\documentclass[prd,aps,showpacs,,nofootinbib,superscriptaddress,preprintnumbers,epsf,psf]{revtex4}
\usepackage[english]{babel}
\usepackage{amsmath}
\usepackage{amsfonts}
\usepackage{amssymb}
\usepackage{pstricks}
\usepackage[dvips]{graphicx}
\usepackage{epsf}
\usepackage{psfrag}
\usepackage{epsfig,graphics}

\newcommand{\widsi}{0.43\columnwidth}
\newcommand{\widss}{0.37\columnwidth}

\newcommand{\kb}{\underline{k}}
\newcommand{\rb}{\underline{r}}

\def\lsim{\raise0.3ex\hbox{$<$\kern-0.75em\raise-1.1ex\hbox{$\sim$}}}
\def\gsim{\raise0.3ex\hbox{$>$\kern-0.75em\raise-1.1ex\hbox{$\sim$}}}

\def\bei{\begin{itemize}}
\def\ei{\end{itemize}}
\def\beqa{\begin{eqnarray}}
\def\bea{\begin{eqnarray}}
\def\eqa{\end{eqnarray}}
\def\beas{\begin{eqnarray*}}
\def\eeas{\end{eqnarray*}}
\def\beqas{\begin{eqnarray*}}
\def\eqas{\end{eqnarray*}}
\def\beq{\begin{equation}}
\def\eq{\end{equation}}
\def\beqd{\begin{displaymath}}
\def\eeqd{\end{displaymath}}
\def\eqd{\end{displaymath}}

\newcommand{\beeq}[1]{
\marginpar{\small\textsf{#1}}
\begin{equation}\label{#1}}
\newcommand{\eeq}{\end{equation}}
\newcommand{\beea}[1]{
\begin{eqnarray}\label{#1}}
\newcommand{\eea}{\end{eqnarray}}

\def\bef{\begin{frame}}

\def\slashchar#1{\setbox0=\hbox{$#1$}
   \dimen0=\wd0
   \setbox1=\hbox{/} \dimen1=\wd1
   \ifdim\dimen0>\dimen1
      \rlap{\hbox to \dimen0{\hfil/\hfil}}
      #1
   \else
      \rlap{\hbox to \dimen1{\hfil$#1$\hfil}}
      /
   \fi}

\def\kb{\underline{k}}

\newcommand{\fV}{f_{3\,\rho}^V}
\newcommand{\fA}{f_{3\,\rho}^A}

\newcommand{\zV}{\zeta_{3}^V}
\newcommand{\zA}{\zeta_{3}^A}

\begin{document}
%\preprint{\hbox{INT-PUB-09-054}}

\title{QCD factorization of exclusive processes beyond leading twist:
\\

$\gamma^*_T \to \rho_T$ impact factor with twist three accuracy}
%\title{Impact factor of $\gamma^{*}\to\rho_{T}$ with twist three accuracy }

\author{I.~V.~Anikin}
\affiliation{Bogoliubov Laboratory of Theoretical Physics, JINR,
             141980 Dubna, Russia}
\author{D.Yu.~Ivanov}
\affiliation{Sobolev Institute of Mathematics, 630090 Novosibirsk, Russia}

\author{B.~Pire}
\affiliation{CPHT, {\'E}cole Polytechnique, CNRS, 91128 Palaiseau Cedex, France}

\author{L.~Szymanowski}
\affiliation{Soltan Institute for Nuclear Studies, PL-00-681 Warsaw, Poland}

\author{S.~Wallon}
\affiliation{LPT, Universit{\'e} Paris-Sud, CNRS, 91405, Orsay, France {\em \&} \\
UPMC Univ. Paris 06, facult\'e de physique, 4 place Jussieu, 75252 Paris Cedex 05, France}

%\date{\today}

\vspace{-.5cm}
\hfill{\hspace*{3cm}CPHT-RR 076.0709, LPT-09-75, INT-PUB-09-054}
\vspace{0cm}

\begin{abstract}

\noindent
We describe a consistent approach to factorization of scattering
amplitudes
for exclusive processes beyond the leading twist approximation. The method
involves the Taylor expansion of the scattering amplitude in the
momentum space around the
dominant light-cone direction and thus  naturally introduces an
appropriate set of non-perturbative correlators which encode  effects
not only of the lowest but also of the higher Fock states of the produced
particle. The reduction
of original set of
correlators to a set of independent ones is achieved with the help of
equations of motion and  invariance of the scattering amplitude under
rotation on the light cone. We compare the proposed method with the covariant method formulated in the
coordinate space, based on the operator product expansion.
We prove the equivalence of two proposed
parametrizations of the $\rho_T$ distribution amplitudes.
As a concrete application,
we compute  the expressions of the
impact factor  for the transition of virtual photon to transversally polarised
$\rho$-meson  up to the twist 3 accuracy within these two quite different methods
and show that they are identical.

\end{abstract}

\maketitle

%\centerline{DRAFT 18/09/09}

\section{Introduction}
\label{Sec_Int}

The study of exclusive reactions in the generalized Bjorken regime has been the scene of significant
progresses in the recent years, thanks to the factorization properties of the leading twist amplitudes \cite{fact}
for deeply virtual Compton scattering and deep exclusive meson production. It however turned out that
transversally polarized $\rho-$meson production did not enter the leading twist controllable case \cite{DGP} but only
the twist 3 more intricate part of the amplitude \cite{MP,AT, Anikin:2002uv}.
This is due to the fact that the leading twist distribution amplitude (DA) of a transversally
polarized vector meson is chiral-odd, and hence decouples from
hard amplitudes at the twist two level, even when another chiral-odd quantity
is involved~\cite{DGP} unless in reactions with more than two final
hadrons \cite{IPST}.
An understanding of the quark--gluon structure of a
transversally polarized vector meson is however an important task of hadronic physics if one cares about studying
confinement dynamics. This quark gluon structure may be described by distribution amplitudes which have been
discussed in great detail \cite{BB,BBtwist4}. On the experimental side, a continuous effort has been devoted to the exploration of $\rho$-meson
 photo and electro-production, from moderate to very large energy \cite{expLow,expHigh}. The kinematical analysis of the final 
$\pi-$meson pair allows then to separate the different helicity amplitudes, hence to measure the transversally polarized $\rho-$meson production amplitude. Although non-dominant for deep electroproduction, this amplitude is by no means negligible at moderately large $Q^2$
 and needs to be understood in terms of QCD. Up to now, experimental information comes from electroproduction on a proton
 or nucleus. Future progress may come from real or virtual photon--photon collisions, which may be accessible either at electron--positron colliders or in ultraperipheral collisions at hadronic colliders, as recently discussed \cite{PSW,IP}.

In the literature there are two approaches to the factorization of the
scattering amplitudes in exclusive processes at leading and higher twists. The first approach \cite{APT,AT}, being
the generalization of the Ellis--Furmanski--Petronzio (EFP) method  \cite{EFP} to the exclusive processes, deals with the factorization in the momentum space around the dominant light-cone
direction. We
shall call it the Light-Cone Collinear Factorization (LCCF). On the other hand, there exists a covariant approach
in coordinate space successfully applied in \cite{BB} for a systematic
description of distribution amplitudes of hadrons carrying different
twists. This approach will be called the Covariant Collinear Factorization approach (CCF). Although being quite
different and using different distribution amplitudes, both
approaches can be applied to the description of the same processes. This fact
calls for verification whether these two descriptions are equivalent and lead
to the same physical consequences.  This can be clarified by
establishing a precise vocabulary between objects appearing in the two
approaches and by comparing physical results obtained with the help of the
two methods. 

The first aim of our paper is to prove that LCCF and CCF are
equivalent methods for the description of exclusive processes. For that we
derive the dictionary between DAs appearing in the LCCF method and
in the CCF one.
We perform
our analysis  within LCCF method in momentum space and use
the invariance
of the scattering amplitude under rotation of the light-cone  vector $n^\mu$,
which we call
 $n$-independence condition.  This method leads to a definitions of relevant soft correlators which are generally not independent ones. The reduction of their number to a minimal set of independent correlators is obtained with the use of equation of motions and of the $n$-independence condition.
We obtain the same number of independent correlators in both
LCCF and in CCF approaches and establish explicit relations between them.

As a concrete application, the second aim of our paper is to calculate
within both methods the impact factor $\gamma^* \to \rho_T$, which is the building block of the
description of the $\gamma^* \, p \to \rho \, p$ and $\gamma^* \gamma^* \to \rho \, \rho$   processes at large $s\,,$ up to twist 3
accuracy and to verify that we get a full consistency between the two results.

 The paper is organized as
follows. In Section \ref{Sec_LCCF} we discuss the general framework of the LCCF method.
In Subsection \ref{SubSec_fact}, we recall basics on factorization within the
LCCF method.
In Subsection  \ref{SubSec_ParamVacuumRho}, we present the parametrization
of the matrix elements relevant to $\rho$-meson production.  In Subsection  \ref{SubSec_EOM} we derive the constraint on these matrix elements coming from the QCD equations of motion. In Subsection  \ref{SubSec_AddSet} we derive additional constraints based on the $n$-independence condition and we then perform the reduction to a minimal set of distribution amplitudes. This results in the dictionary given in subsection  \ref{SubSec_dictionary}.
In Section \ref{Sec_Impact}, we compute the $\gamma^* \to \rho_T$ impact factor. After recalling the necessary definitions and kinematics in Subsection \ref{SubSec_ImpactGeneral}, we firstly perform the calculation in the LCCF framework in Subsection \ref{SubSec_ImpactLCCF} , then in the CCF framework in Subsection \ref{SubSec_ImpactCCF}. We compare the two approaches in subsection \ref{SubSec_ImpactCompare}. Section \ref{Sec_Conclusion} presents our conclusions. A few
appendices present the calculational details needed to complete the proofs.
Partial results of this paper have been briefly  presented in \cite{usSHORT,usProceedings}.

\section{Factorization of exclusive processes in the Light-Cone Collinear approach}
 \label{Sec_LCCF}

\subsection{Factorization beyond leading twist}
 \label{SubSec_fact}

\begin{figure}[h]
\psfrag{l}[cc][cc]{$\ell$}
\psfrag{lm}[cc][cc]{}
\psfrag{q}[cc][cc]{$\gamma^*$}
\psfrag{H}[cc][cc]{$H$}
\psfrag{S}[cc][cc]{$\Phi$}
\psfrag{Hg}[cc][cc]{$H_\mu$}
\psfrag{Sg}[cc][cc]{$\Phi^\mu$}
\psfrag{k}[cc][cc]{}
\psfrag{rmk}[cc][cc]{}
\psfrag{rho}[cc][cc]{$\rho$}
\begin{tabular}{cccc}
\includegraphics[width=5.8cm]{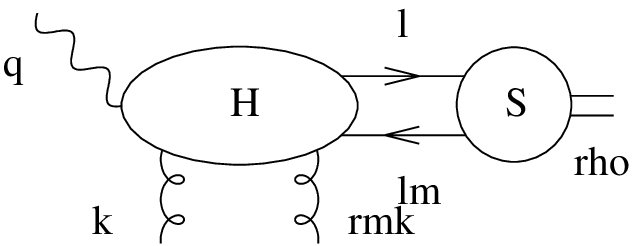}&\hspace{.2cm}
\raisebox{1.2cm}{+}&\hspace{.3cm}\includegraphics[width=5.8cm]{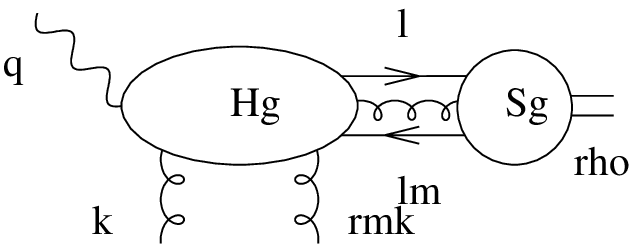}
&\hspace{.2cm}\raisebox{1.2cm}{$+ \cdots$}
\end{tabular}
\caption{2- and 3-parton correlators attached to a hard scattering amplitude in the  example of the $\gamma^* \to \rho$ impact factor, where vertical lines are hard $t-$ channel gluons in the color
singlet state.}
\label{Fig:NonFactorized}
\end{figure}

Let us start with the most general form of the exclusive amplitude for the hard process $A \to \rho \, B$ (where $A$ and $B$ denotes initial and final states in kinematics where a hard scale allows a partonic interpretation) which we are interested in, written in
 the momentum representation and in axial
gauge, as
\begin{eqnarray}
\label{GenAmp}
{\cal A}=
\int d^4\ell \, {\rm tr} \biggl[ H(\ell) \, \Phi (\ell) \biggr]+
\int d^4\ell_1\, d^4\ell_2\, {\rm tr}\biggl[
H_\mu(\ell_1, \ell_2) \, \Phi^{\mu} (\ell_1, \ell_2) \biggr] + \ldots \,,
\end{eqnarray}
where $H$ and $H_\mu$ are the coefficient functions
with two parton legs and three parton  legs,
respectively, as illustrated in Fig.\ref{Fig:NonFactorized} on the example of $\gamma^*\to\rho$
impact factor, which is defined in full length in Sec.\ref{Sec_Impact}.
In (\ref{GenAmp}), the soft parts are given by the
Fourier-transformed two or three partons correlators which are matrix elements of non-local operators.
 We consider the leading asymptotics of $1/Q$ expansion,
separately for the cases of longitudinally (twist 2) and transversely
polarized (twist 3) meson production.
The amplitude (\ref{GenAmp}) is not factorized yet because the hard and soft parts are related by
the four-dimensional integration in the momentum space and by the summation over
the Dirac indices.

To factorize the amplitude, we first choose the dominant direction around which
we intend to decompose our relevant momenta and  we Taylor expand the hard part.
Let $p$ and $n$ be the conventionally called ``plus" and  ``minus" light-cone vectors, respectively, normalized as $p \cdot n =1\,.$
We carry out an expansion of $\ell$ in the basis defined by the $p$ and $n$ light-cone vectors:
\begin{eqnarray}
\label{k}
\ell_{i\, \mu} = y_i\,p_\mu  + (\ell_i\cdot p)\, n_\mu + \ell^\perp_{i\,\mu} ,
\quad y_i=\ell_i\cdot n ,
\end{eqnarray}
and make the following replacement of the integration measure in (\ref{GenAmp}):
\begin{eqnarray}
\label{rep}
d^4 \ell_i \longrightarrow d^4 \ell_i \, dy_i \, \delta(y_i-\ell\cdot n) .
\end{eqnarray}
Afterwards, the hard part coefficient function $H(\ell)$ has to be decomposed around
the dominant ``plus" direction:
\begin{eqnarray}
\label{expand}
H(\ell) = H(y p) + \frac{\partial H(\ell)}{\partial \ell_\alpha} \biggl|_{\ell=y p}\biggr. \,
(\ell-y\,p)_\alpha + \ldots
\end{eqnarray}
where $(\ell-y\,p)_\alpha \approx \ell^\perp_\alpha$ up to twist 3.
One can see that the above-mentioned steps (\ref{k})-(\ref{expand}) do not yet allow us
 to factorize collinearly the amplitude in the momentum
space since
the $l^\perp$ dependence of the hard part is an excursion out of
the collinear framework. To obtain a factorized amplitude, one performs an
 integration by parts
to replace  $\ell^\perp_\alpha$ by $\partial^\perp_\alpha$ acting on
\nolinebreak
~the soft correlator.
 This leads to new operators
${\cal O}^\perp$ which contain
transverse derivatives, such as $\bar \psi \, \partial^\perp \psi $,
and thus
to the necessity of considering additional DAs
$\Phi^\perp (l)$.
This procedure accomplishes the factorization of the amplitude in momentum
space. Factorization in the Dirac space can be achieved by
the Fierz decomposition. For example, in the case of two fermions,  one should project out the Dirac matrix
$\psi_\alpha (0) \, \bar\psi_\beta(z)$ which appears in the soft part of the amplitude on the relevant $\Gamma$ matrices.
%%%%%%%%%%%%%%%%%%%%%%%%%%%%%%%%%%
\begin{figure}[h]
\psfrag{rho}[cc][cc]{$\rho$}
\psfrag{k}[cc][cc]{}
\psfrag{rmk}[cc][cc]{}
\psfrag{l}[cc][cc]{$\ell$}
\psfrag{q}[cc][cc]{}
%\psfrag{q}[cc][cc]{$q$}
\psfrag{lm}[cc][cc]{}
\psfrag{H}[cc][cc]{$ H_{q \bar{q}}$}
\psfrag{S}[cc][cc]{$ \Phi_{q \bar{q}}$}
\begin{tabular}{ccccc}
\epsfig{file=HSqq_rhofact.eps,width=4.5cm}
& \ \raisebox{.9cm}{$\longrightarrow $} \
&
\psfrag{lm}[cc][cc]{\raisebox{.2cm}{$\quad \,\,\, \, \, \Gamma \ \,\, \Gamma$}}
\epsfig{file=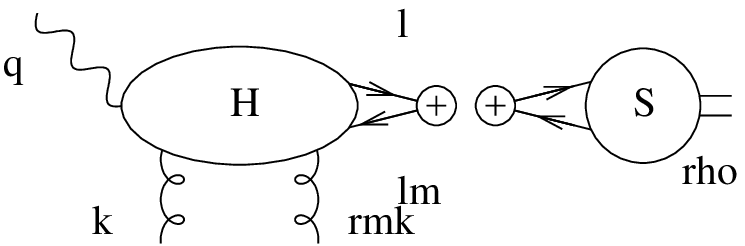,width=5.5cm}
&\raisebox{.9cm}{$+$}
&
\psfrag{H}[cc][cc]{{$H^\perp_{q \bar{q}}$}}
\psfrag{S}[cc][cc]{\scalebox{1}{$\Phi^\perp_{ q \bar{q}}$}}
\psfrag{lm}[cc][cc]{\raisebox{.2cm}{$\quad \,\,\, \, \, \Gamma \ \,\, \Gamma$}}
\hspace{-.2cm}
\epsfig{file=FiertzHSqq_rhofact.eps,width=5.5cm}
\end{tabular}
\caption{Factorization of 2-parton contributions in the  example of the $\gamma^* \to \rho$ impact factor.}
\label{Fig:Factorized2body}
\end{figure}
%%%%%%%%%%%%%%%%%%%%%%%%%%%%%%%%%%%%%
\begin{figure}[h]
\psfrag{rho}[cc][cc]{$\rho$}
%\psfrag{k}[cc][cc]{$k$}
%\psfrag{rmk}[cc][cc]{$k$}
%\psfrag{l}[cc][cc]{$l$}
%\psfrag{q}[cc][cc]{$q$}
\psfrag{k}[cc][cc]{}
\psfrag{rmk}[cc][cc]{}
\psfrag{l}[cc][cc]{}
\psfrag{q}[cc][cc]{}
 \psfrag{Hg}[cc][cc]{$H_{q \bar{q}g}$}
 \psfrag{Sg}[cc][cc]{$\!\Phi_{q \bar{q}g}$}
 \psfrag{lm}[cc][cc]{}
 \psfrag{H}[cc][cc]{$H_{q \bar{q}g}$}
 \psfrag{S}[cc][cc]{$ \Phi_{q \bar{q}}$}
 \psfrag{S}[cc][cc]{$\Phi_{ q \bar{q}g}$}
 \scalebox{.85}{\begin{tabular}{ccc}
 \hspace{-.5cm}\epsfig{file=HSqqg_rhofact.eps,width=7cm}
&
\quad \raisebox{1.4cm}{$\longrightarrow $} \quad
&
\psfrag{lm}[cc][cc]{\raisebox{.2cm}{$\quad \ \, \,\,\, \, \, \, \Gamma \ \ \, \, \Gamma$}}
 \raisebox{.1cm}{\epsfig{file=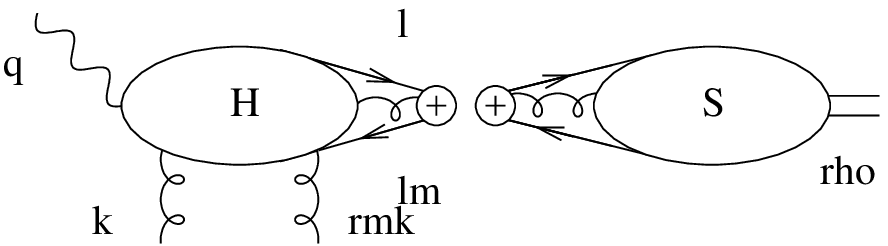,width=9cm}}
 \end{tabular}}
\caption{Factorization of 3-parton contributions in the  example of the $\gamma^* \to \rho$ impact factor.}
\label{Fig:Factorized3body}
\end{figure}
%%%%%%%%%%%%%%%%%%%%%%%%%%%%%%%%%%%%%%%%%%%%%%%%%%%%%%%%%%%%%%%%%%%%%%%%%%%
Thus, after all these stages, the amplitude takes the simple factorized form\footnote{Despite the fact that these formulae are given here up to twist 3, the method can be  extended to higher twist contributions.},
\begin{eqnarray}
\label{GenAmpFac23}
\vspace{-.4cm}{\cal A}=
\int\limits_{0}^{1} dy \,{\rm tr} \left[ H_{q \bar{q}}(y) \, \Gamma \right] \, \Phi_{q \bar{q}}^{\Gamma} (y)
+
\int\limits_{0}^{1} dy \,{\rm tr} \left[ H^{\perp\mu}_{q \bar{q}}(y) \, \Gamma \right] \, \Phi^{\perp\Gamma}_{{q \bar{q}}\,\mu} (y)
+\int\limits_{0}^{1} dy_1\, dy_2 \,{\rm tr} \left[ H_{q \bar{q}g}^\mu(y_1,y_2) \, \Gamma \right] \, \Phi^{\Gamma}_{{q \bar{q}g}\,\mu} (y_1,y_2) \,,
\end{eqnarray}
in which
the two first terms in the r.h.s correspond to the two
parton contribution and the last one to the three body contribution.
This is illustrated symbolically in the example of the $\gamma^* \to \rho$ impact factor in Fig.\ref{Fig:Factorized2body} for 2-parton contributions and in Fig.\ref{Fig:Factorized3body}
for 3-parton contributions.

Alternatively, combining the two last terms together in order to emphasize the fact they
both originate from the Taylor expansion based on the covariant derivative, this factorization can be written as
\begin{eqnarray}
\label{GenAmpFacCov}
&&{\cal A}=
\int\limits_{0}^{1} dy \,{\rm tr} \left[ H(y) \, \Gamma \right] \, \Phi^{\Gamma} (y)
+\int\limits_{0}^{1} dy_1\, dy_2 \,{\rm tr} \left[ H^\mu(y_1,y_2) \, \Gamma \right] \, \Phi^{\Gamma}_\mu (y_1,y_2) .
\end{eqnarray}

For  definiteness\footnote{In the following, the notations $| \rho \rangle$ or $|  V \rangle$ will be used when the specific nature of the vector meson does not matter.}, let us focus on the $\rho$-meson production case where the soft parts of the amplitude
read
\begin{eqnarray}
\label{soft}
\Phi^{\Gamma} (y) &=& \int\limits^{+\infty}_{-\infty} \frac{d\lambda}{2\pi} \, e^{-i\lambda y}
\langle \rho(p) | \bar\psi(\lambda n)\,\Gamma \,\psi(0)| 0 \rangle
\\
\Phi^{\Gamma}_\rho (y_1,y_2)&=& \int\limits^{+\infty}_{-\infty} \frac{d\lambda_1 d\lambda_2}{4\pi^2} \,
e^{-i\lambda y_1\lambda_1-i(y_2-y_1)\lambda_2}
\langle \rho(p) | \bar\psi(\lambda_1 n)\,\Gamma \, i \, \stackrel{\longleftrightarrow}
{D^T_{\rho}}(\lambda_2 n)\, \psi(0)| 0 \rangle \,.
\nonumber
\end{eqnarray}
where 
\beq
\label{defD}
i\stackrel{\rightarrow}{D}_\mu=i\stackrel{\rightarrow}{\partial}_\mu+\,g \, A_\mu\,.
\eq
Eq.(\ref{soft}) supplemented by the appropriate choice of the Fierz matrices defines the set of non-perturbative correlators relevant for the description of the $\rho-$meson, which we will now discuss.

\subsection{Parametrizations of vacuum--to--$\rho$-meson matrix elements up to twist 3}
\label{SubSec_ParamVacuumRho}

In this section, we introduce the parametrizations of
the vacuum--to--$\rho$-meson matrix elements needed when calculating the process of exclusive $\rho-$production. As a concrete example, we shall below calculate the
 $\gamma^*_T\to\rho_T$ impact factor. Since we will
 follow two different approaches for our
calculations, it is instructive to
present two ways for parametrizing the corresponding matrix elements.

\subsubsection{LCCF parametrization}
\label{SubSubSec_ParamVacuumRhoLCCF}

We insist on the fact that in LCCF approach,
 the coordinates $z_i$ in the
 parametrizations
have to be proportional to
the light-cone vector $n$. This is in contrast to the CCF approach where $z$ lies on the light cone but does not correspond
to any fixed light-cone direction.
The transverse polarization of the $\rho-$meson is defined by the conditions
(at twist 3, $p_\rho \sim p$)
\beq
\label{pol_RhoTdef}
e_T \cdot n=e_T \cdot p=0\,,
\eq
i.e. $e_T$ has only a $\perp$ component, while $e_L$ has no $\perp$ component.
Now, we introduce the parametrizations of
the vacuum--to--$\rho$-meson matrix elements needed for the calculation, for example,
of the $\gamma^*_T\to\rho_T$ impact factor.
Keeping all the terms up to the twist-$3$ order
with the axial (light-like) gauge, $n \cdot A=0$,
the matrix elements of quark--antiquark non--local operators
can be written
in terms of the light-cone basis vectors as (here, $z=\lambda n$)
\begin{eqnarray}
\label{par1v}
&&\langle \rho(p_\rho)|\bar\psi(z)\gamma_{\mu} \psi(0)|0\rangle
\stackrel{{\cal F}_1}{=}
m_\rho\,f_\rho \left[ \varphi_1(y)\, (e^*\cdot n)p_{\mu}+\varphi_3(y)\, e^*_{T\mu}\right],
\\
\label{par1a}
&&\langle \rho(p_\rho)|
\bar\psi(z)\gamma_5\gamma_{\mu} \psi(0) |0\rangle
\stackrel{{\cal F}_1}{=}
m_\rho\,f_\rho \, i\varphi_A(y)\, \varepsilon_{\mu\alpha\beta\delta}\,
e^{*\alpha}_{T}p^{\beta}n^{\delta} \,,
\end{eqnarray}
where the corresponding flavour matrix has been omitted\footnote{The normalization in (\ref{par1v},\ref{par1a}) thus corresponds to  a meson which would be a one flavour quark--antiquark state $|V \rangle = |f \, \bar{f}\rangle$,
with for example $\langle V(p)|\bar\psi_f(z)\gamma_{\mu} \psi_f(0)|0\rangle
\stackrel{{\cal F}_1}{=}
m_V\,f_V [ \cdots ]$.}, and where we use $\varepsilon^{0123}=-\varepsilon_{0123}=1$ and $\gamma_5 = i \, \gamma^0\,\gamma^1 \, \gamma^2\, \gamma^3.$
For the sake of conciseness, we denote
$\stackrel{{\cal F}_1}{=}$
 the Fourier transformation with measure
\begin{eqnarray}
\int_{0}^{1}\, dy \,\text{exp}\left[iy\,p\cdot z\right],
\end{eqnarray}
where $z=\lambda n$. The momentum fraction $y$ ($\bar y \equiv 1-y$) corresponds to the quark (antiquark).
Note that the decomposition over the $\gamma$-matrix basis
has been taken in the form:
\begin{eqnarray}
-\langle \psi\, \bar\psi \rangle = \frac{1}{4} \langle \bar{\psi}\, \gamma_\mu \, \psi \rangle \, \gamma^\mu +
\frac{1}{4} \langle \bar{\psi}\, \gamma_5\gamma_\mu \, \psi \rangle \, \gamma^\mu\gamma_5 + \ldots ,
\end{eqnarray}
in such a way that the minus sign in front of the axial term is absorbed into the axial correlators.
The matrix elements of the quark--antiquark operators with transverse derivatives are
parametrized according to
\begin{eqnarray}
\label{par1.1v}
&&\langle \rho(p_\rho)|
\bar\psi(z)\gamma_{\mu}
i\stackrel{\longleftrightarrow}
{\partial^T_{\alpha}} \psi(0)|0 \rangle
\stackrel{{\cal F}_1}{=}m_\rho\,f_\rho \,
\varphi_1^T(y) \, p_{\mu} e^*_{T\alpha}
\\
\label{par1.1a}
&&\langle \rho(p_\rho)| \bar\psi(z)\gamma_5\gamma_{\mu}
i\stackrel{\longleftrightarrow}
{\partial^T_{\alpha}} \psi(0) |0\rangle
\stackrel{{\cal F}_1}{=}m_\rho\,f_\rho \,
i\varphi_A^T (y) \, p_{\mu}\, \varepsilon_{\alpha\lambda\beta\delta}\,
e_T^{*\lambda} p^{\beta}\,n^{\delta}\,,
\end{eqnarray}
where we introduced
$\stackrel{\longleftrightarrow}
{\partial_{\rho}}=\frac{1}{2}(\stackrel{\longrightarrow}
{\partial_{\rho}}-\stackrel{\longleftarrow}{\partial_{\rho}})$
which is the standard antisymmetric derivative.
The DAs $\varphi_1$, $\varphi_3$, $\varphi_A$ satisfy the normalization conditions
\beq
\label{normPhi}
\int\limits^1_0 \,dy\;\varphi_1(y) = 1\,, \int\limits^1_0\,dy\; \varphi_3(y) 
= 1 \quad \mbox{ and } \quad \int\limits^1_0 \,dy\;(y-\bar{y}) \, \varphi_A(y) = \frac{1}{2}\,.
\eq

 In the same way, the matrix elements of quark--gluon non-local operators can be
parametrized as
\begin{eqnarray}
\label{par1.2}
&&\langle \rho(p_\rho)|
\bar\psi(z_1)\gamma_{\mu}g A_{\alpha}^T(z_2) \psi(0) |0\rangle
\stackrel{{\cal F}_2}{=}m_\rho\,\fV \,
B(y_1,y_2;y_g)\, p_{\mu} e^{*}_{T\alpha},
\nonumber\\
&&\langle \rho(p_\rho)|
\bar\psi(z_1)\gamma_5\gamma_{\mu} g A_{\alpha}^T(z_2) \psi(0) |0\rangle
\stackrel{{\cal F}_2}{=}m_\rho\,\fA \,
i \, D(y_1,y_2;y_g)\, p_{\mu} \, \varepsilon_{\alpha\lambda\beta\delta}\,
e_T^{*\lambda}p^{\beta}n^{\delta},
\end{eqnarray}
where the momentum fractions $y_1$, $\bar y_2$ and $y_g$ correspond to the quark, antiquark and gluon, respectively.
The symbol $\stackrel{{\cal F}_2}{=}$ now stands for (here, $z_i=\lambda_i \, n$)
\begin{eqnarray}
\int\limits_{0}^{1}\, dy_1 \,dy_2 \, dy_g \,
\delta(y_2-y_1-y_g) \,
\text{exp}\left[ iy_1\,p\cdot z_1
+iy_g\,p\cdot z_2 \right] .
\end{eqnarray}
In the {\it r.h.s.} of (\ref{par1.2}), it is useful to perform
the integration over the gluon fraction $y_g$ (which then equals $y_2-y_1$). Afterwards, the parametrizations (\ref{par1.2}) take the forms:
\begin{eqnarray}
\label{Correlator3BodyV}
&&\langle \rho(p_\rho)|
\bar\psi(z_1)\gamma_{\mu}g A_{\alpha}^T(z_2) \psi(0) |0\rangle
\stackrel{{\cal F}_2}{=}m_\rho\,\fV \,
B(y_1,y_2)\, p_{\mu} e^*_{T\alpha},
\\
\label{Correlator3BodyA}
&&\langle \rho(p_\rho)|
\bar\psi(z_1)\gamma_5\gamma_{\mu} g A_{\alpha}^T(z_2) \psi(0) |0\rangle
\stackrel{{\cal F}_2}{=}m_\rho\,\fA \,
i D(y_1,y_2)\, p_{\mu} \, \varepsilon_{\alpha\lambda\beta\delta} \,
e^{* \, \lambda}_T \, p^{\beta}n^{\delta}\,,
\end{eqnarray}
with the symbol $\stackrel{{\cal F}_2}{=}$ implying
\begin{eqnarray}
\int\limits_{0}^{1} dy_1 \,\int\limits_{0}^{1} dy_2 \,
\text{exp}\left[ iy_1\,p\cdot z_1+i(y_2-y_1)\,p\cdot z_2 \right] \,.
\end{eqnarray}
Note that the positivity of the gluon light-cone momentum fraction
imposes that quark--gluon parameterizing functions have the form
\begin{eqnarray}
B(y_1,y_2) \stackrel{def}{=} {\cal B}(y_1,y_2; y_2-y_1) \,\theta(y_1 \leq y_2\leq 1), \quad
D(y_1,y_2) \stackrel{def}{=} {\cal D}(y_1,y_2; y_2-y_1) \,\theta(y_1 \leq y_2\leq 1) \,.
\end{eqnarray}

As we already mentioned by writing Eq.(\ref{GenAmpFacCov}), it is also natural to introduce the following objects:
\begin{eqnarray}
\label{cov}
&&\langle \rho(p_\rho)|
\bar\psi(z_1)\gamma_{\mu}i\stackrel{\longleftrightarrow}
{D^T_{\alpha}}(z_2) \psi(0) |0\rangle
\stackrel{{\cal F}_2}{=}m_\rho \, f_\rho\,
\tilde B(y_1,y_2)\, p_{\mu} e^*_{T\alpha},
\nonumber\\
&&\langle \rho(p_\rho)|
\bar\psi(z_1)\gamma_5\gamma_{\mu} i\stackrel{\longleftrightarrow}
{D^T_{\alpha}}(z_2) \psi(0) |0\rangle
\stackrel{{\cal F}_2}{=}m_\rho \, f_\rho\,
i \tilde D(y_1,y_2) \, p_{\mu}\, \varepsilon_{\alpha\lambda\beta\delta}\,
e^{\lambda *}_{T}\, p^{\beta} \, n^{\delta}\,,
\end{eqnarray}
where these parameterizing functions are now equal to
\begin{eqnarray}
\label{defBDcov}
&&\tilde B(y_1,y_2)=\frac{1}{2}\biggl(
\varphi_1^T(y_1)+\varphi_1^T(y_2)\biggr)\delta(y_1-y_2)+\zV \, B(y_1,y_2) ,
\nonumber\\
&&\tilde D(y_1,y_2)=\frac{1}{2}\biggl(
\varphi_A^T(y_1)+\varphi_A^T(y_2)\biggr)\delta(y_1-y_2)+\zA \, D(y_1,y_2) \,,
\end{eqnarray}
with the dimensionless coupling constants
\beq
\label{defZeta}
\zV = \frac{\fV}{f_\rho} \quad {\rm and} \quad \zA = \frac{\fA}{f_\rho}\,.
\eq
Note that the function $\varphi_1$ corresponds to
the twist-$2$, and
functions $B$ and $D$ to the genuine (dynamical) twist-$3$,
while functions $\varphi_3$, $\varphi_A, \varphi_1^T$,
$\varphi_A^T$ (or alternatively $\tilde B$ and  $\tilde D$)
contain both parts: kinematical (\`a la
Wandzura-Wilczek, noted WW) twist-$3$ and genuine (dynamical) twist-$3$.

In (\ref{par1v})--(\ref{par1a}), the functions  $\varphi_1$,
 $\varphi_3$,  $\varphi_A$,  $\varphi_1^T$ and  $\varphi_A^T$ parameterizing the two-particle correlators
obey the following symmetry properties:
\begin{eqnarray}
\label{sym1}
\varphi_1(y)=\varphi_1(1-y), \,  \varphi_3(y)=\varphi_3(1-y), \,
\varphi_A(y)=-\varphi_A(1-y), \, \varphi_1^T(y)=-\varphi_1^T(1-y),\,
\varphi_A^T(y)=\varphi_A^T(1-y)\,.
\end{eqnarray}
These symmetry properties result from  $G$-conjugation (or $C$-conjugation for neutral mesons).
At the same time, the symmetry properties of the functions
parameterizing the quark--gluon correlators are:
\begin{eqnarray}
\label{sym2}
{\cal B}(y_1,y_2;y_g)=-{\cal B}(1-y_2,1-y_1;y_g), \quad
{\cal D}(y_1,y_2;y_g)={\cal D}(1-y_2,1-y_1;y_g) .
\end{eqnarray}
Notice that, in the case of three-particle functions, $G$-conjugation
involves the replacement: $y_1 \leftrightarrow\bar y_2$, while the gluon
fraction $y_g$ remains invariant under $G$-conjugation.

\subsubsection{CCF parametrization}
\label{SubSubSec_ParamVacuumRhoCCF}

We recall and rewrite (doing standard fields transformations) the original CCF parametrizations of the $\rho$ DAs~\cite{BB},
adapting them to our case when vector meson is produced in the final state.
The formula for the axial-vector  correlator reads
\beq
\label{BBA}
\langle \rho(p_\rho)|\bar \psi(z) \, [z,\, 0] \, \gamma_\mu \gamma_5 \psi(0)|0\rangle =
\frac{1}{4}f_\rho\,m_\rho\,\varepsilon_\mu^{\;\;\,e^*_T\,p\,z}     \int\limits_0^1\,dy\,e^{iy(p \cdot z)}\,g_\perp^{(a)}(y)\;,
\eeq
where we denote\footnote{Note that, as already emphasized, our sign convention for the antisymmetric tensor is $\epsilon^{0123}=1$,  opposite to the
one used in Ref.\cite{BB}. The corresponding sign change is taken here into account.}
\beq
\label{defEpsilonCompact}
\varepsilon_\mu^{\;\;\,e^*_T\,p\,z}= \varepsilon_\mu^{\,\,\,\alpha \beta \gamma} e^*_{T \alpha} \,p_\beta \, z_\gamma\,,
\eq
and where 
\beq
\label{defWilson}
[z_1, \, z_2] = P \exp \left[ i g \int\limits^1_0 dt \, (z_1-z_2)_\mu A^\mu(t \,z_1 +(1-t)\,z_2)    \right]
\eq
is the Wilson line, defined 
in accordance with the convention (\ref{defD}).
The transverse vector $e_T$ is orthogonal to the light-cone vectors $p$ and $z$. Neglecting mass effects, i.e. up to twist 3 level, it is decomposed as follows
\beq
\label{pol_Rho}
e_{T \mu}=e_\mu -p_\mu \frac{e \cdot z}{p \cdot z}-z_\mu \frac{e \cdot p}{p \cdot z} \, ,
\eeq
where $e$ is the meson polarization vector. 
% Thus in the CCF parametrization the 
% notion of "transverse" is different with respect to the one defined by 
% Eq.(\ref{pol_RhoTdef}) since as we discuss later in sec.\ref{SubSec_ImpactCCF}  the coordinate $z$ on the 
% light-cone and 
% the light-cone vector $n$ point in two different directions.
Since as we will discuss later $n$ can be arbitrary, and since the concrete definition of $n$ influences
the definition of transverse polarization, it is useful to remove the dependence on $e_T$ in correlation functions. For that,
 we use Eq.(\ref{pol_Rho}) and rewrite the original CCF 
parametrization in terms of the full meson polarization vector $e$. This is already done for the axial-vector correlator (\ref{BBA})
 since due to the properties of fully antisymmetric tensor 
$\epsilon_{\mu\nu\alpha\beta}$ one can use in the r.h.s. of (\ref{BBA}) the full meson polarization vector
$e$ instead of $e_T$.

The definition of 2-parton vector  correlator of a
$\rho$-meson  can be written in the form
\beq
\label{BBV1}
\langle \rho(p_\rho)|\bar \psi(z) \, [z,\, 0] \, \gamma_\mu  \psi(0)|0\rangle = f_\rho\,m_\rho\int\limits_0^1\,dy\,e^{iy(p\cdot z)}\left[
p_\mu\,\frac{e^*\cdot z}{p\cdot z}\phi_{\parallel}(y) +
e^*_{T\mu}\,g_\perp^{(v)}(y)-z_\mu \frac{m^2}{2} \frac{e^* \cdot z}{(p \cdot z)^2}  g_3(y)
  \right]\,.
\eq
All distribution amplitudes describing two particle correlators are normalized to unity
\beq
\label{norm}
\int\limits^1_0 dy \left\{\phi_\parallel, g^{(a)}_\perp, g^{(v)}_\perp, g_3\right\}(y)=1 \, .
\eeq
Using relations (\ref{pol_Rho}, \ref{norm}) and integration by parts one can rewrite the
vector correlator (\ref{BBV1}) in the form
\beq
\label{BBV2}
\langle \rho(p_\rho)|\bar \psi(z) \, [z,\, 0] \, \gamma_\mu  \psi(0)|0\rangle 
= f_\rho\,m_\rho\int\limits_0^1\,dy\,e^{iy(p\cdot z)}\left[
-i \, p_\mu\,(e^*\cdot z)\,h(y) +
e^*_{\mu}\,g_\perp^{(v)}(y) + i z_\mu \frac{m^2}{2} \frac{e^* \cdot z}{p \cdot z}  \, \bar{h}(y)
  \right]\;,
\eq
where we introduce the auxiliary functions
\bea
\label{defh}
h(y)=\int\limits^y_0 dv \left(\phi_\parallel(v)-g^{(v)}_\perp(v)\right) \,, \\
\bar h(y)=\int\limits^y_0 dv \left(g_3(v)-g^{(v)}_\perp(v)\right) \, .
\eea
Note that the r.h.s of (\ref{BBV2}) now only involves the full polarization vector $e$, as was noted above for the axial correlator. The last term in the r.h.s. of (\ref{BBV2}) contributes to the physical amplitude
starting from the twist 4 level only, therefore we will neglect it in the following.

%%%%%%%%%%%%%%%%%%%%%%%%%%%%%%%%

For quark--antiquark--gluon correlators (up to twist 3 level) the parametrizations of Ref.\cite{BB} have the forms\footnote{Note that in those definition  $f_\rho$, $\fV$ and $\fA$ have dimension of mass. This is agreement with Ref.\cite{BBtwist4} but differ from Ref.\cite{BB} in which $\fV$ and $\fA$ have dimension of mass square.}
\bea
&&
\hspace{-1cm}\langle \rho(p_\rho)|\bar \psi(z)[z,t\, z]\gamma_\alpha g \, G_{\mu\nu}(t\, z)[t\,z,0] \psi(0)|0 \rangle =
-i p_\alpha [p_\mu e^*_{\perp \nu}-p_\nu e^*_{\perp \mu} ] m_\rho \, \fV \int D \alpha \, V(\alpha_1,\alpha_2)
e^{i(p \cdot z)(\alpha_1+t\,\alpha_g)} \, , \label{GV}\\
&&
\hspace{-1cm}\langle \rho(p_\rho)|\bar \psi(z)[z,t\, z]\gamma_\alpha\gamma_5 g \, \tilde G_{\mu\nu}(t\, z)[t\,z,0] \psi(0)|0 \rangle =
- p_\alpha [p_\mu e^*_{\perp \nu}-p_\nu e^*_{\perp \mu} ] m_\rho \,\fA \int D \alpha \, A(\alpha_1,\alpha_2)
e^{i(p \cdot z)(\alpha_1+t\,\alpha_g)} \,,\label{GA}
\eea
where $\alpha_1$, $\alpha_2$, $\alpha_g$ correspond to momentum fractions of quark, antiquark and gluon respectively inside the $\rho-$meson, 
\beq
\label{defDalpha}
\int D \alpha =\int\limits^1_0 d\alpha_1\int\limits^1_0 d\alpha_2 \int\limits^1_0 d\alpha_g\,
\delta(1-\alpha_1-\alpha_2-\alpha_g)
\eq
and $\tilde G_{\mu\nu}=-{1\over 2}\epsilon_{\mu\nu\alpha\beta}G^{\alpha\beta}.$
These three partons DAs are normalized as follows
\beq
\int D \alpha \,(\alpha_1-\alpha_2)\, V(\alpha_1,\alpha_2)=1 \, , \quad
\int D \alpha \, A(\alpha_1,\alpha_2)=1 \, .
\eeq
In what follows we will work in the axial gauge $A\cdot n=0$, $n^2=0$.  In this gauge the gluon field can be
expressed in terms of field strength as follows
\beq
\label{axial}
A_\alpha (y)=\int\limits^\infty_0 d\sigma \, e^{-\epsilon \,\sigma} n^\beta G_{\alpha\beta}(y+\sigma n) \, 
\eeq
which implies that
 the
($\bar q \,A\,q$ ) correlators involving the gluon field $A$ reads
\begin{eqnarray}
\label{BBVg}
&&\hspace{-.3cm}\langle \rho(p_\rho)|\bar \psi(z)\gamma_\mu g A_\alpha(tz)\psi(0)|0\rangle =
-p_\mu \, e^*_{T\alpha}m_\rho \,f^V_{3\rho}\int
\frac{D \alpha}{\alpha_g} \,e^{i(p\cdot z)(\alpha_1+t\,\alpha_g)}\,
V(\alpha_1,\alpha_2)\,, \\
\label{BBAg}
&&\hspace{-.3cm}\langle \rho(p_\rho)|\bar \psi(z)\gamma_\mu\gamma_5 g
A_\alpha(tz)\psi(0)|0\rangle =
-i p_\mu \frac{\varepsilon_\alpha^{\;\;z\,p\,e^*_T}}{(p\cdot z)}m_\rho \, f^A_{3\rho}\int
\frac{ D \alpha}{\alpha_g}\, e^{i(p\cdot z)(\alpha_1+t\alpha_g)}\,
A(\alpha_1,\alpha_2)\,.
\end{eqnarray}

\subsection{Equations of motion}
\label{SubSec_EOM}

The correlators introduced above are not independent, since they
are constrained
 by the QCD equations of motion for the field
operators entering them (see, for example, \cite{AT}).
In the simplest case of fermionic fields, they follow from the
vanishing of matrix elements
$\langle (i  %\stackrel{\rightarrow}
{\hat D}(0) \psi(0))_\alpha\, \bar \psi_\beta(z)\rangle = 0$ and
$\langle  \psi_\alpha(0)\, i %\stackrel{\rightarrow}
({\hat D}(z)\bar \psi(z))_\beta \rangle
= 0\,$
due to the Dirac equation, then projected on different Fierz structure.

Let us start with the QCD equation of motion written for the fermion field $\psi(0)$:
\begin{eqnarray}
\label{QCDeom}
\langle i\stackrel{\rightarrow}{\slashchar{D}}(0) \psi(0)\, \bar \psi(z)\rangle = 0,
\end{eqnarray}
where
$\langle\ldots\rangle$ denote arbitrary hadron states which we here specify as
$\langle\rho | \ldots |0\rangle$.
Also, we stress that, in (\ref{QCDeom}), the fermion fields $\psi$ and $\bar\psi$ should
be understood as fields with free Dirac indices.
Then, we first focus on the quark--antiquark part of (\ref{QCDeom}) which can be written as
\begin{eqnarray}
\label{quarkpart}
\int d^4z \,e^{-iyp\cdot z -i\bar y p\cdot x}
\biggl\{\langle i\stackrel{\rightarrow}
{\slashchar{\partial}^{\,x}}_{\!\!\!\!L} \psi(x)\, \bar \psi(z)\rangle + \langle i\stackrel{\rightarrow}
{\slashchar{\partial}^{\,x}}_{\!\!\!\!T} \psi(x)\, \bar \psi(z)\rangle
\biggr\} \biggl|_{x=0}\biggr. .
\end{eqnarray}
Here, we separate out the longitudinal derivatives from the transverse ones. Working with the longitudinal
derivative contribution, we get
\begin{eqnarray}
\label{longder}
i\gamma_\rho  \int d^4z\, e^{-iyp\cdot z -i\bar y p\cdot x} \frac{\partial^L}{\partial x_\rho}
\langle \psi(x)\, \bar \psi(z)\rangle \biggl|_{x=0}\biggr. =
\bar y \, \slashchar{p} \, \int d^4z\, e^{-iyp\cdot z -i\bar y p\cdot x}
\langle \psi(x)\, \bar \psi(z)\rangle \biggl|_{x=0}\biggr. ,
\end{eqnarray}
where an integration by parts has been used. Let us now decompose $\langle \psi(x)\, \bar \psi(z)\rangle$ over
the $\gamma$-basis (the Fierz decomposition):
\begin{eqnarray}
\label{Firtz}
- \langle \psi(x)\, \bar \psi(z)\rangle =\frac{1}{4} \langle \bar\psi(z)\gamma_\alpha \psi(x)\rangle \gamma^\alpha +
\frac{1}{4} \langle \bar\psi(z) \gamma_5 \gamma_\alpha \psi(x)\rangle \gamma^\alpha \gamma_5\,.
\end{eqnarray}
With (\ref{Firtz}), after the use of the parametrization of the relevant correlators,
one gets for the longitudinal derivative contribution (see, (\ref{longder})):
\begin{eqnarray}
\frac{\bar y}{4} \, \slashchar{p} \, \slashchar{e}^T \, \varphi_3(y) + i\, \frac{\bar y}{4} \,\slashchar{p} \,\slashchar{a}^T \,\gamma_5 \, \varphi_A(y) =
- \frac{i}{4}\sigma_{p \, e^T} \biggl\{ \bar y \, \varphi_3(y) + \bar y \, \varphi_A(y) \biggr\},
\end{eqnarray}
where again we introduced the short-hand notations:
\begin{eqnarray}
\sigma_{p \, e^T}= \sigma_{\alpha\, \beta} \, p^{\alpha} \, e_T^{*\beta}, \quad
a_{T \rho} = \varepsilon_{\rho e^*_T p n}\,.
\end{eqnarray}
We thus have for the longitudinal derivative contribution:
\begin{eqnarray}
\label{longderiv}
\int d^4z \,e^{-iyp\cdot z -i\bar y p\cdot x}
\,\langle i\stackrel{\rightarrow}
{\slashchar{\partial}}_L \psi(x)\, \bar \psi(z)\rangle \biggl|_{x=0}\biggr.=
- \frac{i}{4}\sigma_{p \, e_T}\,m_\rho f_\rho\, \biggl\{ \bar y \, \varphi_3(y) + \bar y \, \varphi_A(y) \biggr\}.
\end{eqnarray}
While, the correlators with the transverse derivatives in (\ref{quarkpart}) can directly be expressed via
the corresponding parameterizing functions with the help of (\ref{par1.1v}) and (\ref{par1.1a}):
\begin{eqnarray}
\label{tranderiv}
\int d^4z \,e^{-iyp\cdot z -i\bar y p\cdot x}
\,\langle i\stackrel{\rightarrow}
{\slashchar{\partial}}_T \psi(x)\, \bar \psi(z)\rangle \biggl|_{x=0}\biggr.=
- \frac{i}{4}\sigma_{p \, e^*_T}\,m_\rho f_\rho\, \biggl\{ \varphi_1^T(y) + \varphi_A^T(y) \biggr\}.
\end{eqnarray}
Therefore, within the WW approximation
(where all genuine twist $3$ are disappeared) the equations of motion takes the following simple form:
\begin{eqnarray}
\varphi_+^T(y)=-\bar y \, \varphi_+^{WW}(y) ,
\end{eqnarray}
where the plus (minus)-combination is defined as
\begin{eqnarray}
\label{saf}
\varphi_{\pm}(y)=\varphi_{\text{``vector''}}(y) \pm \varphi_{\text{``axial''}}(y).
\end{eqnarray}

Let us now take into account  the quark--gluon correlators. Using (\ref{par1.2}), one can obtain that
\begin{eqnarray}
- \gamma_\rho \langle A^{T\rho}(0)\, \psi(0)\, \bar\psi(z) \rangle =
\frac{1}{4} \gamma_\rho
\biggl\{ \langle \bar\psi(z)\, \gamma_\alpha\, A^{T\rho}(0)\, \psi(0) \rangle  \gamma_\alpha +
\langle \bar\psi(z)\, \gamma_5 \gamma_\alpha\, A^{T\rho}(0)\, \psi(0) \rangle  \gamma_\alpha \gamma_5
\biggr\} ,
\end{eqnarray}
where
\begin{eqnarray}
&&\langle \bar\psi(z)\, \gamma_\alpha\, g \, A^T_\rho(0)\, \psi(0) \rangle =m_\rho \, \fV \,
p_\alpha\, e^*_{T\rho} \, \int\limits_{0}^{1} \, dy_1\, dy_2 \, e^{iy_1p\cdot z} \, B(y_1,y_2) ,
\nonumber\\
&&\langle \bar\psi(z)\, \gamma_5\gamma_\alpha\, g \, A^T_\rho(0)\, \psi(0) \rangle =m_\rho \, \fA \,
i\,p_\alpha\, a_{T\rho} \, \int\limits_{0}^{1} \, dy_1\, dy_2 \, e^{iy_1p\cdot z} \, D(y_1,y_2).
\end{eqnarray}
Thus, combining the quark--antiquark and quark--gluon correlators, one derives the following relation:
\begin{eqnarray}
\label{em_rho0}
&&\int\limits_{0}^{1} dx \left( \tilde B(y,x)
+ \tilde D(y,x) \right) =
-\bar y \,\varphi_+(y) ,
\end{eqnarray}
where one used the notations (\ref{defBDcov}).

In a similar way, we can derive the relation associated with the
$C$-conjugated equation:
\begin{eqnarray}
\langle \psi(0)\, \bar \psi(z)\, i\stackrel{\leftarrow}{\slashchar{D}}(0) \rangle = 0 .
\end{eqnarray}
which reads
\begin{eqnarray}
\label{em_rho01}
&&\int\limits_{0}^{1} dx \left( \tilde B(x,y)
- \tilde D(x,y) \right) = y \,\varphi_-(y) .
\end{eqnarray}
Combining the relations (\ref{em_rho0}) and (\ref{em_rho01}) with the use of
(\ref{defBDcov}) and (\ref{saf}),
 we obtain
\begin{eqnarray}
\label{em_rho1}
 &&\bar{y}_1 \, \varphi_3(y_1) +  \bar{y}_1 \, \varphi_A(y_1)  +  \varphi_1^T(y_1)  +\varphi_A^T(y_1)\nonumber \\
&&=-\int\limits_{0}^{1} dy_2 \left[ \zV \, B(y_1,\, y_2) +\zA \, D(y_1,\, y_2) \right] \,
\end{eqnarray}
and
\begin{eqnarray}
\label{em_rho2}
 && y_1 \, \varphi_3(y_1) -  y_1 \, \varphi_A(y_1)  -  \varphi_1^T(y_1)  +\varphi_A^T(y_1) \nonumber\\
&&=-\int\limits_{0}^{1} dy_2 \left[ -\zV \, B(y_2,\, y_1) +\zA\, D(y_2,\, y_1) \right] \,.
\end{eqnarray}
Note that Eq.(\ref{em_rho2}) can be obtained by the replacement $y_1 \to \bar{y}_1$ in (\ref{em_rho1}) and the use of symmetry properties (\ref{sym1}, \ref{sym2}).

\subsection{Additional set of equations}
\label{SubSec_AddSet}

\subsubsection{Light-cone factorization direction arbitrariness}
\label{SubSubSec_n-indep}

Contrarily to
the light-cone vector $p$ related to the out-going meson momentum,
 the second light-cone vector $n$ (with $p \cdot n=1$), required for the parametrization of the
needed
correlators introduced in section \ref{SubSubSec_ParamVacuumRhoLCCF} %(\ref{CorrelatorV}-\ref{CorrelatorDerA}),
is arbitrary.
%Let us emphasize that the crucial point of this approach is that
The physical observables do not depend on the specific choice of $n$,
thus the scattering amplitudes should be $n-$independent.
For any specific process, there is a natural choice for $n$, which one may denote as $n_0.$ For instance, in forward $e-p$ collision, the proton momentum defines $n_0.$ 
More generally, one may expand an arbitrary choice of $n$ as \cite{EFP, AT}
\beq
\label{n_general}
n_\mu = \alpha \, p_\mu + \beta \, n_{0\mu} + n^\perp_{\mu}\,,
\eq
with the two constraints 
\beq
\label{constraints}
p \cdot n=1 \quad {\rm and }  \quad n^2=0\,,
\eq
which fixes the coefficients
$\beta=1$ and $\alpha=-n_\perp^2/2\,.$
The light-cone vector $n$ is thus parametrized by its transverse components $n_\perp\,.$

Let us now analyse the various source of $n-$dependence. First, it enters the definition
of the non-local correlators
introduced in Sec.\ref{SubSubSec_ParamVacuumRhoLCCF} through
the light-like separation $z=\lambda \, n$.
%
% intering the non-local correlators
%introduced in Sec.\ref{SubSubSec_ParamVacuumRhoLCCF}
%involves an arbitrary light-like vector $n$ with $n \cdot p=1$
%which already appeared in the Sudakov decomposition (\ref{k}).
These correlators are defined in the axial light-like gauge $n \cdot A=0\,,$
which allows to get rid of  Wilson lines.
Second, it determines the notion of transverse polarization of the $\rho\,.$
Last, $n$ inters the Sudakov decomposition
(\ref{k}) which defines the transverse parton momentum involved in the collinear factorization.
Note that this notion of parton transverse
momentum should not be confused with the notion of transverse momenta
of external particles (e.g. in the case of $\gamma^* \to \rho$ impact factor to be discussed in Sec.\ref{Sec_Impact}, entering in $\gamma^* \, N(p_2) \to \rho(p) \, N$ process, the $t-$channel gluons have a transverse momentum determined within another Sudakov basis defined by the external light-cone momenta $p_1=p$ and incoming nucleon momentum $p_2$).

This $n-$independence principle leads
to additional non-trivial constraints between the non-perturbative correlators entering the factorized amplitude. It was crucial
for the understanding of inclusive structure functions properties
at
 the twist three level  \cite{EFP} and its relevance for  some exclusive processes was pointed out in \cite{AT}. We show now that this  condition expressed at the level of the {\em full amplitude} of any process can be reduced to a set of conditions involving only the soft correlators. The obtained equations are process independent and do not assume a priori any Wandzura-Wilczek approximation. The strategy
for deriving these equations
%that we will detail on a specific example below
relies on the power of the Ward identities
to relate firstly amplitudes  with different number of legs and secondly higher order coefficients in the Taylor expansion (\ref{expand}) to lower order ones.

%%%%%%%%%%%%%%%%%%%%%%%%%%%%%%%%%%%%%%%%
In the case of processes involving $\rho_T$ production up to twist 3 level, we will now derive
the equations
\begin{eqnarray}
\label{ninV}
&&\frac{d}{dy_1}\varphi_1^T(y_1)+\varphi_1(y_1)-\varphi_3(y_1)+\zV\int\limits_0^1\,\frac{dy_2}{y_2-y_1}
 \left( B(y_1,y_2)+B(y_2,y_1) \right)=0\,,  \\
\label{ninA}
&&\frac{d}{dy_1}\varphi_A^T(y_1)-\varphi_A(y_1)+\zA\int\limits_0^1\,\frac{dy_2}{y_2-y_1}
 \left(D(y_1,y_2)+D(y_2,y_1)
   \right)=0 \,.
\end{eqnarray}

The $n-$independence of ${\cal A}$ for an arbitrary fixed polarization vector $e$ is expressed by the condition
\begin{equation}
\label{n-indA}
\frac{d}{dn_{\perp}^{\mu}}\;{\cal A}=0\,,
\end{equation}
which we write in the form
\beqa
\label{n-indA_dev}
\frac{d}{dn_{\perp}^{\mu}}\;{\cal A}
=\frac{\partial\,n^\alpha}{\partial\, n_{\perp}^{\mu}}\,
\frac{\partial {\cal A}}{\partial\,n^\alpha} 
+\frac{\partial (e^*\cdot n)}{\partial n_{\perp}^{\mu}}
\,\frac{\partial {\cal A}}{\partial \,(e^*\cdot n)} =
[ -n_{\perp \,\mu} p^\alpha 
+ g_{\perp \,\mu}^{ \alpha}]
\frac{\partial}{\partial\,n^\alpha}\;{\cal A}+e^{*}_{\perp \,\mu} 
\,\frac{\partial {\cal A}}{\partial \,(e^*\cdot n)}\,.
%&=&
%\frac{\partial}{\partial\,n_\perp^\mu}{\cal A}^{axial}=0
\eqa
Let us emphasize the fact that although $n$ fixes the gauge, 
the hard part does not depend on this 
 gauge fixing vector, as we will show below
after the technical derivation of Eqs.(\ref{ninV}, \ref{ninA}). It means that the variation of $n$
only affects the $n-$dependence related to the definition of transverse momentum $\ell_\perp$
and of transverse $\rho$ polarization.

We also note that the appearance of the total  derivative in Eqs.(\ref{n-indA}, \ref{n-indA_dev}) may be interpreted as a (vector) analog of the renormalization group (RG) invariance equation when the
dependency on the renormalization parameter coming from various sources
 cancel.  One can view this as a RG-like flow in the space of light-cone
directions of contributions to the amplitude where the polarization vector plays the
 role of a beta function.

 The scattering amplitude ${\cal A}$ receives contributions from the vector correlators, which result into ${\cal A}^{vector}$, and from the axial vector correlators, which lead to ${\cal A}^{axial}$ part of ${\cal A}$.
Due to different parity properties of the vector and the axial-vector correlators, the condition  (\ref{n-indA_dev}) means effectively two separate conditions:
\begin{equation}
\label{n-ind_V}
\frac{d}{dn_{\perp}^{\mu}}\;{\cal A}^{vector}=0\;
\end{equation}
and
\begin{equation}
\label{n-ind_A}
\frac{d}{dn_{\perp}^{\mu}}\;{\cal A}^{axial}=0\;.
\end{equation}

The dependence of ${\cal A}$ on the vector $n_\perp$ is obtained through the dependence of
${\cal A}$
 on the full vector $n$. This dependence on $n$ is different in ${\cal A}^{axial}$ and in ${\cal A}^{vector}$ parts.

\noindent
The dependence of ${\cal A}^{axial}$ on the vector $n$ 
enters only through the expression ${\varepsilon}^{p\,n\,\beta\,\gamma}$ involving the contraction with the momentum $p$ and in which the indices $\beta$ and $\gamma$ are contracted with some other vectors. Thus the condition (\ref{n-ind_A}) is equivalent to
\begin{eqnarray}
\label{nIndA}
\frac{\partial\,n^\alpha}{\partial\, n_{\perp}^{\mu}}\,\frac{\partial}{\partial\,n^\alpha}\;{\cal A}^{axial} =
[ -n_{\perp}^\mu p^\alpha + g_{\perp}^{ \alpha\,\mu}]\frac{\partial}{\partial\,n^\alpha}\;{\cal A}^{axial}
=
\frac{\partial}{\partial\,n_\perp^\mu}{\cal A}^{axial}=0
\end{eqnarray}
where  we took into account the peculiar dependence of ${\cal A}^{axial}$ on $n$ discussed above. This will lead at the level of DAs to the equation (\ref{ninA}).

In the vector part, the dependence with respect to $n$ is identified
by rewriting the  polarization vector for transversally polarized  $\rho$
with the help of
the identity
\begin{equation}
\label{epsilonRho}
e_{T}^\mu = e^\mu - p^\mu\, e\cdot n\,,
\end{equation}
since $e \cdot p =0\,.$
Thus the dependence of ${\cal A}^{vector}$ on the vector $n$ enters only through
 the scalar product $e^*\cdot n$, and Eq.(\ref{n-ind_V}) can be written as
\begin{equation}
\label{nIndVint}
\frac{d}{dn_{\perp}^{\mu}}\;{\cal A}^{vector}= e_T^{*\,\mu}\,\frac{\partial}{\partial\,(e^*\cdot n)}\;{\cal A}^{vector}=0\,
\end{equation}
which results in
\begin{equation}
\label{nIndV}
\frac{\partial}{\partial\,(e^*\cdot n)}\;{\cal A}^{vector}=0\,
\end{equation}
from which follows Eq.(\ref{ninV}).
\def\li{.2\columnwidth}
\def\si{\hspace{.9cm}}
\def\sci{\hspace{1.7cm}}
\psfrag{u}{\footnotesize$\hspace{-0cm}y$}
\psfrag{d}{\footnotesize$\hspace{-.2cm}-\bar{y}$}
\begin{figure}[h]
 \scalebox{1}{\begin{tabular}{cccc}
 \hspace{0.cm}\raisebox{0cm}{\epsfig{file=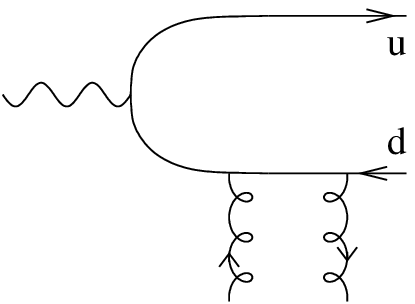,width=\li}} & \si
\psfrag{u}{}
\psfrag{d}{}
 \raisebox{0cm}{\epsfig{file=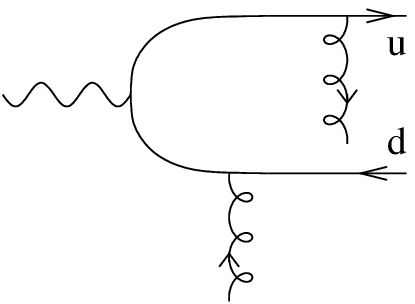,width=\li}} & \si
\psfrag{u}{}
\psfrag{d}{}
\raisebox{1cm}{\epsfig{file=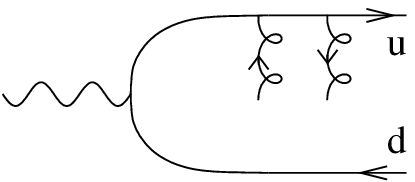,width=\li}}  & \si
\psfrag{u}{}
\psfrag{d}{}
 \raisebox{0cm}{\epsfig{file=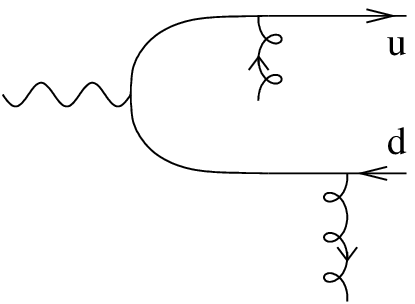,width=\li}}
\\
\\
\hspace{1.2cm} (a) & \sci (b) & \sci (c) & \sci (d)
\\
\\
\\
\psfrag{u}{}
\psfrag{d}{}
 \raisebox{0cm}{\epsfig{file=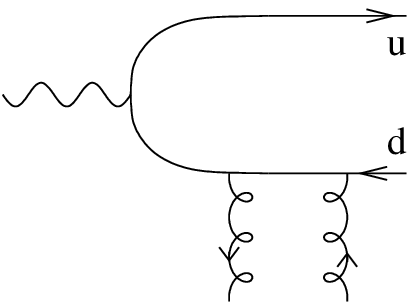,width=\li}}  & \si
\psfrag{u}{}
\psfrag{d}{}
 \raisebox{1cm}{\epsfig{file=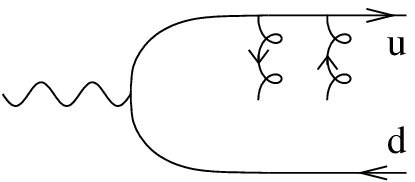,width=\li}} \\
\\
 \hspace{1.2cm}  (e) & \sci (f)
 \end{tabular}}
\caption{The 6 hard diagrams attached to the 2-parton correlators, which contribute to the   $\gamma^* \to \rho$ impact factor, with momentum flux of external line, along $p_1$ direction.
These drawing implicitly 
assume that the two right-hand side spinor lines are closed on the 
 the two possible Fierz structures $\slashchar{p}$
or $\slashchar{p} \, \gamma^5$.}
\label{Fig:NoDer2}
\end{figure}
%%%%%%%%%%%%%%%%%%%%%%%%%%%
\def\li{.2\columnwidth}
\def\si{\hspace{.9cm}}
\def\sci{\hspace{1.7cm}}
\psfrag{i}{}
\psfrag{u}{\footnotesize$\hspace{-0cm}y$}
\psfrag{d}{\footnotesize$\hspace{-.25cm}-\bar{y}$}
\psfrag{m}{}
\begin{figure}[h]
 \scalebox{1}{\begin{tabular}{cccc}
 \hspace{0.cm}\raisebox{0cm}{\epsfig{file=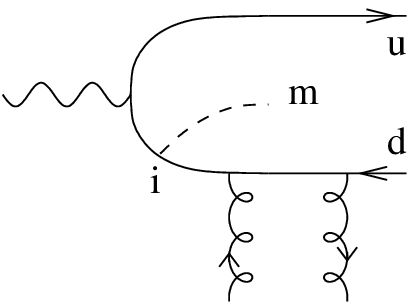,width=\li}} & \si
\psfrag{u}{}
\psfrag{d}{}
 \raisebox{0cm}{\epsfig{file=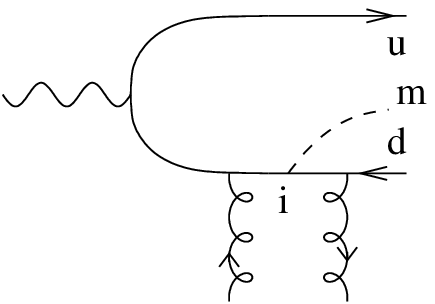,width=\li}} & \si
\psfrag{u}{}
\psfrag{d}{}
\raisebox{0cm}{\epsfig{file=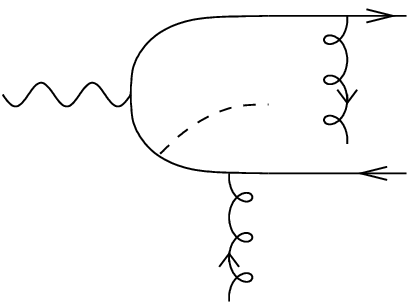,width=\li}}  & \si
\psfrag{u}{}
\psfrag{d}{}
 \raisebox{0cm}{\epsfig{file=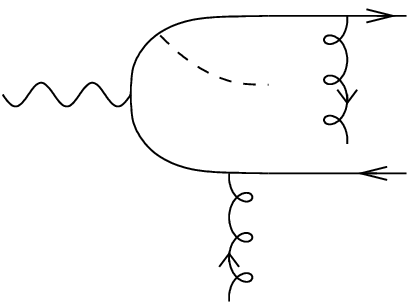,width=\li}}
\\
\\
\hspace{1.2cm} (a1) & \sci (a2) & \sci (b1) & \sci (b2)
\\
\\
\\
\psfrag{u}{}
\psfrag{d}{}
\hspace{0.cm}\raisebox{1cm}{\epsfig{file=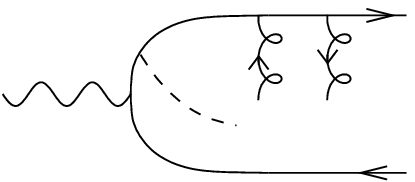,width=\li}} & \si
\psfrag{u}{}
\psfrag{d}{}
 \raisebox{1cm}{\epsfig{file=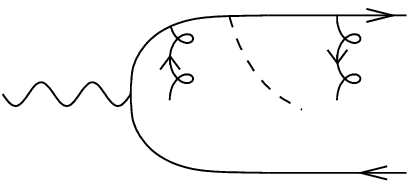,width=\li}} & \si
\psfrag{u}{}
\psfrag{d}{}
\raisebox{0cm}{\epsfig{file=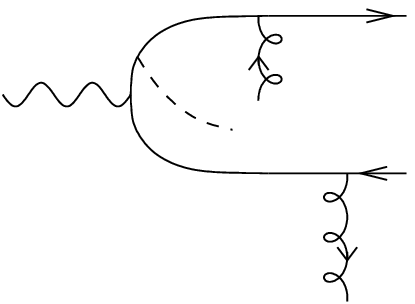,width=\li}}  & \si
\psfrag{u}{}
\psfrag{d}{}
 \raisebox{0cm}{\epsfig{file=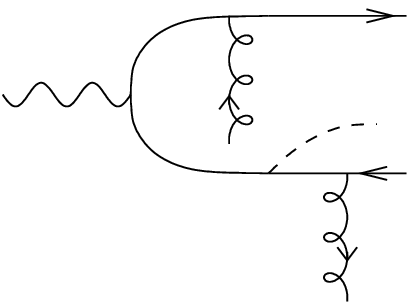,width=\li}}
\\
\\
\hspace{1.2cm} (c1) & \sci (c2) & \sci (d1) & \sci (d2)
\\
\\
\\
\psfrag{u}{}
\psfrag{d}{}
\hspace{0.cm}\raisebox{0cm}{\epsfig{file=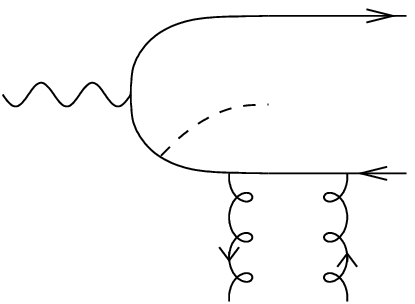,width=\li}} & \si
\psfrag{u}{}
\psfrag{d}{}
 \raisebox{0cm}{\epsfig{file=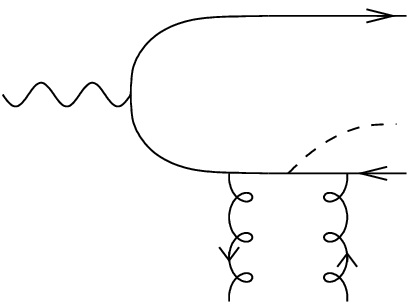,width=\li}} & \si
\psfrag{u}{}
\psfrag{d}{}
\raisebox{1cm}{\epsfig{file=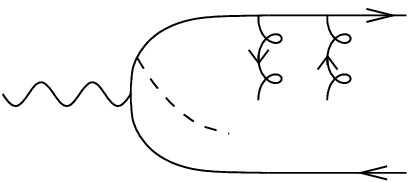,width=\li}}  & \si
\psfrag{u}{}
\psfrag{d}{}
 \raisebox{1cm}{\epsfig{file=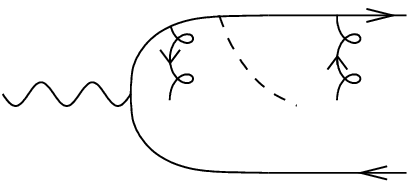,width=\li}}
\\
\\
\hspace{1.2cm} (e1) & \sci (e2) & \sci (f1) & \sci (f2)
 \end{tabular}}
\caption{The 12 contributions arising from the first derivative of the 6 hard diagrams attached to the 2-parton correlators, which contribute to the   $\gamma^* \to \rho$ impact factor, with momentum flux of external line, along $p_1$ direction.}
\label{Fig:Der2}
\end{figure}
%%%%%%%%%%%%%%%%    FIGURE 6    %%%%%%%%%%%%%%%%%%%%%%%
\begin{figure}[tb]
\psfrag{u}{\hspace{-.1cm}$ y_1$}
\psfrag{d}{\raisebox{.05cm}{$\hspace{-.35cm} -\bar{y}_2$}}
\psfrag{m}{$y_2-y_1$}
\psfrag{i}{}
\def\widtlH{0.245\columnwidth}
\def\widtlS{0.24\columnwidth}
\scalebox{1}{\begin{tabular}{ccc}
\hspace{-0.5cm}\epsfig{file=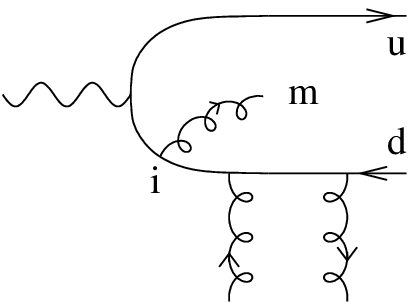,width=\widtlH}&\hspace{1cm}
\epsfig{file=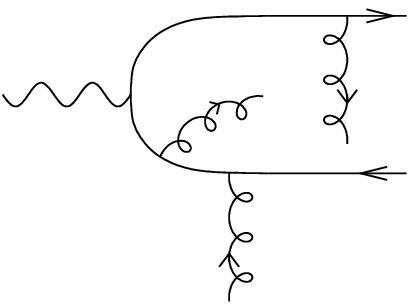,width=\widtlH}
& \hspace{1cm}\raisebox{.07cm}{\epsfig{file=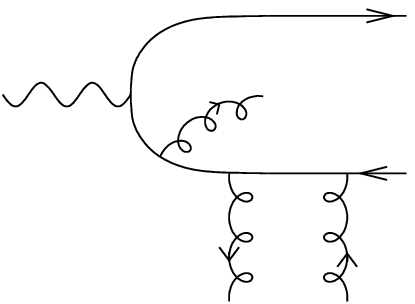,width=\widtlS}}\\
\\
\hspace{1.5cm} (aG1) &   \hspace{1.5cm} (bG1) &   \hspace{1.5cm} (eG1) \\
\\
\\
\psfrag{u}{}
\psfrag{d}{}
\psfrag{m}{}
\psfrag{i}{}
\hspace{-0.5cm}\raisebox{-.03cm}{\epsfig{file=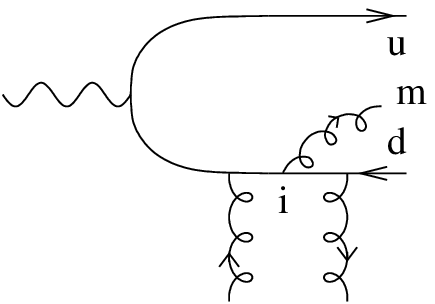,width=\widtlH}} &\hspace{1cm}
\raisebox{-.16cm}{\epsfig{file=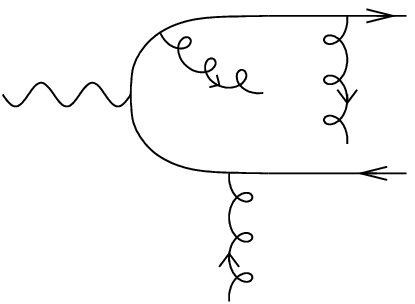,width=\widtlH}} & \hspace{1cm}\raisebox{-.1cm}{\epsfig{file=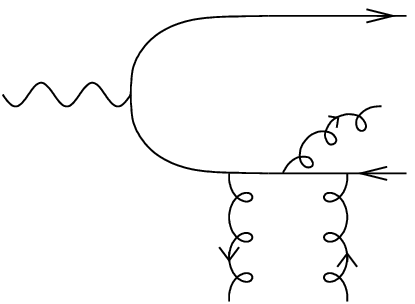,width=\widtlS}}
\\
\\
\hspace{1.5cm} (aG2) & \hspace{1.5cm}  (bG2)  &  \hspace{3cm} (eG2) \\
\\
\\
\hspace{-0.5cm}\epsfig{file=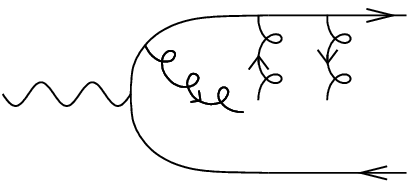,width=\widtlH} &\hspace{1cm}
\raisebox{-1.23cm}{\epsfig{file=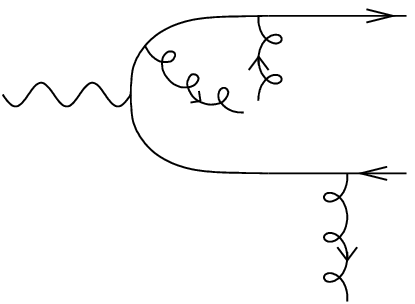,width=\widtlH}}
& \hspace{1cm}\raisebox{.04cm}{\epsfig{file=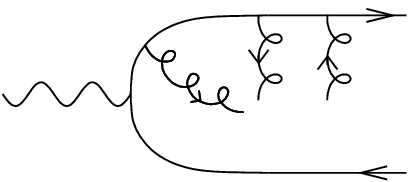,width=\widtlS}}\\
\\
\\
\hspace{1.5cm} (cG1)  & \hspace{1.5cm} (dG1)  & \hspace{3cm} (fG1) \\
\\
\\
\hspace{-0.5cm}\epsfig{file=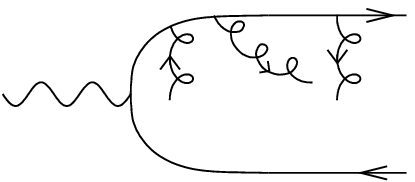,width=\widtlH} &\hspace{1cm}
\raisebox{-1.24cm}{\epsfig{file=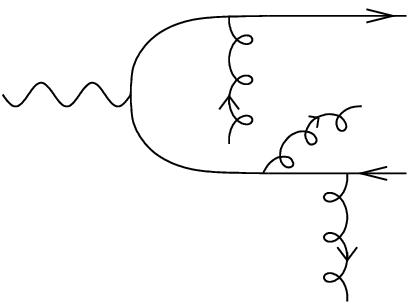,width=\widtlH}}  &\hspace{1cm}\raisebox{.03cm}{\epsfig{file=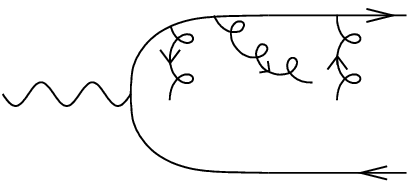,width=\widtlS}}\\
\hspace{1.5cm} (cG2)  & \hspace{1.5cm} (dG2)  &  \hspace{3cm} (fG2) \\
\end{tabular}}
\caption{The 12 ''Abelian`` (i.e. without triple gluon vertex) type contributions from  the hard scattering amplitude attached to the 3-parton correlators for the   $\gamma^* \to \rho$ impact factor, with momentum flux of external line, along $p_1$ direction.}
\label{Fig:3Abelian}
\end{figure}
%%%%%%%%%%%%%%%%%  FIGURE 7  %%%%%%%%%%%%%%%%%%%%%
\def\li{.2\columnwidth}
\def\si{\hspace{.9cm}}
\def\sci{\hspace{1.7cm}}
\psfrag{i}{}
\psfrag{u}{\footnotesize\hspace{-.1cm}$ y_1$}
\psfrag{d}{\raisebox{.05cm}{\footnotesize$\hspace{-.35cm} -\bar{y}_2$}}
% \psfrag{u}{\footnotesize$\hspace{-.3cm}y_1$}
% \psfrag{d}{\footnotesize$\hspace{-.3cm}-\bar{y}_2$}
\psfrag{m}{$\hspace{-.3cm}y_2-y_1$}
\begin{figure}[htb]
 \scalebox{1}{\begin{tabular}{cccc}
 \hspace{0.cm}\raisebox{0cm}{\epsfig{file=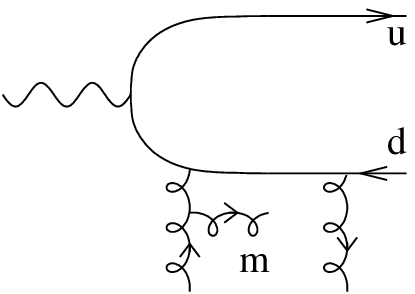,width=\li}} & \si
\psfrag{u}{}
\psfrag{d}{}
\psfrag{m}{}
 \raisebox{0cm}{\epsfig{file=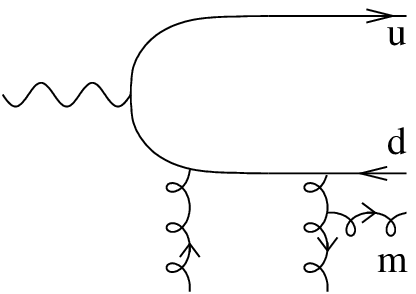,width=\li}} & \si
\psfrag{u}{}
\psfrag{d}{}
\psfrag{m}{}
\raisebox{0cm}{\epsfig{file=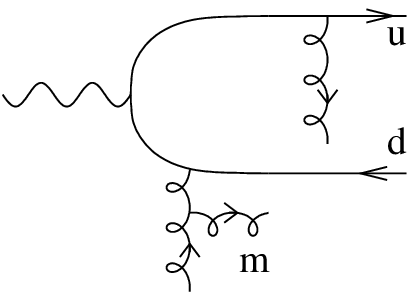,width=\li}}  & \si
\psfrag{u}{}
\psfrag{d}{}
 \psfrag{m}{}
 \raisebox{0.38cm}{\epsfig{file=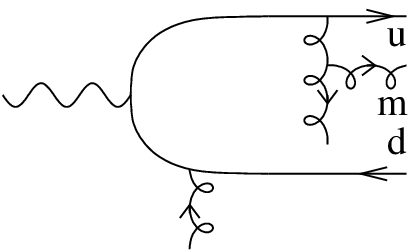,width=\li}}
\\
\\
\hspace{1.2cm} (atG1) & \sci (atG2) & \sci (btG1) & \sci (btG2)
\\
\\
\\
\psfrag{u}{}
\psfrag{d}{}
\psfrag{m}{}
\hspace{0.cm}\raisebox{1cm}{\epsfig{file=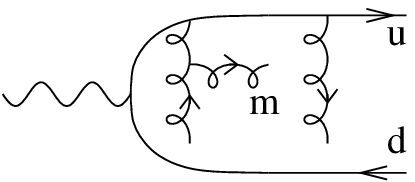,width=\li}} & \si
\psfrag{u}{}
\psfrag{d}{}
\psfrag{m}{}
 \raisebox{1cm}{\epsfig{file=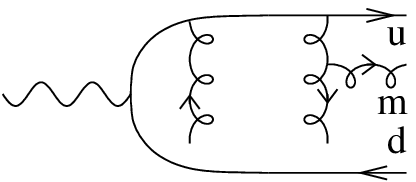,width=\li}} & \si
\psfrag{u}{}
\psfrag{d}{}
\psfrag{m}{}
\raisebox{0.395cm}{\epsfig{file=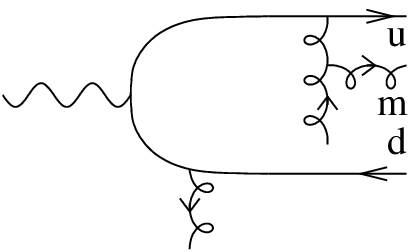,width=\li}}  & \si
\psfrag{u}{}
\psfrag{d}{}
\psfrag{m}{}
 \raisebox{0.05cm}{\epsfig{file=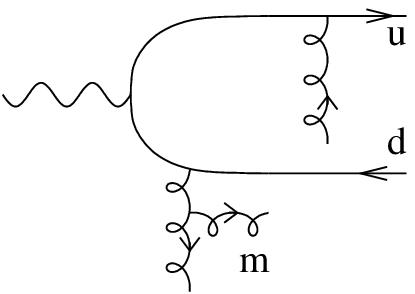,width=\li}}
\\
\\
\hspace{1.2cm} (ctG1) & \sci (ctG2) & \sci (dtG1) & \sci (dtG2)
\\
\\
\\
\psfrag{u}{}
\psfrag{d}{}
\psfrag{m}{}
\hspace{0.cm}\raisebox{0cm}{\epsfig{file=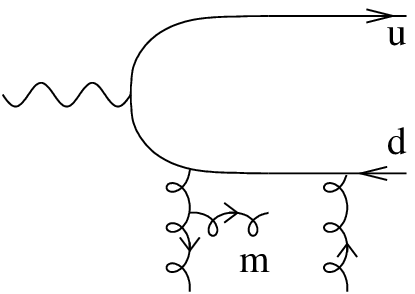,width=\li}} & \si
\psfrag{u}{}
\psfrag{d}{}
\psfrag{m}{}
 \raisebox{0cm}{\epsfig{file=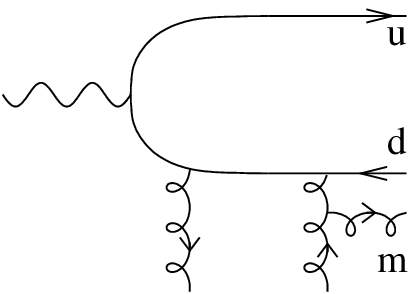,width=\li}} & \si
\psfrag{u}{}
\psfrag{d}{}
\psfrag{m}{}
\raisebox{.9cm}{\epsfig{file=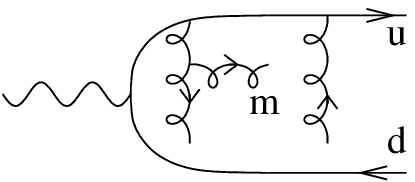,width=\li}}  & \si
\psfrag{u}{}
\psfrag{d}{}
\psfrag{m}{}
 \raisebox{.9cm}{\epsfig{file=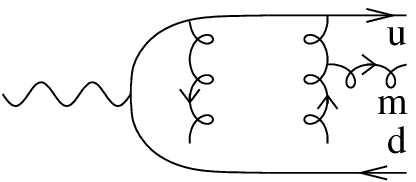,width=\li}}
\\
\\
\hspace{1.2cm} (etG1) & \sci (etG2) & \sci (ftG1) & \sci (ftG2)
 \end{tabular}}
\caption{The 12 ''non-Abelian`` -(with one triple gluon vertex) contributions from the hard scattering amplitude attached to the 3-parton correlators, for the   $\gamma^* \to \rho$ impact factor, with momentum flux of external line, along $p_1$ direction.}
\label{Fig:3NonAbelian}
\end{figure}
%%%%%%%%%%%%%%%%%%%%%    FIGURE 8    %%%%%%%%%%%%%%%%%
\def\li{.2\columnwidth}
\def\si{\hspace{.9cm}}
\def\sci{\hspace{1.7cm}}
\psfrag{i}{}
\psfrag{u}{\footnotesize$\hspace{-.1cm}y_1$}
\psfrag{d}{\footnotesize$\hspace{-.3cm}-\bar{y}_2$}
\psfrag{m}{\footnotesize$\hspace{-.1cm}y_2-y_1$}
\begin{figure}[htb]
 \scalebox{1}{\begin{tabular}{cccc}
 \hspace{0.cm}\raisebox{0cm}{\epsfig{file=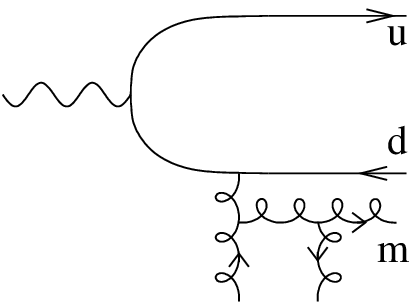,width=\li}} & \si
\psfrag{u}{}
\psfrag{d}{}
\psfrag{m}{}
 \raisebox{0cm}{\epsfig{file=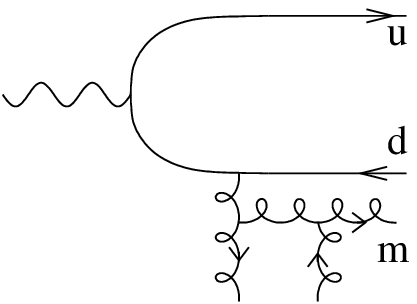,width=\li}} & \si
\psfrag{u}{}
\psfrag{d}{}
\psfrag{m}{}
\raisebox{1cm}{\epsfig{file=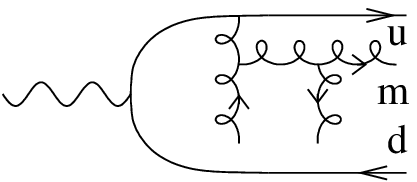,width=\li}}  & \si
\psfrag{u}{}
\psfrag{d}{}
\psfrag{m}{}
 \raisebox{1cm}{\epsfig{file=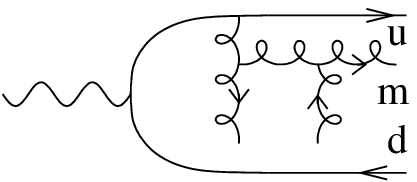,width=\li}}
\\
\\
\hspace{1.2cm} (gttG1) & \sci (gttG2) & \sci (httG1) & \sci (httG2)
 \end{tabular}}
\caption{The 4 ''non-Abelian`` -(with two triple gluon vertices) contributions from the hard scattering amplitude attached to the 3-parton correlators, for the   $\gamma^* \to \rho$ impact factor, with momentum flux of external line, along $p_1$ direction.}
\label{Fig:3NonAbelianTwo}
\end{figure}
%%%%%%%%%%%%%%%%%%%%%%%%%%%%%%%%%%%%%%%%%%%

We will now derive equations  (\ref{ninV}, \ref{ninA}) using as a tool the explicit example of the $\gamma^* \to \rho$ impact factor. We want here to insist on the fact that
the proof is independent of the specific process under consideration, and only rely
on general arguments based on Ward identities. For the $\gamma^* \to \rho$ impact factor, which will be computed in details in section \ref{Sec_Impact}, one needs to consider 2-parton contributions both without transverse derivative (illustrated by diagrams of Fig.\ref{Fig:NoDer2}) and with transverse derivative (see Fig.\ref{Fig:Der2}),
as well as 3-parton contributions (see Figs.\ref{Fig:3Abelian}, \ref{Fig:3NonAbelian}, \ref{Fig:3NonAbelianTwo}). 
Note that these drawing implicitly 
assume that the two right-hand side spinor lines are closed on the 
 the two possible Fierz structures $\slashchar{p}$
or $\slashchar{p} \, \gamma^5$ involved in the correlators of $\rho-$meson DAs.

In the color space, each of those diagrams can be projected
in two parts, characterized by the two Casimir invariants $C_F$ and $N_c$.
The equations (\ref{ninV}, \ref{ninA})
%(\ref{ninV}) and (\ref{ninA})
are obtained by
considering the consequence of the $n-$independence on
 the contribution to the $C_F$
color structure.
The $n-$independence condition applied to the $N_c$ structure is automatically satisfied and does not lead to new constraints, as we show in Appendix \ref{Ap:Nc}.

We start with  the derivation of Eq.(\ref{ninV}), which corresponds  to the vector correlator contributions with $C_F$ invariant.
%within $k_T$-factorization.
The 3-parton ($q \bar{q} g$) contribution  and the 2-parton contribution involving
$\Phi^\perp$ to ${\cal A}$
can be reduced to the convolution of the leading order hard 2-parton contributions with linear combination of correlators,
 thanks to the use of the Ward identity.
%The most involved use of these identities occurs in the case of
%the 3-parton correlator.

In the case of the 3-parton vector correlator (\ref{Correlator3BodyV}), due to (\ref{epsilonRho}) the dependency on
$n$ enters linearly and only through the scalar product $e^* \cdot n\,.$
 Thus, the action on the amplitude of the derivative $d/dn_\perp$ involved in (\ref{nIndV}) can be extracted by the replacement $e_\alpha^* \to -p_\alpha\,,$ which means in practice that the Feynman rule (using  conventions of Ref.\cite{DDT} for computing the $T$ matrix element) $g \, t^a \, \gamma^\alpha \,e^*_\alpha$
 entering the coupling of the gluon
inside the hard part should be replaced by $-g \, t^a \, \gamma^\alpha \, p_\alpha\,.$
Then, using the Ward identity for the hard part, it reads
\begin{eqnarray}
\label{Ward}
(y_1-y_2){\rm tr} \left[ H^{\rho}_{q \bar{q} g}(y_1,y_2) \, p_\rho
\, \slashchar{p}\right] =
{\rm tr} \left[H_{q \bar{q}}(y_1) \, \slashchar{p}\right]-{\rm tr} \left[H_{q \bar{q}}(y_2) \, \slashchar{p}\right]\,,
\nonumber
\end{eqnarray}
which can be seen graphically as
\psfrag{yq}{{\footnotesize $\hspace{-.2cm}y_1$}}
\psfrag{dy}{\raisebox{.04cm}{\footnotesize $\hspace{-.3cm}y_2-y_1$}\raisebox{.6cm}{$\!\!\mu$}}
\psfrag{yb}{{\footnotesize $\hspace{-.2cm}1-y_2$}}
%\hspace{-0.2cm}\raisebox{.9cm}{$(y_1-y_2)$}\
\beqa
\psfrag{yq}{{ $\hspace{-.2cm}y_1$}}
%\psfrag{dy}{\raisebox{.04cm}{ $\hspace{-.3cm}y_2-y_1$}\raisebox{.6cm}{$\!\!\mu$}}
\psfrag{yb}{{ $\hspace{-.2cm}1-y_2$}}
&&\centerline{\hspace{-2cm}
\raisebox{1.8cm}{$ p_\mu \left[ \quad
\raisebox{-1cm}{\includegraphics[width=3.2cm]{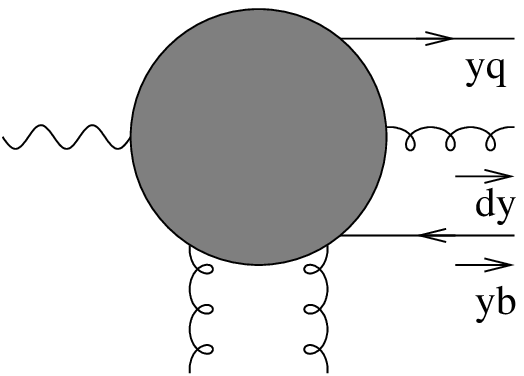}}\hspace{.6cm} \right]$}
} \hspace{-2cm}
\nonumber \\
&&\hspace{0cm}\raisebox{0cm}{$=\ \displaystyle \frac{1}{y_1-y_2} \left[
\psfrag{yq}{{\footnotesize $\hspace{-.2cm}y_1$}}
\psfrag{yb}{{\footnotesize $\hspace{-.2cm}1-y_1$}}
\hspace{0.cm}\raisebox{-1cm}{\includegraphics[width=3.cm]{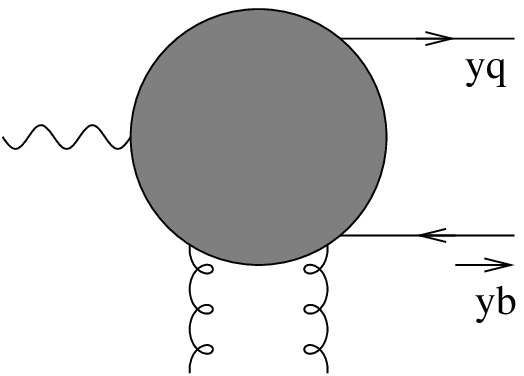}}\hspace{.8cm}
%\raisebox{0cm}{\displaystyle -}
\displaystyle - \hspace{.8cm}
\psfrag{yq}{{\footnotesize $\hspace{-.2cm}y_2$}}
\psfrag{yb}{{\footnotesize $\hspace{-.3cm}1-y_2$}}
\raisebox{-1cm}{\includegraphics[width=3.cm]{RWardIF.eps}}\hspace{0.4cm}\right]\hspace{-.4cm}$}
%\caption{Reduction of 3-parton correlators to 2-parton correlators through Ward identity.}
\label{Fig:WardIF}
\eqa
as we will  show below with more details, and will
give the last term of r.h.s of Eq.(\ref{ninV}).
The proof can be settled easily relying on a graphical rule in order to use the collinear Ward identity.
Indeed within the conventions of \cite{DDT},
the collinear Ward identity can be symbolically written as
\beq
\label{WardCol}
\psfrag{p}{$\hspace{-.3cm}y_1 \, p$}
\psfrag{q}{$y_2 \, p$}
\psfrag{g}{$\gamma^\mu$}
\psfrag{f}{$\!(y_2-y_1) \, p$}
p_\mu \quad  \raisebox{-.5cm}{\epsfig{file=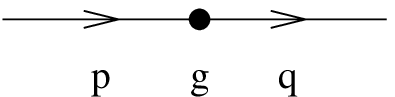,width=3cm}}=\frac{1}{y_2-y_1}\left[\,
\psfrag{t}{$y_2 \, p$}
 \raisebox{-.5cm}{\epsfig{file=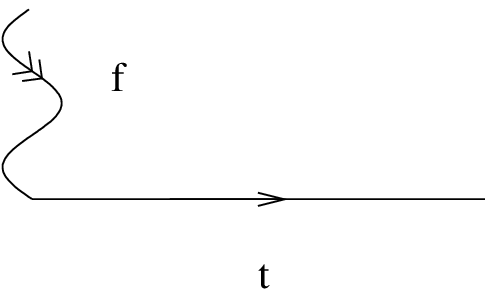,width=3cm}} -
\psfrag{t}{$y_1 \, p$}
\psfrag{f}{$\hspace{-1.5cm}(y_2-y_1)\, p$}
\raisebox{-.5cm}{\epsfig{file=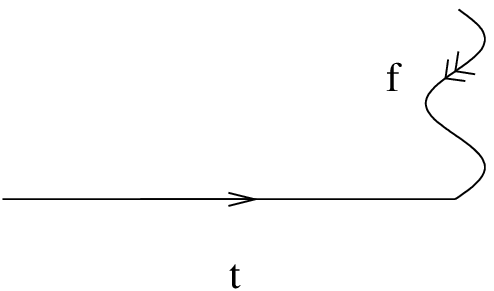,width=3cm}}\, \right]
\eq
where each fermionic line is a propagator.  The wavy lines with double arrows are there in order to fulfill momentum conservation, since the {\em incoming} momentum is $y_1 \, p$ while the {\em outgoing} momentum is $y_2 \, p\,.$

\def\si{\footnotesize}
\def\ti{0.32\columnwidth}
\def\ri{-2.5cm}
Let us consider the 3-parton {\em ''Abelian''} diagrams, illustrated in Fig.\ref{Fig:3Abelian}.
Applying the Ward identity (\ref{WardCol}) to graphs (aG1) and (aG2) gives
\beqa
% \psfrag{i}{$\mu$}
% \psfrag{u}{$y_1$}
% \psfrag{d}{$y_2-1$}
% \psfrag{m}{$y_2-y_1$}
% \psfrag{yq}{{ $\hspace{-.2cm}y_1$}}
% \psfrag{dy}{\raisebox{.04cm}{ $\hspace{-.3cm}y_2-y_1$}\raisebox{.6cm}{$\!\!\mu$}}
\psfrag{yb}{{ $\hspace{-.2cm}1-y_2$}}
&&\scalebox{1}
 {\raisebox{1.8cm}{$ \displaystyle - (y_2-y_1) p_\mu \quad \left[
\psfrag{i}{$\mu$}
\psfrag{u}{$y_1$}
\psfrag{d}{$y_2-1$}
\psfrag{m}{$y_2-y_1$}
%\psfrag{yq}{{ $\hspace{-.2cm}y_1$}}
%\psfrag{yb}{{ $\hspace{-.2cm}1-y_1$}}
\hspace{0.4cm}\raisebox{-1.4cm}{\includegraphics[width=4.5cm]{aG1.eps}}\hspace{1.5cm}
%\raisebox{0cm}{\displaystyle -}
\displaystyle + \hspace{.8cm}
%\psfrag{yq}{{ $\hspace{-.2cm}y_2$}}
%\psfrag{yb}{{ $\hspace{-.3cm}1-y_2$}}
\raisebox{-1.4cm}{\includegraphics[width=4.5cm]{aG2.eps}}\hspace{1.4cm}\right]$}}\nonumber\\
\nonumber\\
%\nonumber\\
&=&
\psfrag{u}{\si \raisebox{.3cm}{$y_1$}}
\psfrag{d}{\si $y_2-1$}
\psfrag{f}{\si $y_1-y_2$}
%%%%%%\scalebox{1}{\begin{tabular}{cccc}
\psfrag{u}{\si \raisebox{0cm}{$y_1$}}
\psfrag{r}{\si $\hspace{-.26cm} y_1-1$}
\psfrag{s}{\si $y_2-1$}
\hspace{.3cm}\raisebox{-2cm}{\epsfig{file=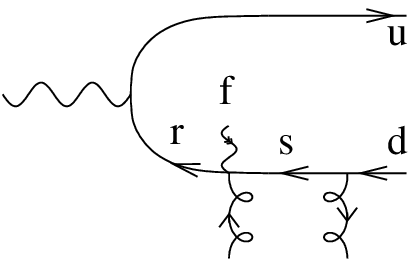,width=\ti}}
\hspace{.8cm} -
\psfrag{r}{\si $\hspace{-.27cm} y_2-1$}
\psfrag{s}{\si $y_2-1$}
 \hspace{0.4cm}\raisebox{-2cm}{\epsfig{file=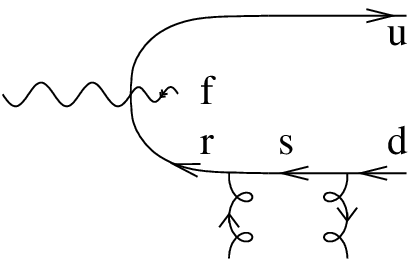,width=\ti}}\nonumber\\
\nonumber\\
\nonumber\\
&+&
\psfrag{u}{\si \raisebox{0cm}{$y_1$}}
\psfrag{r}{\si $\hspace{-.1cm} y_1-1$}
\psfrag{s}{\si $\hspace{-.3cm} y_1-1$}
\psfrag{d}{\si $y_2-1$}
\psfrag{f}{\si $y_1-y_2$}
\hspace{.3cm}\raisebox{-2cm}{\epsfig{file=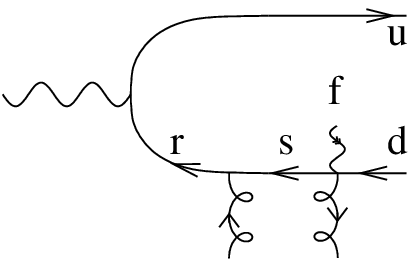,width=\ti}}
\hspace{.8cm}  -
\psfrag{r}{\si $\hspace{-.2cm} y_1-1$}
\psfrag{s}{\si $\hspace{-.1cm} y_2-1$}
\hspace{0.4cm}\raisebox{-2cm}{\epsfig{file=aG1Ward1.eps,width=\ti}}\nonumber\\
\nonumber\\
\nonumber\\
&=&
\psfrag{u}{\si \raisebox{.3cm}{$y_1$}}
\psfrag{d}{\si $y_1-1$}
\psfrag{r}{\si $\hspace{-.2cm} y_1-1$}
\psfrag{s}{\si $y_1-1$}
\psfrag{f}{\si $y_1-y_2$}
\hspace{.3cm}\raisebox{-2.7cm}{\epsfig{file=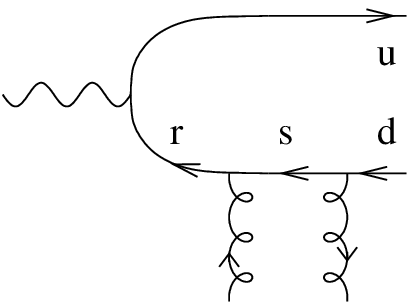,width=\ti}}
\hspace{.8cm}  -
\psfrag{u}{\si \raisebox{.3cm}{$y_2$}}
\psfrag{d}{\si $y_2-1$}
\psfrag{r}{\si $\hspace{-.2cm} y_2-1$}
\psfrag{s}{\si $\hspace{-.1cm}y_2-1$}
 \hspace{0.4cm}\raisebox{-2.7cm}{\epsfig{file=af.eps,width=\ti}} \hspace{1cm} \,,
 %%%%  \end{tabular}}            
%\caption{The collinear Ward identity applied to the 3-parton "Abelian" contributions (aG1) and (aG2).}
\label{Ward3body_aD}
\eqa
where the indicated momentum fractions correspond to flow along the momentum $p_1$ of the $\rho-$meson.
The last line of Eq.(\ref{Ward3body_aD})
has been  obtained after cancellation of the first and fourth
term in the second equality, and the two remaining diagrams have been relabelled (in the first term of the last line, one does $y_2-1 \to y_1-1$ for the outgoing antiquark, and in the second term, one does $y_1 \to y_2$ for the outgoing quark), as far as the external lines are concerned, after using the fact that this does not change its internal structure.

The same identity applies for each couple of graphs (bG1, bG2),  (cG1, cG2),  (dG1, dG2),  (eG1, eG2) and  (fG1, fG2). This leads to the following identity
\def\sih{\hspace{-0.05cm}}
\def\sibh{\hspace{-0.05cm}}
\def\li{.2\columnwidth}
\def\hi{.2cm}
\def\si{\hspace{0.1cm}}
\def\ri{0.9cm}
\def\sib{\hspace{0.1cm}}
\def\sci{\hspace{1.7cm}}
\psfrag{i}{\tiny$\!\!\!\!\mu$}
\psfrag{m}{}
\psfrag{u}{}
\psfrag{d}{}
%\psfrag{u}{\footnotesize$\hspace{-.3cm}y$}
%\psfrag{d}{\footnotesize$\hspace{-.3cm}y-1$}
\def\wi{.11\columnwidth}
\beqa
&&
-\int\limits^1_0 dy_1 \int\limits^1_0 dy_2 \, B(y_1, y_2)
 \nonumber \\
&&\hspace{-.3cm}
\raisebox{1.8cm}{$ \displaystyle  \times \, p_\mu \left[
\scalebox{1}{\begin{tabular}{cccccccccccc}
&
\sib
\psfrag{m}{\tiny$y_2-y_1$}
\psfrag{u}{\tiny$y_1$}
\psfrag{d}{\tiny$y_2-1$}
\hspace{-.9cm}\epsfig{file=aG1.eps,width=\wi}
&\si \raisebox{\ri}{+}
&\sib \epsfig{file=bG1.eps,width=\wi}
&\si \raisebox{\ri}{+}
 &\sib\raisebox{0cm}{\epsfig{file=eG1.eps,width=\wi}}
&\si \raisebox{\ri}{+}
&\psfrag{i}{}
\sib\raisebox{0.05cm}{\epsfig{file=aG2.eps,width=\wi}}
&\si \raisebox{\ri}{+}
&\sib \epsfig{file=bG2.eps,width=\wi}
&\si \raisebox{\ri}{+} &\sib\raisebox{0cm}{\epsfig{file=eG2.eps,width=\wi}}\\
\\
\si \raisebox{\ri}{+}
&
\sib\raisebox{0.6cm}{\epsfig{file=cG1.eps,width=\wi}}
&\si \raisebox{\ri}{+}
& \sib\raisebox{0.06cm}{\epsfig{file=dG1.eps,width=\wi}}
&\si \raisebox{\ri}{+}
& \si\raisebox{.6cm}{\epsfig{file=fG1.eps,width=\wi}}
&\si \raisebox{\ri}{+}
& \sib\raisebox{.6cm}{\epsfig{file=cG2.eps,width=\wi}}
&\si \raisebox{\ri}{+}
&\sib\raisebox{0.06cm}{\epsfig{file=dG2.eps,width=\wi}}
&\si \raisebox{\ri}{+}  &\sib\raisebox{.6cm}{\epsfig{file=fG2.eps,width=\wi}}
\end{tabular}}
\right] $}\nonumber\\
&&
=
\int\limits^1_0 dy_1 \int\limits^1_0 dy_2\, \displaystyle \frac{B(y_1, y_2)}{y_2-y_1}
\nonumber \\
&&\hspace{-.3cm}\times \left\{\left[
\scalebox{1}{\begin{tabular}{ccccccccccc}
\psfrag{u}{\tiny $\hspace{-.5cm}y_1$}
\psfrag{d}{\tiny $\hspace{-.5cm}y_1-1$}
\sibh\epsfig{file=a.eps,width=\wi}
&\sih \raisebox{\ri}{+}
&\sibh \epsfig{file=b.eps,width=\wi}
&\sih \raisebox{\ri}{+}
 &\sibh\raisebox{.55cm}{\epsfig{file=c.eps,width=\wi}}
&\sih \raisebox{\ri}{+}
&\psfrag{i}{}
\sibh\raisebox{0.0cm}{\epsfig{file=d.eps,width=\wi}}
&\sih \raisebox{\ri}{+}
&\sibh \epsfig{file=e.eps,width=\wi}
&\sih \raisebox{\ri}{+} &\sibh\raisebox{.55cm}{\epsfig{file=f.eps,width=\wi}}
\end{tabular}}
 \right]- (y_1 \leftrightarrow y_2) \right\} \nonumber \\
&&
=
\int\limits^1_0 dy_1 \int\limits^1_0 \displaystyle\frac{dy_2}{y_2-y_1}\, [B(y_1, y_2)+B(y_2, y_1)]
 \nonumber \\
\def\si{\hspace{-0.05cm}}
\def\sib{\hspace{-0.05cm}}
&&\hspace{-.3cm}\times \left[
\scalebox{1}{\begin{tabular}{ccccccccccc}
\psfrag{u}{\tiny $\hspace{-.5cm}y_1$}
\psfrag{d}{\tiny $\hspace{-.5cm}y_1-1$}
\sib\epsfig{file=a.eps,width=\wi}
&\si \raisebox{\ri}{+}
&\sib \epsfig{file=b.eps,width=\wi}
&\si \raisebox{\ri}{+}
 &\sib\raisebox{.55cm}{\epsfig{file=c.eps,width=\wi}}
&\si \raisebox{\ri}{+}
&\psfrag{i}{}
\sib\raisebox{0.0cm}{\epsfig{file=d.eps,width=\wi}}
&\si \raisebox{\ri}{+}
&\sib \epsfig{file=e.eps,width=\wi}
&\si \raisebox{\ri}{+} &\sib\raisebox{.55cm}{\epsfig{file=f.eps,width=\wi}}
\end{tabular}}
 \right]
%\caption{The collinear Ward identity applied to the whole 3-parton ``non-Abelian'' contributions.}
\label{Ward3body}
\eqa
where
the last line is obtained after performing the change of variable $y_1 \leftrightarrow y_2$ in  the second term. Note that the hard part given by diagrams inside Eq.(\ref{Ward3body}) is convoluted with the last term of Eq.(\ref{ninV}).

%%%%

A similar treatment of
2-parton correlators with transverse derivative
whose contributions can be viewed as 3-parton processes with vanishing gluon  momentum leads to
\def\sih{\hspace{-0.05cm}}
\def\sibh{\hspace{-0.05cm}}
\def\li{.2\columnwidth}
\def\hi{.2cm}
\def\si{\hspace{0.1cm}}
\def\ri{0.9cm}
\def\sib{\hspace{0.1cm}}
\def\sci{\hspace{1.7cm}}
\psfrag{i}{\tiny$\!\!\!\!\mu$}
\psfrag{m}{}
\psfrag{u}{}
\psfrag{d}{}
%\psfrag{u}{\footnotesize$\hspace{-.3cm}y$}
%\psfrag{d}{\footnotesize$\hspace{-.3cm}y-1$}
\def\wi{.11\columnwidth}
\beqa
&&
-%\mathop{lim}_{y_2 \to y_1}
 \int\limits^1_0 dy_1 \, \int\limits^1_0 dy_2 \,
\delta(y_1-y_2) \,\varphi_1^T(y_1)
   \nonumber \\
&&\hspace{-.3cm}
\raisebox{1.8cm}{$ \displaystyle  \times \, p_\mu \left[
\scalebox{1}{\begin{tabular}{cccccccccccc}
&
\sib
\psfrag{m}{\tiny$y_2-y_1$}
\psfrag{u}{\tiny$y_1$}
\psfrag{d}{\tiny$y_2-1$}
\hspace{-.9cm}\epsfig{file=a1.eps,width=\wi}
&\si \raisebox{\ri}{+}
&\sib \epsfig{file=b1.eps,width=\wi}
&\si \raisebox{\ri}{+}
 &\sib\raisebox{0cm}{\epsfig{file=e1.eps,width=\wi}}
&\si \raisebox{\ri}{+}
&\psfrag{i}{}
\sib\raisebox{0.05cm}{\epsfig{file=a2.eps,width=\wi}}
&\si \raisebox{\ri}{+}
&\sib \epsfig{file=b2.eps,width=\wi}
&\si \raisebox{\ri}{+} &\sib\raisebox{0cm}{\epsfig{file=e2.eps,width=\wi}}\\
\\
\si \raisebox{\ri}{+}
&
\sib\raisebox{0.6cm}{\epsfig{file=c1.eps,width=\wi}}
&\si \raisebox{\ri}{+}
& \sib\raisebox{0.06cm}{\epsfig{file=d1.eps,width=\wi}}
&\si \raisebox{\ri}{+}
& \si\raisebox{.6cm}{\epsfig{file=f1.eps,width=\wi}}
&\si \raisebox{\ri}{+}
& \sib\raisebox{.6cm}{\epsfig{file=c2.eps,width=\wi}}
&\si \raisebox{\ri}{+}
&\sib\raisebox{0.06cm}{\epsfig{file=d2.eps,width=\wi}}
&\si \raisebox{\ri}{+}  &\sib\raisebox{.6cm}{\epsfig{file=f2.eps,width=\wi}}
\end{tabular}}
\right] $}\nonumber\\
&&
=\int\limits^1_0 dy_1 \, \int\limits^1_0 dy_2 \,
\delta(y_1-y_2) \, \displaystyle\frac{\varphi_1^T(y_1)}{y_2-y_1}
\nonumber \\
&&\hspace{-.3cm}\times \left\{\left[
\scalebox{1}{\begin{tabular}{ccccccccccc}
\psfrag{u}{\tiny $\hspace{-.5cm}y_1$}
\psfrag{d}{\tiny $\hspace{-.5cm}y_1-1$}
\sibh\epsfig{file=a.eps,width=\wi}
&\sih \raisebox{\ri}{+}
&\sibh \epsfig{file=b.eps,width=\wi}
&\sih \raisebox{\ri}{+}
 &\sibh\raisebox{.55cm}{\epsfig{file=c.eps,width=\wi}}
&\sih \raisebox{\ri}{+}
&\psfrag{i}{}
\sibh\raisebox{0.0cm}{\epsfig{file=d.eps,width=\wi}}
&\sih \raisebox{\ri}{+}
&\sibh \epsfig{file=e.eps,width=\wi}
&\sih \raisebox{\ri}{+} &\sibh\raisebox{.55cm}{\epsfig{file=f.eps,width=\wi}}
\end{tabular}}
 \right]- (y_1 \leftrightarrow y_2) \right\} \nonumber
\eqa
\beqa
%&&
&&=
\int\limits^1_0 dy_1 \frac{d}{d y_1} \varphi_1^T(y_1)
 \nonumber \\
\def\si{\hspace{-0.05cm}}
\def\sib{\hspace{-0.05cm}}
&&\hspace{-1cm}\times \left[
\scalebox{1}{\begin{tabular}{ccccccccccc}
\psfrag{u}{\tiny $\hspace{-.5cm}y_1$}
\psfrag{d}{\tiny $\hspace{-.5cm}y_1-1$}
\sib\epsfig{file=a.eps,width=\wi}
&\si \raisebox{\ri}{+}
&\sib \epsfig{file=b.eps,width=\wi}
&\si \raisebox{\ri}{+}
 &\sib\raisebox{.55cm}{\epsfig{file=c.eps,width=\wi}}
&\si \raisebox{\ri}{+}
&\psfrag{i}{}
\sib\raisebox{0.0cm}{\epsfig{file=d.eps,width=\wi}}
&\si \raisebox{\ri}{+}
&\sib \epsfig{file=e.eps,width=\wi}
&\si \raisebox{\ri}{+} &\sib\raisebox{.55cm}{\epsfig{file=f.eps,width=\wi}}
\end{tabular}}
 \right]\,,
%\caption{The collinear Ward identity applied to the whole 3-parton ``non-Abelian'' contributions.}
\label{Ward2body}
\eqa
where the last line is obtained after integration
by part.
This leads to the convolution of the first term of the l.h.s of Eq.(\ref{ninV}) with the $[ \cdots ]$ part in (\ref{Ward2body}).

The  second term, with $\varphi_1$,  of the l.h.s of Eq.(\ref{ninV})
originates from the 2-parton vector correlator and corresponds to the contribution for the longitudinally
polarized $\rho$ with
$e_L \sim p.$ The third term with $\varphi_3$ corresponds to the contribution of the same correlator for
the polarization vector of  $\rho_T$ written as in Eq.(\ref{epsilonRho}).
To get Eq.(\ref{ninV}), we used the fact that each individual term obtained above when expressing the $n-$independence condition involve the
{\em same} 2-parton hard part, convoluted with the Eq.(\ref{ninV}) through an integration over $y_1\,.$ The arguments used above, based on the collinear Ward identity, are clearly independent of the detailed structure of this resulting 2-parton hard part. Therefore, we deduce from this that Eq.(\ref{ninV}) itself should be satisfied.

A similar treatment for axial correlators leads to Eq.(\ref{ninA}). To prove this, we start from Eq.(\ref{nIndA}) and we note that
the parametrizations of matrix elements of correlators with
axial-vector currents (\ref{par1.1a}, \ref{par1.2}, \ref{Correlator3BodyA})  involve
the quantity
\begin{equation}
\label{pepsilon}
 p^\mu\,\varepsilon^{\alpha\,e^*_T\,p\,n}
\end{equation}
in which the index $\alpha$ is contracted with the matrix $\gamma^\alpha$
appearing  in the vertex of gluon emission in the hard part and the
momentum $p^\mu$ is contracted with the Fierz  matrix $\gamma_\mu\gamma^5$
corresponding in the hard part to the meson vertex.
First let us note that in the
expression (\ref{pepsilon}) one can replace $e^*_T$ by the full
polarization vector $e^*$, i.e.
\begin{equation}
\label{pepsilon2}
p^\mu\, \varepsilon^{\alpha\,e^*_T\,p\,n} =
p^\mu\, \varepsilon^{\alpha\,e^*\,p\,n}\;.
 \end{equation} 
Secondly, the inspection of the quantity (\ref{pepsilon}) or (\ref{pepsilon2}) leads to
the conclusion that in order to use the Ward identities in
a similar way as it was done in the vector part we need
to interchange in (\ref{pepsilon}) the indices $\mu\;\leftrightarrow\;\alpha$.
It is done with the help of the Schouten identity, which for our peculiar
case means that
\begin{equation}
\label{schoten}
  p^\mu\,\varepsilon^{\alpha\,e^*_T\,p\,n} =
p^\alpha\,\varepsilon^{\mu\,e^*_T\,p\,n} \;.
\end{equation}
After that,
the momentum $p^\alpha$ acts on the gluon vertex in the hard part, so
the consequences of the $n-$independence of the axial part of the
impact factor maybe  derived in exactly the same way as we did above in the case of vector
correlators,  since the vector $\varepsilon^{\mu\,e^*_T\,p\,n}$ is completely factorized. One then obtains Eq.(\ref{ninA})
from Eq.(\ref{ninV}) after the replacements
$B \to D, \, \varphi_3 \to \varphi_A, \, \varphi^T_1 \to \varphi^T_A$ and $\varphi_1 \to 0$
since there is no counterpart of the twist 2 DA
$\varphi_1$ for the axial part.
\\

Since we rely on the $n-$independence of the amplitude, one may wonder about the effect of the gauge choice, which is fixed by $n\,,$ on the hard part.
The QCD Ward identities require the vanishing of the amplitude in which polarization vector of a gluon is replaced by its momentum provided all other partons are on the mass shell. In the framework of the $k_T-$factorization (see Sec.\ref{Sec_Impact}), the $t-$channel gluons are off the mass-shell. Therefore the replacement of the $s-$channel gluon  polarization vector by its momentum leads to the vanishing of scattering amplitude up to terms proportional to $k_\perp^2/s$ where $k_\perp$ are transverse momenta of $t-$channel gluons.
From the point of view of the $t-$channel, the gauge invariance of the impact-factor means that it should vanish when the transverse momentum of any $t-$channel gluon vanishes. To achieve this property it is necessary to include in a consistent way
not only DAs with lowest Fock state containing only quarks but also those involving
quarks and gluon, as we will show in detail in Sec.\ref{Sec_Impact}.

In practice, we here
 check this invariance by contracting the $s-$channel emitted gluon vertex in the hard part with the momentum, which
in collinear factorization is proportional to the $\rho-$meson momentum, which leads to simplifications in the use of (collinear) Ward identities.

In order to prove this,
one should first project on the various color Casimir structure. In the case of the impact factor (see Sec.\ref{Sec_Impact}), this means to distinguish $N_c$ and $C_F$ terms. In this case, $C_F$ terms arise from
2-partons diagrams and from  3-partons diagrams where the emitted gluon
is attached to a quark line, while $N_c$ terms are obtained from  3-partons diagrams where the emitted gluon is attached to a quark line only between 2 $t-$channel exchanged gluons or from diagrams involving at least one triple gluon vertex.

The method is almost identical with the one used above when deriving the $n-$independence equations.
Consider first the case of hard 3-partons  diagrams entering the vector part of Fierz decomposition.  
We
 contract the $s-$channel emitted gluon vertex  with its momentum, which
is proportional to  $p_\mu\,.$ The next step is to use the same method as the one used in Eq.(\ref{Ward3body}), except that the DA $B$ is not involved here. One thus finally gets 
two groups of 6 diagrams (which differ by the labeling of outgoing quarks, one being $y_1$ and $\bar{y}_1$ for the quark and antiquark, respectively, and the other one being 
$y_2$ and $\bar{y}_2$ for the quark and antiquark respectively).
Since we started here from the consideration of the $\gamma^*_T \to \rho_T$ transition, each of these 6 hard contributions, due to the appearance of the remaining $\slashchar{p}$  from the Fierz structure,  thus now  encodes the hard part of the transition $\gamma^*_T \to \rho_L\,.$ This transition vanishes in our kinematics, which leads to the conclusion that this hard part is gauge invariant.
The same treatment can be applied to the hard diagrams with derivative insertion displayed in 
Fig.\ref{Fig:Der2}, since this insertion corresponds to the peculiar limit of vanishing ``gluon'' momentum.

The proof for the axial part of the Fierz decomposition goes along the same line. The only difference lays on the appearance of the $\slashchar{p} \, \gamma^5$ structure, which corresponds to a meson  $ b_1$ with quantum numbers $J^{PC}=1^{+-}$ instead of $1^{--}$ for $\rho$, leading finally 
to the hard part of the transition $\gamma^*_T \to b_{1\, L}\,$ which vanishes in our kinematics.

In the case of contribution proportional to $N_c$, one can prove that these hard terms are also gauge invariant.
 This is proven in Appendix \ref{Ap:Nc}. The reason is the same as the one which led to the conclusion that $N_c$ terms do not lead to additional $n-$independence condition.

Although our implementation of factorization and $n-$independence condition is illustrated here on the particular example of the impact factor at twist 3, we expect
that this procedure is more general and that the above method can be applied for other
exclusive processes, for which the key tool is still the collinear Ward identity. This means in particular that each building block (soft and hard part, for each structure which lead to the introduction of a DA) are separately gauge invariant. This fact simplifies dramatically
the use of the $n-$independence principle.

\subsubsection{A minimal set of non-perturbative correlators}
\label{SubSubSec_DiffEq}

We now solve the previous equations, namely
 the two equation of motions
(\ref{em_rho1},\ref{em_rho2})
and the two equations (\ref{ninV},\ref{ninA}) coming from the $n-$independence.
This effectively
reduces the set of 7 DAs to the set of 3 independent  DAs $\varphi_1,$  $B,$  $D.$

To start with we represent the distributions $\varphi_3(y)$, $\varphi_A(y)$,
$\varphi^T_1(y)$ and  $\varphi^T_A(y)$ generically denoted as $\varphi(y)$
as the sums
\begin{equation}
\label{WWgen}
\varphi(y)=\varphi^{WW}(y)+\varphi^{gen}(y)\;, \;\;\;\;\;  \varphi(y)=\;\varphi_3(y),\;
 \varphi_A(y),\; \varphi^T_1(y),\; \varphi^T_A(y)\,,
\end{equation}
where $\varphi^{WW}(y)$ and $\varphi^{gen}(y)$ are contributions
in the so called Wandzura-Wilczek approximation and the genuine twist-3
contributions, respectively.

The Wandzura-Wilczek contributions are solutions of
Eqs.~(\ref{em_rho1}, \ref{em_rho2}, \ref{ninV}, \ref{ninA}) with vanishing 3-parton
distributions
 $B(y_1,y_2)$ and $D(y_1,y_2)$, i.e. which satisfy the equations
\begin{equation}
\label{em_rho1WW}
 \bar{y}_1 \, \varphi^{WW}_3(y_1) +  \bar{y}_1 \, \varphi^{WW}_A(y_1)
+  \varphi_1^{T\;WW}(y_1)  +\varphi_A^{T\;WW}(y_1)
=0 \,
\end{equation}
\begin{equation}
\label{em_rho2WW}
 y_1 \, \varphi^{WW}_3(y_1) -  y_1 \, \varphi^{WW}_A(y_1)
-  \varphi_1^{T\;WW}(y_1)  +\varphi_A^{T\;WW}(y_1)
=0 \,.
\end{equation}
\begin{equation}
\label{ninVWW}
\frac{d}{dy_1}\varphi_1^{T\;WW}(y_1)=-\varphi_1(y_1) + \varphi^{WW}_3(y_1)\,,\;\;\;\;\;\;\;
\frac{d}{dy_1}\varphi_A^{T\;WW}(y_1)=\varphi^{WW}_A(y_1) \,.
\end{equation}
By adding and subtracting Eqs.~(\ref{em_rho1WW},\ref{em_rho2WW}) together
with the use of Eqs.~(\ref{ninVWW}) one obtains equations which involve only
$\varphi^{WW}_3$ and $\varphi^{WW}_A$
\begin{equation}
\label{WW3A}
\frac{d}{dy_1}\varphi^{WW}_3(y_1)=
-(\bar y_1 -y_1)\frac{d}{dy_1}\varphi^{WW}_A(y_1)
\;, \;\;\;\;\;\;
2\varphi_1(y_1)= \frac{d}{dy_1}\varphi^{WW}_A(y_1) +(\bar y_1 -y_1)\frac{d}{dy_1}\varphi^{WW}_3(y_1)
\end{equation}
and which solutions, satisfying
the normalization conditions
\beq
\label{normalizationWW}
\int\limits_0^1dy\,\varphi_3^{WW}(y)=1 \quad {\rm and} \quad
\int\limits_0^1dy\,\varphi_A^{WW}(y)=1\,,
\eq
read
\begin{equation}
\label{WWA3}
\varphi^{WW}_A(y_1)= \frac{1}{2}\left[
\int\limits_0^{y_1}\,\frac{dv}{\bar v}\varphi_1(v) -
\int\limits_{y_1}^1\,\frac{dv}{ v}\varphi_1(v)   \right]\;, \;\;\;\;\;\;\;
\varphi^{WW}_3(y_1)= \frac{1}{2}\left[
\int\limits_0^{y_1}\,\frac{dv}{\bar v}\varphi_1(v) +
\int\limits_{y_1}^1\,\frac{dv}{ v}\varphi_1(v)   \right]\;.
\end{equation}
These expressions and Eqs.~(\ref{em_rho1WW}, \ref{em_rho2WW}) give finally
the remaining solutions $\varphi^{T \;WW}_A$ and $ \varphi^{T \;WW}_1$
\begin{equation}
\label{WWT}
\varphi^{T\;WW}_A(y_1)= \frac{1}{2}\left[-\bar y_1
\int\limits_0^{y_1}\,\frac{dv}{\bar v}\varphi_1(v) -
y_1 \int\limits_{y_1}^1\,\frac{dv}{ v}\varphi_1(v)   \right]\;, \;\;\;\;
\varphi^{T\;WW}_1(y_1)= \frac{1}{2}\left[
-\bar y_1 \int\limits_0^{y_1}\,\frac{dv}{\bar v}\varphi_1(v) +
y_1 \int\limits_{y_1}^1\,\frac{dv}{ v}\varphi_1(v)   \right]\;.
\end{equation}
We note that these two WW results (\ref{WWA3})  were obtained in Ref.~\cite{Ali:1993vd}
when considering the transition form factor $B \to \rho \, \gamma\,.$ 
The  distributions $\varphi^{gen}$ carrying the genuine twist-3 contributions
satisfy the equations

\begin{equation}
\label{em_rho1gen}
 \bar{y}_1 \, \varphi^{gen}_3(y_1) +  \bar{y}_1 \, \varphi^{gen}_A(y_1)
+  \varphi_1^{T\;gen}(y_1)  +\varphi_A^{T\;gen}(y_1)
=-\int\limits_{0}^{1} dy_2 \left[ \zV \, B(y_1,\, y_2)
+\zA \, D(y_1,\, y_2) \right] \,,
\end{equation}
\begin{equation}
\label{em_rho2gen}
  y_1 \, \varphi^{gen}_3(y_1) -  y_1 \, \varphi^{gen}_A(y_1)  -
\varphi_1^{T\;gen}(y_1)  +\varphi_A^{T\;gen}(y_1)
=-\int\limits_{0}^{1} dy_2 \left[ -\zV \, B(y_2,\, y_1)
+\zA\, D(y_2,\, y_1) \right] \,,
\end{equation}
\begin{eqnarray}
\label{ninVgen}
&&\frac{d}{dy_1}\varphi_1^{T\;gen}(y_1)=\varphi^{gen}_3(y_1)
-\zV\int\limits_0^1\,\frac{dy_2}{y_2-y_1}
 \left( B(y_1,y_2)+B(y_2,y_1) \right)\,, \nonumber \\
&&\frac{d}{dy_1}\varphi_A^{T\;gen}(y_1)=\varphi_A^{gen}(y_1)-\zA\int\limits_0^1\,\frac{dy_2}{y_2-y_1}
 \left(D(y_1,y_2)+D(y_2,y_1)
   \right) \,.
\end{eqnarray}

Similarly as in the WW case, one can obtain equations without
distributions $\varphi^{T\;gen}$ by adding and subtracting
Eqs.~(\ref{em_rho1gen}, \ref{em_rho2gen}) together with the use
of (\ref{ninVgen})
\begin{eqnarray}
\label{gen3A}
&&\frac{d}{dy_1}\varphi_3^{gen}(y_1)+(\bar y_1
-y_1)\frac{d}{dy_1}\varphi_A^{gen}(y_1)
\nonumber \\
&&=
4\,\zA \int\limits_0^1\frac{dy_2}{y_2-y_1} D^{(+)}(y_1,y_2)-
2\,\zV\frac{d}{dy_1}\int\limits_0^1dy_2 \,B^{(-)}(y_1,y_2)-
2\,\zA\frac{d}{dy_1}\int\limits_0^1dy_2 \,D^{(+)}(y_1,y_2)\;,
\end{eqnarray}
\begin{eqnarray}
\label{genA3}
&&\frac{d}{dy_1}\varphi_A^{gen}(y_1)
+(\bar y_1 -y_1)\frac{d}{dy_1}\varphi_3^{gen}(y_1)
\nonumber \\
&&=
4\,\zV \int\limits_0^1\frac{dy_2}{y_2-y_1}B^{(+)}(y_1,y_2)-
2\,\zV \frac{d}{dy_1}\int\limits_0^1dy_2 \,B^{(+)}(y_1,y_2)-
2\,\zA\frac{d}{dy_1}\int\limits_0^1dy_2 \,D^{(-)}(y_1,y_2)\;,
\end{eqnarray}
where
$O^{(\pm)}(y_1,y_2)=O(y_1,y_2)\pm O(y_2,y_1)$ for $O=B,\;D$. Let us note that
Eq.~(\ref{genA3}) can be obtained from Eq.~(\ref{gen3A}) and
by the interchange
$\varphi^{gen}_A \;\leftrightarrow \;\varphi^{gen}_3$ and $B \;\leftrightarrow \;D$.

From Eqs.~(\ref{gen3A}, \ref{genA3})
supplemented by the boundary conditions ${\cal B}(y,y)=0={\cal D}(y,y)$ (see later in Section \ref{SubSec_dictionary})
one gets in a straightforward although somehow
tedious way the
equation for $\varphi_3^{gen}$
\begin{eqnarray}
\label{gen3prime}
&&\frac{d}{dy_1} \varphi_3^{gen}(y_1)=
-\frac{1}{2}\left( \frac{1}{y_1} + \frac{1}{\bar y_1}  \right)
\nonumber \\
&& \left\{  \zV \left[
y_1 \int\limits_{y_1}^1\,dy_2\,\frac{d}{dy_1}{ B}(y_1,y_2) -
\bar y_1 \int\limits_0^{y_1}\frac{d}{dy_1}{ B}(y_2,y_1) +
 (\bar y_1 -y_1)\left(
\int\limits_{y_1}^1\,dy_2\,\frac{{ B}(y_1,y_2)}{y_2-y_1}+
\int\limits_0^{y_1}\,dy_2\,\frac{{ B}(y_2,y_1)}{y_2-y_1}\right)
 \right]\right.
\nonumber \\
&& \left.
+
\zA\left[ y_1 \int\limits_{y_1}^1\,dy_2\,\frac{d}{dy_1}{ D}(y_1,y_2) +
\bar y_1 \int\limits_0^{y_1}\frac{d}{dy_1}{ D}(y_2,y_1) -
 \int\limits_{y_1}^1\,dy_2\,\frac{{ D}(y_1,y_2)}{y_2-y_1}-
\int\limits_0^{y_1}\,dy_2\,\frac{{ D}(y_2,y_1)}{y_2-y_1} \right]\right\}\;,
\end{eqnarray}
which properly normalized solution can be written in the form
\begin{equation}
\label{gen3solution}
\varphi_3^{gen}(y)= \frac{1}{2}\left( -\int\limits_y^1\,\frac{dy_1}{y_1}
+ \int\limits_0^1 \,\frac{dy_1}{\bar y_1}    \right) \left\{ ... \right\}\;,
\end{equation}
in which both integrals act on the expression inside $\left\{ ... \right\}$
on the r.h.s of Eq.({\ref{gen3prime}}). The expression (\ref{gen3solution})
can be  simplified after changing the order of the nested integrals.
After performing this task we obtain that
\begin{eqnarray}
\label{Resphi3}
&&\varphi_3^{gen}(y)=
\\
&&-\frac{1}{2}\int\limits^1_{y} \frac{du}{u}
\biggl[
\int\limits^u_0 dy_2 \frac{d}{du} (\zV { B}-\zA { D})(y_2,\, u)
-\int\limits^1_u \frac{dy_2}{y_2-u}  (\zV { B}-\zA { D})(u, \,y_2)
-\int\limits^u_0 \frac{dy_2}{y_2-u}  (\zV { B}-\zA { D})(y_2, \, u)
\biggr]
\nonumber \\
&&-\frac{1}{2}\int\limits^{y_1}_{0}  \frac{du}{\bar{u}}
\biggl[
\int\limits^1_u dy_2 \frac{d}{du} (\zV { B}+\zA { D})(u, \,y_2)
-\int\limits^1_u \frac{dy_2}{y_2-u}  (\zV { B}+\zA { D})(u, \,y_2)
-\int\limits^u_0 \frac{dy_2}{y_2-u}  (\zV { B}+\zA { D})(y_2, \, u)
\biggr]\,. \nonumber
\end{eqnarray}
Finally, the solution for $\varphi_1^{T\;gen}$ is obtained from the first
Eq.~(\ref{ninVgen}) and (\ref{Resphi3})
\begin{equation}
\label{Resphi1T}
\varphi_1^{T \, gen}(y) =\int\limits^y_0 du \, \varphi_3^{gen}(u) - \zV \int\limits^y_0 dy_1 \int\limits^1_y dy_2 \frac{B(y_1,\, y_2)}{y_2-y_1}\,.
\end{equation}
The corresponding expressions for $\varphi^{gen}_A(y)$ and
   $\varphi_A^{T\, gen}(y)$ are obtained from Eq.(\ref{Resphi3}) and (\ref{Resphi1T}) by the substitutions:
\begin{eqnarray}
\label{ResphiA}
\varphi_{3}^{gen}(y) &\stackrel{\zV B \,\leftrightarrow \,\zA D}{\longleftrightarrow}& \varphi_{A}^{gen}(y)\, ,\\
\label{ResphiAT}
\varphi_{1}^{T\, gen}(y) &\stackrel{\zV B \,\leftrightarrow \,\zA D}{\longleftrightarrow}& \varphi_{A}^{T\, gen}(y)\,.
\end{eqnarray}
In conclusion of this section, we explicitly succeeded in representing our results
(\ref{WWA3}),  (\ref{WWT}),  (\ref{Resphi3}),  (\ref{Resphi1T}), (\ref{ResphiA})  (\ref{ResphiAT}) in terms of
 3 independent DAs: the twist 2 DA $\varphi_1$ and the twist 3  DAs $ B\,, D\,.$

\subsection{Dictionary}
\label{SubSec_dictionary}

For comparison of  expressions (\ref{Correlator3BodyV}, \ref{Correlator3BodyA}) with the definitions (\ref{BBVg}, \ref{BBAg})
we perform the change of variables $z\,\to\, z_1$, $tz\,\to\, z_2$, $\alpha_d \,\to\,y_1$ and $\alpha_u=1-y_2$, i.e. $\alpha_g=y_2-y_1$. It results
in the following
identification of the 3-parton DAs in LCCF and CCF approaches
\begin{eqnarray}
\label{DictB}
 B(y_1,\,y_2)&=&-\frac{V(y_1, \, 1-y_2)}{y_2-y_1} \\
\label{DictD}
D(y_1,\,y_2)&=&-\frac{A(y_1, \, 1-y_2)}{y_2-y_1}\,.
\end{eqnarray}
From (\ref{DictB}, \ref{DictD}) and Ref.\cite{BB} follows the boundary conditions
$B(y, y)=0=D(y, y)\,.$

Taking in Eqs.(\ref{BBV1}, \ref{BBV2}) the coordinate $z$ along the light-cone vector $n$, $z=\lambda \,n$,
permits the identification of the vector DAs in Eq.(\ref{par1v}):
\begin{eqnarray}
\label{relBBvector}
&&\varphi_1(y)=
\phi_{\parallel}(y) ,
\quad
\varphi_3(y)=
 g_\perp^{(v)}(y) \,,
\end{eqnarray}
and of the axial DA in Eq.~(\ref{par1a})
\begin{eqnarray}
\label{relBBaxial}
&&\varphi_A(y) =
-\frac{1}{4} \, \frac{\partial g_\perp^{(a)}(y)}{\partial y}\,.
\end{eqnarray}
We checked also the validity of Eqs.(\ref{relBBvector}, \ref{relBBaxial}) directly by the use of our explicit solutions (\ref{WWA3}, \ref{Resphi3}, \ref{ResphiA}) and expressions for $g_\perp^{(v)}$ and $g_\perp^{(a)}$ given Ref.~\cite{BB} in terms of $\phi_\parallel$, $V$ and $A$ DAs. This non trivial
check can be done with the help of methods similar to those used in Appendix \ref{Ap:Comparison} when comparing the results of calculation of the $\gamma^* \to \rho_T$ impact factor in LCCF and CCF approaches.

%%%%%%%%%%%%%%%%%

\section{$\gamma^* \to \rho_T$ Impact factor up to  twist three accuracy}
\label{Sec_Impact}
\subsection{General recall on impact factor representation and kinematics}
\label{SubSec_ImpactGeneral}

The $\gamma^* \to \rho $ impact factor enters the description of high energy reactions  in the $k_{T}$ factorization approach. As an example, one may study the
reactions
\begin{eqnarray}
\label{prgg}
\gamma^*(q)+\gamma^*(q^\prime)\to \rho_T(p_1)+\rho(p_2)
\end{eqnarray}
or
\begin{eqnarray}
\label{prgP}
\gamma^*(q)+N\to \rho_T(p_1)+N
\end{eqnarray}
where the virtual photons carry large squared  momenta $q^2=-Q^2$
($q'^2=-Q'^2$) $\gg \Lambda^2_{QCD}$\,, and
the Mandelstam variable $s$ obeys the condition
$s\gg Q^2,\,Q^{\prime\, 2}, -t \simeq \rb^2$. The hard scale which justifies the applicability of
perturbative QCD is set by $Q^2$ and $Q'^2$ and/or by $t.$
Neglecting meson masses, one considers for reaction (\ref{prgg}) the light-cone vectors  $p_1$ and $p_2$
as the vector meson momenta  ($2\,p_1\cdot p_2=s$).

\begin{figure}
\psfrag{A}[cc][cc]{\raisebox{-.7cm}{$\kappa$}\rotatebox{-70}{$\underbrace{\rule{1.1cm}{0pt}}$}}
\begin{tabular}{cc} \epsfig{file=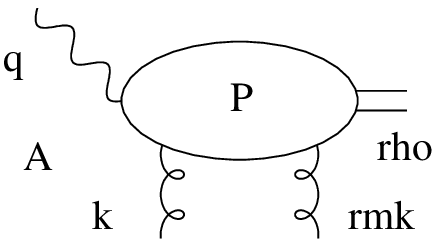,width=\widss} &\qquad \epsfig{file=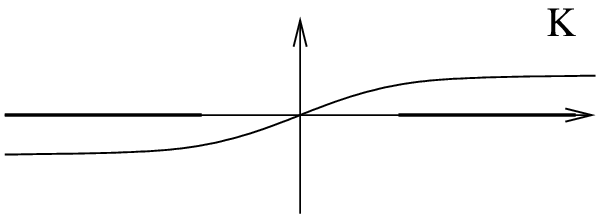,width=\widss}
\end{tabular}
\caption{a: $\gamma^* \to \rho$ impact factor. b: $\kappa$ integration contour entering the definition of the $\gamma^* \to \rho$ impact factor.}
\label{Fig:impactRhoANDkappaint}
\end{figure}

 In this Sudakov light-cone basis, transverse Euclidian momenta
are denoted with underlined letters. The virtual photon momentum $q$  reads
\beq
\label{momq}
q=p_1-\frac{Q^2}{s}\, p_2\,.
\eq
The impact representation of the scattering amplitude for the reaction (\ref{prgg})  is \cite{kT, Collins:1991ty, Levin:1991ry}
\begin{eqnarray}
\label{BFKLamforward}
%\text
%Im\,{\cal A}
{\cal M}=\frac{i s}{(2\pi)^2}\!\!
\int\frac{d^2\kb}{\kb^2} \Phi^{ab}_1(\kb,\,\rb-\kb) \!\!
\int\frac{d^2\kb'}{\kb'^2} \Phi^{ab}_2(-\kb',\,-\rb+\kb') \!\!\!\!\!
\int\limits_{\delta-i\infty}^{\delta+i\infty} \frac{d\omega}{2\pi i}
\biggl(\frac{s}{s_0}\biggr)^\omega G_\omega (\kb,\kb',\rb)
\end{eqnarray}
where  $G_\omega$ is the 4-gluons Green function which obeys the BFKL equation \cite{bfkl}. $G_\omega$ reduces to
\beq
G_\omega^{Born}=
\frac{1}{\omega} \,\delta^2(\kb -\kb') \frac{\kb^2}{(\rb-\kb)^2}
\eq
 within the
Born approximation.
We focus here on the $\gamma^* \to \rho $ impact factor $\Phi$  of the  subprocess
\begin{eqnarray}
\label{ggsubpr}
   g(k_1,\varepsilon_{1})+\gamma^*(q)\to g(k_2, \varepsilon_{2})+\rho_T(p_1) \,,
\end{eqnarray}
illustrated in Fig.\ref{Fig:impactRhoANDkappaint}a,
in the  kinematical
region where virtualities of the photon, $Q^2$, and $t-$channel gluons
$k_\perp^2$, are of the same order, $Q^2\sim k_\perp^2$, and much larger than
$\Lambda_{QCD}$.
\psfrag{P}[cc][cc]{$\Large\Phi$}
\psfrag{q}[cc][cc]{$q$}
\psfrag{k}[cc][cc]{$\, \, k_1$}
\psfrag{rmk}[cc][cc]{$k_2$}
\psfrag{rho}[cc][cc]{$\rho$}
\psfrag{K}[cc][cc]{$\kappa$}
It is
 the integral of the  S-matrix element
 ${\cal S}^{\gamma^*_T\, g\to\rho_T\, g}_\mu$ with respect to the Sudakov component of the t-channel $k$
momentum along $p_2\,.$ Closing the contour of integration below shows that the impact factor can be equivalently be written (see Fig.\ref{Fig:impactRhoANDkappaint}b)
% \begin{figure}
% \psfrag{A}[cc][cc]{\raisebox{-.7cm}{$\kappa$}\rotatebox{-70}{$\underbrace{\rule{1.1cm}{0pt}}$}}
% \epsfig{file=kappaint.eps,width=\widss}
% \caption{$\kappa$ integration contour entering the definition of the $\gamma^* \to \rho$ impact factor.}
% \label{Fig:kappaint}
% \end{figure}
as the integral of the $\kappa$-channel discontinuity of the  S-matrix element
 ${\cal S}^{\gamma^*_T\, g\to\rho_T\, g}_\mu\,:$
% of this subprocess (\ref{ggsubpr}):
\begin{eqnarray}
\label{imfac}
\Phi^{\gamma^*\to\rho}(\kb,\,\rb-\kb)=e^{\gamma^*\mu}\, \frac{1}{2s}\int\limits^{+\infty}_{-\infty}\frac{d\kappa}{2\pi}
\,  {\cal S}^{\gamma^*\, g\to\rho\, g}_\mu(\kb,\,\rb-\kb)
=
 e^{\gamma^*\mu}\, \frac{1}{2s}\int\limits^{+\infty}_0\frac{d\kappa}{2\pi}
\, \hbox{Disc}_\kappa \,  {\cal S}^{\gamma^*\, g\to\rho\, g}_\mu(\kb,\,\rb-\kb)
\,,
\end{eqnarray}
where $\kappa=(q+k_1)^2$ denotes the Mandelstam variable $s$ for the subprocess (\ref{ggsubpr}), as illustrated in Fig.\ref{Fig:impactRhoANDkappaint}a.

Note that the two reggeized
gluons have so-called non-sense polarizations $\varepsilon_1=\varepsilon_2^*=p_2\sqrt{2/s}\,.$

Considering the forward limit for simplicity,
the gluon momenta reduce to
\begin{eqnarray}
\label{gmom}
k_1=\frac{\kappa+Q^2+\kb^2}{s} p_2 + k_\perp, \quad
k_2=\frac{\kappa+\kb^2}{s} p_2 + k_\perp, \quad
k_1^2=k_2^2=k^2_\perp=-\kb^2\,.
\end{eqnarray}
Finally, let us note that  when writing (\ref{gmom}) we took an exact kinematics for the fraction of momentum along $p_2.$ This kinematics naturally extends the usual Regge kinematics in the case where
$t$-channel momentum transfer along $p_2$ is allowed, which corresponds to the skewed
kinematics which is typical of GPD studies. In usual computation
of impact factors used in $k_T$-factorization, one usually makes the approximation that these two fractions are exactly opposite. Here we make such a choice in order to introduce skewness effects in a correct manner since these terms will contribute at the twist 3 order we are interested in. Note that within $k_T$-factorization, the description of impact factor for produced hadron described within QCD collinear approach requires a modification of
twist counting due to the off-shellness of the $t-$channel partons. Therefore, when here we say "up to twist 3" we only mean  twist counting from the point of view of the collinear factorization of the produced $\rho-$meson, and not of the whole amplitude, e.g. $\gamma^* \, p \to \rho \, p$ or  $\gamma^* \, \gamma^* \to \rho \, \rho\,.$

In order to describe the collinear factorization of $\rho$-production inside the impact factor (\ref{imfac}),
we note that the kinematics of the general approach discussed in section \ref{Sec_LCCF} is related to our present kinematics
for the impact factor (\ref{imfac})
by setting $p=p_1$, while a natural choice for $n$ is obtained by setting
$n=n_0=p_2/(p_1 \cdot p_2)$ (this latter choice for $n$, though natural, is somehow arbitrary as we discussed above in section \ref{SubSec_AddSet}).

We will now distinguish and make a comparative analysis of
two different approaches:
LCCF and CCF.
We will show that these two results are actually fully equivalent to each other, when using the dictionary \ref{SubSec_dictionary}.

%%%%%%%%%%%%%%%%%%%%

\subsection{Calculation based on the Light-Cone Collinear Factorization approach}
\label{SubSec_ImpactLCCF}

\subsubsection{$\gamma^*_L \to \rho_L$ transition}
\label{SubSec_LL}

  \begin{figure}
\psfrag{u}{\footnotesize$\hspace{-.3cm}y_1 \, p_1$}
\psfrag{d}{\footnotesize$\hspace{-.3cm}-\bar{y}_1 \, p_1$}
\psfrag{a}{$k_1$}
\psfrag{c}{$\hspace{-1cm} y_1  \,p_1-q$}
\psfrag{e}{$\hspace{-.7cm}-k_2 -\bar{y}_1  \,p_1$}
\psfrag{f}{$\hspace{-.3cm}k_2 $}
\psfrag{q}{$q$}
 \raisebox{0cm}{\epsfig{file=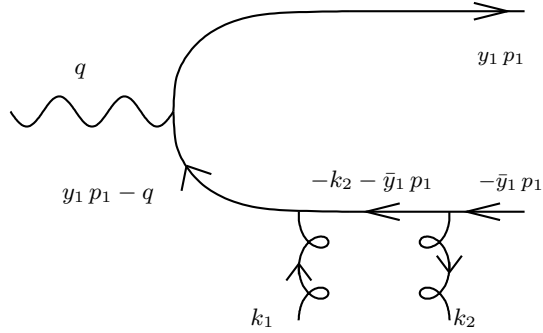,width=7cm}}
\caption{The detailed structure of the diagram (a).}
\label{Fig:a}
\end{figure}
Working within the LCCF, we first recall the calculation of  the $\gamma^*_L \to \rho_L$ transition which receives contribution only from the diagrams with  quark--antiquark
correlators\footnote{Hereafter, except for final results, we perform the computation for a meson which would be a one flavour quark--antiquark state.
Its wave function is then restored at the very end.}.
It is given by contributions from the $p_\mu$ term  of the correlators (\ref{par1v}) of the twist 2. Higher order corrections would start at twist 4,
which is below our accuracy. The computation of the corresponding impact factor is standard \cite{ginzburg}. It involves the computation of the 6 diagrams of Fig.\ref{Fig:NoDer2}.
The longitudinal polarization of the virtual photon reads, in the Sudakov basis,
\beq
\label{eGL}
e_{\gamma L}^\mu= \frac{1}Q \left(p_1^\mu + \frac{Q^2}{s}\, p_2^\mu \right)\,,
\eq
while the momentum of the $\rho$ reads
\beq
\label{p_rho}
p_{\rho}^\mu=  p_1^\mu + \frac{m_\rho^2}{s}\, p_2^\mu \,,
\eq
and the longitudinal polarization of the $\rho$ is
\beq
\label{e_rhoL}
e_L^\mu \equiv e_{\rho L}^\mu= \frac{1}{m_\rho} \left(p_1^\mu - \frac{m_\rho^2}{s}\, p_2^\mu \right)\,.
\eq
Consider for example the diagram (a) of Fig.\ref{Fig:NoDer2},
 as illustrated in Fig.\ref{Fig:a}.
Computing the corresponding $S$-matrix element, the corresponding contribution to the impact factor reads
\beq
\label{Sa}
\Phi_a=-e_q \, \frac{1}4 \frac{2}s (-i) \,  f_\rho \, m_\rho \, g^2 \, \frac{\delta^{ab}}{2 \, N_c}\frac{1}{2s}\int\limits^1_0 dy\int \frac{d \kappa}{2 \pi}\frac{Tr [\slashchar{e}_{\gamma L} \, (y\, \slashchar{p}_1-\slashchar{q}) \, \slashchar{p}_2 \, (\slashchar{k}_2+ \bar{y} \slashchar{p}_1) \, \slashchar{p}_2 \, \slashchar{p}_1]}{[(y \, p_1-q)^2+i \eta] [ (k_2+\bar{y} p_1)^2+i \eta]} \varphi_1(y)\,,
\eq
 where the factor $1/4$ is reminiscent from the Fierz identity, the factor $2/s$ comes from the normalization of the non-sense polarizations. Finally, the color factor $\frac{\delta^{ab}}{2 \, N_c}$ is due to the fact that when summing over
the color of the $t-$channel gluons, the net color coefficient for $\gamma^* \gamma^* \to \rho \rho$ should be $(N_c^2-1)/(4 N_c^2)$ (due a Fierz factor $1/N_c$
when factorizing each of the two $\rho$ DAs). Among the two propagators, only the second one, involving $(k_2+\bar{y} p_1)^2+i\eta = \kappa \, \bar{x} - \kb^2 \, x + i \eta$ has a pole in $\kappa$, contributing when closing the contour below
(therefore contributing to the discontinuity).  The result is then easily obtained after extracting the corresponding residue.
Diagram (c) provide the same contribution, since it can be obtained from (a) by the replacement $x \leftrightarrow \bar{x}$.
Diagrams (b) and (d) vanishes for this twist 2 transition. Diagrams contributes only when closing the $\kappa$ contour above.
 Finally, the net result for the  $\gamma_L^* \to \rho_L$ impact factor is, after taking into account the $\rho^0$ wave function
\beq
\label{gammaLrhoL}
 \Phi^{\gamma^*_L\to\rho_L}(\kb^2)=\frac{2 \, e \, g^2 \, f_\rho}{\sqrt{2} \, Q} \frac{\delta^{ab}}{2 \, N_c}\int\limits^1_0 dy \, \varphi_1(y) \frac{\kb^2}{y \, \bar{y} \, Q^2+\kb^2} \,.
\eq
Note that diagrams (a) and (c) of Fig.\ref{Fig:NoDer2} are the only diagrams which contribute when computing the hard part by closing the $\kappa$ contour below.

\subsubsection{$\gamma^*_T \to\rho_T$ transition}
\label{SubSec_TT}

We now concentrate on the $\gamma^*_T \to\rho_T$ transition, which impact factor will be one of the main results of this paper.
The 2-parton
 contribution contains the terms arising from the diagram Fig.\ref{Fig:NoDer2}, where
the quark--antiquark correlators have no transverse derivative,
and from the diagrams Fig.\ref{Fig:Der2}, where the quark--antiquark correlators stand with a
transverse derivatives. The computation of the diagrams of Fig.\ref{Fig:NoDer2} for the $\gamma^*_T \to\rho_T$ transition goes along the same line as for the twist 2 $\gamma^*_L \to\rho_L$
transition discussed above. The practical trick used for computing the contributions of Fig.\ref{Fig:Der2} is
the Ward identity
\beq
\label{WardDer}
\psfrag{d}[cc][cc]{\raisebox{-.63cm}{$\displaystyle \frac{\partial}{\partial p_\mu}$}}
\psfrag{e}[cc][cc]{$=$}
\psfrag{p}[cc][cc]{$p$}
\psfrag{g}[cc][cc]{$ \gamma^\mu$}
 \begin{tabular}{cc}\epsfig{file=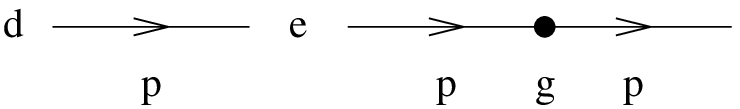,width=\widsi} &  \raisebox{.8cm}{\quad where \ \raisebox{-.4cm}{\epsfig{file=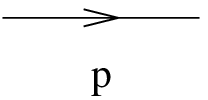,width=1cm}}\raisebox{-.1cm}{$\, =\,\displaystyle \frac{1}{m-\slash \!\!\! p-i \epsilon}\,,$}}
\end{tabular}
\eq
where lines denotes fermionic propagators. This
 leads to an additional Feynman rule when inserting a derivative. The corresponding insertions are denoted with dashed lines in Fig.\ref{Fig:Der2}.

%%%%%%%%%%%%b1%%%%%%%%%%%%%
\begin{figure}
\psfrag{u}{\footnotesize$\hspace{-.3cm}y_1 \, p_1$}
\psfrag{d}{\footnotesize$\hspace{-.3cm}-\bar{y}_1 \, p_1$}
\psfrag{a}{$k_1$}
\psfrag{b}{$\hspace{-1cm}k_1-\bar{y}_1  \,p_1$}
\psfrag{c}{$\hspace{-1cm}k_1-\bar{y}_1  \,p_1$}
\psfrag{e}{$\hspace{-.5cm}k_2 +y_1  \,p_1$}
\psfrag{f}{$\hspace{-.3cm}k_2 $}
\psfrag{q}{$q$}
 \raisebox{0cm}{\epsfig{file=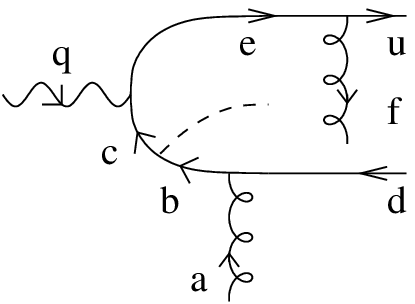,width=7cm}}
\caption{The detailed structure of the diagram (b1).}
\label{Fig:b1}
\end{figure}
Consider for example the diagram (b1) of Fig.\ref{Fig:NoDer2}, illustrated in Fig.\ref{Fig:b1}.
Computing the corresponding $S$-matrix element for the vector part, the corresponding contribution to the impact factor reads
\beq
\label{Sb1V}
\Phi^V_{b1}\!=\!-e_q \, \frac{1}4 \frac{2}s (-i) \,  g^2 \,  f_\rho \, m_\rho \, \frac{\delta^{ab}}{2 \, N_c}\frac{1}{2s}\!\int\limits^1_0 \!dy\!\int \frac{d \kappa}{2 \pi}\frac{Tr [\slashchar{e}_\gamma \, (\slashchar{k}_1-\bar{y} \, \slashchar{p}_1) \, \slashchar{e}^*_T \, (\slashchar{k}_1- \bar{y} \slashchar{p}_1) \, \slashchar{p}_2 \, \slashchar{p}_1 \, \slashchar{p}_2  (\slashchar{k}_2+y \, \slashchar{p}_1)]}{[(k_1-\bar{y} p_1)^2+i \eta]^2 [ (k_2+\bar{y} p_1)^2+i \eta]}\, \varphi_3^T(y)\,,
\eq
This part of the impact factor receives (identical) contributions when closing the $\kappa$ integration contour either from above or from below, which reads
\beq
\label{Sb1VRes}
\Phi^V_{b1}=-\frac{e_q\, g^2}2\,  f_\rho \, m_\rho\, \frac{\delta^{ab}}{2 \, N_c} \int\limits^1_0 dy\, y \frac{-e^*_T \cdot e_\gamma (y \, \bar{y} \, Q^2 + \kb^2)+2 \, e^*_T \cdot k \, e^*_T \cdot k  (1-2 \, y)}{(Q^2 \, y \, \bar{y}+ \kb^2)^2} \varphi_3^T(y)\,.
 \eq
The computation of the axial part is similar (note the $i/4$ factor from Fierz)
\beq
\label{Sb1A}
\Phi^A_{b1}\!=\!-e_q \, \frac{i}4 \frac{2}s (-i)  g^2   f_\rho \, m_\rho \frac{\delta^{ab}}{2 \, N_c}\frac{1}{2s}\!\!\int\limits^1_0 \!\!dy\!\!\int \!\!\frac{d \kappa}{2 \pi}\frac{Tr [\slashchar{e}_\gamma \, (\slashchar{k}_1-\bar{y} \, \slashchar{p}_1) \, \gamma_\alpha \, (\slashchar{k}_1- \bar{y} \slashchar{p}_1) \, \slashchar{p}_2 \, \slashchar{p}_1 \,\gamma_5 \, \slashchar{p}_2  (\slashchar{k}_2+y \, \slashchar{p}_1)]}{[(k_1-\bar{y} p_1)^2+i \eta]^2 [ (k_2+\bar{y} p_1)^2+i \eta]} \epsilon^\alpha_{\,\, e^*_T p n}  \, \varphi_3^T(y)\,,
\eq
and leads to
\beq
\label{Sb1ARes}
\Phi^A_{b1}=-\frac{e_q\, g^2}2 \,  f_\rho \, m_\rho \, \frac{\delta^{ab}}{2 \, N_c} \int\limits^1_0 dy \, y \frac{-e^*_T \cdot e_\gamma (y \, \bar{y} \, Q^2 - \kb^2)+2 \, e^*_T \cdot k \, e^*_T \cdot k }{(Q^2 \, y \, \bar{y}+ \kb^2)^2} \varphi_A^T(y)\,.
 \eq
The contributions of 3-parton correlators are of two types, the first one being of "Abelian" type (without triple gluon vertex, see Fig.\ref{Fig:3Abelian}) and the second involving non-Abelian coupling with one triple gluon vertex (see Fig.\ref{Fig:3NonAbelian}) or two (see Fig.\ref{Fig:3NonAbelianTwo}).
Let us first consider the "Abelian" class. They involve two kind of Casimir invariants:
\beqa
\label{colorAbelian}
&&\hspace{-.4cm}\frac{1}{N_c}\,Tr(t^c \, t^a \, t^b \, t^c)\!\!=\!\!C_F \, \frac{\delta^{ab}}{2 \, N_c}\equiv C_a \, \frac{\delta^{ab}}{2\, N_c}\!:  \mbox{(aG1), (cG1), (eG1), (fG1)} \\
&&\hspace{-.4cm}\frac{1}{N_c} \,Tr(t^c \, t^a \, t^c \, t^b)\!\!=\!\!\left(\!\!C_F-\frac{N_c}2\!\!\right) \! \frac{\delta^{ab}}{2\, N_c}\! \equiv C_b \, \frac{\delta^{ab}}{2\, N_c}\!:  \mbox{(bG1), (dG1), (aG2), (cG2), (bG2), (dG2), (eG2), (fG2)},
\nonumber
\eqa
where the $1/N_c$ comes from the Fierz coefficient when factorizing the quark--antiquark state in color space.
%%%%%%%%%%%%%%%%%%%%%%atG1%%%%%%%%%%%%%
\begin{figure}
\psfrag{u}{\footnotesize$\hspace{-.3cm}y_1 \, p_1$}
\psfrag{d}{\footnotesize$\hspace{-.3cm}-\bar{y}_2 \, p_1$}
\psfrag{m}{$\hspace{-.4cm}(y_2-y_1) \, p_1$}
\psfrag{a}{$k_1$}
\psfrag{b}{$\hspace{-.8cm}y_2  \,p_1-q$}
\psfrag{c}{$\hspace{-.8cm}y_1  \,p_1-q$}
\psfrag{e}{$\hspace{-.5cm}-k_2 -\bar{y}_2  \,p_1$}
\psfrag{f}{$k_2 $}
\psfrag{q}{$q$}
\psfrag{i}{}
 \raisebox{0cm}{\epsfig{file=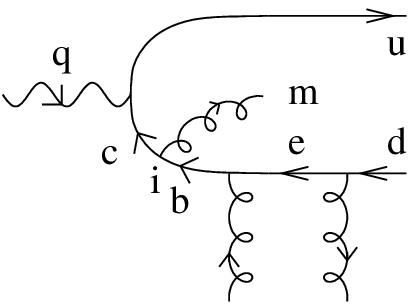,width=8cm}}
\caption{The detailed structure of the diagram (aG1).}
\label{Fig:aG1}
\end{figure}
Again, to illustrate the method, we consider the peculiar diagram (aG1) of Fig.\ref{Fig:3Abelian}, illustrated in Fig.\ref{Fig:aG1}. The vector contribution reads
\beqa
\label{SaG1V}
\Phi^V_{aG1}&=&-e_q \, \frac{1}4 \frac{2}s (i) \,  g^2  \,  f_\rho \, m_\rho \, \frac{\delta^{ab}}{2 \, N_c}\frac{1}{2s}\int\limits^1_0 \! dy_1 \, dy_2\int \frac{d \kappa}{2 \pi}\frac{Tr [\slashchar{e}_\gamma \, (y_1 \, \slashchar{p}_1-\slashchar{q}) \, \slashchar{e}^*_T \, (y_2 \, \slashchar{p}_1-\slashchar{q}) \,\slashchar{p}_2 \, (\slashchar{k}_2+ \bar{y}_2 \, \slashchar{p}_1) \, \slashchar{p}_2 \, \slashchar{p}_1]}{[(y_1 \, p_1-q)^2+i \eta][(y_2 \, p_1-q)^2+i \eta] [ (k_2+\bar{y}_2 \, p_1)^2+i \eta]}
\nonumber \\
&\times&  \, B(y_1, \, y_2)\,,
\eqa
and equals
\beq
\label{SaG1VRes}
\Phi^V_{aG1}=-\frac{e_q\, g^2}2  \,  f_\rho \, m_\rho \, \frac{\delta^{ab}}{2 \, N_c} \int\limits^1_0 dy_1 \, dy_2\, \frac{e^*_T \cdot e_\gamma}{\bar{y}_1 \, Q^2} B(y_1,\,y_2)\,.
 \eq
The corresponding axial contribution reads
\beqa
\label{SaG1A}
\Phi^A_{aG1}&=&-e_q \, \frac{i}4
\frac{2}s (i) \,  g^2 \,  f_\rho \, m_\rho \, \frac{\delta^{ab}}{2 \, N_c}\frac{1}{2s}\int\limits^1_0 \! dy_1 \, dy_2 \int \frac{d \kappa}{2 \pi}\frac{Tr [\slashchar{e}_\gamma \, (y_1 \, \slashchar{p}_1-\slashchar{q}) \,  \gamma_\alpha \, (y_2 \, \slashchar{p}_1-\slashchar{q}) \,\slashchar{p}_2 \, (\slashchar{k}_2+ \bar{y}_2 \, \slashchar{p}_1) \, \slashchar{p}_2 \, \slashchar{p}_1]}
{[(y_1 \, p_1-q)^2+i \eta][(y_2 \, p_1-q)^2+i \eta] [ (k_2+\bar{y}_2 \, p_1)^2+i \eta]} \nonumber \\
&\times&
\epsilon^\alpha_{\,\, e^*_T p n} \, D(y_1, \, y_2)\,,
\eqa
and equals
\beq
\label{SaG1ARes}
\Phi^A_{aG1}=-\frac{e_q\, g^2}2  \,  f_\rho \, m_\rho \, \frac{\delta^{ab}}{2 \, N_c} \int\limits^1_0 dy_1 \,dy_2\,  \frac{e^*_T \cdot e_\gamma}{\bar{y}_1 \, Q^2} D(y_1,\,y_2)\,.
 \eq
Consider now the
 ''non-Abelian`` diagrams of Fig.\ref{Fig:3NonAbelian}, involving a single triple gluon vertex.
They involve two kind of color structure:
\beqa
\label{color3NonAbelian}
&&\hspace{-.4cm} \frac{2}{N_c^2-1} (-i)\,Tr(t^c \, t^b \, t^d) \, f^{cad}= \frac{N_c}2 \,\frac{1}{C_F} \frac{\delta^{ab}}{2\,N_c}\,:  \mbox{ (atG1), (dtG1), (etG1), (btG2), (ctG2), (ftG2)} \\
&&\hspace{-.4cm}
\frac{2}{N_c^2-1} (-i)\,Tr(t^c \, t^d \, t^b) \, f^{cad}= -\frac{N_c}2 \,\frac{1}{C_F} \frac{\delta^{ab}}{2\,N_c}\,:  \mbox{ (ctG1), (btG1), (ftG1), (atG2), (dtG2), (etG2)} \,,
\nonumber
\eqa
where the $2/(N_c^2-1)$ comes from the Fierz coefficient when factorizing the quark--antiquark gluon state in color space.
%%%%%%%%%%%%%%%%%%%%%%atG1%%%%%%%%%%%%%
\begin{figure}
\psfrag{u}{\footnotesize$\hspace{-.3cm}y_1 \, p_1$}
\psfrag{d}{\footnotesize$\hspace{-.3cm}-\bar{y}_2 \, p_1$}
\psfrag{m}{$\hspace{-.4cm}(y_2-y_1) \, p_1$}
%\psfrag{i}{\raisebox{.2cm}{$\mu$}}
\psfrag{i}{}
\psfrag{a}{$k_1$}
\psfrag{b}{$\hspace{-1.7cm}k_1+(y_1-y_2)  \,p_1$}
\psfrag{c}{$\hspace{-.8cm}y_1  \,p_1-q$}
\psfrag{e}{$\hspace{-.5cm}-k_2 -\bar{y}_2  \,p_1$}
\psfrag{f}{$k_2 $}
\psfrag{q}{$q$}
 \raisebox{0cm}{\epsfig{file=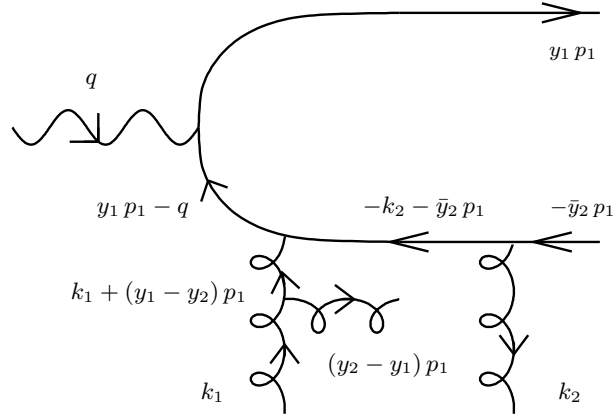,width=8cm}}
\caption{The detailed structure of the ''non-Abelian`` (with one triple gluon vertex) diagram (atG1).}
\label{Fig:atG1}
\end{figure}
 Let us consider the diagram (atG1) of Fig.\ref{Fig:3NonAbelian}, illustrated in Fig.\ref{Fig:atG1}. We denote as
\beq
\label{defPropAxial}
d^{\nu \rho}(k)=g^{\nu \rho}-\frac{k^\nu n^\rho+k^\rho n^\nu}{k \cdot n}
\eq
the numerator of the gluon propagator in axial gauge, and
\beq
\label{defV}
V_{\mu_1 \, \mu_2\, \mu_3}(k_1,\, k_2,\, k_3)=(k_1-k_2)_{\mu_1} \, g_{\mu_1 \mu_2} + \cdots
\eq
the momentum part of the 3-gluon vertex, where $k_i$ are incoming, labeled in the counter-clockwise direction.
The contribution of the diagram (atG1) then reads, for the vector DA,
\beqa
\label{SatG1V}
\Phi^V_{atG1}&\!\!\!=&\!\!-e_q \, \frac{1}4 \frac{2}s \frac{(-i) N_c}{2 \, C_F} \,  g^2 \, m_\rho \, f_\rho \, \frac{\delta^{ab}}{2 \, N_c}\frac{1}{2s}\!\int\limits^1_0 \! dy_1 \, dy_2 \!\!\int \!\frac{d \kappa}{2 \pi}Tr [\slashchar{e}_\gamma \, (y_1 \slashchar{p}_1-\slashchar{q}) \, \gamma_\nu \, (\slashchar{k}_2+\bar{y}_2 \, \slashchar{p}_1) \,\slashchar{p}_2 \, \slashchar{p}_1] \nonumber \\
&\times& \frac{d^{\nu \rho}(k_1+(y_1-y_2)p_1) \, V_{\rho \lambda \alpha}(-k_1-(y_1-y_2)p_1,\,k_1,\,(y_1-y_2)p_1)}{[(y_1 \, p_1-q)^2+i \eta][(k_1+(y_1-y_2) \, p_1)^2+i \eta] [ (k_2+\bar{y}_2 \, p_1)^2+i \eta]}\,  p_2^\lambda \,e^{*\alpha}_T \, e_{\gamma}^\mu \, B(y_1, \, y_2)\,.
\eqa
Note that for this diagram, as well as for all ''non-Abelian`` diagrams, one can easily check that only the $g^{\nu \rho}$ part
 of  (\ref{defPropAxial}) contributes.

Closing the $\kappa$ contour above or below gives for the vector DA part of the diagram (atG1) the result
\beq
\label{SatG1VRes}
\Phi^V_{atG1}=-\frac{e_q\, g^2}2\, m_\rho \, f_\rho \, \frac{\delta^{ab}}{2 \, N_c} \frac{N_c}{C_F} \int\limits^1_0 \! dy_1 \, dy_2 \, \frac{(y_1 - y_2) \, \bar{y}_2 }
{\bar{y}_1 \, (\bar{y}_1 \, \kb^2 + \bar{y}_2 \, (y_2-y_1) \, Q^2)}
e^*_T \cdot e_\gamma \, B(y_1,\,y_2)\,.
 \eq
Similarly, the contribution of the diagram (atG1) reads, for the axial DA,
\beqa
\label{SatG1A}
\Phi^A_{atG1}&\!\!\!=&\!\!-e_q \, \frac{i}4 \frac{2}s \frac{(-i) N_c}{2 \, C_F} \,  g^2 \,  m_\rho \, f_\rho \, \frac{\delta^{ab}}{2 \, N_c}\frac{1}{2s}\!\int\limits^1_0 \! dy_1 \, dy_2 \!\!\int \!\frac{d \kappa}{2 \pi}Tr [\slashchar{e}_\gamma \, (y_1 \slashchar{p}_1-\slashchar{q}) \, \gamma_\nu \, (\slashchar{k}_2+\bar{y}_2 \, \slashchar{p}_1) \,\slashchar{p}_2 \, \slashchar{p}_1\, \gamma_5] \nonumber \\
&\times& \frac{d^{\nu \rho}(k_1+(y_1-y_2)p_1) \, V_{\rho \lambda \alpha}(-k_1-(y_1-y_2)p_1,\,k_1,\,(y_1-y_2)p_1)}{[(y_1 \, p_1-q)^2+i \eta][(k_1+(y_1-y_2) \, p_1)^2+i \eta] [ (k_2+\bar{y}_2 \, p_1)^2+i \eta]} \,  p_2^\lambda \,\epsilon^\alpha_{\,\, e^*_T p n}  \, D(y_1, \, y_2)\,,
\eqa
and closing the $\kappa$ contour above or below gives
\beq
\label{SatG1ARes}
\Phi^A_{atG1}=-\frac{e_q\, g^2}2 \, m_\rho \, f_\rho \, \frac{\delta^{ab}}{2 \, N_c} \frac{N_c}{C_F} \int\limits^1_0 \! dy_1 \, dy_2 \, \frac{(y_1 - y_2) \, \bar{y}_2 }
{\bar{y}_1 \, (\bar{y}_1 \, \kb^2 + \bar{y}_2 \, (y_2-y_1) \, Q^2)}
e^*_T \cdot e_\gamma \, D(y_1,\,y_2)\,.
 \eq
%%%%%%%%%%%%%%%%%%%%%gttG1%%%%%%%%%%%%%%%%%%
\begin{figure}
\psfrag{u}{\footnotesize$\hspace{-.3cm}y_1 \, p_1$}
\psfrag{d}{\footnotesize$\hspace{-.3cm}-\bar{y}_2 \, p_1$}
\psfrag{m}{$\hspace{.2cm}(y_2-y_1) \, p_1$}
\psfrag{a}{$k_1$}
\psfrag{b}{$\hspace{-2.3cm}(1+y_1-y_2)  \,p_1-q$}
\psfrag{c}{$\hspace{-.8cm}y_1  \,p_1-q$}
\psfrag{e}{$\hspace{-.5cm}k_2 +(y_2-y_1)  \,p_1$}
\psfrag{f}{$k_2 $}
\psfrag{q}{$q$}
\psfrag{i}{}
 \raisebox{0cm}{\epsfig{file=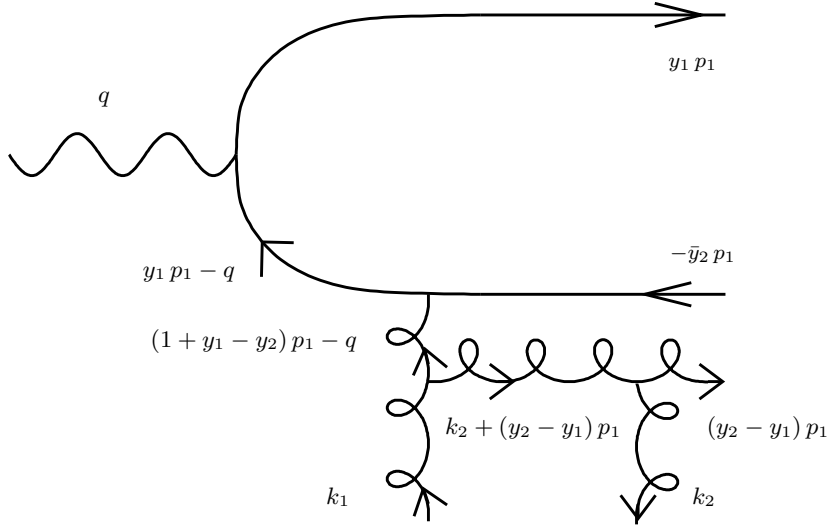,width=10cm}}
\caption{The detailed structure of the diagram (gttG1).}
\label{Fig:gttG1}
\end{figure}
We
consider now the
 ''non-Abelian`` diagrams of Fig.\ref{Fig:3NonAbelianTwo}, involving two triple gluon vertices.
They all involve the color structure
\beq
\label{color3NonAbelianTwo}
-\frac{2}{N_c^2-1}Tr[t^c \, t^d] f^{cea} \, f^{edb}= \frac{N_c}{C_F} \, \frac{\delta^{ab}}{2 \, N_c}\,.
\eq
For illustration, let us consider the diagram (gttG1) of Fig.\ref{Fig:3NonAbelianTwo}, illustrated in Fig.\ref{Fig:gttG1}. It reads, for the vector DA,
\beqa
\label{SgttG1V}
\Phi^V_{gttG1}&\!\!\!=&\!\!-e_q \, \frac{1}4 \frac{2}s \frac{(-i) N_c}{C_F} \,  g^2 \, m_\rho \, f_\rho \, \frac{\delta^{ab}}{2 \, N_c}\frac{1}{2s}\!\int\limits^1_0 \! dy_1 \, dy_2 \!\!\int \!\frac{d \kappa}{2 \pi}Tr [\slashchar{e}_\gamma \, (y_1 \slashchar{p}_1-\slashchar{q}) \, \gamma_\nu \,  \slashchar{p}_1 ]\, d^{\nu \rho}(-q+(1+y_1-y_2) \, p_1) \nonumber \\
&\times& \frac{V_{\rho \lambda \alpha}(q-(1+y_1-y_2) \, p_1,\,k_1,\,-k_2+(y_1-y_2)p_1)\, d^{\alpha \beta}(k_2+(y_2-y_1)p_1)}
{[(y_1 \, p_1-q)^2+i \eta][(-q+(1+y_1-y_2) \, p_1)^2+i \eta] [ (k_2+(y_2-y_1) \, p_1)^2+i \eta]}\nonumber \\
&\times &
V_{\beta \tau \delta}(k_2+(y_2-y_1)p_1,\, -k_2,\, (y_1-y_2) p_1)\,  p_2^\lambda \,p_2^\tau \,e^{*\delta}_T \,  B(y_1, \, y_2)\,.
\eqa
It equals, when closing the $\kappa$ contour below on the single pole coming from the third propagator,
\beq
\label{SgttG1VRes}
\Phi^V_{gttG1}=-\frac{e_q\, g^2}2  \, m_\rho \, f_\rho \, \frac{\delta^{ab}}{2 \, N_c} \frac{N_c}{C_F} \frac{1}{Q^2}
\int\limits^1_0 \! dy_1 \, dy_2 \, \frac{B(y_1,\,y_2)}{\bar{y}_1 }
e^*_T \cdot e_\gamma \,.
 \eq
The axial DA contribution from the diagram (gttG1) reads
\beqa
\label{SgttG1A}
\Phi^V_{gttG1}&\!\!\!=&\!\!-e_q \, \frac{i}4 \frac{2}s \frac{(-i) N_c}{C_F} \,  g^2  \, m_\rho \, f_\rho\, \frac{\delta^{ab}}{2 \, N_c}\frac{1}{2s}\!\int\limits^1_0 \! dy_1 \, dy_2 \!\!\int \!\frac{d \kappa}{2 \pi}Tr [\slashchar{e}_\gamma \, (y_1 \slashchar{p}_1-\slashchar{q}) \, \gamma_\nu \,  \slashchar{p}_1 \, \gamma_5]\, d^{\nu \rho}(-q+(1+y_1-y_2) \, p_1) \nonumber \\
&\times& \frac{V_{\rho \lambda \alpha}(q-(1+y_1-y_2) \, p_1,\,k_1,\,-k_2+(y_1-y_2)p_1)\, d^{\alpha \beta}(k_2+(y_2-y_1)p_1)}
{[(y_1 \, p_1-q)^2+i \eta][(-q+(1+y_1-y_2) \, p_1)^2+i \eta] [ (k_2+(y_2-y_1) \, p_1)^2+i \eta]}\nonumber \\
&\times &
V_{\beta \tau \sigma}(k_2+(y_2-y_1)p_1,\, -k_2,\, (y_1-y_2) p_1)\,  p_2^\lambda \,p_2^\tau  \,\epsilon^\sigma_{\, \, e^*_T p n} \, B(y_1, \, y_2)\,.
\eqa
It equals, when closing the $\kappa$ contour below on the single pole coming from the third propagator,
\beq
\label{SgttG1ARes}
\Phi^A_{gttG1}=-\frac{e_q\, g^2}2  \, m_\rho \, f_\rho \, \frac{\delta^{ab}}{2 \, N_c} \frac{N_c}{C_F} \frac{1}{Q^2}
\int\limits^1_0 \! dy_1 \, dy_2 \, \frac{D(y_1,\,y_2)}{\bar{y}_1 }
e^*_T \cdot e_\gamma \,.
 \eq
%%%%%%%%%%%%%%%%%%%%%%
All other diagrams of each class can be computed according to the previous examples.

In order to present now the full result in a compact form,
we decompose the result impact factor into spin-non-flip and spin-flip
part. The non-flip part is proportional to
\beq
\label{defNonFlip}
T_{n.f.}=-(e_\gamma \cdot e^*_T)\,,
\eq
whereas
the spin-flip part involves
\beq
\label{defFlip}
T_{f.}=\frac{(e_\gamma \cdot k_\perp)(e^*_T \cdot k_\perp)}{\kb^2}+\frac{(e_\gamma \cdot e^*_T)}{2}\,.
\eq
We label the  transverse polarizations as
\beqa
\label{CircularPol}
\epsilon^{(+)}& \equiv& \epsilon^{(R)} =\displaystyle-\frac{i}{\sqrt{2}}\left[e_1+i \, e_2\right]=\displaystyle-\frac{i}{\sqrt{2}}\,(0,1,i,0) \,,\\
\epsilon^{(-)}&\equiv& \epsilon^{(L)}=\displaystyle\frac{i}{\sqrt{2}}\left[e_1-i \, e_2\right]=\displaystyle\frac{i}{\sqrt{2}}\,(0,1,-i,0)\,.
\eqa
They satisfy
\beq
\label{propHelicite}
\epsilon^{(\lambda)\, *}=\epsilon^{(-\lambda)}\,.
\eq
and
\beq
\label{propHeliciteOrtho}
\epsilon^{(+)} \, \epsilon^{(+)\, *}=\epsilon^{(+)} \, \epsilon^{(-)}=-1 \,,\quad {\rm and } \quad
\epsilon^{(-)} \, \epsilon^{(-)\, *}=\epsilon^{(-)} \, \epsilon^{(+)}=-1\,.
\eq
In this basis,
\beq
\label{gperphelicite}
\epsilon^{(+)}_\mu \, \epsilon^{(+)\, *}_\nu+\epsilon^{(-)}_\mu \, \epsilon^{(-)\, *}_\nu=\epsilon^{(+)}_\mu \, \epsilon^{(-)}_\nu+\epsilon^{(-)}_\mu \, \epsilon^{(+)}_\nu
%=e_{1\, \mu} \,e_{1\, \nu}+ e_{2\, \mu} \,e_{2\, \nu}
=-g_{\perp \, \mu\nu}\,.
\eq
Decomposing the impact factor as the sum of spin-non-flip and spin-flip contributions
\beq
\label{impactNonFlipFlip}
\Phi^{\gamma^*_T\to\rho_T}(\kb^2)=\Phi_{n.f.}^{\gamma^*_T\to\rho_T}(\kb^2) \, T_{n.f.}+\Phi_{f.}^{\gamma^*_T\to\rho_T}(\kb^2) \, T_{f} \,,
\eq
and introducing the notations
\beq
\label{CabAlpha}
\alpha=\kb^2/Q^2 \quad {\rm and} \quad C^{ab} = -\frac{e\, g^2 m_\rho f_\rho}{\sqrt{2}\,Q^2}\frac{\delta^{ab}}{2 \, N_c}
\eq
one obtain the following results for the two bodies contribution
\bea
 \label{NonFlip2}
\Phi_{ n.f. 2}^{\gamma^*_T\to\rho_T}(\kb^2)&=&
\frac{C^{ab}}{2}\,\frac{1}{C_F}\, C_F \int\limits^1_0 dy_1 \,
 \left\{\frac{\left(2 y_1-1\right) \varphi _1^T\left(y_1\right)+2 \, y_1
   \left(1-y_1\right) \varphi _3\left(y_1\right)+\varphi
   _A^T\left(y_1\right)}{y_1 \, \left(1-y_1\right)} \right. \nonumber \\
&&\left.-\frac{2 \, \alpha
   \left(\alpha+2  \, y_1 \left(1-y_1\right) \right) \left(\left(2
   y_1-1\right) \varphi _1^T\left(y_1\right)+\varphi
   _A^T\left(y_1\right)\right)}{y_1 \, \left(1-y_1\right)
   \left(\alpha+ \, y_1\left(1-y_1\right) \right)^2}\right\}
\eea
and
\bea
 \label{Flip2}
&&\Phi_{ f. 2}^{\gamma^*_T\to\rho_T}(\kb^2)=
\frac{C^{ab}}{2} \frac{1}{C_F}\, C_F \int\limits^1_0 dy_1 \,
 \frac{4 \, \alpha}{\left( \alpha + \left( 1 - {y_1} \right) \,{y_1} \right) ^2}
\left[\left( 1 - 2\,{y_1} \right) \,
         \varphi_1^T(y_1)
+
      \varphi_A^T({y_1}) \right]\,.
 \eea
The three bodies contribution reads
\bea
\label{NonFlip3}
&&\hspace{-.2cm}\Phi_{n.f. 3}^{\gamma^*_T\to\rho_T}(\kb^2)=
C^{ab}\,\frac{1}{C_F}  \int\limits^1_0 dy_1 \!\int\limits^1_0 dy_2
\left\{\frac{ y_1 \, \zA \,D\left(y_1,y_2\right)}{\alpha+ \left(1-y_1\right)
   y_1}
   \left(\frac{\alpha \left(N_c-2\, C_F \right)}{\left(y_1-y_2+1\right) \alpha+ \, y_1
   \left(1-y_2\right)}+\frac{\alpha N_c \left(1-y_1\right)
   }{y_2 \, \alpha+ y_1
   \left(y_2-y_1\right)}\right)\right. \nonumber \\
&&\hspace{-.2cm}\left.-\frac{y_1 \, \zV \, B\left(y_1,y_2\right) }{\alpha+ \left(1-y_1\right)
   y_1} \left(\frac{\alpha
   \left(2\, C_F -N_c\right) \left(2 y_1-1\right)}{\left(y_1-y_2+1\right)
   \alpha+ y_1
   \left(1-y_2\right)}+\frac{\alpha N_c \left(1-y_1\right)
   }{y_2 \,\alpha + y_1
   \left(y_2-y_1\right)}\right)\right. \\
&&\hspace{-.2cm}\left.+\left(\zV \, B\left(y_1,y_2\right)+\zA  \, D\left(y_1,y_2\right)\right)
   \left(\frac{2 \, C_F \, y_1}{\alpha+ \left(1-y_1\right)
   y_1}-\frac{1}{1-y_1}\left[\frac{N_c \left(1-y_2\right) \left(y_1-y_2\right)
   Q^2}{\left(1-y_1\right) \alpha+ \left(1-y_2\right)
   \left(y_2-y_1\right)}+C_F+N_c\right]\right)\right\}\nonumber
\eea
and
\bea
\label{Flip3}
&&\Phi_{ f. 3}^{\gamma^*_T\to\rho_T}(\kb^2)=
\frac{C^{ab}}{2}\, \frac{1}{C_F}
\frac{ 4\,\alpha \,y_1}{\alpha +
    y_1 \, \left( 1 - {y_1} \right)}\,
    \left( \frac{2 \, C_F- N_c}
       {\alpha\,\left( 1 + {y_1} - {y_2} \right)  + y_1\,\left( 1 - {y_2} \right) } -
      \frac{{N_c}}{\alpha\,{y_2} + y_1\,\left( -{y_1} + {y_2} \right) } \right) \nonumber \\
&& \times
\left[\zA \,D({y_1},{y_2})\,\left( 1 + {y_1} - {y_2} \right)  +
     \zV\, B({y_1},{y_2})\,\left( 1 - {y_1} - {y_2} \right)  \right] \,.
\eea
The full result for the impact factor reads,
 after several simplications due to the use of the equation of motion and the symmetrical properties of 2 and 3-parton correlators, as
\bea
\label{NonFlip}
&&\Phi_{n.f.}^{\gamma^*_T\to\rho_T}(\kb^2) \nonumber
\\
&=&\hspace{-.2cm}\,
\frac{C^{ab}}{2}\,
\left\{-2   \,\int dy_1 \frac{
 \left(\alpha+2  \, y_1\left(1-y_1\right)\right) \alpha} { y_1  \left(1-y_1\right)  \left(\alpha+ \, y_1\left(1-y_1\right)
   \right)^2}\left[\left(2 y_1-1\right)  \varphi
   _1^T(y_1)+ \varphi _A^T(y_1)\right] \right. \nonumber \\
&&\hspace{-.7cm}+2  \int dy_1 \,  dy_2 \left[\zV\,  B\left(y_1,y_2\right)\!-\!\zA \,  D\left(y_1,y_2\right)\right] \frac{y_1\left(1-y_1\right) \alpha}{\alpha+ \, y_1 \left(1-y_1\right)} \!\left[\frac{2 -N_c/C_F}{\alpha  \left(y_1-y_2+1\right)+ \, y_1 \left(1-y_2\right)}\right. \nonumber\\
&&\hspace{-.4cm}\left.-\frac{N_c}{C_F} \frac{1}{y_2 \, \alpha+ \, y_1 \left(y_2-y_1\right)}\right] -2  \int dy_1 \, dy_2 \left[\zV  B\left(y_1,y_2\right)+\zA  D\left(y_1,y_2\right)\right] \left[\frac{2 + N_c/C_F}{ 1-y_1}\right. \nonumber\\
&&\left.\left.+\frac{y_1}{\alpha+ y_1\left(1-y_1\right)}  \left(\frac{ \left(2 -N_c/C_F\right) \, y_1 \, \alpha}{\alpha \left(y_1-y_2+1\right)+ y_1   \left(1-y_2\right)}-2 \right)\right. \right. \nonumber\\
&&\left.\left.+\frac{N_c}{C_F}\frac{ \left(y_1-y_2\right) \left(1-y_2\right)}{1-y_1}\frac{1}{\alpha\left(1-y_1\right)+ \left(y_2-y_1\right) \left(1-y_2\right)}\right]\right\}
\eea
and
\bea
\label{Flip}
&&\hspace{-.2cm}\Phi_{f.}^{\gamma^*_T\to\rho_T}(\kb^2)=\frac{C^{ab}}{2}\,
  \left\{4  \! \int \! dy_1\frac{\alpha}{\left(\alpha+
  \, y_1 \left(1-y_1\right) \right)^2}\left[ \varphi _A^T(y_1)-  \left(2 y_1-1\right)  \varphi
   _1^T(y_1)\right] \right.\nonumber \\
&&-\,4 \!\int \! dy_1 \, dy_2\frac{ y_1 \, \alpha}{\alpha+ \, y_1 \left(1-y_1\right)} \left[\zA  D\left(y_1,y_2\right)
   \left(-y_1+y_2-1\right)+\zV  B\left(y_1,y_2\right) \left(y_1+y_2-1\right)\right] \nonumber \\
&&\left.\times \left[\frac{(2-N_c/C_F)}{ \alpha \left(y_1-y_2+1\right)+ \, y_1
   \left(1-y_2\right)} -\frac{N_c}{C_F} \frac{1}{y_2 \, \alpha+ y_1
   \left(y_2-y_1\right)}\right]\right\}\,.
\eea
The gauge invariance of the considered impact factor requires a special
attention. The $\gamma^*\to\rho_T$
impact factor is constructed in such a manner that it should vanish when
$\kb^2=0$. This fact is a consequence of the gauge invariance of the
impact factor.
From our final formulas (\ref{NonFlip})
and (\ref{Flip}), it is obvious to check that $\Phi_{f.}$ and
$\Phi_{n.f.}$ indeed vanish when $\kb^2=0$ since
$T_{n.f.}$ is a phase for flip transition, which is regular in the $\kb^2
\to 0$ limit.
 The vanishing of the "Abelian'', i.e. proportional to $C_F$ part of
(\ref{NonFlip})
is particularly subtle since it appears as  a consequence of the equations
of motions
(\ref{em_rho1}, \ref{em_rho2}).
Because of that some comments can be useful.
Let us note that the sum of Eq.~(\ref{em_rho1}) multiplied by $y_1$ and of
Eq.~(\ref{em_rho2}) multiplied by
$\bar y_1$ takes the form
\begin{eqnarray}
\label{eomcancellation1}
&&2y_1\bar y_1 \varphi_3(y_1)+(y_1-\bar y_1)\varphi_1^T(y_1)
+\varphi_A^T(y_1)
\\
&& = -y_1\int\limits_0^1dy_2\left[ \zeta_3^VB(y_1,y_2) +
\zeta_3^AD(y_1,y_2)   \right]
 -\bar y_1\int\limits_0^1dy_2\left[ -\zeta_3^VB(y_2,y_1) +
\zeta_3^AD(y_2,y_1)   \right]\;,
\nonumber
\end{eqnarray}
from which, after integration over $y_1$ of the both sides of
(\ref{eomcancellation1}) multiplied by
$\frac{1}{y_1\bar y_1}$, we derive the equality
\begin{equation}
\label{eomcancellation2}
\int\limits_0^1\frac{dy_1}{y_1\bar y_1}\left(
2y_1\bar y_1 \varphi_3(y_1)+(y_1-\bar y_1)\varphi_1^T(y_1)
+\varphi_A^T(y_1)   \right)
= - \int\limits_0^1dy_1 \int\limits_0^1dy_2\;\frac{2}{\bar y_1}\left[
\zeta_3^VB(y_1,y_2)+\zeta_3^AD(y_1,y_2)  \right]\;.
\end{equation}
Now, by inspecting the expression (\ref{NonFlip2}) in the limit $\alpha
\to 0$ we see that only the
 first term in $\left\{ ...\right\}$ survives and it has a form of the l.h.s
of expression (\ref{eomcancellation2}).
Similarly, by inspecting the expression (\ref{NonFlip3}) in the limit
$\alpha \to 0$ we see, that only  the last line
of this expression survives in the limit $\alpha \to 0$ and that the
resulting expression
 coincides with the r.h.s of  (\ref{eomcancellation2}). Consequently, the
non-vanishing terms cancel out due to the
relation (\ref{eomcancellation2}).
 At the same time, the vanishing of non-Abelian ($\sim N_c$) part
of (\ref{NonFlip3}) is the result of direct cancellation of the
non-vanishing contributions of
 the diagrams (bG1), (dG1), (aG2), (cG2), (bG2), (dG2), (eG2), (fG2) of Fig.\ref{Fig:3Abelian} (see Eq.(\ref{colorAbelian})) with the corresponding ones coming from
diagrams of Fig.\ref{Fig:3NonAbelian} and Fig.\ref{Fig:3NonAbelianTwo} containing
 triple-gluon vertices.
Thus, the expression for the $\gamma^*\to\rho_T$ impact factor has finally
a gauge-invariant form only
provided the genuine twist $3$ contributions have been taken into account,
hidden in
formula (\ref{NonFlip}) when writing $\Phi_{n.f.}$
 by the fact that we have used e.o.m., which explicitly relate 2 and 3
particles correlators.

We end up this section with a comment about the problem of end-point
singularities. Such singularities does
not occur both in WW approximation and in full twist-3 order
approximation. First, the flip contribution (\ref{Flip})
obviously does not have any end-point singularity.
 The potential end-point singularity for the non-flip contribution
(\ref{NonFlip}) is spurious since
$\varphi _A^T(x_1),$ $\varphi _1^T(x_1)$
vanishes at $x_1=0,1$ (this is enough to justify the regularity of the
result in the WW approximation),
as well as $B(x_1,x_2)$ and  $D(x_1,x_2).$

\subsection{Calculation based on the Covariant Collinear Factorization}
\label{SubSec_ImpactCCF}

We now calculate the impact factor using the CCF parametrization of Ref.\cite{BB} for vector meson
DAs.
Let us outline basic ideas behind our calculation. We need to express the impact factor in terms of 
hard coefficient functions and soft parts parametrized by light-cone matrix elements. The standard technique 
here is an operator product expansion on the light cone, $z^2\to 0$, which naturally gives the leading term in the power counting and leads to the described above factorized structure. Unfortunately we do not have an operator definition for an impact factor, and therefore, we have to rely in our actual calculation on the perturbation theory.
The primary complication here is that $z^2\to 0$ limit of any single diagram is given in terms of  
light-cone matrix elements without any Wilson line insertion between the quark and gluon operators, like
$$
\langle V(p_V)|\bar \psi(z)\gamma_\mu \psi(0)|0 \rangle \quad {\rm and}\quad  \langle V(p_V)|\bar \psi(z)\gamma_\mu A_\alpha (t\, z) \psi(0)|0 \rangle \, ,
$$
we will call conventionally such objects as perturbative correlators. 
Actually we need to combine together contributions of quark--antiquark and quark--antiquark gluon diagrams in order to obtain a final gauge invariant result.

One should stress that despite working in the axial gauge one can not neglect completely an effect coming from the Wilson
lines since the  two light-cone vectors $z$ and $n$ are not equal to each other and thus, generically, Wilson lines are not equal to unity. Nevertheless in the axial gauge the contribution of each additional parton costs one extra power 
of $1/Q$, therefore a calculation can be
organized in a simple iterative manner expanding the Wilson line. 
At twist three level it is enough to consider the first two terms of such expansion 
\beq
\label{wl}
[z,0]=1+i \,g \int\limits^1_0 dt \, z^\alpha A_\alpha (z t)+{\cal O}(A^2) \, .
\eeq
For instance, the quark--antiquark vector correlator can be written, for the general case $z^2\neq 0$, as
\beq
\label{vec}
\langle V(p_V)|\bar \psi(z)\gamma_\mu \psi(0)|0 \rangle=\langle V(p_V)|\bar \psi(z)\gamma_\mu [z,0] \psi(0)|0 \rangle
-ig\int\limits^1_0 dt \langle V(p_V)|\bar \psi(z)\gamma_\mu z^\alpha A_\alpha (zt)\psi(0)|0 \rangle \, ,
\eeq
where we formally inserted the Wilson line in the r.h.s and performed its approximate subtraction according to (\ref{wl}). Then using relation (\ref{axial}), we express the gluon field operator in the second term of (\ref{vec}) in terms of field strength, which gives us the $\langle V(p_V)|\bar \psi(z)\gamma_\mu G_{\alpha\beta}(t \,z)\psi(0)|0 \rangle$
correlator. For the later we apply again the procedure of Wilson lines insertion (and its approximate subtraction)  
\bea
\label{vec1}
& \langle V(p_V)|\bar \psi(z)\gamma_\mu \psi(0)|0 \rangle=\langle V(p_V)|\bar \psi(z)\gamma_\mu [z,0] \psi(0)|0 \rangle & 
\nonumber \\
&
-ig\int\limits^1_0 \int\limits^\infty_0 dt \,d\sigma  \, e^{-\epsilon \,\sigma}\langle V(p_V)|\bar \psi(z)[z,z\,t+n\,\sigma]\gamma_\mu z^\alpha n^\beta G_{\alpha\beta} (z\,t+n\,\sigma)[z \,t+n\,\sigma,0]\psi(0)|0 \rangle + \dots \, , &
\eea
where $\dots$ stands for the correlators with more than one gluon field. 

Such correlator naturally appears (after Fierz decomposition) in the expression for the impact factor generated by the leading order diagrams of perturbation theory. Now we proceed to the extraction of leading $1/Q$ asymptotic. It is achieved, due to the dimensional counting reasons, by the substitution of the off light-cone correlators by their 
light-cone limit where $z \,t+n\,\sigma\propto z$, $z^2\to 0$. 
In this limit, using the CCF parametrization of section \ref{SubSubSec_ParamVacuumRhoCCF} for the light-cone correlators, one can deduce after some transformations that
\bea
\label{vc-no-wl}
&\langle \rho(p_\rho)|\bar \psi(z)\gamma_\mu \psi(0)|0 \rangle|_{z^2\to0}=& 
\nonumber\\
&f_\rho \, m_\rho \left[-i \,p_\mu (e^* \cdot z)
\int\limits^1_0 dy \, e^{i y (p \cdot z)} (h(y)-\tilde h(y))
+ e^*_{\mu} \int\limits^1_0 dy \, e^{i y (p \cdot z)} g^{(v)}_\perp (z)
\right] + \dots \, ,&
\eea    
where $\dots$ stands for the contributions vanishing at twist 3 level, and 
\beq
\label{defHtilde}
\tilde h(y)=\zeta_3^V\int\limits^y_0d\alpha_1\int\limits^{\bar y}_0d\alpha_2\frac{V(\alpha_1,\alpha_2)}{\alpha_g^2}
\, ,
\eq
The physical polarization vector satisfies $e \cdot p_\rho=0$ (or $e \cdot p=0$ since $p_\rho = p$ up to twist 3). On the other hand, 
the polarization vector of transversely polarized meson is chosen to be orthogonal to the light-cone vectors
fixed by the external kinematics: 
 $e \cdot n_0=0$. But one should take into account that the $e_T$ vector defined by (\ref{pol_Rho}) 
has a non-vanishing scalar product with the vector $n_0$,
\beq
\label{scalar_e.n}
e_T \cdot n_0=-\frac{e \cdot z}{p\cdot z} \, .
\eq
This relation was used to derive (\ref{vc-no-wl}).

Note that the $z \, t+n \, \sigma\propto z$ condition
means actually that the vector $z$ (which is an internal integration variable for the impact factor) is approaching during this limiting procedure the direction of the light-cone vector $n$, $z\propto n$. One should mention, to avoid any misunderstanding, that it does not mean that we must put, say in Eqs. like (\ref{vc-no-wl}), the $e \cdot z$ scalar product 
equal to zero. What we actually do when performing the $1/Q$ power expansion is a Taylor expansion of scalar functions $F(p\cdot z,z^2)$, 
which depend generically on the two variables $p\cdot z$ and $z^2$,  with respect to the variable $z^2$, whereas any scalar product of $z$ with other vectors should remain intact. 

Performing a similar sequence of steps we obtain the following result for the axial-vector correlator
at the twist 3 level
\bea
\label{ax-vc-no-wl}
&\langle \rho(p_\rho)|\bar \psi(z)\gamma_\mu\gamma_5 \psi(0)|0 \rangle|_{z^2\to0}=&
\nonumber\\
&\frac{1}4 f_\rho\, m_\rho \left[\epsilon_{\mu\alpha\beta\gamma} \, e^{*\alpha}p^\beta z^\gamma
\int\limits^1_0 dy \,e^{i y (p \cdot z)} (g^{(a)}_\perp(y)-\tilde g^{(a)}_\perp(y))
+ \epsilon_{\mu\alpha\beta\gamma} e^{*\alpha}p^\beta n^\gamma p \cdot  z
\int\limits^1_0 dy \, e^{i y (p \cdot z)} \tilde g^{(a)}_\perp(y) 
 \right]   \, ,&
\eea
with
\beq
\label{def_gaTilde}
\tilde g^{(a)}_\perp(y)=4\, \zeta_3^A\int\limits^y_0d\alpha_1\int\limits^{\bar y}_0d\alpha_2\frac{A(\alpha_1,\alpha_2)}{\alpha_g^2}\,.
\eq
Comparing the obtained results (\ref{vc-no-wl}), (\ref{ax-vc-no-wl}) for the perturbative correlators  with initial parametrizations (\ref{BBV2}), (\ref{BBA}) we see that at twist 3-level the net effect of the Wilson line is
just some renormalization of the $h$ function in the case of vector correlator, whereas for the axial-vector we obtain in addition to the function $g^a_\perp$ renormalization a new Lorentz structure, the last term in (\ref{ax-vc-no-wl}).
Nevertheless, we found that the last term in (\ref{ax-vc-no-wl}) produces at the end a zero contribution to impact factor.
  
Let us now discuss gluonic diagrams which involves quark--antiquark gluon correlators, like 
$\langle \rho(p_\rho)|\bar \psi(z)\gamma_\mu A_\alpha(w) \psi(0)|0 \rangle$. 
Applying our procedure one can easily show that at twist 3 level 
\bea
\label{3b}
 \langle \rho(p_\rho)|\bar \psi(z)\gamma_\mu g A_\alpha(w) \psi(0)|0 \rangle|_{w\propto z, z^2\to0}&=&
-m_\rho \, \fV \,p_\mu e^{*}_{T\alpha} \int  D\underline \alpha \, \frac{V(\alpha_1,\alpha_2)}{\alpha_g}
e^{i(p\cdot z)\alpha_1+i(p\cdot w)\alpha_g} \, , \\
\langle \rho(p_\rho)|\bar \psi(z)\gamma_\mu\gamma_5 g A_\alpha(w) \psi(0)|0 \rangle|_{w\propto z, z^2\to0}&=&
-i \, m_\rho \, \fA \,p_\mu \,\epsilon_{\alpha\beta\gamma\delta} \,n^\beta p^\gamma e^{*\delta}_{T} \int  D\underline \alpha \, \frac{A(\alpha_1,\alpha_2)}{\alpha_g}
e^{i(p\cdot z)\alpha_1+i(p\cdot w)\alpha_g} \, ,
\nonumber
\eea
see Eq.(\ref{BBVg},\ref{BBAg}).
Note that the first nontrivial effects induced by the Wilson line insertion start for such perturbative correlators at the level
of twist 4 only. Therefore taking into account such 3-partons contributions is quite straightforward: one needs to calculate, projected in accordance with (\ref{3b}), diagrams describing the production of collinear on-shell quark--antiquark gluon state. 

One comment is here in order. Perturbative expansion generates, among others, 
diagrams where the gluon field is attached not to the internal part of the diagrams but to the "external" 
quark (or antiquark) lines. Such diagrams, in accordance to the logic of collinear factorization, should be 
factorized in terms of not 3-parton but 2-parton correlators. Quasi-collinear gluon radiation appears at
large distances and corresponding subprocess should be factorized not in the hard coefficient but is included 
in the soft part of the process, described by the 2-parton quark--antiquark correlator.

The internal variable $z$ (and $w$) integration can be reduced to
the Fourier integrals
$$
\int d^4 z \, e^{i(l\cdot z)}=(2\pi)^4\delta^4(l) \, , \quad \int d^4 z\, z_\alpha \, e^{i(l\cdot z)}=-i (2\pi)^4\frac{\partial }{\partial l^\alpha}\delta^4(l)\, ,
$$
where $l$ stays here for some combination of the external and internal momenta\footnote{The subsequent integration
over the internal momenta with delta function derivative is done using integration by parts.}.
The corresponding intermediate calculations do not contain principle difficulties both for the case of 2-partons and 3-partons contributions.  Note that the contributions computed here have the same hard part as the one of the section \ref{SubSec_ImpactLCCF}, except for 2-partons contributions with derivatives discussed in that section  which have no counterpart here. In what follows, we simply give the
final results. Then we will discuss in details    an important issue related with the restoration of the gauge invariance for the final result. 

We use the notations
$$
\alpha=\frac{\kb^2}{Q^2}\, , \quad
c_f=\frac{N^2}{N^2-1}
%\, , \quad \zeta_3^V=\frac{f^V}{f} \, , \quad  \zeta^A=\frac{f^A}{m f} \, .
$$
For the 3-parton contributions we obtain the result 
\beq
\Phi^3=-\frac{e \, g^2 \, m_\rho \, f_\rho}{\sqrt{2} Q^2}\frac{\delta^{ab}}{2 N_c}
\left\{\Phi^{q\bar q g}(\alpha)+\Delta \Phi^3\right\}\,,
\eeq
with
\bea
\label{3BodyDelta}
&&\Delta \Phi^{3}=-\frac{T_{n.f.}}{2}\int \frac{Dz}{\bar z_1\bar z_2 z_g} \, 
\left\{
\zeta_3^V V(z_1,z_2)(z_1-z_2) + \zeta_3^A A(z_1,z_2)(\bar z_1+\bar z_2)
\right\} \, 
\eea
and
\bea
\label{3BodyCov}
&&\Phi^{q\bar q g}(\alpha)=\int Dz \frac{2\alpha  }{z_1 z_2 z_g^2} \\
&&\times
\left\{
(\zeta_3^V V(z_1,z_2) + \zeta_3^A A(z_1,z_2)) \, T_{n.f.}\left(
\frac{(1-c_f)\bar z_g z_1}{\alpha \bar z_g+z_1z_2}
-\frac{c_f z_g^2}{\alpha\bar z_1+z_2 z_g}
\right.
-\frac{(z_2-\bar z_1 c_f) z_1z_2}{\bar z_1(\alpha+z_1 \bar z_1)}
-\frac{(z_1-\bar z_2 c_f)\bar z_2}{(\alpha+z_2 \bar z_2)}
\right) \nonumber \\
&& +\left.
(\zeta_3^V V(z_1,z_2) - \zeta_3^A A(z_1,z_2)) \, 2z_1 T_{f.} \! \!
\left(
\frac{(1-c_f)\bar z_g^2}{\alpha\bar z_g+z_1z_2}
-\frac{c_f z_g\bar z_1}{\alpha\bar z_1+z_2z_g}
-\frac{c_f z_g\bar z_2}{\alpha\bar z_2+z_1z_g}
+\frac{c_f\bar z_1-z_2}{\alpha+z_1\bar z_1}
+\frac{c_f\bar z_2-z_1}{\alpha+z_2\bar z_2}
\right)
\right\}\,,
\nonumber
\eea
where $D z$ is defined according to (\ref{defDalpha})
and where we have used symmetry properties of gluon distribution amplitudes under the exchange of quark momentum
fractions:
$V(z_1,z_2)=-V(z_2,z_1)$, $A(z_1,z_2)=A(z_2,z_1)$.

Let us discuss now the 2-parton contributions. We obtain 
\beq
\label{Phi2}
\Phi^2=-\frac{e \, g^2 \, m_\rho \, f_\rho}{\sqrt{2} Q^2}\frac{\delta^{ab}}{2 N_c}\left\{\Phi^{q\bar q}(\alpha)+\Delta \Phi^2\right\}\,,
\eq
with 
\beq
\label{2bodyCov}
\Phi^{q\bar q}(\alpha)=\int\limits^1_0 dz
\left\{
T_{n.f.}\Phi^+(z)\frac{\alpha(\alpha+2z\bar z)}{z\bar z (\alpha+z\bar z)^2}
+T_{f.}\Phi^-(z)\frac{2\alpha}{ (\alpha+z\bar z)^2}\,,
\right\}
\eq
where
\beq
\label{defPhi+-}
\Phi^\pm(\alpha)=(2z-1)\left[
h(z)-\tilde h(z)
\right]\pm \frac{g^{(a)}_\perp(z)-\tilde g^{(a)}_\perp(z)}{4}\,,
\eq
whereas for $\Delta \Phi^2$ term we get
\beq
\Delta \Phi^2=T_{n.f.} \int\limits^1_0 dz
\left\{ g^{(v)}_\perp(z)-\frac{\Phi^+(z)}{2z\bar z}
\right\} \, .
\eeq

Note that $\Phi^{q\bar q}(\alpha)$ and $\Phi^{q\bar q g}(\alpha)$ vanish in the limit $\alpha\to 0$,
whereas $\Delta \Phi^{2}$ and $\Delta \Phi^{3}$ do not depend on $\alpha$.  
Now we need to demonstrate that $\Delta \Phi^{2}$ and $\Delta \Phi^{3}$ cancel each other. It guarantees the property of the impact factor, $\Phi(\alpha=0)=0$, which is directly related to the gauge invariance.

One can now separate from $\Delta \Phi^2$ the contribution $\Delta \Phi^2_{a}$ that is due to functions $\tilde h(z)$ and $\tilde g^a_\perp(z)$, which originates in our method from the Wilson lines insertion procedure, 
\beq
\label{delta-phi2}
\Delta \Phi^2=\Delta \Phi^2_{a}+\Delta \Phi^2_{b} \,,
\eeq
where
\beq
\label{defPhia}
\Delta \Phi^2_{a}=T_{n.f.} \int\limits^1_0 \frac{dz}{2z\bar z}
\left\{ (2z-1)\tilde h(z)+\frac{\tilde g^{(a)}_\perp(z)}{4}
\right\} \, .
\eq
To calculate $\Delta \Phi^2_{a}$, it is convenient to present $\tilde h(z)$ and $\tilde g^{(a)}_\perp(z)$ in the form
$$
\tilde h(u)=\zeta_3^V\int\limits^1_0 dt \int Dz\,\delta(u-z_1-z_g t)\frac{V(z_1,z_2)}{z_g}\, ,\quad 
\tilde g^{(a)}_\perp(z)=4\,\zeta_3^A\int\limits^1_0 dt \int Dz\,\delta(u-z_1-z_g t)\frac{A(z_1,z_2)}{z_g}\, .
$$  
Using this relation one can easily found that 
$$
\Delta \Phi^2_{a}=\frac{T_{n.f.}}{2}\int Dz \left\{ \zeta_3^V\frac{V(z_1,z_2)}{z_g^2}\ln\frac{z_1 \bar z_1}{z_2 \bar z_2}+\zeta_3^A\frac{A(z_1,z_2)}{z_g^2}\ln\frac{\bar z_1 \bar z_2}{z_1 z_2}
\right\} \, .
$$

The second term in (\ref{delta-phi2}) can be reduced to the form
$$
\Delta \Phi^2_{b}=\frac{T_{n.f.}}{2}\int\limits^1_0 dz
\left\{ \ln(z)\, g^{\uparrow \downarrow}(z)+\ln(\bar z)\, g^{ \downarrow \uparrow }(z)
\right\} \, .
$$
where \cite{BB}
$$
g^{\uparrow \downarrow}(z)=g_\perp^{(v)}(z)+\frac{1}{4}\frac{d}{dz}g_\perp^{(a)}(z)\, ,\quad 
g^{ \downarrow\uparrow}(z)=g_\perp^{(v)}(z)-\frac{1}{4}\frac{d}{dz}g_\perp^{(a)}(z)\, .
$$
Then, we separate the WW and genuine twist 3 contributions to $\Delta \Phi^2_{b}$,
$$
\Delta \Phi^2_{b}=\Delta \Phi^{2\, WW}_{b}+\Delta \Phi^{2\, gen}_{b}\, ,
$$
in accordance with 
$$
g^{\uparrow \downarrow}(z)=g^{\uparrow \downarrow\, WW}(z)+g^{\uparrow \downarrow\, gen}(z)\, ,
\quad 
g^{\uparrow \downarrow}(z)=g^{\uparrow \downarrow\, WW}(z)+g^{\uparrow \downarrow\, gen}(z)\, .
$$
Using the explicit expressions for these functions in the WW limit
$$
g^{\uparrow \downarrow\, WW}(z)=\int\limits^1_z \frac{du}{u}\phi_\parallel (u) \, , \quad
g^{ \downarrow\uparrow\, WW}(z)=\int\limits^z_0 \frac{du}{\bar u}\phi_\parallel (u) \, ,
$$
one can easily found that the WW contribution to $\Delta \Phi^{2}_{b}$ vanishes, 
$$\Delta \Phi^{2\, WW}_{b}=0\, .$$
Then, using results of Ref.\cite{BB}, which allow to express $g^{\uparrow \downarrow\, gen}(z)$ and $g^{\downarrow\uparrow \, gen}(z)$ in terms of 3-partons DAs, 
we found after some transformations that
$$
\Delta \Phi^{2\, gen}_{b}=\frac{T_{n.f.}}{2}\int Dz \left\{ \zeta_3^V\frac{V(z_1,z_2)}{z_g}\left(\frac{z_1-z_2}{\bar z_1\bar z_2}-\ln\frac{z_1 \bar z_1}{z_2\bar z_2}\frac{1}{z_g}\right)+\zeta_3^A\frac{A(z_1,z_2)}{z_g}\left(\frac{\bar z_1+\bar z_2}{\bar z_1\bar z_2}-\ln\frac{\bar z_1 \bar z_2}{z_1 z_2}\frac{1}{z_g}\right)
\right\} \, .
$$
These results mean that
\beq
\label{DeltaPhi2Gen}
\Delta \Phi^2=\frac{T_{n.f.}}{2}\int \frac{Dz}{z_g \bar z_1\bar z_2} \left\{ \zeta_3^V V(z_1,z_2)(z_1-z_2)+\zeta_3^A A(z_1,z_2)(\bar z_1+\bar z_2)
\right\}  \, ,
\eq
and thus that the constant terms of 2-parton (\ref{DeltaPhi2Gen}) and 3- parton (\ref{3BodyDelta})  contributions cancel each other
$$
\Delta \Phi^2+\Delta \Phi^3=0 \, .
$$

Finally, the impact factor is given as a sum of two contributions
\beq
\label{decomposePhi2and3}
\Phi(\alpha)=-\frac{e \, g^2 \, m_\rho \, f_\rho}{\sqrt{2} Q^2}\frac{\delta^{ab}}{2 N_c}\left\{\Phi^{q\bar q}(\alpha)+\Phi^{q\bar q g}(\alpha)\right\} \,
\eq
where $\Phi^{q\bar q}(\alpha)$ and $\Phi^{q\bar q g}(\alpha)$ are given in eqs.  (\ref{2bodyCov}) and (\ref{3BodyCov}).

\subsection{Comparison of the two computations}
\label{SubSec_ImpactCompare}

The above results for the  $\gamma^* \to \rho$ impact factor were obtained
based on the LCCF and CCF method, and look at first sight very different.
As a testing ground of the validity of the dictionary elaborated in section \ref{SubSec_dictionary}, it is interesting to show the exact equivalence between the two results. Since the detailed proof is rather involved and technical, this is done with
whole details in Appendix \ref{Ap:Comparison}.

\section{Conclusion}
\label{Sec_Conclusion}

In conclusion, we presented a general formalism which allows us  to include in a systematic way higher twist effects. The general scheme was exemplified on the study of the
 $rho$-meson production up to twist 3 accuracy. We did this using two
different
methods: the LCCF method in the momentum space and the CCF method in the
coordinate space. The crucial point in the comparison of these two methods
was the use of Lorentz invariance constraints formulated as the
$n$-independence of the scattering amplitude within LCCF method, which leads to the
necessity of taking into account the contribution of 3-parton correlators.
We derived the dictionary between the DAs appearing in the two methods.
Due to different calculational techniques used by the LCCF and CCF methods, we
performed the calculation of the $\gamma^* \rho$ impact factor in two different ways; we
checked that they lead to the same result.  This does not preclude the solution of the well known
end-point singularity problem \cite{MP, GK} which may still require a separate treatment. The phenomenological application to our result to the HERA data \cite{expHigh} will be discussed in a separate publication \cite{usPhen}.

\section*{Acknowledgments}

We thank  V.~M.~Braun, R.~Kirschner, G.~Korchemsky and M.~Segond  for discussions. Special thanks to O.~V.~Teryaev for many useful comments.
 This work is partly supported by the ECO-NET program, contract
18853PJ, the French-Polish scientific agreement Polonium, the grant
ANR-06-JCJC-0084, the RFBR (grants 09-02-01149,
 08-02-00334, 08-02-00896), the NSh-1027.2008.2 grant and
the Polish Grant N202 249235. LS acknowledges the support of INT in Seattle during the final stage of this project.

%%%%%%%%%%%%

\eject

\appendix

\renewcommand{\theequation}{\Alph{section}.\arabic{equation}}
\section*{Appendices}

\section{$n-$independence constraint for the $N_c$ structure}
\label{Ap:Nc}

\def\sai{\footnotesize}
In this appendix we show that the $n-$independence
constraint is automatically fulfilled for $N_c$
contributions.
According to Eq.(\ref{colorAbelian}),
one should consider the ''Abelian`` diagrams (bG1), (dG1), (aG2), (cG2), (bG2), (dG2), (eG2), (fG2), and from Eq.(\ref{color3NonAbelian})
and Eq.(\ref{color3NonAbelianTwo}) all the non-Abelian diagrams.

\begin{figure}
\psfrag{i}{$\mu$}
\psfrag{a}{$k$}
\psfrag{b}{$\ell_g$}
\psfrag{c}{$\hspace{-.5cm}k-\ell_g$}
\includegraphics[width=3cm]{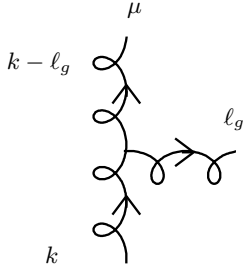}
\caption{3-gluon vertex with one open index $\mu.$}
\label{Fig:vertex3Ward}
\end{figure}

As a preliminary, let us consider a triple gluon vertex  with a $t-$channel gluon of momentum $k$ (which will stand for $k_1$ or $-k_2$ in the all cases discussed below), saturated with a non-sense polarization $p_2$, and a real gluon of momentum $\ell_g$ and polarization $\epsilon_g$, leaving the index $\mu$ open, as illustrated in Fig.\ref{Fig:vertex3Ward}.
It reads
\beq
\label{vertex3Ward}
V^\mu_{\,\,\, \lambda \alpha}(-k+\ell_g,\, k,\, -\ell_g)  \epsilon_g^\alpha \, p_2^\lambda = -(2 \, k -\ell_g) \, \epsilon_g \, p_2^\mu +(k+\ell_g)_\mu (p_2  \epsilon_g)-(2 \, \ell_g -k) p_2 \, \epsilon_g^\mu
\eq
which reduces, after making the replacement $\epsilon_g \to \ell_g$ and using the fact that
in our kinematics $k$ has no component along $p_1$, to
 \beq
\label{vertex3Ward1}
V^\mu_{\,\,\, \lambda \alpha}(-k+\ell_g,\, k,\, -\ell_g) \cdot \ell_g^\alpha \, p_2^\lambda = -(2 \, k \cdot \ell_g)  \, p_2^\mu -(\ell_g \cdot p_2)\, (\ell_g -k)^\mu=-\left[k^2 -(k-\ell_g)^2\right] \, p_2^\mu - (\ell_g \cdot p_2) \, (\ell_g -k)^\mu
\eq
 where in the last line we used the fact that $\ell_g$ in on mass-shell. Hereafter, we will symbolically denote with the index $p_2$ the first term in the r.h.s of Eq.(\ref{vertex3Ward1})
and with W the second term (having in mind for this second term further use of Ward identities).

We now apply the same method based on the Ward identity which led to Eq.(\ref{Ward3body_aD}), now for graphs (bG1) and (bG2). This gives\footnote{The signs in l.h.s of Eqs.(\ref{Ward3body_b}, \ref{Ward_btG1}, \ref{Ward_btG1})
are related to the signs in front of the corresponding $N_c$ coefficients, see Eqs.(\ref{colorAbelian}, \ref{color3NonAbelian}).}
\def\tti{0.2\columnwidth}
\def\ttiB{0.22\columnwidth}
\def\rri{-1.6cm}
\def\ssi{\footnotesize}
\beqa
% \psfrag{i}{$\mu$}
% \psfrag{u}{$y_1$}
% \psfrag{d}{$y_2-1$}
% \psfrag{m}{$y_2-y_1$}
% \psfrag{yq}{{ $\hspace{-.2cm}y_1$}}
% \psfrag{dy}{\raisebox{.04cm}{ $\hspace{-.3cm}y_2-y_1$}\raisebox{.6cm}{$\!\!\mu$}}
\psfrag{yb}{{\sai $\hspace{-.2cm}1-y_2$}}
&&\scalebox{1}
 {\raisebox{1.8cm}{$ \displaystyle - (y_2-y_1) \, p_\mu \quad \left[
\psfrag{i}{\sai$\mu$}
\psfrag{f}{}
\psfrag{a}{}
\psfrag{u}{\sai$y_1$}
\psfrag{e}{\sai$y_1$}
\psfrag{d}{\sai$\hspace{-.3cm}-\bar{y}_2$}
\psfrag{r}{\sai\hspace{-.3cm}$-\bar{y}_1$}
\psfrag{b}{\sai\hspace{-.4cm}\raisebox{.1cm}{$-\bar{y}_1$}}
\psfrag{m}{\sai\hspace{-.4cm}$y_2-y_1$}
%\psfrag{yq}{{ $\hspace{-.2cm}y_1$}}
%\psfrag{yb}{{ $\hspace{-.2cm}1-y_1$}}
\hspace{0.4cm}\raisebox{-1.4cm}{\includegraphics[width=4.2cm]{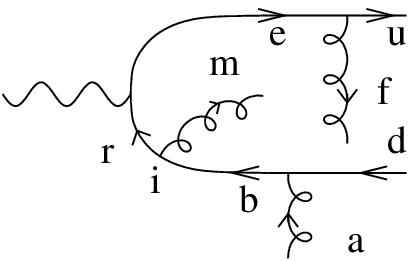}}\hspace{1.5cm}
%\raisebox{0cm}{\displaystyle -}
\displaystyle + \hspace{.8cm}
%\psfrag{yq}{{ $\hspace{-.2cm}y_2$}}
%\psfrag{yb}{{ $\hspace{-.3cm}1-y_2$}}
\psfrag{m}{\sai$\hspace{-.5cm} y_2-y_1$}
\raisebox{-1.4cm}{\includegraphics[width=4.2cm]{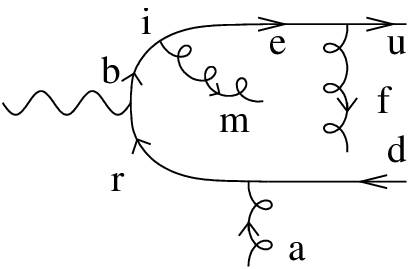}}\hspace{1.4cm}\right]$}}\nonumber\\
&&
\psfrag{u}{ \raisebox{0cm}{\sai$ y_1$}}
\psfrag{d}{ \raisebox{0cm}{\sai$\hspace{-.3cm}-\,\bar{y}_2$}}
\psfrag{f}{\ssi \hspace{-.5cm} \sai$y_1-y_2$}
\scalebox{1}{\begin{tabular}{cccccccc}
\hspace{-.8cm}=&
\psfrag{r}{\ssi \hspace{-.4cm} \raisebox{-.6cm}{$-\,\bar{y}_1$}}
\psfrag{s}{\ssi \hspace{-.3cm} \raisebox{.1cm}{$y_1$}}
\hspace{-.3cm}\raisebox{\rri}{\epsfig{file=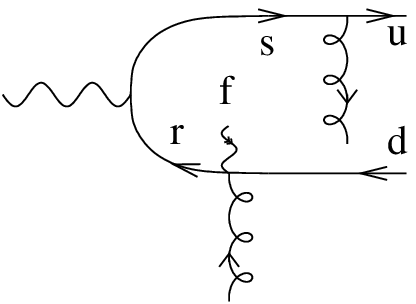,width=\tti}}
&\hspace{-.9cm}
$\displaystyle -$&
\psfrag{f}{\ssi \hspace{-.5cm}  \raisebox{-.2cm}{$y_1-y_2$}}
\psfrag{r}{\ssi \hspace{-.6cm} \raisebox{-.6cm}{$-\,\bar{y}_2$}}
\psfrag{s}{\ssi \raisebox{.1cm}{$y_1$}}
 %\hspace{-3.2cm}
\raisebox{\rri}{ \hspace{-2.2cm}\epsfig{file=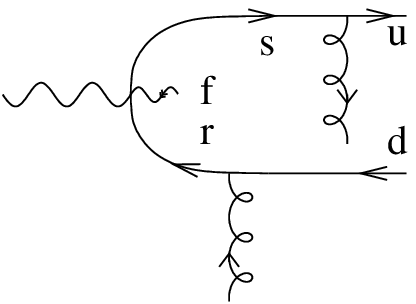,width=\tti}}
& \hspace{-1.2cm}
+&
\psfrag{f}{\ssi \hspace{-.5cm}  \raisebox{-.2cm}{$y_1-y_2$}}
\psfrag{r}{\ssi \hspace{-.6cm} \raisebox{-.6cm}{$-\,\bar{y}_2$}}
\psfrag{s}{\ssi \raisebox{.1cm}{$y_1$}}
\hspace{-0.1cm}
\raisebox{\rri}{\epsfig{file=bG1Ward2.eps,width=\tti}}
&\hspace{0cm}
$\displaystyle -$&
\psfrag{f}{\ssi \hspace{-.9cm}  \raisebox{0cm}{$y_1-y_2$}}
\psfrag{r}{\ssi \hspace{-.6cm} \raisebox{-.6cm}{$-\,\bar{y}_2$}}
\psfrag{s}{\ssi \raisebox{.1cm}{$y_2$}}
\hspace{-.1cm}
\raisebox{\rri}{\epsfig{file=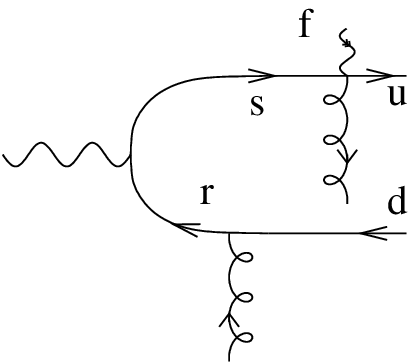,width=\tti}}
\\
\\
\hspace{-.5cm}=&
\psfrag{u}{\ssi \raisebox{0cm}{$ y_1$}}
\psfrag{d}{ \raisebox{0cm}{\sai$\hspace{-.3cm}-\,\bar{y}_2$}}
\psfrag{r}{\ssi \hspace{-.4cm} \raisebox{-.6cm}{$-\,\bar{y}_1$}}
\psfrag{s}{\ssi \hspace{-.3cm} \raisebox{.1cm}{$y_1$}}
\hspace{0cm}\raisebox{\rri}{\epsfig{file=bG1Ward1.eps,width=\tti}}
&\hspace{.8cm} $\displaystyle -$&
\psfrag{u}{\ssi \raisebox{0cm}{$ y_1$}}
\psfrag{d}{ \raisebox{0cm}{\sai$\hspace{-.3cm}-\,\bar{y}_2$}}
\psfrag{f}{\ssi \hspace{-.9cm}  \raisebox{0cm}{$y_1-y_2$}}
\psfrag{r}{\ssi \hspace{-.6cm} \raisebox{-.6cm}{$-\,\bar{y}_2$}}
\psfrag{s}{\ssi \raisebox{.1cm}{$y_2$}}
\hspace{.5cm}
\raisebox{\rri}{\epsfig{file=bG2Ward2.eps,width=\tti}}
               \end{tabular}}
%\caption{The collinear Ward identity applied to the 3-parton "Abelian" contributions (aG1) and (aG2).}
\label{Ward3body_b}
\eqa
The $p_2$ term obtained after applying the identity (\ref{vertex3Ward1}) to the diagrams (btG1) and (btG2) gives, respectively
\beq
\label{Ward_btG1}
\displaystyle - (y_2-y_1) \, p_\mu \
\psfrag{q}{}
\psfrag{i}{\sai\raisebox{.1cm}{$\mu$}}
\psfrag{u}{\sai$y_1$}
\psfrag{f}{}%{$k'$}
\psfrag{a}{}%{$k$}
\psfrag{b}{}
\psfrag{e}{\sai$y_1$}
\psfrag{c}{\hspace{-.35cm}\sai$-\bar{y}_1$}
%\psfrag{d}{\sai$-\bar{y}_2$}
\psfrag{d}{ \raisebox{0cm}{\sai$\hspace{-.3cm}-\,\bar{y}_2$}}
\psfrag{m}{\sai$y_2-y_1$}
%\psfrag{yq}{{ $\hspace{-.2cm}y_1$}}
%\psfrag{yb}{{ $\hspace{-.2cm}1-y_1$}}
\hspace{0cm}\raisebox{-2cm}{\includegraphics[width=4.2cm]{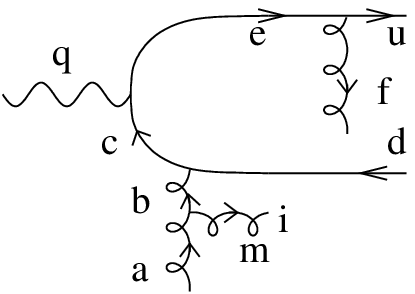}}
\hspace{0.5cm}
\raisebox{-2cm}{\rotatebox{90}{$\underline{\rule{3cm}{0pt}}$}}_{\displaystyle ''p_2''}
\hspace{0cm}\displaystyle = \, -\left[   1-\frac{k_1^2}{(k_1-\ell_g)^2}\right] \hspace{0.2cm}
\psfrag{u}{\sai$y_1$}
\psfrag{d}{ \raisebox{0cm}{\sai$\hspace{-.3cm}-\,\bar{y}_2$}}
\psfrag{f}{\sai$\hspace{-.2cm}y_1-y_2$}
\psfrag{s}{\sai$y_1$}
\psfrag{r}{\sai\hspace{-1cm}$-\bar{y}_1$}
\raisebox{-2cm}{\includegraphics[width=4.2cm]{bG1Ward1.eps}}
\eq
and
\beq
\label{Ward_btG2}
\hspace{0.2cm}\displaystyle  (y_2-y_1) \, p_\mu \
\psfrag{q}{}
\psfrag{i}{\raisebox{.1cm}{\sai$\mu$}}
\psfrag{u}{\raisebox{.5cm}{\sai$y_1$}}
\psfrag{e}{\raisebox{.5cm}{\sai$y_2$}}
\psfrag{f}{}%{$k'$}
\psfrag{a}{}%{$k$}
\psfrag{b}{}
\psfrag{r}{\hspace{-.35cm}\sai$-\bar{y}_2$}
%\psfrag{r}{\ssi \hspace{-.4cm} \raisebox{.2cm}{\sai$-\,\bar{y}_2$}}
\psfrag{c}{\sai\hspace{-.35cm}$\bar{y}_1$}
\psfrag{d}{\sai$-\bar{y}_2$}
\psfrag{m}{\sai$y_2-y_1$}
%\psfrag{yq}{{ $\hspace{-.2cm}y_1$}}
%\psfrag{yb}{{ $\hspace{-.2cm}1-y_1$}}
\hspace{0cm}\raisebox{-1.5cm}{\includegraphics[width=4.2cm]{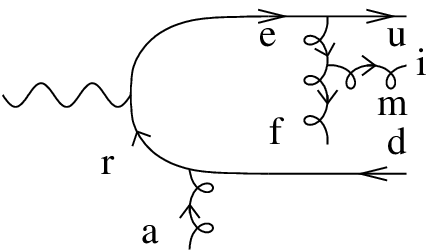}}\hspace{0.5cm}
\raisebox{-1.6cm}{\rotatebox{90}{$\underline{\rule{2.7cm}{0pt}}$}}_{\displaystyle ''p_2''}  %&
\hspace{-0.1cm} \displaystyle = \left[   1-\frac{k_2^2}{(k_2+\ell_g)^2}\right]%&
\hspace{0.2cm}
\psfrag{u}{\sai$y_1$}
\psfrag{d}{\sai$-\bar{y}_2$}
\psfrag{f}{\raisebox{.3cm}{\sai$\hspace{-.2cm}y_1-y_2$}}
\psfrag{r}{\ssi \hspace{-1cm} \raisebox{-.6cm}{\sai$-\,\bar{y}_2$}}
\psfrag{s}{\hspace{0cm}{\sai$y_2$}}
\raisebox{-2cm}{\includegraphics[width=4.2cm]{bG2Ward2.eps}}
\eq
In the perturbative Regge limit we are considering here, the terms $k^2/(k-\ell_g)^2$ and $k'^2/(k'+\ell_g)^2$ can be neglected, since they are of the order of $k^2/s\,,$ and therefore  Eqs.(\ref{Ward3body_b}, \ref{Ward_btG1}, \ref{Ward_btG2})
exhibit an explicit cancellation.

Turning now to the case of graphs (dG1) and (dG2) one gets, using again the Ward identity,
\beqa
% \psfrag{i}{$\mu$}
% \psfrag{u}{$y_1$}
% \psfrag{d}{$y_2-1$}
% \psfrag{m}{$y_2-y_1$}
% \psfrag{yq}{{ $\hspace{-.2cm}y_1$}}
% \psfrag{dy}{\raisebox{.04cm}{ $\hspace{-.3cm}y_2-y_1$}\raisebox{.6cm}{$\!\!\mu$}}
\psfrag{yb}{{\sai $\hspace{-.2cm}1-y_2$}}
&&\scalebox{1}
 {\raisebox{1.8cm}{$ \displaystyle - (y_2-y_1) \, p_\mu \quad \left[
\psfrag{a}{}
\psfrag{f}{}
\psfrag{i}{\sai$\mu$}
\psfrag{u}{\sai$y_1$}
\psfrag{d}{\ssi \raisebox{0cm}{\hspace{-.3cm}$-\,\bar{y}_2$}}
\psfrag{b}{\hspace{-.2cm}\sai\raisebox{.1cm}{$y_2$}}\psfrag{r}{\hspace{-.35cm}\sai$-\bar{y}_2$}
\psfrag{e}{\sai\hspace{-.2cm}$y_1$}
\psfrag{m}{\sai\hspace{-.4cm}$y_2-y_1$}
%\psfrag{yq}{{ $\hspace{-.2cm}y_1$}}
%\psfrag{yb}{{ $\hspace{-.2cm}1-y_1$}}
\hspace{0.4cm}\raisebox{-1.4cm}{\includegraphics[width=4.2cm]{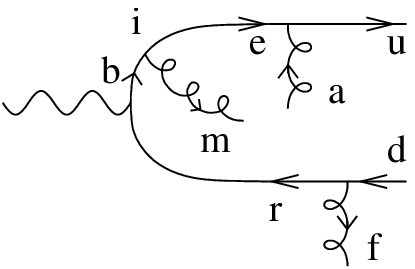}}\hspace{1.5cm}
%\raisebox{0cm}{\displaystyle -}
\displaystyle + \hspace{.8cm}
%\psfrag{yq}{{ $\hspace{-.2cm}y_2$}}
%\psfrag{yb}{{ $\hspace{-.3cm}1-y_2$}}
\psfrag{b}{\hspace{0cm}\sai\raisebox{.3cm}{$y_1$}}
\psfrag{r}{\hspace{-.35cm}\sai$-\bar{y}_1$}
\psfrag{m}{\sai\raisebox{.2cm}{$\hspace{-.3cm} y_2-y_1$}}
\raisebox{-1.4cm}{\includegraphics[width=4.2cm]{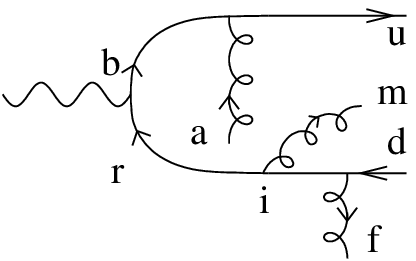}}\hspace{1.4cm}\right]$}}\nonumber\\
&&
\psfrag{u}{ \raisebox{.1cm}{\sai$ y_1$}}
\psfrag{d}{\ssi \raisebox{0cm}{\hspace{-.3cm}$-\,\bar{y}_2$}}
\psfrag{f}{\ssi \hspace{-.5cm} \sai$y_1-y_2$}
\scalebox{1}{\begin{tabular}{cccccccc}
\hspace{-.5cm}=&
\psfrag{a}{}
\psfrag{o}{}
\psfrag{u}{\ssi \raisebox{0cm}{$ y_1$}}
\psfrag{d}{\ssi \raisebox{0cm}{\hspace{-.3cm}$-\,\bar{y}_2$}}
\psfrag{e}{\ssi \hspace{-.4cm} \raisebox{0cm}{$y_1$}}
\psfrag{r}{\ssi \hspace{-.6cm} \raisebox{0cm}{$-\,\bar{y}_1$}}
\hspace{0cm}\raisebox{-1.4cm}{\epsfig{file=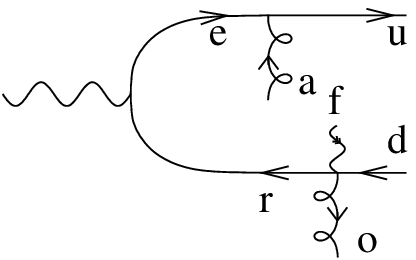,width=\ttiB}}
&\hspace{.8cm} $\displaystyle -$&
\psfrag{a}{}
\psfrag{o}{}
\psfrag{u}{\ssi \raisebox{0cm}{$ y_1$}}
\psfrag{d}{\ssi \raisebox{0cm}{\hspace{-.3cm}$-\,\bar{y}_2$}}
\psfrag{e}{\ssi \hspace{-.4cm} \raisebox{0cm}{$y_2$}}
\psfrag{f}{\ssi \hspace{-.9cm}  \raisebox{0cm}{$y_1-y_2$}}
\psfrag{r}{\ssi \hspace{-.6cm} \raisebox{0cm}{$-\,\bar{y}_2$}}
\psfrag{s}{\ssi \raisebox{.1cm}{$y_2$}}
\hspace{.5cm}
\raisebox{-1.4cm}{\epsfig{file=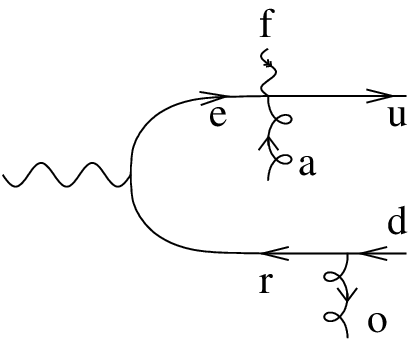,width=\ttiB}}
               \end{tabular}}
%\caption{The collinear Ward identity applied to the 3-parton "Abelian" contributions (aG1) and (aG2).}
\label{Ward3body_d}
\eqa
The $p_2$ term obtained after applying the identity (\ref{vertex3Ward1}) to the diagrams (dtG1) and (dtG2) gives, respectively
\beq
\label{Ward_dtG1}
%\hspace{-.6cm}\raisebox{0.2cm}{$ \displaystyle  (y_2-y_1) \, p_\mu$}
\hspace{0.2cm} \displaystyle  (y_2-y_1) \, p_\mu
\psfrag{q}{}
\psfrag{i}{\raisebox{.1cm}{\sai $\mu$}}
\psfrag{u}{\sai $y_1$}
\psfrag{f}{}%{$k'$}
\psfrag{a}{}%{$k$}
\psfrag{b}{}
\psfrag{e}{\sai$y_2$}
\psfrag{c}{\sai\hspace{-.35cm}$-\bar{y}_2$}
\psfrag{d}{\ssi \raisebox{0cm}{\hspace{-.3cm}$-\,\bar{y}_2$}}
\psfrag{m}{\sai$y_2-y_1$}
%\psfrag{yq}{{ $\hspace{-.2cm}y_1$}}
%\psfrag{yb}{{ $\hspace{-.2cm}1-y_1$}}
\hspace{0.4cm}\raisebox{-1.6cm}{\includegraphics[width=4.5cm]{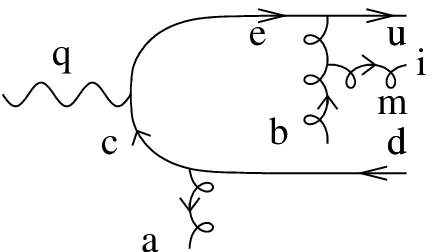}}\hspace{.4cm}\raisebox{-1.7cm}{\rotatebox{90}{$\underline{\rule{2.7cm}{0pt}}$}}_{\displaystyle ''p_2''}  %\hspace{-.3cm}\raisebox{0cm}{$ \displaystyle =  \, \left[   1-\frac{k^2}{(k-\ell_g)^2}\right]$}&
\hspace{-.3cm}\displaystyle =  \, \left[   1-\frac{k_1^2}{(k_1-\ell_g)^2}\right]
\psfrag{a}{}
\psfrag{o}{}
\psfrag{u}{\ssi \raisebox{0cm}{$ y_1$}}
\psfrag{d}{\ssi \raisebox{0cm}{\hspace{-.3cm}$-\,\bar{y}_2$}}
\psfrag{e}{\ssi \hspace{-.4cm} \raisebox{0cm}{$y_2$}}
\psfrag{f}{\ssi \hspace{-.4cm}  \raisebox{0cm}{$y_1-y_2$}}
\psfrag{r}{\ssi \hspace{-.6cm} \raisebox{0cm}{$-\,\bar{y}_2$}}
\psfrag{s}{\ssi \raisebox{.1cm}{$y_2$}}
\hspace{0.3cm}\raisebox{-1.4cm}{\epsfig{file=dG1Ward2.eps,width=\ttiB}}\,\,
\eq
and
\beq
\label{Ward_dtG2}
%\hspace{-.6cm}\raisebox{-.1cm}{$ \displaystyle - (y_2-y_1) \, p_\mu$}
 \hspace{0cm}\displaystyle - (y_2-y_1) \, p_\mu
\psfrag{q}{}
\psfrag{i}{\sai\raisebox{.1cm}{$\mu$}}
\psfrag{u}{\sai\raisebox{0cm}{$y_1$}}
\psfrag{f}{}%{$k'$}
\psfrag{a}{}%{$k$}
\psfrag{b}{}
\psfrag{e}{\sai$y_1$}
\psfrag{c}{\sai\hspace{-.35cm}$-\bar{y}_1$}
\psfrag{d}{\ssi \raisebox{0cm}{\hspace{-.3cm}$-\,\bar{y}_2$}}
\psfrag{m}{\sai$y_2-y_1$}
%\psfrag{yq}{{ $\hspace{-.2cm}y_1$}}
%\psfrag{yb}{{ $\hspace{-.2cm}1-y_1$}}
\hspace{0.4cm}\raisebox{-2.1cm}{\includegraphics[width=4.5cm]{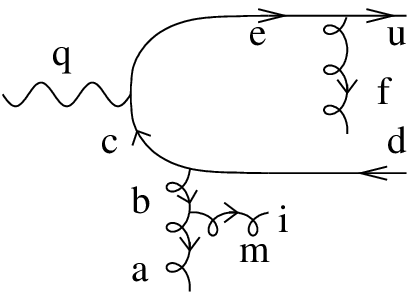}}\hspace{.4cm}
\raisebox{-2.1cm}{\rotatebox{90}{$\underline{\rule{3.2cm}{0pt}}$}}_{\displaystyle ''p_2''}
\hspace{-0.3cm}\displaystyle =  \, - \left[   1-\frac{k_2^2}{(k_2+\ell_g)^2}\right]
\psfrag{a}{}
\psfrag{o}{}
\psfrag{f}{\ssi \hspace{-.4cm}  \raisebox{0cm}{$y_1-y_2$}}
\psfrag{u}{\sai$y_1$}
\psfrag{d}{\ssi \raisebox{0cm}{\hspace{-.3cm}$-\,\bar{y}_2$}}
\psfrag{e}{\ssi \hspace{-.4cm} \raisebox{0cm}{$y_1$}}
\psfrag{r}{\ssi \hspace{-.6cm} \raisebox{0cm}{$-\,\bar{y}_1$}}
\hspace{0.3cm}\raisebox{-1.4cm}{\epsfig{file=dG2Ward1.eps,width=\ttiB}}\ \ .
\eq
Contributions (\ref{Ward3body_d}), (\ref{Ward_dtG1}) and (\ref{Ward_dtG2})
add to zero.

The proofs for (aG2) goes along the same line and reads
\beq
\raisebox{-.4cm}{
\begin{tabular}{ccc}
\hspace{-0.5cm}$\displaystyle - (y_2-y_1) \, p_\mu$
\psfrag{yb}{{\sai $\hspace{-.2cm}1-y_2$}}
\psfrag{a}{}
\psfrag{f}{}
\psfrag{i}{\raisebox{.7cm}{\hspace{-.3cm}\sai$\mu$}}
\psfrag{u}{\sai$y_1$}
\psfrag{d}{\ssi \raisebox{0cm}{\hspace{-.3cm}$-\,\bar{y}_2$}}
\psfrag{c}{\hspace{-.35cm}\sai$-\bar{y}_2$}
\psfrag{e}{\raisebox{0cm}{\hspace{-.1cm}\sai$-\bar{y}_1$}}
\psfrag{r}{\hspace{-.35cm}\sai$-\bar{y}_1$}
\psfrag{m}{\raisebox{.2cm}{\sai\hspace{-.4cm}$y_2-y_1$}}
%\psfrag{yq}{{ $\hspace{-.2cm}y_1$}}
%\psfrag{yb}{{ $\hspace{-.2cm}1-y_1$}}
\hspace{0cm}\raisebox{-1.6cm}{\includegraphics[width=4.3cm]{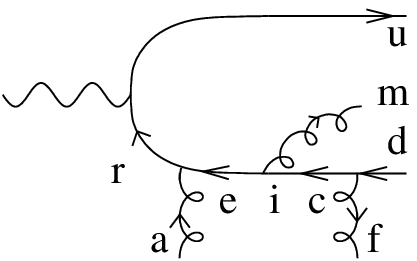}}
&
\psfrag{u}{ \raisebox{0cm}{\sai$ y_1$}}
\psfrag{d}{ \raisebox{0cm}{\sai$\hspace{-.3cm}-\,\bar{y}_2$}}
\psfrag{f}{\ssi \hspace{-.5cm} \sai$y_1-y_2$}
\hspace{.15cm}
$=$
\psfrag{u}{\ssi \raisebox{0cm}{$ y_1$}}
\psfrag{d}{\ssi \raisebox{0cm}{\hspace{-.3cm}$-\,\bar{y}_2$}}
\psfrag{r}{\ssi \hspace{-.6cm} \raisebox{-.6cm}{$-\,\bar{y}_1$}}
\psfrag{s}{\ssi \hspace{-.4cm} \raisebox{0cm}{$-\,\bar{y}_1$}}
\hspace{0.1cm}\raisebox{-1.4cm}{\epsfig{file=aG2Ward1.eps,width=\ttiB}}
\hspace{.8cm}
&\hspace{-0.6cm}$-$
\psfrag{u}{\ssi \raisebox{.1cm}{$ y_1$}}
\psfrag{d}{\ssi \raisebox{0cm}{\hspace{-.3cm}$-\,\bar{y}_2$}}
\psfrag{f}{\ssi \hspace{-.4cm}  \raisebox{0cm}{$y_1-y_2$}}
\psfrag{r}{\ssi \hspace{-.6cm} \raisebox{-.6cm}{$-\,\bar{y}_1$}}
\psfrag{s}{\ssi \raisebox{0cm}{\hspace{-.4cm} $-\,\bar{y}_2$}}
\hspace{0cm}
\raisebox{-1.4cm}{\epsfig{file=aG1Ward1.eps,width=\ttiB}}
\end{tabular}}\hspace{.1cm} ,
\label{Ward3body_a}
\eq
The $p_2$ term obtained after applying the identity (\ref{vertex3Ward1}) to
the diagrams (atG1) and (atG2) gives, respectively
\beq
\label{Ward_atG1}
%\hspace{-.6cm}\raisebox{0.2cm}{$ \displaystyle  (y_2-y_1) \, p_\mu$}
\hspace{0.2cm} \displaystyle  (y_2-y_1) \, p_\mu
\psfrag{q}{}
\psfrag{i}{\raisebox{.1cm}{\sai $\mu$}}
\psfrag{u}{\sai $y_1$}
\psfrag{f}{}%{$k'$}
\psfrag{a}{}%{$k$}
\psfrag{b}{}
\psfrag{e}{\sai$-\bar{y}_2$}
\psfrag{c}{\sai\hspace{-.35cm}$-\bar{y}_1$}
\psfrag{d}{\ssi \raisebox{0cm}{\hspace{-.3cm}$-\,\bar{y}_2$}}
\psfrag{m}{\hspace{-0.2cm}\sai$y_2-y_1$}
%\psfrag{yq}{{ $\hspace{-.2cm}y_1$}}
%\psfrag{yb}{{ $\hspace{-.2cm}1-y_1$}}
\hspace{0.4cm}\raisebox{-2.1cm}{\includegraphics[width=4.5cm]{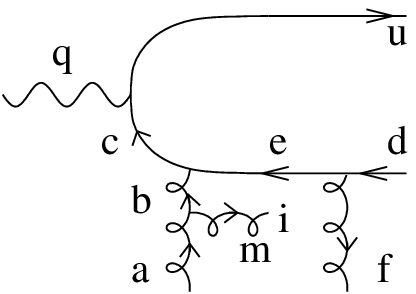}}
\hspace{.4cm}
\raisebox{-2.1cm}{\rotatebox{90}{$\underline{\rule{3.2cm}{0pt}}$}}_{\displaystyle ''p_2''}
%\hspace{-.3cm}\raisebox{0cm}{$ \displaystyle =  \, \left[   1-\frac{k^2}{(k-\ell_g)^2}\right]$}&
\hspace{-.3cm}\displaystyle =  \, \left[   1-\frac{k_1^2}{(k_1-\ell_g)^2}\right]
\psfrag{a}{}
\psfrag{o}{}
\psfrag{u}{\ssi \raisebox{0cm}{$ y_1$}}
\psfrag{d}{\ssi \raisebox{0cm}{\hspace{-.3cm}$-\,\bar{y}_2$}}
\psfrag{f}{\ssi \hspace{-.4cm}  \raisebox{0cm}{$y_1-y_2$}}
\psfrag{r}{\ssi \hspace{-.8cm} \raisebox{-0.4cm}{$-\,\bar{y}_1$}}
\psfrag{s}{\ssi \raisebox{0cm}{{\hspace{-.2cm}$-\,\bar{y}_2$}}}
\hspace{0.3cm}\raisebox{-1.4cm}{\epsfig{file=aG1Ward1.eps,width=\ttiB}}\,\,
\eq
and
\beq
\label{Ward_atG2}
%\hspace{-.6cm}\raisebox{-.1cm}{$ \displaystyle - (y_2-y_1) \, p_\mu$}
 \hspace{0cm}\displaystyle - (y_2-y_1) \, p_\mu
\psfrag{q}{}
\psfrag{i}{\sai\raisebox{.1cm}{$\mu$}}
\psfrag{u}{\sai\raisebox{0cm}{$y_1$}}
\psfrag{f}{}%{$k'$}
\psfrag{a}{}%{$k$}
\psfrag{b}{}
\psfrag{e}{\hspace{-.2cm}\sai$-\bar{y}_1$}
\psfrag{c}{\sai\hspace{-.35cm}$-\bar{y}_1$}
\psfrag{d}{\ssi \raisebox{0cm}{\hspace{-.3cm}$-\,\bar{y}_2$}}
\psfrag{m}{\hspace{-.1cm}\sai$y_2-y_1$}
%\psfrag{yq}{{ $\hspace{-.2cm}y_1$}}
%\psfrag{yb}{{ $\hspace{-.2cm}1-y_1$}}
\hspace{0.4cm}\raisebox{-2cm}{\includegraphics[width=4.5cm]{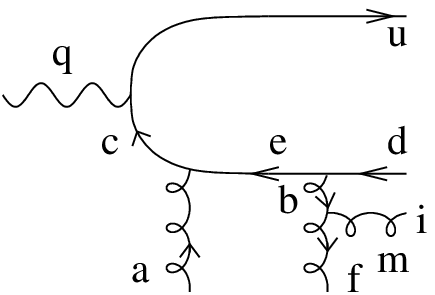}}\hspace{.4cm}
\raisebox{-2.1cm}{\rotatebox{90}{$\underline{\rule{3.2cm}{0pt}}$}}_{\displaystyle ''p_2''}
\hspace{-0.3cm}\displaystyle =  \, - \left[   1-\frac{k_2^2}{(k_2+\ell_g)^2}\right]
\psfrag{a}{}
\psfrag{o}{}
\psfrag{f}{\ssi \hspace{-.4cm}  \raisebox{0cm}{$y_1-y_2$}}
\psfrag{u}{\sai$y_1$}
\psfrag{d}{\ssi \raisebox{0cm}{\hspace{-.3cm}$-\,\bar{y}_2$}}
\psfrag{e}{\ssi \hspace{-.4cm} \raisebox{0cm}{$y_1$}}
\psfrag{s}{\ssi \raisebox{0cm}{{\hspace{-.3cm}$-\,\bar{y}_2$}}}
\psfrag{r}{\ssi \hspace{-.8cm} \raisebox{-0.3cm}{$-\,\bar{y}_1$}}
\hspace{0.3cm}\raisebox{-1.4cm}{\epsfig{file=aG2Ward1.eps,width=\ttiB}}\ \ .
\eq
Contributions (\ref{Ward3body_a}), (\ref{Ward_atG1}) and (\ref{Ward_atG2})
add to zero in the Regge limit.

%%%%%%%%%%%%%%%
The proofs for (cG2) goes along the same line and reads
\beq
%\begin{tabular}{ccc}
\hspace{-0.2cm}\displaystyle - (y_2-y_1) \, p_\mu \
\psfrag{yb}{{\sai $\hspace{-.2cm}1-y_2$}}
\psfrag{a}{}
\psfrag{f}{}
\psfrag{i}{\raisebox{0.1cm}{\hspace{-.05cm}\sai$\mu$}}
\psfrag{u}{\raisebox{.5cm}{\sai$y_1$}}
\psfrag{d}{\ssi \raisebox{0cm}{\hspace{-.3cm}$-\,\bar{y}_2$}}
\psfrag{b}{\raisebox{0.2cm}{\hspace{-.15cm}\sai$y_2$}}
\psfrag{e}{\raisebox{0.5cm}{\hspace{-.3cm}\sai$y_2$}}
\psfrag{r}{\raisebox{0.5cm}{\hspace{-.15cm}\sai$y_1$}}
\psfrag{m}{\raisebox{.1cm}{\sai\hspace{-.4cm}$y_2-y_1$}}
%\psfrag{yq}{{ $\hspace{-.2cm}y_1$}}
%\psfrag{yb}{{ $\hspace{-.2cm}1-y_1$}}
\hspace{0cm}\raisebox{-.8cm}{\includegraphics[width=4.3cm]{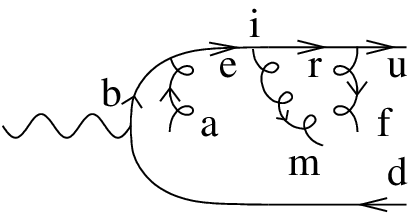}}
%&
\hspace{.15cm}
=
\psfrag{f}{\ssi \hspace{-.5cm} \sai$y_1-y_2$}
\psfrag{o}{}
\psfrag{u}{\ssi \raisebox{0cm}{$ y_1$}}
\psfrag{d}{\ssi \raisebox{0cm}{\hspace{-.3cm}$-\,\bar{y}_2$}}
\psfrag{e}{\raisebox{0.4cm}{\hspace{-.3cm}\sai$y_2$}}
\psfrag{r}{\ssi \hspace{-.2cm} \raisebox{0cm}{$y_1$}}
\hspace{0.1cm}\raisebox{-.7cm}{\epsfig{file=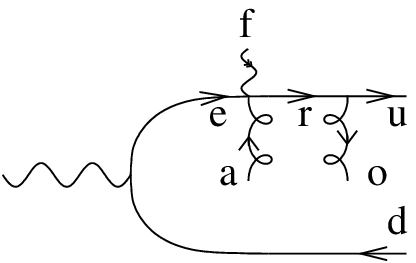,width=\ttiB}}
\hspace{.8cm}
%&
\hspace{-0.6cm}- \ \
\psfrag{o}{}
\psfrag{u}{\ssi \raisebox{0cm}{$ y_1$}}
\psfrag{d}{\ssi \raisebox{0cm}{\hspace{-.3cm}$-\,\bar{y}_2$}}
\psfrag{f}{\ssi \hspace{-.9cm}  \raisebox{0cm}{$y_1-y_2$}}
\psfrag{e}{\raisebox{0.4cm}{\hspace{-.3cm}\sai$y_2$}}
\psfrag{r}{\ssi \hspace{-.2cm} \raisebox{0cm}{$y_2$}}
\psfrag{s}{\ssi \raisebox{.1cm}{$y_2$}}
\hspace{0cm}
\raisebox{-.7cm}{\epsfig{file=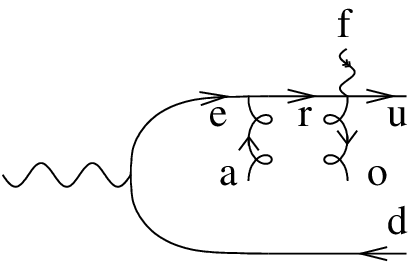,width=\ttiB}}
%\end{tabular}}
\hspace{.5cm} ,
\label{Ward3body_c}
\eq
The $p_2$ term obtained after applying the identity (\ref{vertex3Ward1}) to the diagrams (ctG1) and (ctG2) gives, respectively
\beq
\label{Ward_ctG1}
%\hspace{-.6cm}\raisebox{0.2cm}{$ \displaystyle  (y_2-y_1) \, p_\mu$}
\hspace{0.2cm} \displaystyle  -(y_2-y_1) \, p_\mu
\psfrag{q}{}
\psfrag{i}{\raisebox{-.1cm}{\sai \hspace{.2cm}$\mu$}}
\psfrag{u}{\sai $y_1$}
\psfrag{f}{}%{$k'$}
\psfrag{a}{}%{$k$}
\psfrag{b}{\raisebox{0.2cm}{\sai\hspace{-.1cm}$y_2$}}
\psfrag{e}{\raisebox{-.4cm}{\sai$y_1$}}
\psfrag{c}{\sai\hspace{-.35cm}$-\bar{y}_1$}
\psfrag{d}{\ssi \raisebox{0cm}{\hspace{-.3cm}$-\,\bar{y}_2$}}
\psfrag{m}{\hspace{-0.4cm}\sai$y_2-y_1$}
%\psfrag{yq}{{ $\hspace{-.2cm}y_1$}}
%\psfrag{yb}{{ $\hspace{-.2cm}1-y_1$}}
\hspace{0.4cm}\raisebox{-.9cm}{\includegraphics[width=4.5cm]{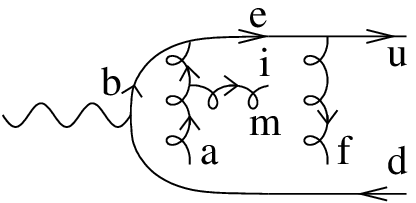}}\hspace{.4cm}\raisebox{-1.1cm}{\rotatebox{90}{$\underline{\rule{2.1cm}{0pt}}$}}_{\displaystyle ''p_2''}  %\hspace{-.3cm}\raisebox{0cm}{$ \displaystyle =  \, \left[   1-\frac{k^2}{(k-\ell_g)^2}\right]$}&
\hspace{-.3cm}\displaystyle =  \,- \left[   1-\frac{k_1^2}{(k_1-\ell_g)^2}\right]
\psfrag{a}{}
\psfrag{o}{}
\psfrag{u}{\ssi \raisebox{0cm}{$ y_1$}}
\psfrag{d}{\ssi \raisebox{0cm}{\hspace{-.3cm}$-\,\bar{y}_2$}}
\psfrag{e}{\ssi \raisebox{0cm}{\hspace{-.2cm}$ y_2$}}
\psfrag{f}{\ssi \hspace{-.4cm}  \raisebox{0cm}{$y_1-y_2$}}
\psfrag{r}{\ssi \hspace{-.2cm} \raisebox{0cm}{$y_1$}}
\psfrag{s}{\ssi \raisebox{0cm}{{\hspace{-.2cm}$-\,\bar{y}_2$}}}
\hspace{0.3cm}\raisebox{-.7cm}{\epsfig{file=cG2Ward1.eps,width=\ttiB}}\,\,
\eq
and
\beq
\label{Ward_ctG2}
%\hspace{-.6cm}\raisebox{-.1cm}{$ \displaystyle - (y_2-y_1) \, p_\mu$}
 \hspace{0cm}\displaystyle  (y_2-y_1) \, p_\mu
\psfrag{q}{}
\psfrag{i}{\sai\raisebox{.1cm}{$\mu$}}
\psfrag{u}{\sai\raisebox{0cm}{$y_1$}}
\psfrag{f}{}%{$k'$}
\psfrag{a}{}%{$k$}
\psfrag{b}{\raisebox{0.2cm}{\sai\hspace{-.1cm}$y_2$}}
\psfrag{e}{\raisebox{-.4cm}{\sai$y_1$}}
\psfrag{c}{}
%\sai\hspace{-.35cm}$-\bar{y}_1$}
\psfrag{d}{\ssi \raisebox{0cm}{\hspace{-.3cm}$-\,\bar{y}_2$}}
\psfrag{m}{\hspace{-.1cm}\sai$y_2-y_1$}
%\psfrag{yq}{{ $\hspace{-.2cm}y_1$}}
%\psfrag{yb}{{ $\hspace{-.2cm}1-y_1$}}
\hspace{0.4cm}\raisebox{-.8cm}{\includegraphics[width=4.5cm]{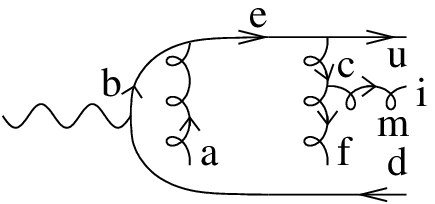}}\hspace{.4cm}
\raisebox{-1.1cm}{\rotatebox{90}{$\underline{\rule{2.1cm}{0pt}}$}}_{\displaystyle ''p_2''}
\hspace{-0.3cm}\displaystyle =  \,  \left[   1-\frac{k_2^2}{(k_2+\ell_g)^2}\right]
\psfrag{a}{}
\psfrag{o}{}
\psfrag{u}{\ssi \raisebox{0cm}{$ y_1$}}
\psfrag{d}{\ssi \raisebox{0cm}{\hspace{-.3cm}$-\,\bar{y}_2$}}
\psfrag{e}{\ssi \raisebox{0cm}{\hspace{-.2cm}$ y_2$}}
\psfrag{f}{\ssi \hspace{-.4cm}  \raisebox{0cm}{$y_1-y_2$}}
\psfrag{r}{\ssi \hspace{-.2cm} \raisebox{0cm}{$y_2$}}
\psfrag{s}{\ssi \raisebox{0cm}{{\hspace{-.2cm}$-\,\bar{y}_2$}}}
\hspace{0.3cm}\raisebox{-.7cm}{\epsfig{file=cG2Ward2.eps,width=\ttiB}}\ \ .
\eq
Contributions (\ref{Ward3body_c}), (\ref{Ward_ctG1}) and (\ref{Ward_ctG2})
add to zero in the Regge limit.

%%%%%%%%%%%%%%%

The proofs for (eG2) goes along the same line and reads
\beq
\raisebox{-.4cm}{
\begin{tabular}{ccc}
\hspace{-0.5cm}$\displaystyle - (y_2-y_1) \, p_\mu$
\psfrag{yb}{{\sai $\hspace{-.2cm}1-y_2$}}
\psfrag{a}{}
\psfrag{f}{}
\psfrag{i}{\raisebox{.7cm}{\hspace{-.3cm}\sai$\mu$}}
\psfrag{u}{\sai$y_1$}
\psfrag{d}{\ssi \raisebox{0cm}{\hspace{-.3cm}$-\,\bar{y}_2$}}
\psfrag{c}{\hspace{-.35cm}\sai$-\bar{y}_2$}
\psfrag{e}{\raisebox{0cm}{\hspace{-.1cm}\sai$-\bar{y}_1$}}
\psfrag{r}{\hspace{-.35cm}\sai$-\bar{y}_1$}
\psfrag{m}{\raisebox{.2cm}{\sai\hspace{-.4cm}$y_2-y_1$}}
%\psfrag{yq}{{ $\hspace{-.2cm}y_1$}}
%\psfrag{yb}{{ $\hspace{-.2cm}1-y_1$}}
\hspace{0cm}\raisebox{-1.6cm}{\includegraphics[width=4.3cm]{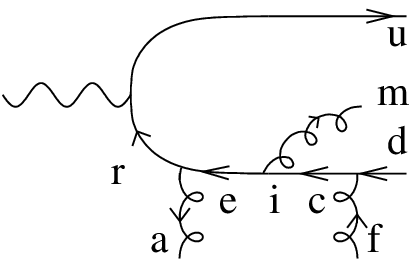}}
&
\psfrag{u}{ \raisebox{0cm}{\sai$ y_1$}}
\psfrag{d}{ \raisebox{0cm}{\sai$\hspace{-.3cm}-\,\bar{y}_2$}}
\psfrag{f}{\ssi \hspace{-.5cm} \sai$y_1-y_2$}
\hspace{.15cm}
$=$
\psfrag{u}{\ssi \raisebox{0cm}{$ y_1$}}
\psfrag{d}{\ssi \raisebox{0cm}{\hspace{-.3cm}$-\,\bar{y}_2$}}
\psfrag{r}{\ssi \hspace{-.8cm} \raisebox{-0.3cm}{$-\,\bar{y}_1$}}
\psfrag{s}{\ssi \hspace{-.4cm} \raisebox{0cm}{$-\,\bar{y}_1$}}
\hspace{0.1cm}\raisebox{-1.4cm}{\epsfig{file=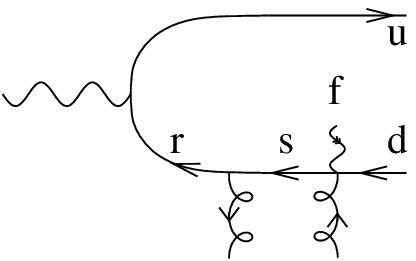,width=\ttiB}}
\hspace{.8cm}
&\hspace{-0.6cm}$-$
\psfrag{u}{\ssi \raisebox{0cm}{$ y_1$}}
\psfrag{d}{\ssi \raisebox{0cm}{\hspace{-.3cm}$-\,\bar{y}_2$}}
\psfrag{f}{\ssi \hspace{-.4cm}  \raisebox{0cm}{$y_1-y_2$}}
\psfrag{r}{\ssi \hspace{-.8cm} \raisebox{-0.3cm}{$-\,\bar{y}_1$}}
\psfrag{s}{\ssi \raisebox{0cm}{\hspace{-.4cm} $-\,\bar{y}_2$}}
\hspace{0cm}
\raisebox{-1.4cm}{\epsfig{file=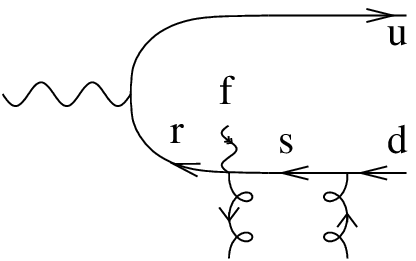,width=\ttiB}}
\end{tabular}}\hspace{.1cm} ,
\label{Ward3body_e}
\eq
The $p_2$ term obtained after applying the identity (\ref{vertex3Ward1}) to the diagrams (etG1) and (etG2) gives, respectively
\beq
\label{Ward_etG1}
%\hspace{-.6cm}\raisebox{0.2cm}{$ \displaystyle  (y_2-y_1) \, p_\mu$}
\hspace{0.2cm} \displaystyle  (y_2-y_1) \, p_\mu
\psfrag{q}{}
\psfrag{i}{\raisebox{.1cm}{\sai \hspace{0cm}$\mu$}}
\psfrag{u}{\sai $y_1$}
\psfrag{e}{\ssi \hspace{-.4cm} \raisebox{0cm}{$-\,\bar{y}_2$}}
\psfrag{f}{}%{$k'$}
\psfrag{a}{}%{$k$}
\psfrag{b}{}
%\psfrag{e}{\raisebox{-.4cm}{\sai$y_1$}}
\psfrag{c}{\sai\hspace{-.35cm}$-\bar{y}_1$}
\psfrag{d}{\ssi \raisebox{0cm}{\hspace{-.3cm}$-\,\bar{y}_2$}}
\psfrag{m}{\hspace{-0.4cm}\sai$y_2-y_1$}
%\psfrag{yq}{{ $\hspace{-.2cm}y_1$}}
%\psfrag{yb}{{ $\hspace{-.2cm}1-y_1$}}
\hspace{0.4cm}\raisebox{-2.1cm}{\includegraphics[width=4.5cm]{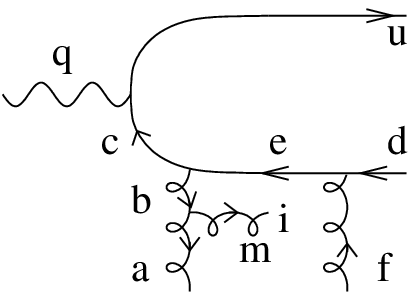}}\hspace{.4cm}\raisebox{-2.2cm}{\rotatebox{90}{$\underline{\rule{3.3cm}{0pt}}$}}_{\displaystyle ''p_2''}  %\hspace{-.3cm}\raisebox{0cm}{$ \displaystyle =  \, \left[   1-\frac{k^2}{(k-\ell_g)^2}\right]$}&
\hspace{-.3cm}\displaystyle =  \, \left[   1-\frac{k_1^2}{(k_1-\ell_g)^2}\right]
\psfrag{a}{}
\psfrag{o}{}
\psfrag{u}{\ssi \raisebox{0cm}{$ y_1$}}
\psfrag{d}{\ssi \raisebox{0cm}{\hspace{-.3cm}$-\,\bar{y}_2$}}
\psfrag{e}{\ssi \raisebox{0cm}{\hspace{-.2cm}$ y_2$}}
\psfrag{f}{\ssi \hspace{-.4cm}  \raisebox{0cm}{$y_1-y_2$}}
\psfrag{r}{\ssi \hspace{-.8cm} \raisebox{-0.3cm}{$-\,\bar{y}_1$}}
\psfrag{s}{\ssi \raisebox{0cm}{{\hspace{-.2cm}$-\,\bar{y}_2$}}}
\hspace{0.3cm}\raisebox{-1.4cm}{\epsfig{file=eG1Ward1.eps,width=\ttiB}}\,\,
\eq
and
\beq
\label{Ward_etG2}
%\hspace{-.6cm}\raisebox{-.1cm}{$ \displaystyle - (y_2-y_1) \, p_\mu$}
 \hspace{0cm}\displaystyle  -(y_2-y_1) \, p_\mu
\psfrag{q}{}
\psfrag{i}{\sai\raisebox{.1cm}{$\mu$}}
\psfrag{u}{\sai\raisebox{0cm}{$y_1$}}
\psfrag{e}{\ssi \hspace{-.4cm} \raisebox{0cm}{$-\,\bar{y}_1$}}
\psfrag{f}{}%{$k'$}
\psfrag{a}{}%{$k$}
\psfrag{b}{}
\psfrag{c}{\sai\hspace{-.35cm}$-\bar{y}_1$}
%\sai\hspace{-.35cm}$-\bar{y}_1$}
\psfrag{d}{\ssi \raisebox{0cm}{\hspace{-.3cm}$-\,\bar{y}_2$}}
\psfrag{m}{\hspace{-.1cm}\sai$y_2-y_1$}
%\psfrag{s}{\ssi \raisebox{0cm}{{\hspace{-.2cm}$-\,\bar{y}_1$}}}
%\psfrag{yq}{{ $\hspace{-.2cm}y_1$}}
%\psfrag{yb}{{ $\hspace{-.2cm}1-y_1$}}
\hspace{0.4cm}\raisebox{-2cm}{\includegraphics[width=4.5cm]{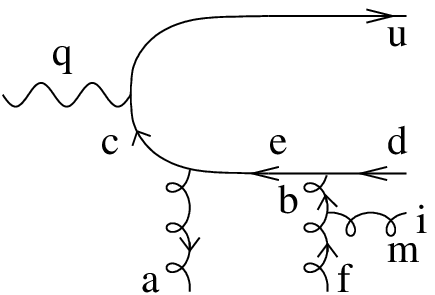}}\hspace{.4cm}
\raisebox{-2.1cm}{\rotatebox{90}{$\underline{\rule{3.1cm}{0pt}}$}}_{\displaystyle ''p_2''}
\hspace{-0.3cm}\displaystyle =  \, - \left[   1-\frac{k_2^2}{(k_2+\ell_g)^2}\right]
\psfrag{a}{}
\psfrag{o}{}
\psfrag{u}{\ssi \raisebox{0cm}{$ y_1$}}
\psfrag{d}{\ssi \raisebox{0cm}{\hspace{-.3cm}$-\,\bar{y}_2$}}
\psfrag{e}{\ssi \raisebox{0cm}{\hspace{-.2cm}$ y_2$}}
\psfrag{f}{\ssi \hspace{-.4cm}  \raisebox{0cm}{$y_1-y_2$}}
\psfrag{r}{\ssi \hspace{-.8cm} \raisebox{-0.3cm}{$-\,\bar{y}_1$}}
\psfrag{s}{\ssi \raisebox{0cm}{{\hspace{-.35cm}$-\,\bar{y}_1$}}}
\hspace{0.3cm}\raisebox{-1.4cm}{\epsfig{file=eG2Ward1.eps,width=\ttiB}}\ \ .
\eq
Contributions (\ref{Ward3body_e}), (\ref{Ward_etG1}) and (\ref{Ward_etG2})
add to zero in the Regge limit.

%%%%%%%%%%%%%%%%%%%%%%%%
Finally, the proofs for (fG2) goes along the same line and reads
\beq
%\begin{tabular}{ccc}
\hspace{-0.2cm}\displaystyle - (y_2-y_1) \, p_\mu \
\psfrag{yb}{{\sai $\hspace{-.2cm}1-y_2$}}
\psfrag{a}{}
\psfrag{f}{}
\psfrag{i}{\raisebox{0.1cm}{\hspace{-.05cm}\sai$\mu$}}
\psfrag{u}{\raisebox{.5cm}{\sai$y_1$}}
\psfrag{d}{\ssi \raisebox{0cm}{\hspace{-.3cm}$-\,\bar{y}_2$}}
\psfrag{b}{\raisebox{0.2cm}{\hspace{-.15cm}\sai$y_2$}}
\psfrag{e}{\raisebox{0.5cm}{\hspace{-.3cm}\sai$y_2$}}
\psfrag{r}{\raisebox{0.5cm}{\hspace{-.15cm}\sai$y_1$}}
\psfrag{m}{\raisebox{.1cm}{\sai\hspace{-.4cm}$y_2-y_1$}}
%\psfrag{yq}{{ $\hspace{-.2cm}y_1$}}
%\psfrag{yb}{{ $\hspace{-.2cm}1-y_1$}}
\hspace{0cm}\raisebox{-.8cm}{\includegraphics[width=4.3cm]{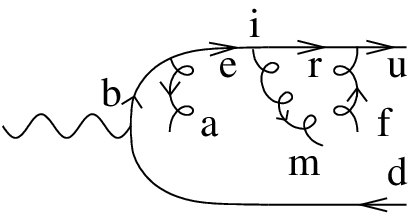}}
%&
\hspace{.15cm}
=
\psfrag{f}{\ssi \hspace{-.5cm} \sai$y_1-y_2$}
\psfrag{o}{}
\psfrag{u}{\ssi \raisebox{0cm}{$ y_1$}}
\psfrag{d}{\ssi \raisebox{0cm}{\hspace{-.3cm}$-\,\bar{y}_2$}}
\psfrag{e}{\raisebox{0.4cm}{\hspace{-.3cm}\sai$y_2$}}
\psfrag{r}{\ssi \hspace{-.2cm} \raisebox{0cm}{$y_1$}}
\hspace{0.1cm}\raisebox{-.7cm}{\epsfig{file=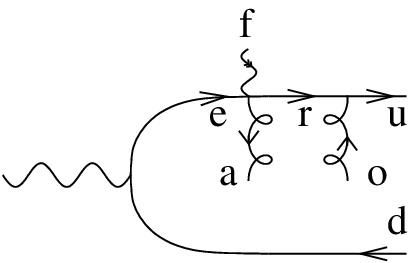,width=\ttiB}}
\hspace{.8cm}
%&
\hspace{-0.6cm}- \ \
\psfrag{o}{}
\psfrag{u}{\ssi \raisebox{0cm}{$ y_1$}}
\psfrag{d}{\ssi \raisebox{0cm}{\hspace{-.3cm}$-\,\bar{y}_2$}}
\psfrag{f}{\ssi \hspace{-.9cm}  \raisebox{0cm}{$y_1-y_2$}}
\psfrag{e}{\raisebox{0.4cm}{\hspace{-.3cm}\sai$y_2$}}
\psfrag{r}{\ssi \hspace{-.2cm} \raisebox{0cm}{$y_2$}}
\psfrag{s}{\ssi \raisebox{.1cm}{$y_2$}}
\hspace{0cm}
\raisebox{-.7cm}{\epsfig{file=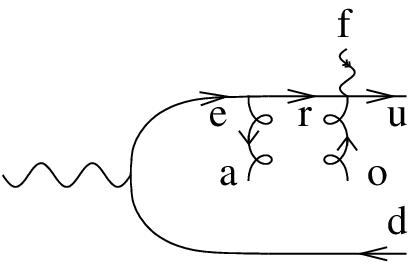,width=\ttiB}}
%\end{tabular}}
\hspace{.5cm} ,
\label{Ward3body_f}
\eq
The $p_2$ term obtained after applying the identity (\ref{vertex3Ward1}) to the diagrams (ftG1) and (ftG2) gives, respectively
\beq
\label{Ward_ftG1}
%\hspace{-.6cm}\raisebox{0.2cm}{$ \displaystyle  (y_2-y_1) \, p_\mu$}
\hspace{0.2cm} \displaystyle  -(y_2-y_1) \, p_\mu
\psfrag{q}{}
\psfrag{i}{\raisebox{-.2cm}{\sai \hspace{0.2cm}$\mu$}}
\psfrag{u}{\sai $y_1$}
\psfrag{e}{\ssi \hspace{-.3cm} \raisebox{-0.4cm}{$y_1$}}
\psfrag{f}{}%{$k'$}
\psfrag{a}{}%{$k$}
\psfrag{b}{\sai\hspace{-.1cm}\raisebox{0.2cm}{$y_2$}}
%\psfrag{e}{\raisebox{-.4cm}{\sai$y_1$}}
\psfrag{c}{}
\psfrag{d}{\ssi \raisebox{0cm}{\hspace{-.3cm}$-\,\bar{y}_2$}}
\psfrag{m}{\hspace{-0.4cm}\sai$y_2-y_1$}
%\psfrag{yq}{{ $\hspace{-.2cm}y_1$}}
%\psfrag{yb}{{ $\hspace{-.2cm}1-y_1$}}
\hspace{0.4cm}\raisebox{-.9cm}{\includegraphics[width=4.5cm]{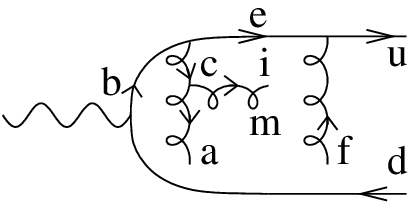}}\hspace{.4cm}\raisebox{-1.4cm}{\rotatebox{90}{$\underline{\rule{2.3cm}{0pt}}$}}_{\displaystyle ''p_2''}  %\hspace{-.3cm}\raisebox{0cm}{$ \displaystyle =  \, \left[   1-\frac{k^2}{(k-\ell_g)^2}\right]$}&
\hspace{-.3cm}\displaystyle =  \,- \left[   1-\frac{k_1^2}{(k_1-\ell_g)^2}\right]
\psfrag{a}{}
\psfrag{o}{}
\psfrag{u}{\ssi \raisebox{0cm}{$ y_1$}}
\psfrag{d}{\ssi \raisebox{0cm}{\hspace{-.3cm}$-\,\bar{y}_2$}}
\psfrag{e}{\ssi \raisebox{0cm}{\hspace{-.2cm}$ y_2$}}
\psfrag{f}{\ssi \hspace{-.4cm}  \raisebox{0cm}{$y_1-y_2$}}
\psfrag{r}{\ssi \hspace{-.2cm} \raisebox{0cm}{$y_1$}}
\psfrag{s}{\ssi \raisebox{0cm}{{\hspace{-.2cm}$-\,\bar{y}_2$}}}
\hspace{0.3cm}\raisebox{-.7cm}{\epsfig{file=fG2Ward1.eps,width=\ttiB}}\,\,
\eq
and
\beq
\label{Ward_ftG2}
%\hspace{-.6cm}\raisebox{-.1cm}{$ \displaystyle - (y_2-y_1) \, p_\mu$}
 \hspace{0cm}\displaystyle  (y_2-y_1) \, p_\mu
\psfrag{q}{}
\psfrag{i}{\sai\raisebox{.1cm}{$\mu$}}
\psfrag{u}{\sai\raisebox{0cm}{$y_1$}}
\psfrag{e}{\ssi \hspace{-.2cm} \raisebox{0cm}{$y_2$}}
\psfrag{f}{}%{$k'$}
\psfrag{a}{}%{$k$}
\psfrag{b}{\sai\hspace{-.1cm}\raisebox{0.2cm}{$y_2$}}
\psfrag{c}{}
%\sai\hspace{-.35cm}$-\bar{y}_1$}
\psfrag{d}{\ssi \raisebox{0cm}{\hspace{-.3cm}$-\,\bar{y}_2$}}
\psfrag{m}{\hspace{-.1cm}\sai$y_2-y_1$}
%\psfrag{s}{\ssi \raisebox{0cm}{{\hspace{-.2cm}$-\,\bar{y}_1$}}}
%\psfrag{yq}{{ $\hspace{-.2cm}y_1$}}
%\psfrag{yb}{{ $\hspace{-.2cm}1-y_1$}}
\hspace{0.4cm}\raisebox{-.9cm}{\includegraphics[width=4.5cm]{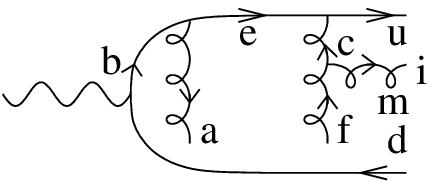}}\hspace{.4cm}
\raisebox{-1.4cm}{\rotatebox{90}{$\underline{\rule{2.3cm}{0pt}}$}}_{\displaystyle ''p_2''}
\hspace{-0.3cm}\displaystyle =  \,  \left[   1-\frac{k_2^2}{(k_2+\ell_g)^2}\right]
\psfrag{a}{}
\psfrag{o}{}
\psfrag{u}{\ssi \raisebox{0cm}{$ y_1$}}
\psfrag{d}{\ssi \raisebox{0cm}{\hspace{-.3cm}$-\,\bar{y}_2$}}
\psfrag{e}{\ssi \raisebox{0cm}{\hspace{-.2cm}$ y_2$}}
\psfrag{f}{\ssi \hspace{-.4cm}  \raisebox{0cm}{$y_1-y_2$}}
\psfrag{r}{\ssi \hspace{-.2cm} \raisebox{0cm}{$y_2$}}
\psfrag{s}{\ssi \raisebox{0cm}{{\hspace{-.35cm}$-\,\bar{y}_1$}}}
\hspace{0.3cm}\raisebox{-.7cm}{\epsfig{file=fG2Ward2.eps,width=\ttiB}}\ \ .
\eq
Contributions (\ref{Ward3body_f}), (\ref{Ward_ftG1}) and (\ref{Ward_ftG2})
add to zero in the Regge limit.

%%%%%%%%%

We now consider the "Ward" part (second term in the l.h.s of Eq.(\ref{vertex3Ward1})),
first for the set (atG1), (dtG1) and (etG2).
The contribution arising from (atG1) reads
\beqa
\label{Ward_atG1_Ward}
%\hspace{-.6cm}\raisebox{0.2cm}{$ \displaystyle  (y_2-y_1) \, p_\mu$}
&&\hspace{0.2cm} \displaystyle  (y_2-y_1) \, p_\mu
\psfrag{q}{}
\psfrag{i}{\raisebox{.2cm}{\sai \hspace{0cm}$\mu$}}
\psfrag{u}{\sai $y_1$}
\psfrag{f}{}%{$k'$}
\psfrag{a}{}%{$k$}
\psfrag{b}{}
\psfrag{e}{\ssi \raisebox{0cm}{\hspace{-.3cm}$-\,\bar{y}_2$}}
\psfrag{c}{\sai\hspace{-.35cm}$-\bar{y}_1$}
\psfrag{d}{\ssi \raisebox{0cm}{\hspace{-.3cm}$-\,\bar{y}_2$}}
\psfrag{m}{\hspace{-0.2cm}\sai$y_2-y_1$}
%\psfrag{yq}{{ $\hspace{-.2cm}y_1$}}
%\psfrag{yb}{{ $\hspace{-.2cm}1-y_1$}}
\hspace{0.4cm}
\raisebox{-1.9cm}{\includegraphics[width=4cm]{atG1_kin.eps}}
\hspace{.4cm}
\raisebox{-2.1cm}{\rotatebox{90}{$\underline{\rule{3cm}{0pt}}$}}_{\displaystyle ''W''}
\hspace{-.3cm}\displaystyle
=  \,- \frac{\ell_g \cdot p_2}{(k_1-\ell_g)^2} (\ell_g-k_1)_\mu
\psfrag{a}{\raisebox{-.2cm}{\sai $\begin{array}{c}\mu \\ y_1-y_2 \end{array}$}}
\psfrag{o}{}
\psfrag{u}{\ssi \raisebox{0cm}{$ y_1$}}
\psfrag{d}{\ssi \raisebox{0cm}{\hspace{-.3cm}$-\,\bar{y}_2$}}
\psfrag{e}{\ssi \raisebox{0cm}{\hspace{-.3cm}$-\,\bar{y}_2$}}
\psfrag{r}{\ssi \hspace{-.2cm} \raisebox{0cm}{$y_2$}}
\psfrag{s}{\ssi \raisebox{0cm}{{\hspace{-.2cm}$-\,\bar{y}_2$}}}
\hspace{0.3cm}\raisebox{-1.6cm}{\epsfig{file=a_kin.eps,width=4cm}}\nonumber \\
&&= \,\frac{\ell_g \cdot p_2}{(k_1-\ell_g)^2}  \left[ \raisebox{-1.4cm}{
\psfrag{f}{\sai \hspace{-.2cm}$y_1-y_2$}
\psfrag{s}{\ssi \raisebox{0cm}{$ y_1$}}
\psfrag{u}{\ssi \raisebox{0cm}{$ y_1$}}
\psfrag{d}{\ssi \raisebox{0cm}{\hspace{-.3cm}$-\,\bar{y}_2$}}
\psfrag{r}{\ssi \raisebox{0cm}{\hspace{-.3cm}$-\,\bar{y}_1$}}
\epsfig{file=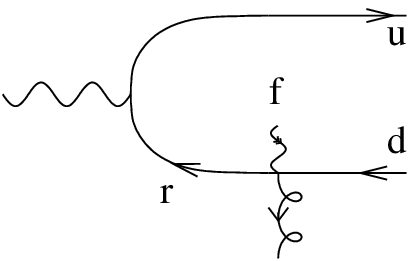,width=\ttiB}}
\ -\ \raisebox{-1.4cm}{
\psfrag{s}{\ssi \raisebox{0cm}{$ y_2$}}
\psfrag{u}{\ssi \raisebox{0cm}{$ y_1$}}
\psfrag{r}{\ssi \raisebox{0cm}{\hspace{-.3cm}$-\,\bar{y}_2$}}
\psfrag{f}{\raisebox{0.2cm}{\sai $y_1-y_2$}}
\psfrag{d}{\ssi \raisebox{0cm}{\hspace{-.3cm}$-\,\bar{y}_2$}}
\epsfig{file=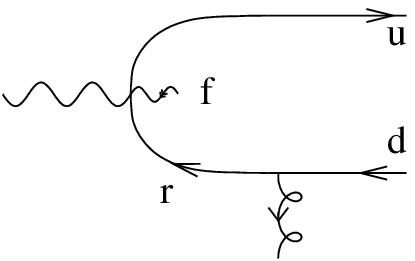,width=\ttiB}} \ \right]\,,
\eqa
while
the contribution arising from (dtG1) is
\beqa
\label{Ward_dtG1_Ward}
%\hspace{-.6cm}\raisebox{0.2cm}{$ \displaystyle  (y_2-y_1) \, p_\mu$}
&&\hspace{-0.1cm} \displaystyle  (y_2-y_1) \, p_\mu
\psfrag{q}{}
\psfrag{i}{\raisebox{0cm}{\sai \hspace{0cm}$\mu$}}
\psfrag{u}{\sai $y_1$}
\psfrag{f}{}%{$k'$}
\psfrag{a}{}%{$k$}
\psfrag{b}{}
\psfrag{e}{\raisebox{0cm}{\sai$y_1$}}
\psfrag{c}{\sai\hspace{-.35cm}$-\bar{y}_1$}
\psfrag{d}{\ssi \raisebox{0cm}{\hspace{-.3cm}$-\,\bar{y}_2$}}
\psfrag{m}{\raisebox{0cm}{\hspace{-0.1cm}\sai$y_2-y_1$}}
%\psfrag{yq}{{ $\hspace{-.2cm}y_1$}}
%\psfrag{yb}{{ $\hspace{-.2cm}1-y_1$}}
\hspace{0.2cm}
\raisebox{-1.6cm}{\includegraphics[width=4.5cm]{dtG1_kin.eps}}
\hspace{.4cm}
\raisebox{-1.7cm}{\rotatebox{90}{$\underline{\rule{2.7cm}{0pt}}$}}_{\displaystyle ''W''}
 \hspace{-.5cm}\displaystyle
 =  \, -\frac{\ell_g \cdot p_2}{(k_1-\ell_g)^2} (\ell_g-k_1)_\mu
 \psfrag{a}{\raisebox{-.1cm}{\sai $\hspace{-.1cm}\mu \hspace{.3cm} y_1-y_2$}}
 \psfrag{o}{}
 \psfrag{u}{\ssi \raisebox{0cm}{$ y_1$}}
 \psfrag{d}{\ssi \raisebox{0cm}{\hspace{-.3cm}$-\,\bar{y}_2$}}
\psfrag{d}{\ssi \raisebox{0cm}{\hspace{-.3cm}$-\,\bar{y}_1$}}
 \psfrag{r}{\ssi \hspace{-.2cm} \raisebox{0cm}{$y_2$}}
 \psfrag{s}{\ssi \raisebox{0cm}{{\hspace{-.2cm}$-\,\bar{y}_2$}}}
 \hspace{0.2cm}\raisebox{-1.6cm}{\epsfig{file=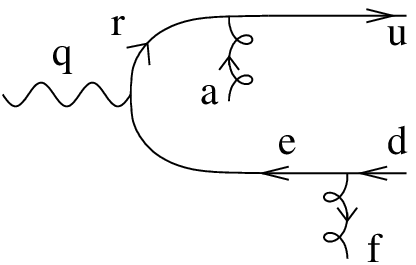,width=3.9cm}}\nonumber \\
&=& \frac{\ell_g \cdot p_2}{(k_1-\ell_g)^2}
 \psfrag{f}{\sai \hspace{.1cm}$y_1-y_2$}
 \psfrag{s}{\ssi \raisebox{0cm}{$ y_1$}}
\psfrag{r}{\ssi \raisebox{0cm}{\hspace{-.3cm}$-\,\bar{y}_2$}}
 \psfrag{u}{\ssi \raisebox{0cm}{$ y_1$}}
 \psfrag{d}{\ssi \raisebox{0cm}{\hspace{-.3cm}$-\,\bar{y}_2$}}
\hspace{.2cm} \raisebox{-1.5cm}{\epsfig{file=WardDd2.eps,width=3.9cm}}
 \ ,
\eqa
which cancels the last term in r.h.s of Eq.(\ref{Ward_atG1_Ward}).
The remaining term is thus the first term of the r.h.s of Eq.(\ref{Ward_atG1_Ward}),
which equals
the contribution arising from (etG2) since
\beqa
\label{Ward_etG2_Ward}
%\hspace{-.6cm}\raisebox{0.2cm}{$ \displaystyle  (y_2-y_1) \, p_\mu$}
&&\hspace{-0.1cm} \displaystyle  -(y_2-y_1) \, p_\mu
\psfrag{q}{}
\psfrag{i}{\raisebox{0cm}{\sai \hspace{0cm}$\mu$}}
\psfrag{u}{\sai $y_1$}
\psfrag{f}{}%{$k'$}
\psfrag{a}{}%{$k$}
\psfrag{b}{}
\psfrag{e}{\raisebox{0cm}{\sai\hspace{-.35cm}$-\bar{y}_1$}}
\psfrag{c}{\sai\hspace{-.35cm}$-\bar{y}_1$}
\psfrag{d}{\ssi \raisebox{0cm}{\hspace{-.3cm}$-\,\bar{y}_2$}}
\psfrag{m}{\raisebox{0cm}{\hspace{-0.1cm}\sai$y_2-y_1$}}
%\psfrag{yq}{{ $\hspace{-.2cm}y_1$}}
%\psfrag{yb}{{ $\hspace{-.2cm}1-y_1$}}
\hspace{0.2cm}
\raisebox{-1.9cm}{\includegraphics[width=4cm]{etG2_kin.eps}}
\hspace{.5cm}
\raisebox{-1.9cm}{\rotatebox{90}{$\underline{\rule{2.7cm}{0pt}}$}}_{\displaystyle ''W''}
 \hspace{-.5cm}\displaystyle
 =  \, \frac{\ell_g \cdot p_2}{(k_1-\ell_g)^2} (\ell_g-k_1)_\mu
 \psfrag{f}{\raisebox{-.1cm}{\sai $\hspace{-.5cm}\mu \hspace{.3cm} y_1-y_2$}}
 \psfrag{o}{}
\psfrag{a}{}
 \psfrag{u}{\ssi \raisebox{0cm}{$ y_1$}}
 \psfrag{d}{\ssi \raisebox{0cm}{\hspace{-.3cm}$-\,\bar{y}_2$}}
 \psfrag{r}{\ssi \hspace{-.2cm} \raisebox{0cm}{$y_2$}}
 \psfrag{s}{\ssi \raisebox{0cm}{{\hspace{-.2cm}$-\,\bar{y}_2$}}}
 \hspace{0.2cm}\raisebox{-1.6cm}{\epsfig{file=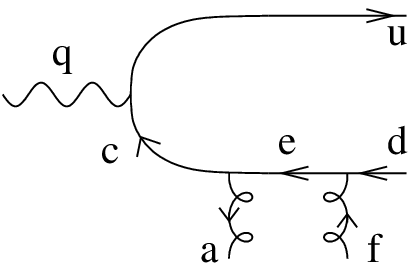,width=3.9cm}}\nonumber \\
&=& \frac{\ell_g \cdot p_2}{(k_1-\ell_g)^2}
 \psfrag{f}{\sai \hspace{-.2cm}$y_1-y_2$}
 \psfrag{s}{\ssi \raisebox{0cm}{$ y_1$}}
\psfrag{r}{\ssi \raisebox{0cm}{\hspace{-.3cm}$-\,\bar{y}_1$}}
 \psfrag{u}{\ssi \raisebox{0cm}{$ y_1$}}
 \psfrag{d}{\ssi \raisebox{0cm}{\hspace{-.3cm}$-\,\bar{y}_2$}}
\hspace{.2cm} \raisebox{-1.5cm}{\epsfig{file=WardDd1.eps,width=3.9cm}}
 \ .
\eqa

Consider now the set (ctG1), (btG1) and (ftG2).
The contribution arising from (ctG1) reads
\beqa
\label{Ward_ctG1_Ward}
%\hspace{-.6cm}\raisebox{0.2cm}{$ \displaystyle  (y_2-y_1) \, p_\mu$}
&&\hspace{0.2cm} \displaystyle  -(y_2-y_1) \, p_\mu
\psfrag{q}{}
\psfrag{i}{\raisebox{-.1cm}{\sai \hspace{.2cm}$\mu$}}
\psfrag{u}{\sai $y_1$}
\psfrag{f}{}%{$k'$}
\psfrag{a}{}%{$k$}
\psfrag{b}{\raisebox{0.2cm}{\sai\hspace{-.1cm}$y_2$}}
\psfrag{e}{\raisebox{-.4cm}{\sai$y_1$}}
\psfrag{c}{\sai\hspace{-.35cm}$-\bar{y}_1$}
\psfrag{d}{\ssi \raisebox{0cm}{\hspace{-.3cm}$-\,\bar{y}_2$}}
\psfrag{m}{\hspace{-0.4cm}\sai$y_2-y_1$}
%\psfrag{yq}{{ $\hspace{-.2cm}y_1$}}
%\psfrag{yb}{{ $\hspace{-.2cm}1-y_1$}}
\hspace{0.4cm}
\raisebox{-.9cm}{\includegraphics[width=4.5cm]{ctG1_kin.eps}}
\hspace{.4cm}
\raisebox{-1.1cm}{\rotatebox{90}{$\underline{\rule{2.1cm}{0pt}}$}}_{\displaystyle ''W''}
\hspace{-.3cm}\displaystyle
=  \, \frac{\ell_g \cdot p_2}{(k_1-\ell_g)^2} (\ell_g-k_1)_\mu
\psfrag{a}{\raisebox{-.2cm}{\sai $\begin{array}{c}\mu \\ y_1-y_2 \end{array}$}}
\psfrag{o}{}
\psfrag{u}{\ssi \raisebox{0cm}{$ y_1$}}
\psfrag{d}{\ssi \raisebox{0cm}{\hspace{-.3cm}$-\,\bar{y}_2$}}
\psfrag{e}{\ssi \raisebox{0cm}{\hspace{0cm}$ y_1$}}
\psfrag{r}{\ssi \hspace{-.2cm} \raisebox{0cm}{$y_2$}}
\psfrag{s}{\ssi \raisebox{0cm}{{\hspace{-.2cm}$-\,\bar{y}_2$}}}
\hspace{0.3cm}\raisebox{-.9cm}{\epsfig{file=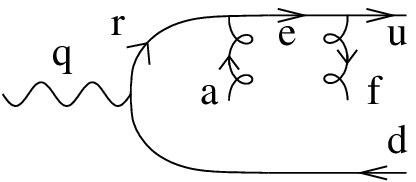,width=4.5cm}}\nonumber \\
&&= \,-\frac{\ell_g \cdot p_2}{(k_1-\ell_g)^2}  \left[ \raisebox{-.7cm}{
\psfrag{f}{\sai \hspace{-.2cm}$y_1-y_2$}
\psfrag{s}{\ssi \raisebox{0cm}{$ y_1$}}
\psfrag{u}{\ssi \raisebox{0cm}{$ y_1$}}
\psfrag{d}{\ssi \raisebox{0cm}{\hspace{-.3cm}$-\,\bar{y}_2$}}
\epsfig{file=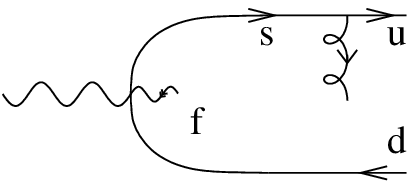,width=\ttiB}}
\,-\, \raisebox{-.7cm}{
\psfrag{s}{\ssi \raisebox{0cm}{$ y_2$}}
\psfrag{u}{\ssi \raisebox{0cm}{$ y_1$}}
\psfrag{f}{\raisebox{0.2cm}{\sai $y_1-y_2$}}
\psfrag{d}{\ssi \raisebox{0cm}{\hspace{-.3cm}$-\,\bar{y}_2$}}
\epsfig{file=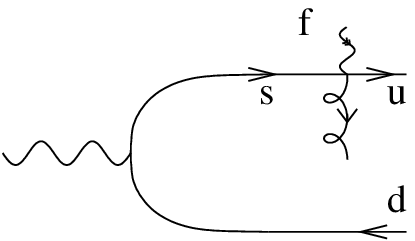,width=\ttiB}} \ \right]\,,
\eqa
while
the contribution arising from (btG1) is
\beq
\label{Ward_btG1_Ward}
%\hspace{-.6cm}\raisebox{0.2cm}{$ \displaystyle  (y_2-y_1) \, p_\mu$}
\hspace{-0.1cm} \displaystyle  -(y_2-y_1) \, p_\mu
\psfrag{q}{}
\psfrag{i}{\raisebox{.1cm}{\sai \hspace{0cm}$\mu$}}
\psfrag{u}{\sai $y_1$}
\psfrag{f}{}%{$k'$}
\psfrag{a}{}%{$k$}
\psfrag{b}{}
\psfrag{e}{\raisebox{0cm}{\sai$y_1$}}
\psfrag{c}{\sai\hspace{-.35cm}$-\bar{y}_1$}
\psfrag{d}{\ssi \raisebox{0cm}{\hspace{-.3cm}$-\,\bar{y}_2$}}
\psfrag{m}{\raisebox{-.1cm}{\hspace{0cm}\sai$y_2-y_1$}}
%\psfrag{yq}{{ $\hspace{-.2cm}y_1$}}
%\psfrag{yb}{{ $\hspace{-.2cm}1-y_1$}}
\hspace{0.1cm}
\raisebox{-1.3cm}{\includegraphics[width=2.8cm]{btG1_kin.eps}}
\hspace{.4cm}
\raisebox{-1.4cm}{\rotatebox{90}{$\underline{\rule{2.1cm}{0pt}}$}}_{\displaystyle ''W''}
 \hspace{-.5cm}\displaystyle
 =  \, \frac{\ell_g \cdot p_2}{(k_1-\ell_g)^2} (\ell_g-k_1)_\mu
 \psfrag{a}{\raisebox{-.2cm}{\sai $\begin{array}{c}\mu \\ y_1-y_2 \end{array}$}}
 \psfrag{o}{}
 \psfrag{u}{\ssi \raisebox{0cm}{$ y_1$}}
 \psfrag{d}{\ssi \raisebox{0cm}{\hspace{-.3cm}$-\,\bar{y}_2$}}
 \psfrag{e}{\ssi \raisebox{0cm}{\hspace{0cm}$ y_1$}}
 \psfrag{r}{\ssi \hspace{-.2cm} \raisebox{0cm}{$y_2$}}
 \psfrag{s}{\ssi \raisebox{0cm}{{\hspace{-.2cm}$-\,\bar{y}_2$}}}
 \hspace{0.1cm}\raisebox{-1.1cm}{\epsfig{file=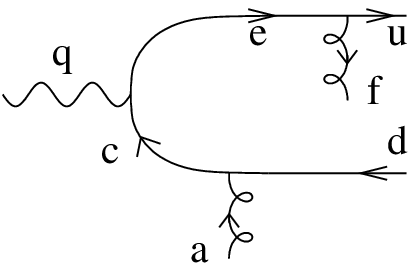,width=2.8cm}}= \frac{\ell_g \cdot p_2}{(k_1-\ell_g)^2}
 \psfrag{f}{\sai \hspace{-.2cm}$y_1-y_2$}
 \psfrag{s}{\ssi \raisebox{0cm}{$ y_1$}}
 \psfrag{u}{\ssi \raisebox{0cm}{$ y_1$}}
 \psfrag{d}{\ssi \raisebox{0cm}{\hspace{-.3cm}$-\,\bar{y}_2$}}
\raisebox{-.5cm}{\epsfig{file=WardUd1.eps,width=2.8cm}}
 \,,
\eq
which cancels the first term in r.h.s of Eq.(\ref{Ward_ctG1_Ward}).
The remaining term is thus the last term of the r.h.s of Eq.(\ref{Ward_ctG1_Ward}),
which equals
the contribution arising from (ftG2) since
\beq
\label{Ward_ftG2_Ward}
%\hspace{-.6cm}\raisebox{0.2cm}{$ \displaystyle  (y_2-y_1) \, p_\mu$}
\hspace{-0.1cm} \displaystyle  (y_2-y_1) \, p_\mu
\psfrag{q}{}
\psfrag{i}{\raisebox{.1cm}{\sai \hspace{0cm}$\mu$}}
\psfrag{u}{\raisebox{.4cm}{\sai $y_1$}}
\psfrag{f}{}%{$k'$}
\psfrag{a}{}%{$k$}
\psfrag{b}{\raisebox{.2cm}{\sai \hspace{-.1cm}$y_2$}}
\psfrag{e}{\raisebox{.4cm}{\sai $y_2$}}
\psfrag{c}{}
\psfrag{d}{\ssi \raisebox{-0.4cm}{\hspace{-.3cm}$-\,\bar{y}_2$}}
\psfrag{m}{\raisebox{-.1cm}{\hspace{0cm}\sai$y_2-y_1$}}
%\psfrag{yq}{{ $\hspace{-.2cm}y_1$}}
%\psfrag{yb}{{ $\hspace{-.2cm}1-y_1$}}
\hspace{0.1cm}
\raisebox{-.5cm}{\includegraphics[width=2.8cm]{ftG2_kin.eps}}
\hspace{.6cm}
\raisebox{-.8cm}{\rotatebox{90}{$\underline{\rule{1.6cm}{0pt}}$}}_{\displaystyle ''W''}
 \hspace{-.5cm}\displaystyle
 =  \,- \frac{\ell_g \cdot p_2}{(k_1-\ell_g)^2} (\ell_g-k_1)_\mu
\psfrag{f}{\raisebox{-.1cm}{\hspace{-.45cm}\sai $\begin{array}{c}\mu \\ \!\!y_1-y_2 \end{array}$}}
 \psfrag{a}{}
 \psfrag{o}{}
\psfrag{u}{\raisebox{.4cm}{\sai $y_1$}}
\psfrag{e}{\raisebox{.4cm}{\sai $y_2$}}
\psfrag{d}{\ssi \raisebox{-0.4cm}{\hspace{-.3cm}$-\,\bar{y}_2$}}
 \psfrag{r}{\ssi \hspace{-.3cm} \raisebox{-0.1cm}{$y_2$}}
 \hspace{0.1cm}\raisebox{-.5cm}{\epsfig{file=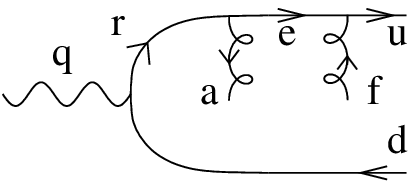,width=2.8cm}}= \frac{\ell_g \cdot p_2}{(k_1-\ell_g)^2}
 \psfrag{f}{\raisebox{.2cm}{\hspace{-.1cm}\sai $y_1-y_2$}}
\psfrag{s}{\ssi \raisebox{0cm}{$ y_2$}}
 \psfrag{u}{\ssi \raisebox{0cm}{$ y_1$}}
\raisebox{-.5cm}{\epsfig{file=WardUd2.eps,width=2.8cm}}
 \,.
\eq

Consider now the set (etG1), (btG2) and (atG2).
The contribution arising from (etG1) reads
\beqa
\label{Ward_etG1_Ward}
%\hspace{-.6cm}\raisebox{0.2cm}{$ \displaystyle  (y_2-y_1) \, p_\mu$}
&&\hspace{0.2cm} \displaystyle  (y_2-y_1) \, p_\mu
\psfrag{q}{}
\psfrag{i}{\raisebox{.2cm}{\sai \hspace{0cm}$\mu$}}
\psfrag{u}{\sai $y_1$}
\psfrag{f}{}%{$k'$}
\psfrag{a}{}%{$k$}
\psfrag{b}{}
\psfrag{e}{\ssi \raisebox{0cm}{\hspace{-.3cm}$-\,\bar{y}_2$}}
\psfrag{c}{\sai\hspace{-.35cm}$-\bar{y}_1$}
\psfrag{d}{\ssi \raisebox{0cm}{\hspace{-.3cm}$-\,\bar{y}_2$}}
\psfrag{m}{\hspace{-0.2cm}\sai$y_2-y_1$}
%\psfrag{yq}{{ $\hspace{-.2cm}y_1$}}
%\psfrag{yb}{{ $\hspace{-.2cm}1-y_1$}}
\hspace{0.4cm}
\raisebox{-1.9cm}{\includegraphics[width=4cm]{etG1_kin.eps}}
\hspace{.4cm}
\raisebox{-2.1cm}{\rotatebox{90}{$\underline{\rule{3cm}{0pt}}$}}_{\displaystyle ''W''}
\hspace{-.3cm}\displaystyle
=  \,- \frac{\ell_g \cdot p_2}{(k_2+\ell_g)^2} (\ell_g+k_2)_\mu
\psfrag{a}{\raisebox{-.2cm}{\sai $\begin{array}{c}\mu \\ y_1-y_2 \end{array}$}}
\psfrag{o}{}
\psfrag{u}{\ssi \raisebox{0cm}{$ y_1$}}
\psfrag{d}{\ssi \raisebox{0cm}{\hspace{-.3cm}$-\,\bar{y}_2$}}
\psfrag{e}{\ssi \raisebox{0cm}{\hspace{-.3cm}$-\,\bar{y}_2$}}
\psfrag{r}{\ssi \hspace{-.2cm} \raisebox{0cm}{$y_2$}}
\psfrag{s}{\ssi \raisebox{0cm}{{\hspace{-.2cm}$-\,\bar{y}_2$}}}
\hspace{0.3cm}\raisebox{-1.6cm}{\epsfig{file=e_kin.eps,width=4cm}}\nonumber \\
&&= \,\frac{\ell_g \cdot p_2}{(k_2+\ell_g)^2}  \left[ \raisebox{-1.4cm}{
\psfrag{f}{\sai \hspace{-.2cm}$y_1-y_2$}
\psfrag{s}{\ssi \raisebox{0cm}{$ y_1$}}
\psfrag{u}{\ssi \raisebox{0cm}{$ y_1$}}
\psfrag{d}{\ssi \raisebox{0cm}{\hspace{-.3cm}$-\,\bar{y}_2$}}
\psfrag{r}{\ssi \raisebox{0cm}{\hspace{-.3cm}$-\,\bar{y}_1$}}
\epsfig{file=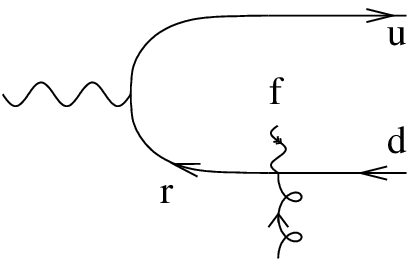,width=\ttiB}}
\ - \ \raisebox{-1.4cm}{
\psfrag{s}{\ssi \raisebox{0cm}{$ y_2$}}
\psfrag{u}{\ssi \raisebox{0cm}{$ y_1$}}
\psfrag{r}{\ssi \raisebox{0cm}{\hspace{-.3cm}$-\,\bar{y}_2$}}
\psfrag{f}{\raisebox{0.2cm}{\sai $y_1-y_2$}}
\psfrag{d}{\ssi \raisebox{0cm}{\hspace{-.3cm}$-\,\bar{y}_2$}}
\epsfig{file=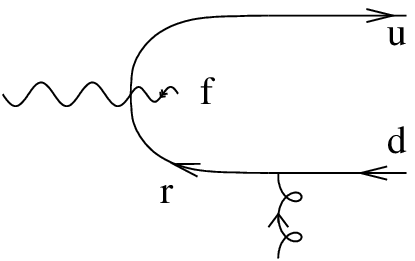,width=\ttiB}} \ \right]\,,
\eqa
while
the contribution arising from (btG2) is
\beqa
\label{Ward_btG2_Ward}
%\hspace{-.6cm}\raisebox{0.2cm}{$ \displaystyle  (y_2-y_1) \, p_\mu$}
&&\hspace{-0.1cm} \displaystyle  (y_2-y_1) \, p_\mu
\psfrag{q}{}
\psfrag{i}{\raisebox{0cm}{\sai \hspace{0cm}$\mu$}}
\psfrag{u}{\sai $y_1$}
\psfrag{f}{}%{$k'$}
\psfrag{a}{}%{$k$}
\psfrag{b}{}
\psfrag{e}{\raisebox{0cm}{\sai$y_1$}}
\psfrag{c}{\sai\hspace{-.35cm}$-\bar{y}_1$}
\psfrag{d}{\ssi \raisebox{0cm}{\hspace{-.3cm}$-\,\bar{y}_2$}}
\psfrag{m}{\raisebox{0cm}{\hspace{-0.1cm}\sai$y_2-y_1$}}
\psfrag{r}{\ssi \raisebox{0cm}{\hspace{-.3cm}$-\,\bar{y}_2$}}
%\psfrag{yq}{{ $\hspace{-.2cm}y_1$}}
%\psfrag{yb}{{ $\hspace{-.2cm}1-y_1$}}
\hspace{0.2cm}
\raisebox{-1.6cm}{\includegraphics[width=4.5cm]{btG2_kin.eps}}
\hspace{.4cm}
\raisebox{-1.7cm}{\rotatebox{90}{$\underline{\rule{2.7cm}{0pt}}$}}_{\displaystyle ''W''}
 \hspace{-.5cm}\displaystyle
 =  \,- \frac{\ell_g \cdot p_2}{(k_2+\ell_g)^2} (\ell_g+k_2)_\mu
 \psfrag{f}{\raisebox{-.1cm}{\sai $\hspace{-.5cm}\mu \hspace{.3cm} y_2-y_1$}}
\psfrag{a}{}
 \psfrag{o}{}
 \psfrag{u}{\ssi \raisebox{0cm}{$ y_1$}}
 \psfrag{d}{\ssi \raisebox{0cm}{\hspace{-.3cm}$-\,\bar{y}_2$}}
\psfrag{d}{\ssi \raisebox{0cm}{\hspace{-.3cm}$-\,\bar{y}_1$}}
 \psfrag{r}{\ssi \hspace{-.2cm} \raisebox{0cm}{$y_2$}}
 \psfrag{s}{\ssi \raisebox{0cm}{{\hspace{-.2cm}$-\,\bar{y}_2$}}}
 \hspace{0.2cm}\raisebox{-1.6cm}{\epsfig{file=b_kin.eps,width=3.9cm}}\nonumber \\
&=& \frac{\ell_g \cdot p_2}{(k_2+\ell_g)^2}
 \psfrag{f}{\sai \hspace{.1cm}$y_1-y_2$}
 \psfrag{s}{\ssi \raisebox{0cm}{$ y_1$}}
\psfrag{r}{\ssi \raisebox{0cm}{\hspace{-.3cm}$-\,\bar{y}_2$}}
 \psfrag{u}{\ssi \raisebox{0cm}{$ y_1$}}
 \psfrag{d}{\ssi \raisebox{0cm}{\hspace{-.3cm}$-\,\bar{y}_2$}}
\hspace{.2cm} \raisebox{-1.5cm}{\epsfig{file=WardDu2.eps,width=3.9cm}}
 \ ,
\eqa
which cancels the last term in r.h.s of Eq.(\ref{Ward_etG1_Ward}).
The remaining term is thus the first term of the r.h.s of Eq.(\ref{Ward_etG1_Ward}),
which equals
the contribution arising from (atG2) since
\beq
\label{Ward_atG2_Ward}
%\hspace{-.6cm}\raisebox{0.2cm}{$ \displaystyle  (y_2-y_1) \, p_\mu$}
\hspace{0cm} \displaystyle  -(y_2-y_1) \, p_\mu
\psfrag{q}{}
\psfrag{i}{\raisebox{.1cm}{\sai \hspace{0cm}$\mu$}}
\psfrag{d}{\ssi \raisebox{0cm}{\hspace{-.3cm}$-\,\bar{y}_2$}}
\psfrag{u}{\sai $y_1$}
\psfrag{f}{}%{$k'$}
\psfrag{a}{}%{$k$}
\psfrag{b}{}
\psfrag{e}{\raisebox{0cm}{\sai \hspace{-.5cm}$-\,\bar{y}_1$}}
\psfrag{c}{\ssi \raisebox{-0.2cm}{\hspace{-.3cm}$-\,\bar{y}_1$}}
\psfrag{m}{\raisebox{-.1cm}{\hspace{0cm}\sai$y_2-y_1$}}
%\psfrag{yq}{{ $\hspace{-.2cm}y_1$}}
%\psfrag{yb}{{ $\hspace{-.2cm}1-y_1$}}
\hspace{0.05cm}
\raisebox{-1.25cm}{\includegraphics[width=2.8cm]{atG2_kin.eps}}
\hspace{.6cm}
\raisebox{-1.4cm}{\rotatebox{90}{$\underline{\rule{2cm}{0pt}}$}}_{\displaystyle ''W''}
 \hspace{-.6cm}\displaystyle
 =  \, \frac{\ell_g \cdot p_2}{(k_2+\ell_g)^2} (\ell_g+k_2)_\mu
\psfrag{f}{\raisebox{-.2cm}{\hspace{-.4cm}\sai $\begin{array}{c}\mu \\ \hspace{-.3cm}y_1-y_2 \end{array}$}}
 \psfrag{a}{}
 \psfrag{o}{}
\psfrag{d}{\ssi \raisebox{0cm}{\hspace{-.3cm}$-\,\bar{y}_2$}}
\psfrag{u}{\sai $y_1$}
\psfrag{e}{\raisebox{0cm}{\sai \hspace{-.5cm}$-\,\bar{y}_1$}}
 \psfrag{r}{\ssi \hspace{-.3cm} \raisebox{-0.1cm}{$-\,\bar{y}_1$}}
 \hspace{0.1cm}\raisebox{-1.1cm}{\epsfig{file=a_kin.eps,width=2.8cm}}= \frac{\ell_g \cdot p_2}{(k_2+\ell_g)^2}
 \psfrag{f}{\raisebox{.1cm}{\hspace{-.1cm}\sai $y_1-y_2$}}
\psfrag{s}{\ssi \raisebox{0cm}{$-\,\bar{y}_1$}}
 \psfrag{u}{\ssi \raisebox{0cm}{$ y_1$}}
\raisebox{-1cm}{\epsfig{file=WardDu1.eps,width=2.8cm}}
 \,.
\eq
Consider now the set (ftG1), (dtG2) and (ctG2).
The contribution arising from (ftG1) reads
\beqa
\label{Ward_ftG1_Ward}
%\hspace{-.6cm}\raisebox{0.2cm}{$ \displaystyle  (y_2-y_1) \, p_\mu$}
&&\hspace{0.2cm} \displaystyle  -(y_2-y_1) \, p_\mu
\psfrag{q}{}
\psfrag{i}{\raisebox{-0.2cm}{\sai \hspace{0.2cm}$\mu$}}
\psfrag{u}{\sai $y_1$}
\psfrag{f}{}%{$k'$}
\psfrag{a}{}%{$k$}
\psfrag{b}{\raisebox{0.2cm}{\hspace{-0.1cm}\sai$y_2$}}
\psfrag{e}{\ssi \raisebox{0.1cm}{\hspace{0cm}$y_1$}}
\psfrag{c}{}
\psfrag{d}{\ssi \raisebox{0cm}{\hspace{-.3cm}$-\,\bar{y}_2$}}
\psfrag{m}{\hspace{-0.4cm}\sai$y_2-y_1$}
%\psfrag{yq}{{ $\hspace{-.2cm}y_1$}}
%\psfrag{yb}{{ $\hspace{-.2cm}1-y_1$}}
\hspace{0.2cm}
\raisebox{-1cm}{\includegraphics[width=4.8cm]{ftG1_kin.eps}}
\hspace{.4cm}
\raisebox{-2.1cm}{\rotatebox{90}{$\underline{\rule{3cm}{0pt}}$}}_{\displaystyle ''W''}
\hspace{-.3cm}\displaystyle
=  \, \frac{\ell_g \cdot p_2}{(k_1-\ell_g)^2} (\ell_g-k_1)_\mu
\psfrag{a}{\raisebox{-.2cm}{\sai $\begin{array}{c}\mu \\ y_2-y_1 \end{array}$}}
\psfrag{o}{}
\psfrag{u}{\ssi \raisebox{0cm}{$ y_1$}}
\psfrag{d}{\ssi \raisebox{0cm}{\hspace{-.3cm}$-\,\bar{y}_2$}}
\psfrag{e}{\ssi \raisebox{0cm}{\hspace{0cm}$y_1$}}
\psfrag{r}{\ssi \hspace{-.2cm} \raisebox{0cm}{$y_2$}}
\psfrag{s}{\ssi \raisebox{0cm}{{\hspace{-.2cm}$-\,\bar{y}_2$}}}
\hspace{0.2cm}\raisebox{-.8cm}{\epsfig{file=f_kin.eps,width=4cm}}\nonumber \\
&&= \,-\frac{\ell_g \cdot p_2}{(k_1-\ell_g)^2}  \left[ \raisebox{-.7cm}{
\psfrag{f}{\sai \hspace{-.2cm}$y_1-y_2$}
\psfrag{s}{\ssi \raisebox{0cm}{$ y_1$}}
\psfrag{u}{\ssi \raisebox{0cm}{$ y_1$}}
\psfrag{d}{\ssi \raisebox{0cm}{\hspace{-.3cm}$-\,\bar{y}_2$}}
\psfrag{r}{\ssi \raisebox{0cm}{\hspace{-.3cm}$-\,\bar{y}_1$}}
\epsfig{file=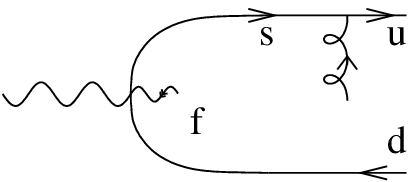,width=\ttiB}}
\ - \ \raisebox{-.7cm}{
\psfrag{s}{\ssi \raisebox{0cm}{$ y_2$}}
\psfrag{u}{\ssi \raisebox{0cm}{$ y_1$}}
\psfrag{r}{\ssi \raisebox{0cm}{\hspace{-.3cm}$-\,\bar{y}_2$}}
\psfrag{f}{\raisebox{0.2cm}{\sai $y_1-y_2$}}
\psfrag{d}{\ssi \raisebox{0cm}{\hspace{-.3cm}$-\,\bar{y}_2$}}
\epsfig{file=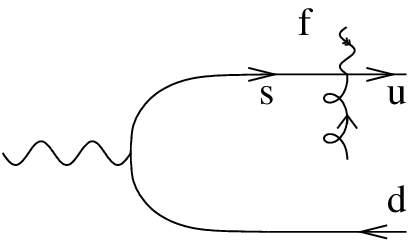,width=\ttiB}} \ \right]\,,
\eqa
while
the contribution arising from (dtG2) is
\beq
\label{Ward_dtG2_Ward}
%\hspace{-.6cm}\raisebox{0.2cm}{$ \displaystyle  (y_2-y_1) \, p_\mu$}
\hspace{-0.1cm} \displaystyle  -(y_2-y_1) \, p_\mu
\psfrag{q}{}
\psfrag{i}{\raisebox{.1cm}{\sai \hspace{0cm}$\mu$}}
\psfrag{u}{\sai $y_1$}
\psfrag{f}{}%{$k'$}
\psfrag{a}{}%{$k$}
\psfrag{b}{}
\psfrag{e}{\raisebox{0cm}{\sai$y_1$}}
\psfrag{c}{\sai\hspace{-.35cm}$-\bar{y}_1$}
\psfrag{d}{\ssi \raisebox{0cm}{\hspace{-.3cm}$-\,\bar{y}_2$}}
\psfrag{m}{\raisebox{-.1cm}{\hspace{0cm}\sai$y_2-y_1$}}
%\psfrag{yq}{{ $\hspace{-.2cm}y_1$}}
%\psfrag{yb}{{ $\hspace{-.2cm}1-y_1$}}
\hspace{0.1cm}
\raisebox{-1.3cm}{\includegraphics[width=2.8cm]{dtG2_kin.eps}}
\hspace{.4cm}
\raisebox{-1.4cm}{\rotatebox{90}{$\underline{\rule{2.1cm}{0pt}}$}}_{\displaystyle ''W''}
 \hspace{-.5cm}\displaystyle
 =  \, \frac{\ell_g \cdot p_2}{(k_1-\ell_g)^2} (\ell_g-k_1)_\mu
 \psfrag{f}{\raisebox{-.1cm}{\sai $\hspace{-.5cm}\mu \hspace{.3cm} y_2-y_1$}}
\psfrag{a}{}
 \psfrag{o}{}
\psfrag{e}{\raisebox{0cm}{\hspace{-.3cm}\sai$-\, \bar{y}_1$}}
 \psfrag{u}{\ssi \raisebox{0cm}{$ y_1$}}
\psfrag{d}{\ssi \raisebox{0cm}{\hspace{-.3cm}$-\,\bar{y}_2$}}
 \psfrag{r}{\ssi \hspace{-.3cm} \raisebox{0.1cm}{$y_1$}}
 \psfrag{s}{\ssi \raisebox{0cm}{{\hspace{-.2cm}$-\,\bar{y}_2$}}}
 \hspace{0.2cm}\raisebox{-1.1cm}{\epsfig{file=d_kin.eps,width=2.8cm}}
= \frac{\ell_g \cdot p_2}{(k_1-\ell_g)^2}
 \psfrag{f}{\sai \hspace{-.2cm}$y_1-y_2$}
 \psfrag{s}{\ssi \raisebox{0cm}{$ y_1$}}
 \psfrag{u}{\ssi \raisebox{0cm}{$ y_1$}}
 \psfrag{d}{\ssi \raisebox{0cm}{\hspace{-.3cm}$-\,\bar{y}_2$}}
\raisebox{-.5cm}{\epsfig{file=WardUu1.eps,width=2.8cm}}
 \,,
\eq
which cancels the first term in r.h.s of Eq.(\ref{Ward_ftG1_Ward}).
The remaining term is thus the last term of the r.h.s of Eq.(\ref{Ward_ftG1_Ward}),
which equals
the contribution arising from (ctG2) since
\beq
\label{Ward_ctG2_Ward}
%\hspace{-.6cm}\raisebox{0.2cm}{$ \displaystyle  (y_2-y_1) \, p_\mu$}
\hspace{0cm} \displaystyle  (y_2-y_1) \, p_\mu
\psfrag{q}{}
\psfrag{i}{\raisebox{.1cm}{\sai \hspace{0cm}$\mu$}}
\psfrag{u}{\raisebox{.4cm}{\sai $y_1$}}
\psfrag{f}{}%{$k'$}
\psfrag{a}{}%{$k$}
\psfrag{b}{\raisebox{.2cm}{\sai \hspace{-.1cm}$y_2$}}
\psfrag{e}{\raisebox{0.13cm}{\sai $y_2$}}
\psfrag{c}{}
\psfrag{d}{\ssi \raisebox{-0.4cm}{\hspace{-.3cm}$-\,\bar{y}_2$}}
\psfrag{m}{\raisebox{-.1cm}{\hspace{0cm}\sai$y_2-y_1$}}
%\psfrag{yq}{{ $\hspace{-.2cm}y_1$}}
%\psfrag{yb}{{ $\hspace{-.2cm}1-y_1$}}
\hspace{0.05cm}
\raisebox{-0.5cm}{\includegraphics[width=2.8cm]{ctG2_kin.eps}}
\hspace{.6cm}
\raisebox{-1.1cm}{\rotatebox{90}{$\underline{\rule{2cm}{0pt}}$}}_{\displaystyle ''W''}
 \hspace{-.6cm}\displaystyle
 =  \,- \frac{\ell_g \cdot p_2}{(k_2+\ell_g)^2} (\ell_g+k_2)_\mu
\psfrag{f}{\raisebox{-.1cm}{\hspace{-.4cm}\sai $\begin{array}{c}\mu \\ \hspace{-.25cm}\raisebox{.1cm}{$y_2-y_1$} \end{array}$}}
 \psfrag{a}{}
 \psfrag{o}{}
\psfrag{u}{\raisebox{.4cm}{\sai $y_1$}}
\psfrag{d}{\ssi \raisebox{-0.4cm}{\hspace{-.3cm}$-\,\bar{y}_2$}}
\psfrag{e}{\raisebox{0.4cm}{\sai $y_2$}}
 \psfrag{r}{\ssi \hspace{-.3cm} \raisebox{-0.1cm}{$y_2$}}
 \hspace{0.1cm}\raisebox{-.6cm}{\epsfig{file=c_kin.eps,width=2.8cm}}= \frac{\ell_g \cdot p_2}{(k_2+\ell_g)^2}
 \psfrag{f}{\raisebox{.1cm}{\hspace{-.1cm}\sai $y_1-y_2$}}
\psfrag{s}{\ssi \raisebox{0cm}{\hspace{-.2cm}$y_2$}}
 \psfrag{u}{\ssi \raisebox{0cm}{$ y_1$}}
\raisebox{-.5cm}{\epsfig{file=WardUu2.eps,width=2.8cm}}
 \,.
\eq
Collecting now the 4 remaining contributions (\ref{Ward_etG2_Ward},\ref{Ward_ftG2_Ward}, \ref{Ward_atG2_Ward}, \ref{Ward_ctG2_Ward}), one
gets
\beq
\label{sumWard}
\psfrag{u}{\ssi \raisebox{0cm}{$ y_1$}}
\psfrag{d}{\ssi \raisebox{0cm}{\hspace{-.3cm}$-\,\bar{y}_2$}}
\psfrag{s}{\ssi \raisebox{0cm}{\hspace{-.2cm}$y_2$}}
\psfrag{r}{\ssi \hspace{-.3cm} \raisebox{-0.1cm}{$-\,\bar{y}_1$}}
 \psfrag{f}{\raisebox{-.3cm}{\hspace{-.4cm}\sai $y_2-y_1$}}
``Ward \ terms``=2 \left[\frac{p_2 \cdot \ell_g}{(k_1-\ell_g)^2}+ \frac{p_2 \cdot \ell_g}{(k_2+\ell_g)^2}  \right] \left[\raisebox{-1.1cm}{\epsfig{file=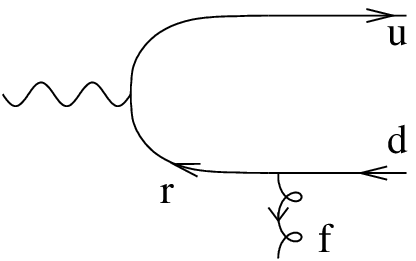,width=2.8cm}}
\,+\,
\psfrag{d}{\ssi \raisebox{-0.5cm}{\hspace{-.2cm}$-\,\bar{y}_2$}}
\raisebox{-.5cm}{\epsfig{file=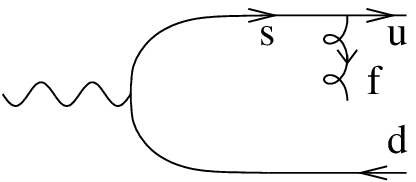,width=2.8cm}} \hspace{.2cm} \right]
\eq
which vanishes since
\beq
\label{vanishSumWard}
Tr \left[\frac{y_1 \slashchar{p}_1- \slashchar{q}}{(y_1 \, p_1 -q)^2} \slashchar{p}_2 \, \slashchar{p}_1 \, \slashchar{e}_\gamma  \right]
+ Tr \left[\slashchar{p}_1 \, \slashchar{p}_2 \,\frac{\slashchar{q}-\bar{y}_2 \, \slashchar{p}_1}{(q-\bar{y}_2 \, p_1)^2}  \, \slashchar{e}_\gamma  \right]
=0
\eq
as expected from the fact that a quark--antiquark collinear pair cannot emit a gluon.
\\

\begin{figure}
\psfrag{i}{$\mu$}
\psfrag{a}{$k$}
\psfrag{f}{$k'$}
\psfrag{g}{$\ell_g$}
\psfrag{t}{}%$\tau$}
\psfrag{l}{}%$\lambda$}
\psfrag{s}{$\sigma$}
\psfrag{b}{$\ell_g$}
\psfrag{c}{$\hspace{-1.3cm}k-k'-\ell_g$}
\includegraphics[width=3cm]{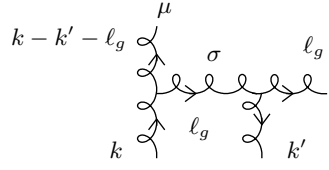}
\caption{Two 3-gluon vertices connected, with one open index $\mu.$}
\label{Fig:vertexTwo3Ward}
\end{figure}

We end up by considering the diagrams (gttG1), (httG1), (gttG2) and (httG2).
As a preliminary, let us consider two triple gluon vertices connected by a gluon line, with $t-$channel gluons of momentum $k$ and $k'$ (with $k=k_1$ (resp. $k=-k_2$) and $k'=k_2$ (resp. $k'=-k_1$) for (httG1) and (gttG1) (resp. (httG2) and gttG2)) saturated with non-sense polarizations $p_2$ and with the gluon momentum $\ell_g$, leaving the index $\mu$ open, as illustrated in Fig.\ref{Fig:vertexTwo3Ward}.
It reads
\beq
\label{vertexTwo3Ward}
V^\mu_{\,\,\, \tau \sigma}(-k+k'+\ell_g,\, k,\, -\ell_g-k') \, V^\sigma_{\,\,\, \lambda \alpha}(\ell_g+k',\, -k',\, -\ell_g) \epsilon_g^\alpha \, p_2^\lambda \, p_2^\tau= -(2 \, k -\ell_g) \, \epsilon_g \, p_2^\mu +(k+\ell_g)_\mu (p_2 \cdot \epsilon_g)-(2 \, \ell_g -k)\cdot p_2 \, \epsilon_g^\mu
\eq
which reduces, after making the replacement $\epsilon_g \to \ell_g$,
%and using the fact that
%in our kinematics $k$ has no component along $p_1$,
to
 \beqa
\label{vertexTwo3Ward1}
&&V^\mu_{\,\,\, \tau \sigma}(-k+k'+\ell_g,\, k,\, -\ell_g-k') \, V^\sigma_{\,\,\, \lambda \alpha}(\ell_g+k',\, -k',\, -\ell_g) \ell_g^\alpha \, p_2^\lambda \, p_2^\tau \nonumber \\
&&=
(\ell_g \cdot p_2) \left[ p_2^\mu  \left(2 \, k \cdot k' + 2 \, \ell_g \cdot (k-k') -(k'+\ell_g)^2\right)  +(\ell_g \cdot p_2)(k'+\ell_g-k)^\mu \right]
\eqa
 where in the last line we used the fact that $\ell_g$ in on mass-shell. Hereafter and as for the case of a single gluon vertex, we will symbolically denote with the index $p_2$ the first term in the r.h.s of Eq.(\ref{vertex3Ward1})
and with W the second term (having in mind for this second term further use of Ward identities).

Let us first consider the ''Ward`` terms. Applying the identity (\ref{vertexTwo3Ward1}) to the diagram (gttG1) leads to
\beqa
\label{Ward_gttG1_Ward}
%\hspace{-.6cm}\raisebox{0.2cm}{$ \displaystyle  (y_2-y_1) \, p_\mu$}
\hspace{0cm} \displaystyle  (y_2-y_1) \, p_\mu
\psfrag{q}{}
\psfrag{i}{\raisebox{.1cm}{\sai \hspace{0cm}$\mu$}}
\psfrag{u}{\raisebox{0cm}{\sai $y_1$}}
\psfrag{f}{}%{$k'$}
\psfrag{a}{}%{$k$}
\psfrag{b}{\raisebox{0cm}{\sai \hspace{-.4cm}$y_1$}}
\psfrag{e}{}
\psfrag{c}{}
\psfrag{d}{\ssi \raisebox{0cm}{\hspace{-.3cm}$-\,\bar{y}_2$}}
\psfrag{m}{\raisebox{-.1cm}{\hspace{.3cm}\sai $\begin{array}{c}\hspace{-.6cm}\mu \\ \hspace{-.25cm}\raisebox{.1cm}{$y_2-y_1$} \end{array}$}}
%\psfrag{yq}{{ $\hspace{-.2cm}y_1$}}
%\psfrag{yb}{{ $\hspace{-.2cm}1-y_1$}}
\hspace{0.05cm}
\raisebox{-1.3cm}{\includegraphics[width=2.8cm]{gttG1_kin.eps}}
\hspace{.9cm}
\raisebox{-1.5cm}{\rotatebox{90}{$\underline{\rule{2.1cm}{0pt}}$}}_{\displaystyle ''W''}
 \hspace{-0.6cm}\displaystyle
 &=&  \,- \frac{(\ell_g \cdot p_2)^2}{(k_1-k_2-\ell_g)^2(k_2+\ell_g)^2} (-k_1+k_2+\ell_g)_\mu
\psfrag{f}{\raisebox{-.3cm}{\hspace{-.4cm}\sai $\begin{array}{c}\mu \\ \hspace{-.25cm}\raisebox{.1cm}{$y_2-y_1$} \end{array}$}}
 \psfrag{a}{}
 \psfrag{o}{}
\psfrag{u}{\raisebox{0cm}{\sai $y_1$}}
\psfrag{d}{\ssi \raisebox{-0.4cm}{\hspace{-.3cm}$-\,\bar{y}_2$}}
\psfrag{e}{\raisebox{0.4cm}{\sai $y_2$}}
 \psfrag{r}{\ssi \hspace{-.5cm} \raisebox{-0cm}{$-\,\bar{y}_1$}}
 \hspace{0.1cm}\raisebox{-1cm}{\epsfig{file=DipoleD.eps,width=2.8cm}}
 \nonumber \\
&=&
\frac{(\ell_g \cdot p_2)^2}{((k_1-k_2-\ell_g)^2(k_2+\ell_g)^2}
 \psfrag{f}{\raisebox{.1cm}{\hspace{-.1cm}\sai $y_1-y_2$}}
\psfrag{d}{\ssi \raisebox{0cm}{\hspace{-.2cm}$-\,\bar{y}_2$}}
 \psfrag{u}{\ssi \raisebox{0cm}{$ y_1$}}
\hspace{.1cm}\raisebox{-.5cm}{\epsfig{file=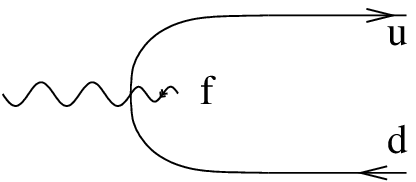,width=2.8cm}}\hspace{.2cm},
\eqa
while the same identity applied to (httG1) gives
\beqa
\label{Ward_httG1_Ward}
%\hspace{-.6cm}\raisebox{0.2cm}{$ \displaystyle  (y_2-y_1) \, p_\mu$}
\hspace{0cm} \displaystyle  (y_2-y_1) \, p_\mu
\psfrag{q}{}
\psfrag{i}{\raisebox{.1cm}{\sai \hspace{0cm}$\mu$}}
\psfrag{u}{\raisebox{0.4cm}{\sai $y_1$}}
\psfrag{f}{}%{$k'$}
\psfrag{a}{}%{$k$}
\psfrag{b}{\raisebox{.2cm}{\sai \hspace{-.2cm}$y_2$}}
\psfrag{e}{}%\raisebox{-0.1cm}{\sai \hspace{-.05cm}$y_2$}}
\psfrag{c}{}
\psfrag{d}{\ssi \raisebox{-0.4cm}{\hspace{-.3cm}$-\,\bar{y}_2$}}
\psfrag{m}{\raisebox{.1cm}{\hspace{.3cm}\sai $\begin{array}{c}\hspace{-.6cm}\mu \\ \hspace{-.25cm}\raisebox{.1cm}{$y_2-y_1$} \end{array}$}}
%\psfrag{yq}{{ $\hspace{-.2cm}y_1$}}
%\psfrag{yb}{{ $\hspace{-.2cm}1-y_1$}}
\hspace{0.05cm}
\raisebox{-.6cm}{\includegraphics[width=2.8cm]{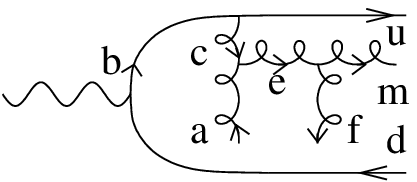}}
\hspace{.9cm}
\raisebox{-1.2cm}{\rotatebox{90}{$\underline{\rule{1.8cm}{0pt}}$}}_{\displaystyle ''W''}
 \hspace{-0.6cm}\displaystyle
 &=&  \,- \frac{(\ell_g \cdot p_2)^2}{(k_1-k_2-\ell_g)^2(k_2+\ell_g)^2} (-k_1+k_2+\ell_g)_\mu
\psfrag{f}{\raisebox{-.3cm}{\hspace{-.4cm}\sai $\begin{array}{c}\mu \\ \hspace{-.25cm}\raisebox{.1cm}{$y_2-y_1$} \end{array}$}}
 \psfrag{s}{\sai \hspace{-.3cm} $y_2$}
\psfrag{u}{\raisebox{0cm}{\sai $y_1$}}
\psfrag{d}{\ssi \raisebox{-0.4cm}{\hspace{-.3cm}$-\,\bar{y}_2$}}
\psfrag{e}{}%\raisebox{0.4cm}{\sai $y_2$}}
 \psfrag{r}{\ssi \hspace{-.5cm} \raisebox{-0cm}{$-\,\bar{y}_1$}}
 \hspace{0.1cm}\raisebox{-.6cm}{\epsfig{file=DipoleU.eps,width=2.8cm}}
 \nonumber \\
&=&
-\frac{(\ell_g \cdot p_2)^2}{((k_1-k_2-\ell_g)^2(k_2+\ell_g)^2}
 \psfrag{f}{\raisebox{.1cm}{\hspace{-.1cm}\sai $y_1-y_2$}}
\psfrag{d}{\ssi \raisebox{0cm}{\hspace{-.2cm}$-\,\bar{y}_2$}}
 \psfrag{u}{\ssi \raisebox{0cm}{$ y_1$}}
\hspace{.1cm}\raisebox{-.5cm}{\epsfig{file=WardNoEm.eps,width=2.8cm}}
\eqa
from which we deduce that the sum of the two contributions (\ref{Ward_gttG1_Ward}) and (\ref{Ward_httG1_Ward}) equals zero.

Applying again the identity (\ref{vertexTwo3Ward1}) to the diagram (gttG2) leads to
\beqa
\label{Ward_gttG2_Ward}
%\hspace{-.6cm}\raisebox{0.2cm}{$ \displaystyle  (y_2-y_1) \, p_\mu$}
\hspace{0cm} \displaystyle  (y_2-y_1) \, p_\mu
\psfrag{q}{}
\psfrag{i}{\raisebox{.1cm}{\sai \hspace{0cm}$\mu$}}
\psfrag{u}{\raisebox{0cm}{\sai $y_1$}}
\psfrag{f}{}%{$k'$}
\psfrag{a}{}%{$k$}
\psfrag{b}{\raisebox{0cm}{\sai \hspace{-.4cm}$y_1$}}
\psfrag{e}{}
\psfrag{c}{}
\psfrag{d}{\ssi \raisebox{0cm}{\hspace{-.3cm}$-\,\bar{y}_2$}}
\psfrag{m}{\raisebox{-.1cm}{\hspace{.3cm}\sai $\begin{array}{c}\hspace{-.6cm}\mu \\ \hspace{-.25cm}\raisebox{.1cm}{$y_2-y_1$} \end{array}$}}
%\psfrag{yq}{{ $\hspace{-.2cm}y_1$}}
%\psfrag{yb}{{ $\hspace{-.2cm}1-y_1$}}
\hspace{0.05cm}
\raisebox{-1.3cm}{\includegraphics[width=2.8cm]{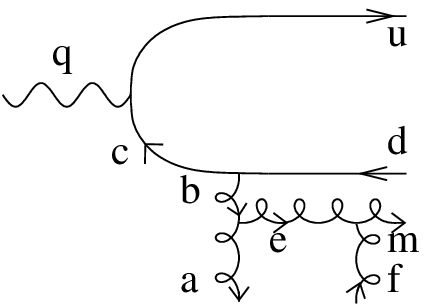}}
\hspace{.9cm}
\raisebox{-1.5cm}{\rotatebox{90}{$\underline{\rule{2.1cm}{0pt}}$}}_{\displaystyle ''W''}
 \hspace{-0.6cm}\displaystyle
 &=&  \,- \frac{(\ell_g \cdot p_2)^2}{(k_1-k_2-\ell_g)^2(k_2+\ell_g)^2} (-k_1+k_2+\ell_g)_\mu
\psfrag{f}{\raisebox{-.3cm}{\hspace{-.4cm}\sai $\begin{array}{c}\mu \\ \hspace{-.25cm}\raisebox{.1cm}{$y_2-y_1$} \end{array}$}}
 \psfrag{a}{}
 \psfrag{o}{}
\psfrag{u}{\raisebox{0cm}{\sai $y_1$}}
\psfrag{d}{\ssi \raisebox{-0.4cm}{\hspace{-.3cm}$-\,\bar{y}_2$}}
\psfrag{e}{\raisebox{0.4cm}{\sai $y_2$}}
 \psfrag{r}{\ssi \hspace{-.5cm} \raisebox{-0cm}{$-\,\bar{y}_1$}}
 \hspace{0.1cm}\raisebox{-1cm}{\epsfig{file=DipoleD.eps,width=2.8cm}}
 \nonumber \\
&=&
\frac{(\ell_g \cdot p_2)^2}{((k_1-k_2-\ell_g)^2(k_2+\ell_g)^2}
 \psfrag{f}{\raisebox{.1cm}{\hspace{-.1cm}\sai $y_1-y_2$}}
\psfrag{d}{\ssi \raisebox{0cm}{\hspace{-.2cm}$-\,\bar{y}_2$}}
 \psfrag{u}{\ssi \raisebox{0cm}{$ y_1$}}
\hspace{.1cm}\raisebox{-.5cm}{\epsfig{file=WardNoEm.eps,width=2.8cm}}\hspace{.2cm},
\eqa
while the same identity applied to (httG2) gives
\beqa
\label{Ward_httG2_Ward}
%\hspace{-.6cm}\raisebox{0.2cm}{$ \displaystyle  (y_2-y_1) \, p_\mu$}
\hspace{0cm} \displaystyle  (y_2-y_1) \, p_\mu
\psfrag{q}{}
\psfrag{i}{\raisebox{.1cm}{\sai \hspace{0cm}$\mu$}}
\psfrag{u}{\raisebox{0.4cm}{\sai $y_1$}}
\psfrag{f}{}%{$k'$}
\psfrag{a}{}%{$k$}
\psfrag{b}{\raisebox{.2cm}{\sai \hspace{-.2cm}$y_2$}}
\psfrag{e}{}
\psfrag{c}{}
\psfrag{d}{\ssi \raisebox{-0.4cm}{\hspace{-.3cm}$-\,\bar{y}_2$}}
\psfrag{m}{\raisebox{.1cm}{\hspace{.3cm}\sai $\begin{array}{c}\hspace{-.6cm}\mu \\ \hspace{-.25cm}\raisebox{.1cm}{$y_2-y_1$} \end{array}$}}
%\psfrag{yq}{{ $\hspace{-.2cm}y_1$}}
%\psfrag{yb}{{ $\hspace{-.2cm}1-y_1$}}
\hspace{0.05cm}
\raisebox{-.6cm}{\includegraphics[width=2.8cm]{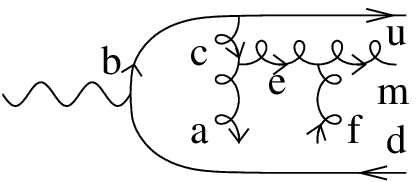}}
\hspace{.9cm}
\raisebox{-1.2cm}{\rotatebox{90}{$\underline{\rule{1.8cm}{0pt}}$}}_{\displaystyle ''W''}
 \hspace{-0.6cm}\displaystyle
 &=&  \,- \frac{(\ell_g \cdot p_2)^2}{(k_1-k_2-\ell_g)^2(k_2+\ell_g)^2} (-k_1+k_2+\ell_g)_\mu
\psfrag{f}{\raisebox{-.3cm}{\hspace{-.4cm}\sai $\begin{array}{c}\mu \\ \hspace{-.25cm}\raisebox{.1cm}{$y_2-y_1$} \end{array}$}}
 \psfrag{s}{\sai \hspace{-.3cm} $y_2$}
\psfrag{u}{\raisebox{0cm}{\sai $y_1$}}
\psfrag{d}{\ssi \raisebox{-0.4cm}{\hspace{-.3cm}$-\,\bar{y}_2$}}
\psfrag{e}{\raisebox{0.4cm}{\sai $y_2$}}
 \psfrag{r}{\ssi \hspace{-.5cm} \raisebox{-0cm}{$-\,\bar{y}_1$}}
 \hspace{0.1cm}\raisebox{-.6cm}{\epsfig{file=DipoleU.eps,width=2.8cm}}
 \nonumber \\
&=&
-\frac{(\ell_g \cdot p_2)^2}{((k_1-k_2-\ell_g)^2(k_2+\ell_g)^2}
 \psfrag{f}{\raisebox{.1cm}{\hspace{-.1cm}\sai $y_1-y_2$}}
\psfrag{d}{\ssi \raisebox{0cm}{\hspace{-.2cm}$-\,\bar{y}_2$}}
 \psfrag{u}{\ssi \raisebox{0cm}{$ y_1$}}
\hspace{.1cm}\raisebox{-.5cm}{\epsfig{file=WardNoEm.eps,width=2.8cm}}
\eqa
from which we again deduce that the sum of the two contributions (\ref{Ward_gttG2_Ward}) and (\ref{Ward_httG2_Ward}) equals zero.
%%%%%%%%%
%%%%%%%%%%

We now consider the ''$p_2$`` terms. Denoting
\beq
\label{defC1}
c_1=
- \frac{(\ell_g \cdot p_2)\, \left(2 \, k_1 \cdot k_2 + 2 \, \ell_g \cdot (k_1-k_2) -(k_2+\ell_g)^2\right)}{(k_1-k_2-\ell_g)^2(k_2+\ell_g)^2}
\eq
and
\beq
\label{defC2}
c_2=-\frac{(\ell_g \cdot p_2)\, \left(2 \, k_1 \cdot k_2 + 2 \, \ell_g \cdot (k_1-k_2) -(-k_1+\ell_g)^2\right)}{(k_1-k_2-\ell_g)^2(-k_1+\ell_g)^2}
\eq
and applying the identity (\ref{vertexTwo3Ward1}) to the diagram (gttG1), one obtains
\beqa
\label{Ward_gttG1_p2}
%\hspace{-.6cm}\raisebox{0.2cm}{$ \displaystyle  (y_2-y_1) \, p_\mu$}
\hspace{0cm} \displaystyle  (y_2-y_1) \, p_\mu
\psfrag{q}{}
\psfrag{i}{\raisebox{.1cm}{\sai \hspace{0cm}$\mu$}}
\psfrag{u}{\raisebox{0cm}{\sai $y_1$}}
\psfrag{f}{}%{$k'$}
\psfrag{a}{}%{$k$}
\psfrag{b}{\raisebox{0cm}{\sai \hspace{-.4cm}$y_1$}}
\psfrag{e}{}
\psfrag{c}{}
\psfrag{d}{\ssi \raisebox{0cm}{\hspace{-.3cm}$-\,\bar{y}_2$}}
\psfrag{m}{\raisebox{-.1cm}{\hspace{.3cm}\sai $\begin{array}{c}\hspace{-.6cm}\mu \\ \hspace{-.25cm}\raisebox{.1cm}{$y_2-y_1$} \end{array}$}}
%\psfrag{yq}{{ $\hspace{-.2cm}y_1$}}
%\psfrag{yb}{{ $\hspace{-.2cm}1-y_1$}}
\hspace{0.05cm}
\raisebox{-1.3cm}{\includegraphics[width=2.8cm]{gttG1_kin.eps}}
\hspace{.9cm}
\raisebox{-1.5cm}{\rotatebox{90}{$\underline{\rule{2.1cm}{0pt}}$}}_{\displaystyle ''p_2''}
 \hspace{-0.6cm}\displaystyle
 &=&  \, c_1 \, p_{2\mu}
\psfrag{f}{\raisebox{-.3cm}{\hspace{-.4cm}\sai $\begin{array}{c}\mu \\ \hspace{-.25cm}\raisebox{.1cm}{$y_2-y_1$} \end{array}$}}
 \psfrag{a}{}
 \psfrag{o}{}
\psfrag{u}{\raisebox{0cm}{\sai $y_1$}}
\psfrag{d}{\ssi \raisebox{-0.4cm}{\hspace{-.3cm}$-\,\bar{y}_2$}}
\psfrag{e}{\raisebox{0.4cm}{\sai $y_2$}}
 \psfrag{r}{\ssi \hspace{-.5cm} \raisebox{-0cm}{$-\,\bar{y}_1$}}
 \hspace{0.1cm}\raisebox{-1cm}{\epsfig{file=DipoleD.eps,width=2.8cm}}\,,
\eqa
while the same identity applied to (httG1) gives
\beqa
\label{Ward_httG1_p2}
%\hspace{-.6cm}\raisebox{0.2cm}{$ \displaystyle  (y_2-y_1) \, p_\mu$}
\hspace{0cm} \displaystyle  (y_2-y_1) \, p_\mu
\psfrag{q}{}
\psfrag{i}{\raisebox{.1cm}{\sai \hspace{0cm}$\mu$}}
\psfrag{u}{\raisebox{0.4cm}{\sai $y_1$}}
\psfrag{f}{}%{$k'$}
\psfrag{a}{}%{$k$}
\psfrag{b}{\raisebox{.2cm}{\sai \hspace{-.2cm}$y_2$}}
\psfrag{e}{}%\raisebox{-0.1cm}{\sai \hspace{-.05cm}$y_2$}}
\psfrag{c}{}
\psfrag{d}{\ssi \raisebox{-0.4cm}{\hspace{-.3cm}$-\,\bar{y}_2$}}
\psfrag{m}{\raisebox{.1cm}{\hspace{.3cm}\sai $\begin{array}{c}\hspace{-.6cm}\mu \\ \hspace{-.25cm}\raisebox{.1cm}{$y_2-y_1$} \end{array}$}}
%\psfrag{yq}{{ $\hspace{-.2cm}y_1$}}
%\psfrag{yb}{{ $\hspace{-.2cm}1-y_1$}}
\hspace{0.05cm}
\raisebox{-.6cm}{\includegraphics[width=2.8cm]{httG1_kin.eps}}
\hspace{.9cm}
\raisebox{-1.2cm}{\rotatebox{90}{$\underline{\rule{1.8cm}{0pt}}$}}_{\displaystyle ''p_2''}
 \hspace{-0.6cm}\displaystyle
 &=& \, c_1 \, p_{2\mu}
\psfrag{f}{\raisebox{-.3cm}{\hspace{-.4cm}\sai $\begin{array}{c}\mu \\ \hspace{-.25cm}\raisebox{.1cm}{$y_2-y_1$} \end{array}$}}
 \psfrag{s}{\sai \hspace{-.3cm} $y_2$}
\psfrag{u}{\raisebox{0cm}{\sai $y_1$}}
\psfrag{d}{\ssi \raisebox{-0.4cm}{\hspace{-.3cm}$-\,\bar{y}_2$}}
\psfrag{e}{}%\raisebox{0.4cm}{\sai $y_2$}}
 \psfrag{r}{\ssi \hspace{-.5cm} \raisebox{-0cm}{$-\,\bar{y}_1$}}
 \hspace{0.1cm}\raisebox{-.6cm}{\epsfig{file=DipoleU.eps,width=2.8cm}}\,,
\eqa
from which we deduce that the sum of the two contributions (\ref{Ward_gttG1_p2}) and (\ref{Ward_httG1_p2}) equals zero, has was already shown in Eq.(\ref{vanishSumWard}).

Applying again the identity (\ref{vertexTwo3Ward1}) to the diagram (gttG2) leads to
\beqa
\label{Ward_gttG2_p2}
%\hspace{-.6cm}\raisebox{0.2cm}{$ \displaystyle  (y_2-y_1) \, p_\mu$}
\hspace{0cm} \displaystyle  (y_2-y_1) \, p_\mu
\psfrag{q}{}
\psfrag{i}{\raisebox{.1cm}{\sai \hspace{0cm}$\mu$}}
\psfrag{u}{\raisebox{0cm}{\sai $y_1$}}
\psfrag{f}{}%{$k'$}
\psfrag{a}{}%{$k$}
\psfrag{b}{\raisebox{0cm}{\sai \hspace{-.4cm}$y_1$}}
\psfrag{e}{}
\psfrag{c}{}
\psfrag{d}{\ssi \raisebox{0cm}{\hspace{-.3cm}$-\,\bar{y}_2$}}
\psfrag{m}{\raisebox{-.1cm}{\hspace{.3cm}\sai $\begin{array}{c}\hspace{-.6cm}\mu \\ \hspace{-.25cm}\raisebox{.1cm}{$y_2-y_1$} \end{array}$}}
%\psfrag{yq}{{ $\hspace{-.2cm}y_1$}}
%\psfrag{yb}{{ $\hspace{-.2cm}1-y_1$}}
\hspace{0.05cm}
\raisebox{-1.3cm}{\includegraphics[width=2.8cm]{gttG2_kin.eps}}
\hspace{.9cm}
\raisebox{-1.5cm}{\rotatebox{90}{$\underline{\rule{2.1cm}{0pt}}$}}_{\displaystyle ''p_2''}
 \hspace{-0.6cm}\displaystyle
 &=&
\, c_2\, p_{2\mu}
\psfrag{f}{\raisebox{-.3cm}{\hspace{-.4cm}\sai $\begin{array}{c}\mu \\ \hspace{-.25cm}\raisebox{.1cm}{$y_2-y_1$} \end{array}$}}
 \psfrag{a}{}
 \psfrag{o}{}
\psfrag{u}{\raisebox{0cm}{\sai $y_1$}}
\psfrag{d}{\ssi \raisebox{-0.4cm}{\hspace{-.3cm}$-\,\bar{y}_2$}}
\psfrag{e}{\raisebox{0.4cm}{\sai $y_2$}}
 \psfrag{r}{\ssi \hspace{-.5cm} \raisebox{-0cm}{$-\,\bar{y}_1$}}
 \hspace{0.1cm}\raisebox{-1cm}{\epsfig{file=DipoleD.eps,width=2.8cm}}
 \,,
\eqa
while the same identity applied to (httG2) gives
\beqa
\label{Ward_httG2_p2}
%\hspace{-.6cm}\raisebox{0.2cm}{$ \displaystyle  (y_2-y_1) \, p_\mu$}
\hspace{0cm} \displaystyle  (y_2-y_1) \, p_\mu
\psfrag{q}{}
\psfrag{i}{\raisebox{.1cm}{\sai \hspace{0cm}$\mu$}}
\psfrag{u}{\raisebox{0.4cm}{\sai $y_1$}}
\psfrag{f}{}%{$k'$}
\psfrag{a}{}%{$k$}
\psfrag{b}{\raisebox{.2cm}{\sai \hspace{-.2cm}$y_2$}}
\psfrag{e}{}
\psfrag{c}{}
\psfrag{d}{\ssi \raisebox{-0.4cm}{\hspace{-.3cm}$-\,\bar{y}_2$}}
\psfrag{m}{\raisebox{.1cm}{\hspace{.3cm}\sai $\begin{array}{c}\hspace{-.6cm}\mu \\ \hspace{-.25cm}\raisebox{.1cm}{$y_2-y_1$} \end{array}$}}
%\psfrag{yq}{{ $\hspace{-.2cm}y_1$}}
%\psfrag{yb}{{ $\hspace{-.2cm}1-y_1$}}
\hspace{0.05cm}
\raisebox{-.6cm}{\includegraphics[width=2.8cm]{httG2_kin.eps}}
\hspace{.9cm}
\raisebox{-1.2cm}{\rotatebox{90}{$\underline{\rule{1.8cm}{0pt}}$}}_{\displaystyle ''p_2''}
 \hspace{-0.6cm}\displaystyle
 &=&
\, c_2 \, p_{2\mu}
\psfrag{f}{\raisebox{-.3cm}{\hspace{-.4cm}\sai $\begin{array}{c}\mu \\ \hspace{-.25cm}\raisebox{.1cm}{$y_2-y_1$} \end{array}$}}
 \psfrag{s}{\sai \hspace{-.3cm} $y_2$}
\psfrag{u}{\raisebox{0cm}{\sai $y_1$}}
\psfrag{d}{\ssi \raisebox{-0.4cm}{\hspace{-.3cm}$-\,\bar{y}_2$}}
\psfrag{e}{\raisebox{0.4cm}{\sai $y_2$}}
 \psfrag{r}{\ssi \hspace{-.5cm} \raisebox{-0cm}{$-\,\bar{y}_1$}}
 \hspace{0.1cm}\raisebox{-.6cm}{\epsfig{file=DipoleU.eps,width=2.8cm}}\,,
\eqa
from which we again deduce that the sum of the two contributions (\ref{Ward_gttG2_p2}) and (\ref{Ward_httG2_p2}) equals zero.

This achieves the proof that the $n-$independence constraint for the $N_c$ terms is automatically fulfilled and does not lead to any new set of equations.

\section{Comparison between LCCF and CCF results}
\label{Ap:Comparison}

Let us first consider the 2-parton spin-flip contribution (\ref{Flip2}).
Using the equation of motions (\ref{em_rho1}, \ref{em_rho2}) we can
express $\varphi_1^T$ and $\varphi_A^T$ in Eq.~(\ref{Flip2}) in terms
of the remaining distributions and we thus obtain from the LCCF approach
\begin{eqnarray}
 \label{Flip2noT}
&&
\Phi_{ f. 2}^{\gamma^*_T\to\rho_T}(\kb^2)=
\frac{C^{ab}}2
 \\ && \hspace{-.3cm}
\times \int\limits_0^1
 \frac{dy_1\;\;4 \, \alpha}{\left( \alpha + \left( 1 - {y_1} \right) \,{y_1} \right) ^2}
\left[ -\frac{1}{2}\left((1-2y_1)^2+1  \right)\varphi_3(y_1)
-(1-2y_1)\varphi_A(y_1) +2y_1\int\limits_0^{y_1}dy_2\,(\zV \,B\, - \zA \,D )(y_2,y_1)    \right]\,.
\nonumber
\end{eqnarray}
Now our aim is to show that the Eq.~(\ref{Flip2noT}) coincides with
the corresponding expression of the CCF approach, i.e. with the flip part of Eq.(\ref{2bodyCov}), which reads
\begin{eqnarray}
\label{DqqFlip}
&&\Phi^{q\bar q}_{f.}(\alpha)=T_f\int\limits_0^1\,
\frac{dz\;2\alpha}{(\alpha +z\bar z)^2}\left\{
(2z-1)\left[
\int\limits_0^z\,dv\,(\phi_{\parallel}(z) - g_\perp^{v}(z))
-\zeta_3^V\int\limits_0^z\,d\alpha_1\int\limits_0^{\bar z}\,d\alpha_2\,
\frac{V(\alpha_1,\alpha_2)}{\alpha_g^2}
\right] \right.
\nonumber \\
&&\left. - \frac{1}{4}g^a_\perp + \zeta_3^A \,\int\limits_0^z\,d\alpha_1\int\limits_0^{\bar z}\,d\alpha_2\,\frac{A(\alpha_1,\alpha_2)}{\alpha_g^2}\;.
\right\}
\end{eqnarray}

We begin with rewriting (\ref{DqqFlip}) using the results of Ref.~\cite{BB}
in such a way that the form      of
resulting expression will be similar to one of the Eq.~(\ref{Flip2noT}).
For that we reexpress in (\ref{DqqFlip}) the
twist-2 distribution $\phi_{\parallel}(z)$ in terms of other DAs.
Let us note that $\phi_{\parallel}(z)$ satisfies equations
\begin{equation}
\label{phi1}
\phi_{\parallel}(z) = -z\frac{d}{dz}\left(g_\perp^{(v)}(z) + \frac{d}{4dz}g^{(a)}_\perp -
g^{\uparrow \downarrow \,gen}  \right)\;,
\end{equation}
or
\begin{equation}
\label{phi2}
\phi_{\parallel}(z) = \bar z\frac{d}{dz}\left(g_\perp^{(v)}(z) -
\frac{d}{4dz}g^{(a)}_\perp -
g^{\downarrow \uparrow \,gen}  \right)\;,
\end{equation}
which follow in an obvious way from the definitions of DAs and their solutions
in the WW approximation. Eqs.~(\ref{phi1}, \ref{phi2}) lead in turn to the
consistency condition
\begin{equation}
\label{ConsCond}
0= \frac{d}{dz}\,g^{(v)}_\perp(z)+\frac{1}{4}(z
- \bar z)\frac{d^2}{dz^2}g^{(a)}_\perp(z) -z \frac{d}{dz}g^{\uparrow \downarrow\,gen}
-\bar z \frac{d}{dz}g^{\downarrow \uparrow\,gen}\;,
\end{equation}
from which it follows  that
\begin{equation}
\label{gasolution}
g^{(a)}_\perp(z)=2 g^{(v)}_\perp(z) +\frac{1}{2}(z-\bar z)\frac{d}{dz}g^{(a)}_\perp(z)
-2z\,g^{\uparrow\downarrow\,gen}- 2\bar z\,g^{\downarrow\uparrow\,gen}
+2\int\limits_0^z\,dv\,g^{\uparrow\downarrow\,gen}(v)- 2
\int\limits_0^z\,dv\,g^{\downarrow\uparrow\,gen}(v)\;.
\end{equation}
This relies on  the boundary conditions
of the different DAs (see
\cite{BB}), which read
\beq
\label{normalizationDAs1}
g^{\uparrow \downarrow}(1)=g^{ \downarrow\uparrow  }(0)=g^{\uparrow \downarrow\,WW}(1)=g^{ \downarrow\uparrow\, WW  }(0)=0\, ,\quad g^{(a)}_\perp(0)=g^{(a)}_\perp(1)=0
\eq
and
\beq
\label{normalizationDAs2}
\int\limits^1_0 du \, g^{\uparrow \downarrow \,WW}(u) = \int\limits^1_0 du \, g^{\uparrow \downarrow}(u)=\int\limits^1_0 du \, g^{ \downarrow \uparrow\,WW}(u) = \int\limits^1_0 du \, g^{\downarrow \uparrow}(u)=1\,.
\eq
Note that they satisfy the symmetrical properties
\beq
\label{symDAS}
g^{\uparrow \downarrow}(u)  \stackrel{u \,\leftrightarrow \,\bar{u}}{\longleftrightarrow}g^{ \downarrow \uparrow}(u)\,.
\eq
The substitution in Eq.~(\ref{DqqFlip}) of $\phi_\parallel(z)$  by the
expression
(\ref{phi1}) and then replacing  resulting $g_\perp^a$'s by the formula
(\ref{gasolution}) leads to the equation
\begin{eqnarray}
\label{DqqFlipTrans}
&&\Phi^{q\bar q}_{f.}(\alpha)=T_f\int\limits_0^1\,
\frac{dz\;2\alpha}{(\alpha +z\bar z)^2}\left\{
-\frac{1}{2}\left((z - \bar z)^2 +1 \right)g^{(v)}_\perp(z) +
\left( 1-2z  \right)\frac{d}{4\,dz}g^{(a)}_\perp(z) \right.
\\
&& \left. +2z^2 g^{\uparrow \downarrow\,gen}(z) -2z\int\limits_0^z\,dv\,
g^{\uparrow \downarrow\,gen}(v)
-\zV(2z-1)\int\limits_0^z\,d\alpha_1
\int\limits_0^{\bar z}\,d\alpha_2\,\frac{V(\alpha_1,\alpha_2)}{\alpha_g^2}
+\zA\int\limits_0^z\,d\alpha_1
\int\limits_0^{\bar z}\,d\alpha_2\,\frac{A(\alpha_1,\alpha_2)}{\alpha_g^2}
\right\}\;,
\nonumber
\end{eqnarray}
in which we observe that the two terms in the first line of
(\ref{DqqFlipTrans}) reproduce -using the vocabulary formulas
(\ref{relBBvector}, \ref{relBBaxial}) - the
first two terms of expression in $[...]$ of Eq.~(\ref{Flip2noT}).

Now consider the expression in the second line of (\ref{DqqFlipTrans}), i.e.
\begin{eqnarray}
\label{DqqNonFlipTrans2line}
&&\Phi^{q\bar q}_{f.}(\alpha)|_{2nd \,line}=T_f\int\limits_0^1\,
\frac{dz\;2\alpha}{(\alpha +z\bar z)^2}\\
&&\hspace{-.4cm} \times \left\{  2z^2 g^{\uparrow \downarrow\,gen}(z) -2z\int\limits_0^z\,dv\,
g^{\uparrow \downarrow\,gen}(v)
-\zV(2z-1)\int\limits_0^z\,d\alpha_1
\int\limits_0^{\bar z}\,d\alpha_2\,\frac{V(\alpha_1,\alpha_2)}{\alpha_g^2}
+\zA\int\limits_0^z\,d\alpha_1
\int\limits_0^{\bar z}\,d\alpha_2\,\frac{A(\alpha_1,\alpha_2)}{\alpha_g^2}
\right\}\;.
\nonumber
\end{eqnarray}
Using results of Ref.~\cite{BB} the formula for
$\Phi^{q\bar q}_{f.2}(\alpha)|_{2nd \,line}$ reads
\begin{eqnarray}
\label{DqqFlipTrans2line2}
&&\Phi^{q\bar q}_{f.}(\alpha)|_{2nd \,line}=T_f\int\limits_0^1\,
\frac{dz\;4\alpha}{(\alpha +z\bar z)^2}\\
&& \hspace{-.4cm}\times \left\{ \zV \!\left[
z^2\left(\int\limits_z^1 \frac{dv}{v}\,H(v) + M(z)   \right)\!
-z\int\limits_0^z \! dv \left(
\int\limits_v^1 \frac{du}{u}\,H(u) + M(v)
\right) \!
-\frac{1}{2}(2z-1)\int\limits_0^z\,d\alpha_1
\int\limits_0^{\bar z}\,d\alpha_2
\frac{V(\alpha_1,\alpha_2)}{(1-\alpha_1-\alpha_2)^2}
\right] \right.
\nonumber \\
&&\left. +\zA\left[
z^2\left(\int\limits_z^1 \frac{dv}{v}\,L(v) + N(z)   \right)
-z\int\limits_0^z dv \left(
\int\limits_v^1 \frac{du}{u}\,L(u) + N(v)
\right)
+\frac{1}{2}\int\limits_0^z\,d\alpha_1
\int\limits_0^{\bar z}\,d\alpha_2
\frac{A(\alpha_1,\alpha_2)}{(1-\alpha_1-\alpha_2)^2}
\right] \right\}\;,
\nonumber
\end{eqnarray}
and after interchanging of integration over the variables
$v$ and $u$ it can be simplified
and takes the form
\begin{eqnarray}
\label{DqqFlipTrans2line3}
&&\Phi^{q\bar q}_{f.}(\alpha)|_{2nd \,line}=T_f\int\limits_0^1\,
\frac{dz\;4\alpha}{(\alpha +z\bar z)^2}\\
&&\times \left\{ \zV \left[
z^2 M(z)
-z\int\limits_0^z dv \left(
H(v) + M(v)
\right)
-\frac{1}{2}(2z-1)\int\limits_0^z\,d\alpha_1
\int\limits_0^{\bar z}\,d\alpha_2
\frac{V(\alpha_1,\alpha_2)}{(1-\alpha_1-\alpha_2)^2}
\right] \right.
\nonumber \\
&&\left. +\,\zA \left[
z^2  N(z)
-z\int\limits_0^z dv \left(L(v) + N(v)
\right)
+\frac{1}{2}\int\limits_0^z\,d\alpha_1
\int\limits_0^{\bar z}\,d\alpha_2
\frac{A(\alpha_1,\alpha_2)}{(1-\alpha_1-\alpha_2)^2}
\right] \right\}\;.
\nonumber
\end{eqnarray}
The expressions $M(v)$, $H(v)$, $L(v)$ and $N(v)$ are the integrals over
the vector and axial-vector 3-partonic DAs
\begin{equation}
\label{M}
M(v)=\frac{d}{dv}\int\limits_0^v\,d\alpha_1\,\int\limits_0^{\bar v}
d\alpha_2\,\frac{V(\alpha_1,\alpha_2)}{1-\alpha_1-\alpha_2}\;,
\end{equation}
\begin{equation}
\label{H}
H(v)=\frac{d}{dv}\int\limits_0^v\,d\alpha_1\,\int\limits_0^{\bar v}
\frac{d\alpha_2}{1-\alpha_1-\alpha_2}\left(\alpha_2 \frac{d}{d\alpha_2}
+ \alpha_1 \frac{d}{d\alpha_1}  \right)V(\alpha_1,\alpha_2)\;,
\end{equation}
\begin{equation}
\label{N}
N(v)=\int\limits_0^v\,d\alpha_1\,\int\limits_0^{\bar v}
\frac{d\alpha_2}{1-\alpha_1-\alpha_2}\left( \frac{d}{d\alpha_2}
+  \frac{d}{d\alpha_1}  \right)A(\alpha_1,\alpha_2)\;,
\end{equation}
\begin{equation}
\label{L}
L(v)=\frac{d}{dv}\int\limits_0^v\,d\alpha_1\,\int\limits_0^{\bar v}
\frac{d\alpha_2}{1-\alpha_1-\alpha_2}\left(\alpha_1 \frac{d}{d\alpha_1}
- \alpha_2 \frac{d}{d\alpha_2}  \right)A(\alpha_1,\alpha_2)\;.
\end{equation}
Using the vocabulary (\ref{DictB}, \ref{DictD}) and calculating the derivatives in
above definitions we put them into the forms
\begin{equation}
\label{ourM}
M(u)=-\int\limits_u^1 \,dv\,B(u,v) + \int\limits_0^u\,dv\,B(v,u)
\;,
\end{equation}
\begin{equation}
\label{ourH}
H(u)=\int\limits_u^1 \,dv\,\left(\frac{1}{v-u}-u\frac{d}{du} \right)B(u,v)
 - \int\limits_0^u\,dv\,\left(\frac{1}{u-v} +\bar u\frac{d}{du} \right)B(v,u)
\;,
\end{equation}
\begin{equation}
\label{ourN}
N(u)=-\int\limits_u^1\,dv\,D(u,v) - \int\limits_0^u\,dv\,D(v,u)+
2\int\limits_u^1dy_2 \int\limits_0^u dy_1 \frac{D(y_1,y_2)}{y_2-y_1}\;,
\end{equation}
\begin{equation}
\label{ourL}
L(u)= -u\int\limits_u^1\,dv\,\frac{d}{du}\,D(u,v)
+\bar u \int\limits_0^u\,dv\,\frac{d}{du}\,D(v,u) -
(\bar u - u)\left(\int\limits_u^1\,\frac{dv}{v-u}\,D(u,v) -
\int\limits_0^u\,\frac{dv}{u-v}\,D(v,u)   \right)\;\,
\end{equation}
from which we calculate integrals appearing in the formula
(\ref{DqqFlipTrans2line3})
\begin{equation}
\label{intM}
-z\int\limits_0^z\,du\,M(u)= z\int\limits_0^z\,dy_1\,\int\limits_z^1\,dy_2
\,B(y_1,y_2)\;
\end{equation}
\begin{equation}
\label{intH}
-z\int\limits_0^z\,du\,H(u)=
z\left[ -\int\limits_z^1\,dy_1\,\int\limits_0^z\,dy_2\,B(y_2,y_1)
\left(\frac{1}{y_1-y_2}+1  \right) +z\int\limits_z^1\,du\,B(z,u)
+\bar z \int\limits_0^z\,du\,B(u,z)   \right]
\end{equation}
\begin{equation}
\label{intN}
-z\int\limits_0^z\,du\,N(u)=
z \int\limits_z^1\,dy_1\,\int\limits_0^z\,dy_2\,D(y_2,y_1)\,
\left(1 - \frac{2(z-y_2)}{y_1-y_2}  \right)
\end{equation}
\begin{equation}
\label{intL}
-z\int\limits_0^z\,du\,L(u)=
z\left[
z\int\limits_z^1\,du\,D(z,u) -\bar z \int\limits_0^z\,du\,D(u,z) +
\int\limits_z^1\,dy_1\,\int\limits_0^z\,dy_2\,D(y_2,y_1)\left(-1 +
\frac{\bar y_2 - y_2}{y_1-y_2}   \right)
\right]\;.
\end{equation}
Finally, after substitution into Eq.~(\ref{DqqFlipTrans2line3}) of the
expressions
(\ref{M}, ... , \ref{intL}) and noting that
\begin{equation}
\label{integralB}
\int\limits_0^z\,d\alpha_1\,\int\limits_0^{\bar z}\,d\alpha_2\,
\frac{V(\alpha_1,\alpha_2)}{(1-\alpha_1-\alpha_2)^2}
= -\int\limits_0^z\,dy_1\int\limits_z^1\,dy_2\,\frac{B(y_1,y_2)}{y_2-y_1}
\end{equation}
and
\begin{equation}
\label{integralD}
\;\;\;\;\;\;
\int\limits_0^z\,d\alpha_1\,\int\limits_0^{\bar z}\,d\alpha_2\,
\frac{A(\alpha_1,\alpha_2)}{(1-\alpha_1-\alpha_2)^2}
= -\int\limits_0^z\,dy_1\int\limits_z^1\,dy_2\,\frac{D(y_1,y_2)}{y_2-y_1}
\end{equation}
and the simplification resulting on the following properties
\begin{equation}
\label{integralBvan}
\int\limits^1_0 dz \, f_{sym}(z) \int\limits_0^z\,dy_1\int\limits_z^1\,dy_2\,\frac{B(y_1,y_2)}{y_2-y_1}=0
\end{equation}
and
\begin{equation}
\label{integralDvan}
\;\;\;\;\;\;
\int\limits^1_0 dz \, f_{antisym}(z)\int\limits_0^z\,dy_1\int\limits_z^1\,dy_2\,\frac{D(y_1,y_2)}{y_2-y_1}=0
\end{equation}
where $f_{sym}$ (resp. $f_{antisym}$) is (anti)symmetric with respect to $z \leftrightarrow \bar{z}\,,$
we obtain that
\begin{equation}
\Phi^{q\bar q}_{f.}(\alpha)|_{2nd \,line}=T_f\int\limits_0^1\,
\frac{dz\;2\alpha}{(\alpha +z\bar z)^2}\left\{ 2z\int\limits_0^z\,du\,\left(
\zV B(u,z) - \zA D(u,z)   \right)\right\}\;,
\end{equation}
which reproduces the last term in Eq.~(\ref{Flip2noT}).
\\

We now proceed our comparison by considering the 3-parton flip contribution.
We should thus compare the last line of (\ref{3BodyCov}), which is the covariant 3-parton flip result, with the corresponding LCCF result (\ref{Flip3}).
This is straightforward after using the dictionary (\ref{DictB}, \ref{DictD}) in the last line of (\ref{3BodyCov}), in order to express it only in terms of $B$ and $D$.
Then, using the symmetrical properties Eq.(\ref{sym2}) ($B(y_1, y_2)$ and $D(y_1, y_2)$
are respectively  antisymmetric and symmetric under the transformation
$y_1 \leftrightarrow \bar{y}_2$), one should rewrite the  prefactors of $B$ in an antisymmetrical form (and correspondingly the prefactor of $D$ in a symmetrical form).
Doing this for both expressions leads to identical expressions.

The same method applies to  $N_c$ 3-parton non-flip contribution. Indeed, comparing the $N_c/C_F=2 \, c_f$ term of Eq.(\ref{NonFlip3}) with the corresponding $c_f$ contribution for
the $T_{n.f.}$ tensor coefficient of (\ref{3BodyCov}) after using the dictionary  (\ref{DictB}, \ref{DictD}) and
restoring the symmetry of the integrands, one get identical expressions.
\\

We now focus on the non-trivial proof that the $C_F$ contribution arising from the non-flip
contributions are identical in LCCF and CCF approaches. Only this part leads to potential violations of gauge invariance, as we saw in CCF approach of  section\ref{SubSec_ImpactCCF}, where it was needed to combine $\Delta \Phi^2$ and $\Delta \Phi^3$.  In LCCF approach, this also requires to combine 2 and 3-parton contributions.
We should thus prove that the 2-parton contribution (\ref{NonFlip2}) supplemented by
the constant term (in color space) of Eq.(\ref{NonFlip3})  is identical with the constant term (in color space) of the  $T_{n.f.}$  coefficient of (\ref{3BodyCov})
supplemented by the
$T_{n.f.}$  coefficient of (\ref{2bodyCov}).

Consider first the 3-parton non-flip contribution with $C_F$ color structure of the LCCF result (\ref{NonFlip3}). It reads
\bea
\label{NonFlip3CF}
&&\Phi_{n.f. 3}^{\gamma^*_T\to\rho_T\, C_F}(\kb^2)=
C^{ab} \int\limits_0^1 dy_1 \int\limits_0^1 dy_2
  \\
&&\hspace{-.3cm}\times \, \left\{ \zV \, B\left(y_1,y_2\right)\left[\frac{-2 \, \alpha \, y_1   \left(2 y_1-1\right)}{\alpha+ \left(1-y_1\right)
   y_1} \frac{1
   }{\left(y_1-y_2+1\right)
   \alpha+ y_1
   \left(1-y_2\right)}+\frac{2 \, y_1}{\alpha+ \left(1-y_1\right)
   y_1}-2/\bar{y}_1 +1/\bar{y}_1\right]\right. \nonumber\\
&&\left.+\,\zA \, D\left(y_1,y_2\right)
\left[\frac{-2 \, \alpha \, y_1   }{\alpha+ \left(1-y_1\right)
   y_1} \frac{1
   }{\left(y_1-y_2+1\right)
   \alpha+ y_1
   \left(1-y_2\right)}+\frac{2 \, y_1}{\alpha+ \left(1-y_1\right)
   y_1}-\frac{2}{\bar{y}_1} +\frac{1}{\bar{y}_1}\right]
\right\}\,.\nonumber
\eea
In this expression, we have written the last term $-1/\bar{y}_1$ of $B$ and $D$ contributions as $-2/\bar{y}_1 +1/\bar{y}_1$
in order to prepare the identification with the CCF approach result. Removing for a moment the  $1/\bar{y}_1$ contribution from Eq.(\ref{NonFlip3CF}) (i.e. last term for both $B$ and $D$ brackets), the obtained expression vanishes in the $\alpha \to 0$ limit (we therefore denote it with an index $van.$), and, after restoring the appropriate symmetry under $y_1 \leftrightarrow \bar{y}_2$ (antisymmetrized
for $B$ and symmetrized for $D$), it reads
\bea
\label{NonFlip3CFvanish}
&&\Phi_{n.f. 3\, van.}^{\gamma^*_T\to\rho_T\, C_F}(\kb^2)=
C^{ab}\,\alpha \int\limits_0^1 d y_1 \int\limits_0^1 d y_2
 \left\{ \zV \, B\left(y_1,y_2\right)\left[\frac{y_1+ \bar{y}_2}{(y_1-y_2)((y_1 + \bar{y}_2)\alpha + y_1 \, \bar{y}_2)} \left( \frac{1}{\bar{y}_2}-\frac{1}{y_1} \right)\right.\right. \nonumber \\
&& \left. \left. -\frac{1}{y_1-y_2} \frac{y_1-\bar{y}_1}{y_1 (\alpha+ y_1 \bar{y}_1)} +\frac{1}{y_1-y_2} \frac{\bar{y}_2-y_2}{\bar{y}_2 (\alpha+ y_2 \bar{y}_2)}-\frac{1}{\bar{y}_1}\frac{1}{\alpha + y_1 \bar{y}_1} +\frac{1}{y_2}\frac{1}{\alpha + y_2 \bar{y}_2} \right]\right. \nonumber \\
&&\left.+\zA \,D\left(y_1,y_2\right)
\left[\frac{y_1+ \bar{y}_2}{(y_1-y_2)((y_1 + \bar{y}_2)\alpha + y_1 \, \bar{y}_2)} \left( \frac{1}{\bar{y}_2}+\frac{1}{y_1} \right)\right.\right. \nonumber \\
&& \left. \left. -\frac{1}{\alpha+ y_1 \bar{y}_1)}\left(\frac{1}{y_1(y_1-y_2)} +\frac{1}{\bar{y}_1}\right)
-\frac{1}{\alpha+ y_2 \bar{y}_2)}\left(\frac{1}{\bar{y}_2(y_1-y_2)} +\frac{1}{y_2}\right)\right]\right. \,.
\eea
The corresponding expression obtained from the CCF approach reads, according to (\ref{3BodyCov}),
\bea
\label{3BodyCovCF}
&&\Phi^{q\bar q g \, C_F}_{n.f.}(\alpha)=\int Dz \frac{2\alpha  }{z_1 z_2 z_g^2}
(\zeta_3^V V(z_1,z_2) + \zeta_3^A A(z_1,z_2)) \,\left(
\frac{\bar z_g z_1}{\alpha \bar z_g+z_1z_2}
-\frac{z_1 z_2^2}{\bar z_1(\alpha+z_1 \bar z_1)}
-\frac{z_1\bar z_2}{\alpha+z_2 \bar z_2}
\right) \nonumber \\
\eea
which after using the dictionary (\ref{DictB}, \ref{DictD}) turns into
\bea
\label{3BodyCovCF1}
&&\Phi^{q\bar q g \, C_F}_{n.f.}(\alpha)=-\int\limits_0^1 d z_1 \int\limits_0^1 d z_2 \, \frac{2\alpha  }{z_1 z_2 (1-z_1-z_2)}
(\zeta_3^V B(z_1,1-z_2) + \zeta_3^A D(z_1,1-z_2)) \\
&&\times \left(
\frac{(z_1+z_2) z_1}{\alpha  (z_1+z_2)+z_1 \, z_2}
-\frac{z_1 \, z_2^2}{\bar z_1(\alpha+z_1 \,\bar z_1)}
-\frac{z_1\,\bar z_2}{\alpha+z_2 \,\bar z_2}\,.
\right) \nonumber \\
\eea
This reads, after restoring the symmetry properties of the integrands,
\bea
\label{3BodyCovCF2}
&&\Phi^{q\bar q g \, C_F}_{n.f.}(\alpha)=\alpha \int\limits_0^1 d y_1   \int\limits_0^1 d y_2 \, \frac{1}{y_1-y_2} \\
&&\hspace{-.5cm}\times
\left\{\zV B(y_1,y_2) \left[
\frac{y_1+ \bar{y}_2}{(y_1 + \bar{y}_2)\,\alpha + y_1 \, \bar{y}_2} \left( \frac{1}{y_1}-\frac{1}{\bar{y}_2} \right)
+\frac{\bar{y}_2}{\bar{y}_1}
\frac{1}
{\alpha+ y_1 \bar{y}_1}
+\frac{y_2}{\bar{y}_2}
\frac{1}
{\alpha+ y_2 \bar{y}_2}
-\frac{y_1}{y_2}
\frac{1}
{\alpha+ y_2 \bar{y}_2}
-\frac{\bar{y}_1}{y_1}
\frac{1}
{\alpha+ y_1 \bar{y}_1}\right]\right.\nonumber \\
&&\hspace{-.5cm}
\left.+\zA D(y_1,y_2) \left[
-\frac{y_1+ \bar{y}_2}{(y_1 + \bar{y}_2)\alpha + y_1 \, \bar{y}_2} \left( \frac{1}{y_1}+\frac{1}{\bar{y}_2} \right)
+\frac{\bar{y}_2}{\bar{y}_1}
\frac{1}
{\alpha+ y_1\, \bar{y}_1}
+\frac{y_2}{\bar{y}_2}
\frac{1}
{\alpha+ y_2 \,\bar{y}_2}
+\frac{y_1}{y_2}
\frac{1}
{\alpha+ y_2 \bar{y}_2}
+\frac{\bar{y}_1}{y_1}
\frac{1}
{\alpha+ y_1\, \bar{y}_1}\right]\right\}\nonumber
\eea
which agrees with the result
(\ref{NonFlip3CFvanish}) after elementary algebra.

We now prove that the 2-parton non-flip $C_F$ LCCF result (\ref{NonFlip2}) agrees with the non-flip $C_F$ CCF contribution $\Phi^{q\bar q}_{n.f.}$,
after supplementing the LCCF result with the 3-parton $1/\bar{y}_1$ term of Eq.(\ref{NonFlip3CF}).

Starting from the LCCF result (\ref{NonFlip2}) and using the equations of motions (\ref{em_rho1}, \ref{em_rho2}) we can
express $\varphi_1^T$ and $\varphi_A^T$ in Eq.~(\ref{Flip2}) in terms
of the remaining distributions, which leads to
\bea
 \label{NonFlip2Modif}
&&\hspace{-.4cm}\Phi_{ n.f. 2}^{\gamma^*_T\to\rho_T}(\kb^2)=
\frac{C^{ab}}2 \int\limits_0^1 \frac{d y_1}{y_1 \, \bar{y}_1}
 \left[-\frac{2 \, \alpha
   \left(\alpha+2  \, y_1 \bar{y}_1 \right)}{
   \left(\alpha+ \, y_1 \, \bar{y}_1\right)^2}
\left\{
-2 \, y_1 \, \bar{y}_1 \, \varphi_3 (y_1) - y_1 \int\limits^1_{y_1} dy_2 \left[\zV \,B(y_1, y_2) + \zA \,D(y_1, y_2)  \right]\right.\right.\nonumber \\
&& \hspace{0cm}\left. \left.
- \bar{y}_1\! \int\limits^{y_1}_0 \!dy_2 \left[-\zV \,B( y_2,y_1) + \zA \,D( y_2,y_1)  \right]\!
\right\}
-y_1 \!\!\int\limits^1_{y_1}\!dy_2\left[\zV \,B(y_1, y_2) + \zA \,D(y_1, y_2)  \right] \right.\nonumber\\
&&\left. - \bar{y}_1 \!\int\limits^{y_1}_0 \!dy_2 \left[-\zV \,B( y_2,y_1) + \zA \,D( y_2,y_1)  \right]
\right].
\eea
Now let us consider the non-flip CCF result  (\ref{2bodyCov}, \ref{defPhi+-}), which reads
\begin{eqnarray}
\label{DqqNonFlip}
&&\Phi^{q\bar q}_{n.f.}(\alpha)=T_{n.f}\int\limits_0^1dz
\frac{\alpha (\alpha+ 2 z \, \bar{z})}{z \, \bar{z} (\alpha +z\bar z)^2}\left\{
(2z-1)\left[
\int\limits_0^z\,dv\,(\phi_\parallel(z) - g_\perp^{(v)}(z))
-\zV\int\limits_0^z\,d\alpha_1\int\limits_0^{\bar z}\,d\alpha_2\,
\frac{V(\alpha_1,\alpha_2)}{\alpha_g^2}
\right] \right.
\nonumber \\
&&\left. + \frac{1}{4}g^{(a)}_\perp - \zA \,\int\limits_0^z \, d\alpha_1\int\limits_0^{\bar z}\,d\alpha_2\,\frac{A(\alpha_1,\alpha_2)}{\alpha_g^2}
\right\}\,.
\end{eqnarray}
The same series of transformations which led from
(\ref{DqqNonFlip}) to (\ref{DqqFlipTrans}) now gives
\begin{eqnarray}
\label{DqqNonFlipTrans}
&&\Phi^{q\bar q}_{n.f.}(\alpha)=T_{n.f.}\int\limits_0^1\,
dz
\frac{2 \, \alpha \,(\alpha+ 2 z \, \bar{z})}{z \, \bar{z} (\alpha +z\bar z)^2}
\left\{
z \, \bar{z} \, g^{(v)}_\perp(z) - z \, \bar{z} \, g^{\uparrow \downarrow\,gen}(z)
+\bar{z} \int\limits_0^z\,dv\,
g^{\uparrow \downarrow\,gen}(v)\right. \\
&&\left.
-\frac{z-\bar{z}}2\,\zV\int\limits_0^z\,d\alpha_1
\int\limits_0^{\bar z}\,d\alpha_2\,\frac{V(\alpha_1,\alpha_2)}{\alpha_g^2}
-\frac{1}{2}\,\zA\int\limits_0^z\,d\alpha_1
\int\limits_0^{\bar z}\,d\alpha_2\,\frac{A(\alpha_1,\alpha_2)}{\alpha_g^2}
\right\}\;.
\nonumber
\end{eqnarray}
Using the dictionary (\ref{relBBvector}) one immediately gets agreement between
the first term of (\ref{NonFlip2Modif}) involving $\varphi_3$  and the first term of
(\ref{DqqNonFlipTrans})  involving $g_\perp^v\,.$
We now consider the remaining term of (\ref{DqqNonFlipTrans}), involving $B$ and $D$ through $g^{\uparrow \downarrow\,gen},$ $g^{\downarrow \uparrow\,gen},$ $V$ and $A\,.$

Using results of Ref.~\cite{BB} the formula for
$\Phi^{q\bar q}_{n.f.}(\alpha)|_{rem.}$ reads, after interchanging the integration over the variables
$v$ and $u$ as we did in order to get (\ref{DqqFlipTrans2line3}),
\begin{eqnarray}
\label{DqqNonFlipTrans2rem}
&&\Phi^{q\bar q}_{n.f.}(\alpha)|_{rem.}=T_{n.f.}\int\limits_0^1\,
dz \,
\frac{2 \, \alpha \,(\alpha+ 2 z \, \bar{z})}{z \, \bar{z} (\alpha +z\bar z)^2}\\
&&\times \left\{ \zV \left[
-z \, \bar{z} \, M(z)
+\bar{z}\int\limits_0^z dv \left(
H(v) + M(v)
\right)
-\frac{1}{2}(2z-1)\int\limits_0^z\,d\alpha_1
\int\limits_0^{\bar z}\,d\alpha_2
\frac{V(\alpha_1,\alpha_2)}{(1-\alpha_1-\alpha_2)^2}
\right] \right.
\nonumber \\
&&\left. +\,\zA \left[
-z \, \bar{z} \, N(z)
+\bar{z}\int\limits_0^z dv \left(L(v) + N(v)
\right)
-\frac{1}{2}\int\limits_0^z\,d\alpha_1
\int\limits_0^{\bar z}\,d\alpha_2
\frac{A(\alpha_1,\alpha_2)}{(1-\alpha_1-\alpha_2)^2}
\right] \right\}\;.
\nonumber
\end{eqnarray}

Finally, after substitution into Eq.~(\ref{DqqNonFlipTrans2rem}) of the
expressions
(\ref{M}, ... , \ref{intL}) and using the properties (\ref{integralB}) and (\ref{integralD}), one obtains
\begin{eqnarray}
\label{DqqNonFlipTrans2remSimplified}
\Phi^{q\bar q}_{n.f.}(\alpha)|_{rem.}=-T_{n.f.}\int\limits^1_0 dz \,
\frac{2 \, \alpha \,(\alpha+ 2 z \, \bar{z})}{z \, (\alpha +z\bar z)^2} \int\limits^z_0 dy_1 \, \left[\zV \,B(y_1, z)-\zA \, D(y_1, z)\right]\,.
\end{eqnarray}
Coming back to the LCCF result, the remaining contribution after removing the $\varphi_3$ term reads, after using the symmetrical properties (\ref{sym2}),
\bea
 \label{NonFlip2ModifSym}
&&\hspace{-.4cm}\Phi_{ n.f. 2 \, rem.}^{\gamma^*_T\to\rho_T}(\kb^2)=
\frac{C^{ab}}2 \int\limits_0^1 \frac{d y_1}{y_1}
 \left\{-\frac{4 \, \alpha
   \left(\alpha+2  \, y_1 \bar{y}_1 \right)}{
   \left(\alpha+ \, y_1 \, \bar{y}_1\right)^2}
 \int\limits_0^{y_1} dy_2 \left[\zV \, B(y_2, y_1) -\zA \, D(y_1, y_2)  \right]\right.\nonumber \\
&& \hspace{-.5cm}\left.
- 2\! \int\limits^{y_1}_0 \!dy_2 \left[-\zV \, B( y_2,y_1) + \zA \,D( y_2,y_1)  \right]\!
\right\}
.
\eea
Now, the remaining term coming from the LCCF $C_F$ non-flip 3-parton contribution (\ref{NonFlip3CF}) which we omitted for the moment is
\bea
\label{NonFlip3CFapart}
&&\Phi_{n.f. 3\, rem.}^{\gamma^*_T\to\rho_T\, C_F}(\kb^2)=
C^{ab}\int\limits_0^1 dy_1 \int\limits_0^1 dy_2
 \, \left\{ \zV \, B\left(y_1,y_2\right)\frac{1}{\bar{y}_1}+\zA \, D\left(y_1,y_2\right)
\frac{1}{\bar{y}_1}
\right\}\,.
\eea
Using the symmetrical properties (\ref{sym2}), this contribution is opposite to the contribution from
the second line of (\ref{NonFlip2ModifSym}) while the first line of (\ref{NonFlip2ModifSym})
is identical to the result (\ref{DqqNonFlipTrans2remSimplified}). This ends the proof of the exact equivalence
between LCCF  and CCF results.

%\end{appendix}


\begin{thebibliography}{99}


\bibitem{fact}
J.~C.~Collins, L.~Frankfurt, M.~Strikman,
%``Factorization for hard exclusive electroproduction of mesons in
%QCD,''
Phys.\ Rev.\ D {\bf 56}, 2982 (1997).
%%CITATION = HEP-PH 9611433;%%

\bibitem{DGP}
  M.~Diehl, T.~Gousset and B.~Pire,
  %``Exclusive electroproduction of vector mesons and transversity
  %distributions,''
  Phys.\ Rev.\  D {\bf 59}, 034023 (1999);
  %%CITATION = PHRVA,D59,034023;%%
  J.~C.~Collins and M.~Diehl,
  %``Transversity distribution does not contribute to hard exclusive
  %electroproduction of mesons,''
  Phys.\ Rev.\  D {\bf 61}, 114015 (2000).
  %%CITATION = PHRVA,D61,114015;%%


  \bibitem{MP}
L.~Mankiewicz and G.~Piller, Phys. Rev. D {\bf 61}, 074013 (2000);




\bibitem{AT}
I.~V.~Anikin and O.~V.~Teryaev,
  %``Genuine twist 3 in exclusive electroproduction of transversely polarized
  %vector mesons,''
  Phys.\ Lett.\  B {\bf 554}, 51 (2003),
  %%CITATION = PHLTA,B554,51;%%
 %``Wandzura-Wilczek approximation from generalized rotational invariance,''
  Phys.\ Lett.\  B {\bf 509}, 95 (2001).
  %%CITATION = PHLTA,B509,95;%%

%\cite{Anikin:2002uv}
\bibitem{Anikin:2002uv}
I.~V.~Anikin and O.~V.~Teryaev,
  %``Non-factorized genuine twist 3 in exclusive electro-production of  vector
  %mesons,''
  Nucl.\ Phys.\  A {\bf 711}, 199 (2002).
  %%CITATION = NUPHA,A711,199;%%



\bibitem{IPST}
D.~Yu.~Ivanov {\em et al.},
%``Probing chiral-odd GPD's in diffractive electroproduction of two
%vector
%mesons,''
Phys.\ Lett.\ B {\bf 550}, 65 (2002);
%%CITATION = PHLTA,B550,65;%%
R.~Enberg, B.~Pire and L.~Szymanowski,
%``Transversity GPD in photo- and electroproduction of two vector
%mesons,''
Eur.\ Phys.\ J.\ C {\bf 47}, 87 (2006).
%[arXiv:hep-ph/0601138].
%%CITATION = EPHJA,C47,87;%%


\bibitem{BB}
  P.~Ball, V.~M.~Braun, Y.~Koike and K.~Tanaka,
  %``Higher twist distribution amplitudes of vector mesons in {QCD}: Formalism
  %and twist three distributions,''
  Nucl.\ Phys.\  B {\bf 529}, 323 (1998);
  %%CITATION = NUPHA,B529,323;%%
   P.~Ball and V.~M.~Braun,
  %``The $\rho$ Meson Light-Cone Distribution Amplitudes of Leading Twist
  %Revisited,''
  Phys.\ Rev.\  D {\bf 54}, 2182 (1996).
  %%CITATION = PHRVA,D54,2182;%%


%\cite{Ball:1998ff}
\bibitem{BBtwist4}
  P.~Ball and V.~M.~Braun,
  %``Higher twist distribution amplitudes of vector mesons in {QCD}: Twist-4
  %distributions and meson mass corrections,''
  Nucl.\ Phys.\  B {\bf 543} (1999) 201.
%  [arXiv:hep-ph/9810475].
  %%CITATION = NUPHA,B543,201;%%



\bibitem{expLow}
S.~A.~Morrow {\it et al.}  [CLAS Collaboration],
  %``Exclusive $\rho^0$ electroproduction on the proton at CLAS,''
  Eur.\ Phys.\ J.\  A {\bf 39}, 5 (2009)
  [arXiv:0807.3834 [hep-ex]].
  %%CITATION = EPHJA,A39,5;%%

 A.~Borissov  [HERMES Collaboration],
  %``Spin density matrix elements from rho0 and phi meson electroproduction at
  %HERMES,''
  AIP Conf.\ Proc.\  {\bf 1105} (2009) 19.
  %%CITATION = APCPC,1105,19;%%

 V.~Y.~Alexakhin {\it et al.}  [COMPASS Collaboration],
  %``Double spin asymmetry in exclusive rho0 muoproduction at COMPASS,''
  Eur.\ Phys.\ J.\  C {\bf 52}, 255 (2007)
  [arXiv:0704.1863 [hep-ex]].
  %%CITATION = EPHJA,C52,255;%%

\bibitem{expHigh}
 A.~Levy,
  %``Exclusive vector meson electroproduction at HERA,''
  arXiv:0711.0737 [hep-ex].
  %%CITATION = ARXIV:0711.0737;%%
 S.~Chekanov {\it et al.}  [ZEUS Collaboration],
  %``Exclusive rho^0 production in deep inelastic scattering at HERA,''
  PMC Phys.\  A {\bf 1}, 6 (2007)
  [arXiv:0708.1478 [hep-ex]].
  %%CITATION = PMCPA,A1,6;%%

\bibitem{IP}
  D.~Yu.~Ivanov, M.~I.~Kotsky and A.~Papa,
  %``The impact factor for the virtual photon to light vector meson
  %transition,''
  Eur.\ Phys.\ J.\  C {\bf 38}, 195 (2004),
  %%CITATION = EPHJA,C38,195;%%
  D.~Yu.~Ivanov and A.~Papa,
  %``Electroproduction of two light vector mesons in the next-to-leading
  %approximation,''
  Nucl.\ Phys.\  B {\bf 732}, 183 (2006) and Eur.\ Phys.\ J.\  C {\bf 49}, 947 (2007).
  %%CITATION = EPHJA,C49,947;%%
  %%CITATION = NUPHA,B732,183;%%

  \bibitem{PSW}
   B.~Pire, L.~Szymanowski and S.~Wallon,
  %``Double diffractive rho-production in gamma* gamma* collisions,''
  Eur.\ Phys.\ J.\  C {\bf 44}, 545 (2005);
  %%CITATION = EPHJA,C44,545;%%
  B.~Pire, M.~Segond, L.~Szymanowski and S.~Wallon,
  %``QCD factorizations in gamma* gamma* --> rho0(L) rho0(L),''
  Phys.\ Lett.\  B {\bf 639}, 642 (2006);
  %%CITATION = PHLTA,B639,642;%%
  R.~Enberg, B.~Pire, L.~Szymanowski and S.~Wallon,
  %``BFKL resummation effects in gamma* gamma* --> rho rho,''
  Eur.\ Phys.\ J.\  C {\bf 45}, 759 (2006)
  [Erratum-ibid.\  C {\bf 51}, 1015 (2007)];
  %%CITATION = EPHJA,C45,759;%%
  M.~Segond, L.~Szymanowski and S.~Wallon,
  %``Diffractive production of two rho0(L) mesons in e+ e- collisions,''
  Eur.\ Phys.\ J.\  C {\bf 52}, 93 (2007).
  %%CITATION = EPHJA,C52,93;%%

  \bibitem{APT}
   I.~V.~Anikin, B.~Pire and O.~V.~Teryaev,
  %``On the gauge invariance of the DVCS amplitude,''
  Phys.\ Rev.\  D {\bf 62} (2000) 071501;
  %%CITATION = PHRVA,D62,071501;%%

\bibitem{EFP}
A.~V.~Efremov and O.~V.~Teryaev,
  %``On Spin Effects In Quantum Chromodynamics,''
  Sov.\ J.\ Nucl.\ Phys.\  {\bf 36}, 140 (1982)
  [Yad.\ Fiz.\  {\bf 36}, 242 (1982)]; E.~V.~Shuryak and A.~I.~Vainshtein,
  %``Theory Of Power Corrections To Deep Inelastic Scattering In Quantum
  %Chromodynamics. 1. Q**2 Effects,''
  Nucl.\ Phys.\  B {\bf 199}, 451 (1982),
  %%CITATION = NUPHA,B199,451;%%
 %E.~V.~Shuryak and A.~I.~Vainshtein,
  %``Theory Of Power Corrections To Deep Inelastic Scattering In Quantum
  %Chromodynamics. 2. Q**4 Effects: Polarized Target,''
  Nucl.\ Phys.\  B {\bf 201}, 141 (1982);
  %%CITATION = NUPHA,B201,141;%%
%\cite{Ellis:1982cd}
%\bibitem{Ellis:1982cd}
R.K.~Ellis {\em et al.},
%  R.~K.~Ellis, W.~Furmanski and R.~Petronzio,
  %``Unraveling Higher Twists,''
  Nucl.\ Phys.\  B {\bf 212} (1983) 29;
  %%CITATION = NUPHA,B212,29;%%
%\bibitem{Efremov:1983eb}
%\nolinebreak
 A.V.~Efremov  and O.V.~Teryaev,
 %``The Transversal Polarization In Quantum Chromodynamics,''
 Sov.\ J.\ Nucl.\ Phys.\  {\bf 39}, 962  (1984);
% [Yad.\ Fiz.\  {\bf 39}, 1517 (1984)].
 %%CITATION = YAFIA,39,1517;%%
O.~V.~Teryaev,
  %``Twist - three in proton nucleon single spin asymmetries,''
  arXiv:hep-ph/0102296;
 A.~V.~Radyushkin and C.~Weiss,
  %``Kinematical twist-3 effects in DVCS as a quark spin rotation,''
  Phys.\ Rev.\  D {\bf 64} (2001) 097504.
  %%CITATION = PHRVA,D64,097504;%%





\bibitem{usSHORT}
 I.~V.~Anikin, D.~Yu.~Ivanov, B.~Pire, L.~Szymanowski and S.~Wallon,
  %``On the description of exclusive processes beyond the leading twist
  %approximation,''
  arXiv:0903.4797 [hep-ph].
  %%CITATION = ARXIV:0903.4797;%%

\bibitem{usProceedings}
I.~V.~Anikin, D.~Yu.~Ivanov, B.~Pire, L.~Szymanowski and S.~Wallon,
  %``QCD factorization beyond leading twist in exclusive rho(T) meson
  %production,''
  arXiv:0904.1482 [hep-ph];
  %%CITATION = ARXIV:0904.1482;%%
%I.~V.~Anikin, D.~Y.~Ivanov, B.~Pire, L.~Szymanowski and S.~Wallon,
  %``gamma* -> rhoT impact factor with twist three accuracy,''
  AIP Conf.\ Proc.\  {\bf 1105} (2009) 390
  [arXiv:0811.2394 [hep-ph]];
  %%CITATION = APCPC,1105,390;%%
%
arXiv:0909.4038 [hep-ph];
%EPS 09
%
%
arXiv:0909.4042 [hep-ph].
%QCD PARIS

\bibitem{DDT}
%\cite{Dokshitzer:1978hw}
  Y.~L.~Dokshitzer, D.~Diakonov and S.~I.~Troian,
  %``Hard Processes In Quantum Chromodynamics,''
  Phys.\ Rept.\  {\bf 58} (1980) 269.
  %%CITATION = PRPLC,58,269;%%


%\cite{Ali:1993vd}
\bibitem{Ali:1993vd}
  A.~Ali, V.~M.~Braun and H.~Simma,
  %``Exclusive radiative B decays in the light cone QCD sum rule approach,''
  Z.\ Phys.\  C {\bf 63} (1994) 437
  [arXiv:hep-ph/9401277].
  %%CITATION = ZEPYA,C63,437;%%


\bibitem{kT}
%\cite{Catani:1990xk}
%\bibitem{Catani:1990xk}
  S.~Catani, M.~Ciafaloni and F.~Hautmann,
  %``GLUON CONTRIBUTIONS TO SMALL x HEAVY FLAVOR PRODUCTION,''
  Phys.\ Lett.\  B {\bf 242} (1990) 97;
  %%CITATION = PHLTA,B242,97;%%
%
%\cite{Catani:1990eg}
%\bibitem{Catani:1990eg}
  S.~Catani, M.~Ciafaloni and F.~Hautmann,
  %``High-energy factorization and small x heavy flavor production,''
  Nucl.\ Phys.\  B {\bf 366} (1991) 135.
  %%CITATION = NUPHA,B366,135;%%

%\cite{Collins:1991ty}
\bibitem{Collins:1991ty}
  J.~C.~Collins and R.~K.~Ellis,
  %``Heavy quark production in very high-energy hadron collisions,''
  Nucl.\ Phys.\  B {\bf 360} (1991) 3.
  %%CITATION = NUPHA,B360,3;%%

%\cite{Levin:1991ry}
\bibitem{Levin:1991ry}
  E.~M.~Levin, M.~G.~Ryskin, Yu.~M.~Shabelski and A.~G.~Shuvaev,
  %``Heavy Quark Production In Semihard Nucleon Interactions,''
  Sov.\ J.\ Nucl.\ Phys.\  {\bf 53} (1991) 657
  [Yad.\ Fiz.\  {\bf 53} (1991) 1059].
  %%CITATION = YAFIA,53,1059;%%




\bibitem{bfkl} E.A.~Kuraev,  L.N.~Lipatov and V.S.~Fadin,
% {\it On the
% Pomeranchuk Singularity in Asymptotically Free Theory},
Phys.~Lett. {\bf B60} (1975) 50-52;
% {\it Multiregge processes in the Yang-Mills
%theory},
Sov.~Phys.~JETP {\bf 44} (1976) 443-451;
%{\it The Pomeranchuk Singularity
% in Non Abelian Gauge Theories},
 Sov.~Phys.~JETP {\bf 45} (1977) 199-204;
Ya.Ya.~Balitskii and L.N.~Lipatov,
% {\it The Pomeranchuk Singularity in Quantum
% Chromodynamics},
Sov.~J.~Nucl.~Phys. {\bf 28} (1978) 822-829.

\bibitem{ginzburg}
  I.~F.~Ginzburg, S.~L.~Panfil and V.~G.~Serbo,
  %``Possibility Of The Experimental Investigation Of The QCD Pomeron In
  %Semihard Processes At The Gamma Gamma Collisions,''
  Nucl.\ Phys.\ B {\bf 284} (1987) 685.
  %%CITATION = NUPHA,B284,685;%%

\bibitem{GK}
  S.~V.~Goloskokov and P.~Kroll,
 %``Vector meson electroproduction at small Bjorken-x and generalized  parton
  %distributions,''
  Eur.\ Phys.\ J.\  {\bf C 42}, 281 (2005);
  %%CITATION = EPHJA,C42,281;%%
  %%%%%%%%%%%%%%%%
  %``The longitudinal cross section of vector meson electroproduction,''
  %Eur.\ Phys.\ J.\  C {\bf 50}, 829 (2007);
  ibid. {\bf C  50}, 829 (2007);
%%%%%%%%%%%%%%%%%%%%
%``The role of the quark and gluon GPDs in hard vector-meson
  %electroproduction,''
  %Eur.\ Phys.\ J.\  C {\bf 53}, 367 (2008).
ibid. {\bf C  53}, 367 (2008).
%\bibitem{DA}
%DA



\bibitem{usPhen}
I.~V.~Anikin, D.~Yu.~Ivanov, B.~Pire, L.~Szymanowski and S.~Wallon,
 in preparation.








\end{thebibliography}
\end{document}